%% file: precphys.tex
\documentclass[final]{cim}

\usepackage{cite,epsfig}

\title{Precision Physics at LEP}

\author{G.~Montagna\from{univ}, 
        O.~Nicrosini\from{univ} \atque
        F.~Piccinini\from{univ} }
        
\instlist{\inst{univ} Dipartimento di Fisica Nucleare e Teorica - Universit\`a 
                      di Pavia, and \\
Istituto Nazionale di Fisica Nucleare - Sezione di Pavia,
                      Italy }

\PACSes{\PACSit{00.01}{First PACS number}
        \PACSit{00.02}{Second PACS number} }
 
\begin{document}

\bibliographystyle{unsrt}  

\maketitle

\begin{center}
To appear in the RIVISTA DEL NUOVO CIMENTO
\end{center}
       
\vskip 24pt 
\hrule

\tableofcontents

\vskip 24pt
\hrule
\vskip 36pt


\input{intro.txi}

\input{sab.txi}

\input{z0.txi}

\input{fits.txi}
\input{lep2.txi}
\input{concl.txi}

\solong

The authors are indebted with several colleagues for having provided figures
that appear in the present paper, in particular with 
G.~Altarelli for Figs.~39 and~40, 
S.~Ambrosanio for Fig.~42, 
W.~Hollik for Fig.~41, 
B.~Pietrzyk for Figs.~35, 67, 68 and~69, 
G.~Quast for  Figs.~1, 33, 36 and~37, and 
D.~Ward for  Figs.~27, 28, 34 and~56. 
The authors are also grateful to F.~Teubert for useful discussions concerning
the propagation of the luminosity error to $Z$-boson parameters. 

The present paper has been written after almost a decade of intense activity on
precision physics at LEP. During these years, the   authors had the occasion to
collaborate with several colleagues: they wish to warmly thank all of them 
for all the stimulating discussions, and the  collaborative environment found.
Without these interactions, most probably this  paper would never be  written. 

Last, but not least, the authors gratefully acknowledge the Italian Physical
Society (SIF) 
for having provided the opportunity of writing the present review paper.  

\appendix

\input{upc.txi}

\input{vp.txi}

\input{scalint.txi}

\bibliography{intro,sab,z0,fits,lep2,upc,vp}


\end{document}

%% file: intro.txi
\section{Introduction}
\label{sect:intro}
According to the present understanding of particle physics,  
the fundamental interactions among elementary particles are 
described by the so-called Standard Model (SM) of the electroweak and 
strong interactions. It is a non abelian gauge theory based 
on the group $SU(2) \otimes U(1)$ for the description of the 
electromagnetic and weak forces in unified form (the so-called 
standard electroweak theory)~\cite{gsw} and the group $SU(3)$ colour 
for the strong interactions (the so-called Quantum 
Chromodynamics or QCD)~\cite{qcd}. 

In the SM the fundamental
constituents of matter are quarks and leptons (matter fields), while 
the quanta mediating the interactions among the matter particles are 
the photon, the $W^\pm$ and $Z^0$ bosons and the gluons (gauge fields). 
Both quarks and leptons interact {\it via} the electroweak force, 
while quarks in addition do participate also in strong interactions. In the 
electroweak sector of the SM, the fundamental fermions are grouped 
into three different generations. 
Each generation contains a charged lepton ($e,\mu,\tau$), an 
associated neutrino ($\nu_e,\nu_\mu,\nu_\tau$), an ``up'' quark 
($u,c,t$) and a ``down'' quark ($d,s,b$). Fermions belonging to 
different generations have identical couplings but are distinguishable by 
their masses. Although the SM does not predict 
the number of sequential generations, the present precision measurements at 
LEP/SLC rule out the existence of a fourth generation with a neutrino of mass 
up to 45~GeV (see Sect.~\ref{sect:detpar}). All the matter particles 
assumed by the SM have been experimentally discovered. In fact, the long-awaited 
{\it top} quark has been recently observed by the CDF and D0 
Collaborations at the TEVATRON, and the value obtained for its mass 
($m_t = 175.6 \pm 5.5$~GeV) is in good agreement with the mass range indirectly 
obtained from LEP/SLC precision measurements (see Sect.~\ref{sect:tqm}). 

In order to realize the unification of electromagnetic and weak interactions, 
the standard electroweak theory predicts the existence of four vector bosons 
as carriers of the electroweak force: $\gamma, W^\pm, Z^0$. The electromagnetic 
interaction is described in terms of the massless photon $\gamma$. Both 
charged and neutral current weak interactions are predicted by the theory and 
described by the exchange of heavy $W^\pm$ and $Z^0$ bosons. The intermediate vector 
bosons $W^\pm$ and $Z^0$, and all the fermions, 
acquire mass by the mechanism of spontaneous symmetry
breaking of the $SU(2) \otimes U(1)$ gauge symmetry, the so-called 
Higgs mechanism~\cite{higgs}. Strictly speaking, no mass value is predicted by the theory; on
the other hand, by making use of low-energy experimental data, it is possible to determine
the vector boson masses and the masses of all the fermions with the exception of the {\it
top} quark, for which high-energy data are necessary. 
As a consequence of the Higgs mechanism, the existence of a neutral scalar 
boson, the Higgs boson, is predicted by the  theory. 
However, also the mass of this particle is a free parameter that has to be determined 
experimentally. At present, the Higgs boson has not been yet experimentally 
observed, and its discovery would  certainly represent one of the most 
important tests of 
the SM. Present negative searches at LEP provide the mass limit 
$m_H > 77$~GeV at 95\%~CL (see Sect.~\ref{sect:hm}).   

The first experimental confirmation of the electroweak theory was the observation
of the neutral currents in neutrino-electron scattering 
experiments~\cite{ncobs}. Since then, important experimental results, such as 
the discovery of the $W^\pm$ and $Z^0$ bosons by the UA1/UA2 Collaborations in 
proton-antiproton collisions~\cite{ua1ua2} and the evidence for the $\gamma$-$Z$ 
interference in $e^+ e^- \to f \bar f$ processes at PEP/PETRA~\cite{pep}, provided  
further support to the SM. With the advent of the electron-positron accelerators  
LEP (Large Electron Positron collider at CERN) and SLC (Stanford Linear Collider at SLAC), 
the modern era of {\it precision tests} of the SM 
was started. At LEP in its first phase (LEP1) and SLC the $Z^0$ bosons 
are copiously produced in a very clean experimental environment, enabling 
the determination of the electroweak observables with impressive precision and 
offering the possibility to test the predictive power of the SM at the level of its 
quantum structure.

\begin{figure}[hbt]
\begin{center}
\epsfig{file=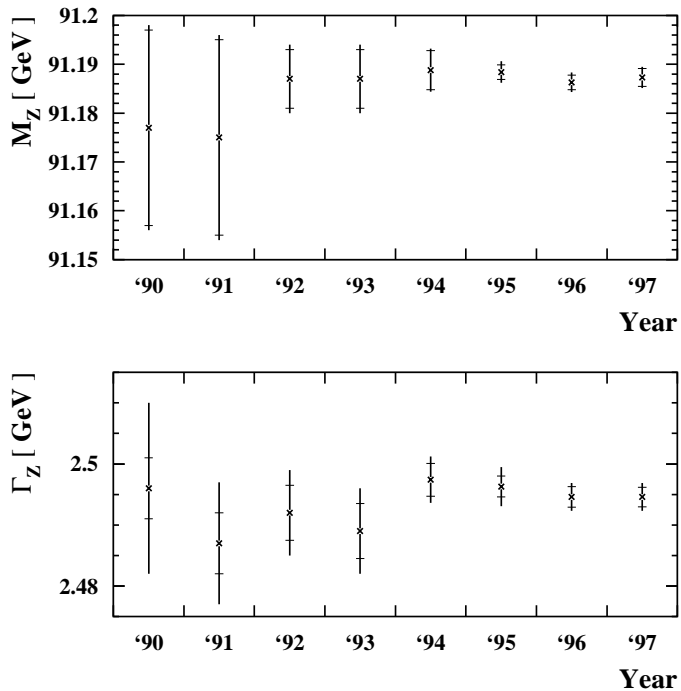, width=10truecm}
\end{center}
\caption{The time evolution of the errors on $M_Z$ and $\Gamma_Z$ (from~\cite{quast97}). }
\label{fig:mzgz_year}
\end{figure}

LEP1 and SLC colliders began their 
operation in the fall 1989. The energies of the electron ($e^-$) and 
positron ($e^+$) beams were chosen to be approximately equal to 45~GeV,    
in such a way that the available centre of mass 
energy was centered around the mass of the 
neutral vector boson of the weak interaction, the $Z^0$ boson ($E_{cm} \simeq 91$~GeV). 
LEP1 was terminated in the fall 1995 in order to allow a second phase in the 
LEP experimental program (LEP2) at energies above the $Z^0$ peak, 
while SLC will continue in data 
taking close to the $Z^0$ resonance still for a few years. The high-energy reactions 
studied at LEP1 by the four experiments ALEPH, DELPHI, L3 and OPAL and at SLC by the 
SLD Collaboration are two-fermion production processes in $e^+ e^-$ collision, 
\idest
\begin{eqnarray*}  
e^+ e^- \to \gamma, Z \to f \bar f \, ,
\end{eqnarray*}
with
\begin{eqnarray}
f \bar f = & & \nu \bar \nu (\nu_e \bar\nu_e, \nu_\mu \bar\nu_\mu, \nu_\tau 
\bar\nu_\tau) \quad \quad \quad \, ({\hbox {\rm invisible final states}} ) , \nonumber \\
& & l^+ l^- (e^+ e^-, \mu^+ \mu^-, \tau^+ \tau^- )
\quad ({\hbox {\rm charged leptons final states}}) , \nonumber \\
& & q \bar q (u \bar u, d \bar d, s \bar s, c \bar c, b \bar b ) \quad \quad \quad \,
\, \, ({\hbox {\rm hadron final states}}) \nonumber \, .
\end{eqnarray}
About 16 millions of $Z^0$ events have been detected at LEP and about 100 thousands at SLC, 
using in the latter polarized electron beams. Thanks to the very large data sample, 
a remarkable precision has been reached in the measurements of the $Z$-boson 
properties and observables, such as mass, total and partial decay widths, 
production cross sections and forward-backward asymmetries. For example, the mass 
of the $Z^0$ boson is at present 
 known with a relative error of $2 \times 10^{-5}$, $M_Z = 91186.7 \pm 
2.0$~MeV; the total $Z^0$ decay width has 
been measured with a relative error of $1 \times 10^{-3}$, \idest 
$\Gamma_Z = 2494.8 \pm 2.5$~MeV~\cite{quast97,dward97}. The evolution with time of the 
experimental uncertainties on $M_Z$ and $\Gamma_Z$ during the years 1989-1997 
is shown in Fig.~\ref{fig:mzgz_year}.  
Three years of data taking at LEP1 (1991, 1993 and 1995) were devoted to a precision scan 
around the resonance, doing measurements also off-peak in the range 88-95~GeV; the rest 
of data taking was performed exclusively on the $Z^0$ peak. Several technological 
ingredients contributed to reach the fantastic performances of the LEP1 program 
summarized in Fig.~\ref{fig:mzgz_year}. For
instance, the method of resonant spin depolarization adopted in the LEP beam energy 
measurement allowed to reach a very precise calibration, of the order of 1~MeV. Quite 
curiously, many important unexpected phenomena affected LEP energy calibration during  
operation: tide effects and field rise due to the passage of the TGV on the nearby railway 
are probably the two most famous examples. The installation of precision luminometers 
also contributed significantly to the success of the LEP program, allowing a luminosity 
measurement at the level of 0.1\% or better, that is unavoidable to make optimal use of 
the high statistics collected at LEP. In parallel with the technological and experimental
progress, a large effort was undertaken by hundreds of theorists 
in the calculation of  radiative corrections. The excellent 
performance of the machine, associated with the above theoretical effort, 
allowed to test the fine structure of the Standard Model of electroweak interactions 
with an unprecedented level of precision ({\it precision physics at LEP}). In this sense, 
precision tests at the $Z^0$ peak can be compared with classical tests of Quantum Electrodynamics
(QED), like $g-2$ and Lamb shift experiments. Besides a 
highly nontrivial test of the standard theory at the level of quantum loops, the 
availability of both accurate measurements and calculations provided the opportunity  
 of putting limitations on models predicting physics beyond the standard model.

After the long data taking at the $Z^0$ pole and a short run at intermediate 
energy in the 1995 fall (LEP1.5 phase), the second phase of LEP (LEP2) started operating
in 1996. The main reason for the energy upgrade of the LEP machine was the precise 
measurement of the properties of the charged vector bosons of the weak interaction, 
the $W^{\pm}$ bosons, through the study of the reaction
\begin{eqnarray*}  
e^+ e^- \to W^+ W^- \to f_1 \bar f_2 f_3 \bar f_4 \, ,
\end{eqnarray*}
with the following possible four-fermion final-states
\begin{eqnarray}
& & l \bar\nu_l \bar{l}' \nu_{l'} (l=e,\mu,\tau) \qquad \qquad \qquad \quad
\, \, \, \, \, 
({\hbox {\rm leptonic final states}} ) , \nonumber \\
& & l \bar\nu_l q_1 \bar q_2 (l=e,\mu,\tau; q=u,d,c,s) \quad \quad 
({\hbox {\rm semi-leptonic final states}} ) , \nonumber \\
& & q_1 \bar{q}_2 q_3 \bar{q}_4 (q=u,d,c,s) \qquad \qquad \qquad
({\hbox {\rm hadronic final states}} ) \nonumber \, .
\end{eqnarray}
The 1996 LEP2 run was split into two parts. The first run was at the centre of 
mass energy 
$\sqrt{s} = 161.3$~GeV, \idest\  0.5~GeV above the nominal $WW$ production threshold,  
while the second one at $\sqrt{s} = 172$~GeV. The current LEP2 schedule foresees running 
during the years 1997-1999 with centre of mass energies from 184~GeV in 1997 to 194~GeV in 
1998-1999. Compared to LEP1, the LEP2 physics potential will be characterized by a
statistical error of the order of 1\%, instead of 0.1\% at LEP1. Nonetheless, also LEP2 
has  to be considered as  a machine for precision physics, since it will allow to 
measure the mass of the $W$ boson with an envisaged precision of 40-50~MeV, that is 
better than the current best determination of $M_W$ at hadron machines, and to 
study in detail the couplings of the $W$'s with the other gauge bosons $\gamma$ and $Z^0$, 
thus directly probing the non-abelian nature of the electroweak theory. Therefore, 
the LEP2 experimental program does constitute a nice completion of the precision studies 
at LEP1 and SLC. Needless to  say, besides the precision measurement of the       
 properties of the charged vector bosons, LEP2 could also provide important pieces of 
information  on the Higgs boson and  physics  beyond the SM.
 
In the recent few years, several reviews on the status of 
precision measurements at LEP1/SLC and of the calculation of radiative corrections in the 
SM have been published in the literature. From an experimental point of view, 
the main aspects of precision tests at LEP1/SLC have been summarized for
instance in 
ref.~\cite{reviewexp}. 
The concepts and results of the calculation of radiative corrections have been 
reviewed for instance in ref.~\cite{reviewth}. Before the start of the LEP1/SLC operations, 
the status of theoretical calculations and computational tools was described 
in refs.~\cite{yrlep86,yrlep88,yrlep189}. In view of the final analysis of the precision data, the 
status of precision calculations was critically reviewed in ref.~\cite{yrwg95}, with special emphasis 
on the uncertainty inherent to the theoretical predictions. The theoretical results and 
related software necessary for the experiments at LEP2 were summarized in 
ref.~\cite{yrlep296}.  

The LEP/SLC physics program is a very wide one, ranging from precision measurements of
vector boson properties, to QCD studies,  searches of new particles, determination of dynamical 
properties of heavy flavours, and so on. 
The main goal of the present review is to give a comprehensive account of 
{\it precision physics at LEP}, 
meant as the whole of precision tests of the electroweak  sector
of the SM and its implications for new physics.  
More precisely, the theoretical apparatus needed for precision 
calculations at LEP/SLC will be described and its link with data analysis 
pointed out, aiming at  emphasizing the intimate and fruitful connection between 
theoretical ideas and experimental results. Too technical details are inevitably omitted;
however, particular care has been devoted to compiling an as  complete as possible   
bibliography, where the interested reader can find additional information. The experimental  data
considered throughout the paper are the ones presented at the 1997 Summer Conferences~\cite{quast97,dward97,lepewwg97}.
 
The review is organized as follows. Section~\ref{sect:sablm} is dedicated to small-angle
Bhabha scattering and its relevance for the luminosity measurement. In Sect.~\ref{sect:z0phys}
the large-angle processes at the $Z^0$ resonance, relevant for the study of the 
$Z$-boson 
properties, are considered, reviewing the theoretical ingredients necessary for 
high-precision calculations of the $Z$-boson observables. It is then shown how the theoretical 
tools developed are used to fit the experimental data, in order to determine the 
{\it top}-quark and Higgs-boson masses and the strong coupling
constant $\alpha_s$,  and eventually establish the existence of new physics 
beyond the SM (Sect.~\ref{sect:fits}). 
The most important issues of electroweak physics at LEP2 are covered in Sect.~\ref{sect:lep2phys},
where two-fermion processes far from the resonance (Sect.~\ref{sect:tfproc}) and four-fermion
processes at and above the $W$-boson pair production threshold (Sect.~\ref{sect:ffproc}) are 
discussed. At last, the conclusions are drawn in Sect.~\ref{sect:concl}.  Three technical
appendices are dedicated to relevant theoretical subjects common to 
various items of LEP physics, namely the universal photonic corrections (Appendix~\ref{sect:upc}),
the vacuum polarization correction (Appendix~\ref{sect:vacpol}) and the scalar integrals 
and dimensional regularization (Appendix~\ref{sect:scalint}).

%% file: sab.txi
\section{Small-Angle Bhabha Scattering and the Luminosity Measurement}
\label{sect:sablm}
\begin{figure}[hbt]
\begin{center}
\epsfig{file=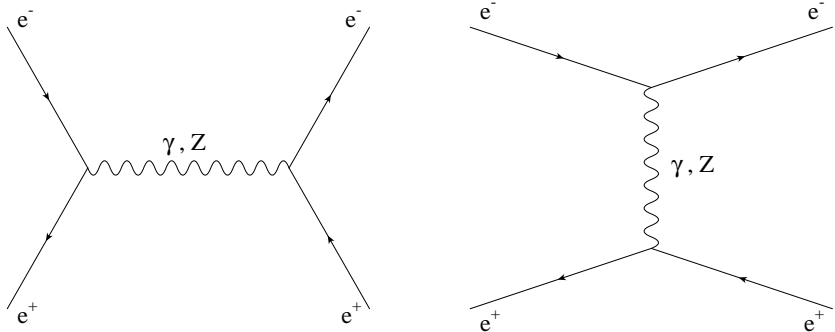, width=11truecm}
\end{center}
\caption{The Feynman diagrams for the tree-level Bhabha scattering cross section. }
\label{fig:bhabhatree}
\end{figure}

The process $e^+ e^- \to e^+  e^-$ (Bhabha scattering) is a peculiar one.
Actually, at a difference from all  the other  two-fermion production
processes,  
that occur as $e^+ e^-$  annihilation, \idest\   are $s$-channel processes, the
amplitude for $e^+ e^- \to e^+  e^-$ receives contributions both  from $s$- and
$t$-channel diagrams (see Fig.~\ref{fig:bhabhatree}). 
In the energy region  typical  of LEP1/SLC ($\sqrt{s} 
\simeq M_Z$, $M_Z$  being  the $Z$-boson  mass) the Bhabha  scattering process
exhibits completely different features, depending on the fact that the final 
state electrons are detected at large or small scattering angles. In the first
case, the amplitude is dominated by $s$-channel subprocesses, and in particular 
by the $e^+ e^- \to Z^0$ annihilation,  in  such a  way that the process is
sensitive to  the $Z$-boson properties. In  the second one, the smaller is the 
scattering angle the larger is the contribution of the $t$-channel 
photon-exchange  diagram,  in such a way that for sufficiently small scattering
angles Bhabha scattering is essentially a pure QED process, substantially
independent of the $Z^0$ physics.
These dynamical features of  Bhabha scattering allow to use it both as a tool
for studying $Z^0$ physics  (large-angle  cross  sections  and 
forward-backward  asymmetries)  and as   a  tool for the high  precision
luminosity monitoring (small-angle cross section).  The first case will be
examined in Sect.~\ref{sect:z0phys},  while the second  one will be 
addressed in the
following. Before starting the discussion  on small-angle  Bhabha scattering,
it is worth noticing that a large amount of work has been  dedicated to  Bhabha
scattering in general. The interested reader is  referred for instance 
to refs.~\cite{cr89} and \cite{bharep96} for a detailed account and a
comprehensive compilation of the relevant literature. 
\subsection{Luminosity Monitoring}
\label{sect:lummon}
The luminosity ${\cal L}$ of a collider is the proportionality constant between the event rate
$d N / dt$ and the corresponding cross section $\sigma$ for any given process, 
according to the relation 
\begin{equation}
{{d N} \over {dt}} = {\cal  L} \sigma , \qquad N = \sigma \int dt {\cal L} = \sigma L  . 
\label{eq:deflum}
\end{equation}
In the practice, an experiment measures the number of events for a given process and, 
by making use of the inverse of eq.~(\ref{eq:deflum}), quotes the experimental cross sections 
as
\begin {equation}
\sigma = {{1} \over   L} N . 
\label{eq:sigexp}
\end{equation}
From eq.~(\ref{eq:sigexp}) it is clear that, in order to fully exploit the experimental
information contained in $N $, the error affecting the luminosity $ L$ must be smaller
than the experimental error affecting $N $.   
The luminosity of a collider depends in a highly non-trivial way on machine and beam parameters. 
It is of course possible to compute it given these parameters, but, in particular for LEP, 
such a determination is affected by an intrinsic uncertainty which is completely unsatisfactory
in view of the extremely high experimental accuracy. This is why the luminosity is monitored 
adopting a different strategy, namely by identifying a process which is  in principle not affected
(or only slightly affected) by {\it unknown} physics, so that its cross section can be computed
within a firmly established theory, and determining  the luminosity {\it via} the
relation 
\begin{equation}
{ L} = {{1} \over {\sigma_{known}}} N . 
\label{eq:lummeas}
\end{equation}
The luminosity determined by eq.~(\ref{eq:lummeas}) is then used for  the determination of 
all the experimental cross sections  {\it via} 
eq.~(\ref{eq:sigexp}) (for  a review on the precision
determinations of the accelerator luminosity in LEP experiments, the interested reader is referred
to~\cite{dvm96} and references therein).   

Such a process does exist, and is the Bhabha scattering $e^+ e^- \to e^+ e^-$ at small scattering
angle. The main reasons why the small-angle Bhabha scattering (SABH) fulfils the requirements
just described are the following: 
\begin{itemize}
\item the SABH process is substantially a QED process, dominated by photon exchange in the 
      $t$-channel; this, in turn, implies that
             \begin{itemize}
             \item its theoretical cross section is dominated by a contribution that is in 
                   principle calculable by means of perturbative QED at arbitrary precision; 
             \item the $Z$-boson exchange contribution to its cross section, {\it via} $Z$-boson
                   annihilation in the $s$ channel, $Z$-boson exchange in the $t$ channel and 
                   $Z$-$\gamma$ interferences, is very small; hence a  detailed knowledge of  
                   the $Z$-boson properties,
                   like  the precise  value of its  mass  and  decay width,  
                   to  be determined  in  large-angle  processes, has a negligible influence    
                   on the luminosity  monitoring; 
             \end{itemize}
\item the SABH cross section is large, and can be rendered much larger than the typical 
      $Z$-boson
      annihilation peak cross sections provided the detection angular region for  the SABH 
      events is sufficiently close to the beam pipe, so that the statistical uncertainty of  
      $N $ in the r.h.s. of eq.~(\ref{eq:lummeas}) can be kept small.  
\end{itemize}

The tree-level differential cross section corresponding to 
$t$-channel photon exchange is given by (see for instance
ref.~\cite{cr89} and references therein)
\begin{equation}
{{d \sigma_0} \over {d  \Omega}} = {{\alpha^2} \over {4 s}} {{2} \over {(1 - \cos \vartheta)^2}}
\left[ 4 + (1 + \cos \vartheta)^2 \right] , 
\end{equation}
where $\vartheta $ is the electron  scattering angle, $\sqrt{s} $ is the total centre of mass (c.m.) 
energy and $\alpha$ is the QED coupling constant. 

The calculation of the theoretical SABH cross section to be inserted into eq.~(\ref{eq:lummeas})
will involve in general the calculation of radiative corrections. These are numerically dominated
by photonic corrections, \idest\  by those corrections obtained by adding real or virtual  photon
lines to the tree-level amplitudes, 
which in turn depend very critically on the phase space integration
region. Hence, in order to exploit eq.~(\ref{eq:lummeas}) for the high  precision luminosity
monitoring, the theoretical cross section must be computed by taking into account carefully the
experimental definition of a SABH event. Skimming over the details of the various SABH Event
Selections (ES's), which vary from experiment to experiment, the general features of SABH event 
are the following: 
\begin{itemize}
\item the forward and backward luminometers cover an angular region of few degrees, starting
      from, say, $1^\circ$; 
\item a SABH event is detected as a signal coincidence of the forward luminometer with the
      backward luminometer;  
\item for technical reasons, asymmetric angular acceptances are adopted; for instance, a narrower
      (N) acceptance on the $e^+$ side and  a wider  (W) acceptance on the $e^-$ side; the 
      results of the NW and WN ES's are averaged; 
\item for  technical reasons, a calorimetric measurement on the final-state electrons is
      performed; this means that experimentally a final-state electron cannot be distinguished 
      by a final-state electron accompanied by collinear photons; 
\item very mild energy/acollinearity cuts are imposed on the scattered electrons, compatible with
      the previous items. 
\end{itemize}  

From the general features of a SABH event, and taking into account that photonic radiation is
mostly soft and/or collinear to the charged lines, the following picture can be drawn: 
\begin{itemize}
\item events with the final-state electrons inside the lu\-mi\-no\-me\-ters,  
      ac\-com\-pa\-nied by soft photons
      or initial-state photons  lost in the beam pipe are detected as good SABH events; 
\item since initial-state radiation (ISR) 
      causes a boost of the c.m. of the reaction, and since the final-state
      electrons cannot be too acollinear, there is a natural cut-off on the amount of ISR 
      allowed; 
\item events with the fi\-nal-state e\-lec\-trons in\-side the lu\-mi\-no\-me\-ters 
      ac\-com\-pa\-nied by col\-li\-near
      pho\-tons  are detected as good SABH events; this means that final-state radiation (FSR) 
      must be integrated over a finite region surrounding the final-state electron; this in 
      turn implies, for a sort of Kinoshita-Lee-Nauenberg (KLN) 
      mechanism~\cite{kln6264}, that for FSR the so-called  
      ``logarithmically enhanced'' corrections 
      are greatly suppressed (more on this in Sect.~\ref{sect:rcsabs}); 
\item initial- or final-state photons can convert into additional fermionic pairs, which for
      most of the events are lost in the beam pipe or collinear to the final-state electrons; 
\item events accompanied by the radiation of virtual photons are degenerate with the elastic
      event;  
\item since in general there is no particle identification in the luminometers, it can happen
      that, for instance, a final-state electron radiates a hard photon in such a way that the
      photon hits the luminometer, while the electron is lost; also such an event is
      in general registered as a good SABH event.   
\end{itemize}

Given  the above picture, it is clear that in order to perform the phase space integration over
all the configurations that correspond to a good SABH event the most versatile tool, \idest\ a
Monte Carlo (MC) integrator and/or event generator, is mandatory. Nonetheless, as a matter of fact the
calculations necessary to compute the theoretical SABH cross section are so sophisticated  that
all kind of information, even analytical or semi-analytical, are precious. A detailed account of
the status of all the theoretical calculations available at present will be given in
Sect.~\ref{sect:sabct}. It is worth noting that the total error affecting the determination of
the luminosity by means of eq.~(\ref{eq:lummeas}) receives contributions both from the
experimental error of the number of events $N$ and from the theoretical error of the SABH
process cross section. Actually, by performing the standard propagation of errors on
eq.~(\ref{eq:lummeas}) one obtains
\begin{equation}
{{\delta L} \over {L}} = {{\delta N} \over {N}} \oplus {{\delta \sigma} \over {\sigma}} , 
\label{eq:lemerr}
\end{equation}
where the first term in the r.h.s. is the experimental error, the last one is the relative
error on the theoretical cross section and $\oplus$ means the sum in quadrature. 
\subsection{Sensitivity of Observables to Luminosity}
\label{sect:sensol}
The LEP experiments determine the $Z$-boson  parameters by means of combined fits to the
measured hadronic and leptonic cross sections and leptonic forward-backward asymmetries (see
Sects.~\ref{sect:z0phys} and \ref{sect:fits} for more details). 
In particular, in order to facilitate the comparison
and the combination of the results, the LEP Collaborations quote, besides other choices, the
following set of nine parameters:  
\begin{equation}
M_Z, \quad \Gamma_Z, \quad \sigma^0_h, \quad R_{e,\mu,\tau}, \quad A_{FB}^{e,\mu,\tau} , 
\end{equation}
where $M_Z$ and  $\Gamma_Z$ are the $Z$-boson mass and total width, respectively, 
$\sigma^0_h$ is the hadronic peak cross section for $Z$-boson exchange only, 
$R_i$ are the ratios of the hadronic width to the $i$-th leptonic width and 
$A_{FB}^{i}$ are  the leptonic forward-backward asymmetries at $\sqrt s  = M_Z$ 
for $Z$-boson exchange only. These parameters are quoted after having been corrected for the
effect of ISR, final-state QED and QCD corrections, 
as well as for $t$-channel and $s$-$t$ interference for $e^+
e^-$ final states. By assuming lepton universality, the three values of $R_i$ and
$A_{FB}^{i}$ are combined into a single $R_l$ and $A_{FB}^{l}$, and the corresponding fit
is a 5-parameter fit. In both  cases, the so called derived parameters, such as for
instance the leptonic width and the number of light neutrinos, are computed by expressing them
as functions of the fundamental parameters. 

\begin{table}
\caption{
Line shape and asymmetry parameters from 5-parameter fits to the data of the
four LEP1 experiments, made with a theoretical luminosity error of 0.16\%, 
0.11\% and 0.06\%.
In the lower part of the Table also derived parameters are listed (from 
ref.~\cite{bharep96}). 
}
\label{tab:sensitivity}
\begin{tabular}{llll}
\hline
theoretical luminosity error   & 0.16\% &  0.11\% & 0.06\% \\
\hline
$M_Z$~[GeV]
   & $ 91.1884\pm0.0022$ & $ 91.1884\pm0.0022$ & $91.1884\pm0.0022$ \\
$\Gamma_Z$~[GeV] 
   & $ 2.4962\pm0.0032$  & $ 2.4962\pm0.0032$  & $2.4961\pm0.0032$  \\
$\sigma^0_h$~[nb] 
   & $ 41.487\pm0.075$   & $ 41.487\pm0.057$   & $41.487\pm0.044$   \\
$R_l$
   & $ 20.788\pm0.032$   & $ 20.787\pm0.032$   & $20.786\pm0.032$   \\
$A_{FB}^{l}$ 
   & $ 0.0173\pm0.0012$  & $ 0.0173\pm0.0012$  & $0.0173\pm0012$    \\
\hline
$\Gamma_{h}$~[GeV]
   & $ 1.7447\pm0.0030$  & $ 1.7447\pm0.0028$  & $1.7446\pm0.0027$ \\
$\Gamma_{l}$~[MeV]
   & $ 83.93\pm0.13$     & $ 83.93\pm0.13$     & $83.93\pm0.12$    \\
$\sigma^0_{l}$~[nb]
   & $ 1.9957\pm0.0044$  & $ 1.9958\pm0.0038$  &$ 1.9959\pm0.0034$ \\
${\Gamma_{h}}/{\Gamma_Z}$~[\%]
   & $69.90\pm0.089$     & $ 69.90\pm0.079$    &$ 69.89\pm0.072$   \\
${\Gamma_{l}}/{\Gamma_Z}$~[\%]
   & $3.362\pm0.0037$    & $ 3.362\pm0.0032$   &$ 3.362\pm0.0028$  \\
$\Gamma_{inv}$~[MeV]
  & $ 499.9\pm2.4$       & $ 499.9\pm2.1$      & $499.9\pm1.9$     \\
${\Gamma_{inv}}/{\Gamma_{l}}$~[\%]
  & $5.956\pm0.030$      & $ 5.956\pm0.024$    &$ 5.956\pm0.020$   \\
$N_{\nu}$
  & $ 2.990\pm0.015$     & $ 2.990\pm0.013$    & $2.990\pm0.011$   \\
\hline
\end{tabular}
\end{table}

In order to understand the effect of the luminosity error on the physical observables, one can
assume, as a first approximation, that the luminosity error affects, among the fundamental
parameters,  the hadronic peak cross
section only, in a way that is determined from eq.~(\ref{eq:sigexp}) by  performing the
standard propagation of errors: 
\begin{equation}
\delta \sigma_h^0 = (\delta \sigma_h^0)_{exp} \oplus \sigma_h^0 {{\delta L} \over {L}} . 
\label{eq:dsh0}
\end{equation}
In eq.~(\ref{eq:dsh0}) the first term in the r.h.s. is the experimental error on  the hadronic
peak cross section, while the second  term is the  error on the cross  section as due to the luminosity
uncertainty. The
influence of the luminosity error on the derived parameters is then determined by their
dependence on $\sigma_h^0$. 

Actually, in the practice the LEP Collaborations combine their results, propagating both the
systematic and statistical uncertainties, and taking into account all the correlations between
the parameters by means of the full covariance matrix. Table~\ref{tab:sensitivity} shows an
example of  the results of such a procedure, by assuming a relative luminosity error of
0.16\%, 0.11\% and 0.06\%, respectively. As can be noticed, the luminosity error affects,
besides the hadronic peak cross section $\sigma_h^0$, almost all the derived parameters.  
\subsection{Radiative Corrections to Small-Angle Bhabha Scattering}
\label{sect:rcsabs}
The radiative corrections to the SABH process are dominated by photonic corrections, \idest\
by those corrections coming from graphs obtained from the tree-level ones by adding real
and/or virtual photon lines, and by the vacuum polarization correction. This last effect is
taken into account by simply using the running QED coupling constant $\alpha (-t)$, where $t$
is the squared four-momentum transfer of the reaction, as shown in Appendix~\ref{sect:vacpol}. In the
present section, some generalities on photonic corrections are reviewed. 
  
Once the infra-red (IR) divergence present in real and virtual corrections separately has been
canceled by properly summing over all the degenerate states, the $n$-th order QED  correction
takes on the following form
\begin{equation}
\sigma^{(n)} = \left( {{\alpha} \over {\pi}} \right)^n \sum_{k=0}^n  a_k^{(n)} {\cal L}^k (Q^2), 
\label{eq:signbha}
\end{equation}
where $\sigma^{(n)}$  is the $n$-th order contribution to the corrected cross section,  
${\cal  L} (Q^2) = L_{Q^2} - 1$ and $L_{Q^2} = \ln ( \vert Q^2 \vert / m_e^2)$ is the so 
called {\it collinear logarithm}, $Q^2$ being a typical scale involved in the process. 
The collinear
logarithm $L_{Q^2}$ originates from the phase space integration of the emitted photon, and in
particular from those configurations in which the emitted photon is almost collinear to the
radiating charged fermion line. In the SABH process, the relevant scale entering the collinear
logarithm is $Q^2  =  -t$, $t$ being the squared four-momentum transfer of  the reaction. 
It is worth noticing that, due to the smallness of the electron
mass $m_e$, $L_{t}$ is of the order of 15 for $-t \simeq 1$~GeV$^2$, \idest\ for values of the
four-momentum transfer typical of small-angle processes at LEP. This is the reason why QED
radiative corrections are numerically very relevant. 

The coefficients $a_k^{(n)}$ in eq.~(\ref{eq:signbha}) are in turn given by the following general
expression
\begin{equation}
a_k^{(n)} = \sum_{j=0}^k b_{kj}^{(n)} l^j , 
\label{eq:akeqbha}
\end{equation}
where $l$ is the so called IR logarithm, $l \simeq \ln ( E / E_\gamma^{max})$, $E$ being  the
beam energy and $E_\gamma^{max}$ being  some cutoff on the maximum photon energy,
respectively. As a  property of QED corrections, the IR logarithm $l$ can become huge
when shrinking the photon phase space. In general, however, such corrections depend critically on the
details of the ES considered.  

The purpose  of the present discussion is to derive a sort of {\it rule of thumb} able to 
provide the order of magnitude of the various QED corrections. To this aim, an approximate
calculation is a useful guideline. As a general warning, however, it is worth noting that the
coefficients $a_k $, due to the presence of the IR logarithm,  are IR sensitive, so that they
can become very large in {\it pathological} situations, \idest\ when very tight cuts on the
emitted photons are imposed. After this {\it caveat}, the corrected cross section $\sigma$ for  the SABH
process in the  leading logarithmic (LL) approximation, obtained for instance by means of the Structure
Function method, taking into account initial- and final-state radiation but  
neglecting convolution effects (see 
Appendix~\ref{sect:upc} for more details) can be written as~\cite{cr89}
\begin{equation}
\sigma \simeq \sigma_0 \varepsilon^\beta . 
\label{eq:sigll}
\end{equation}
The symbols in eq.~(\ref{eq:sigll}) have the following meaning: $\sigma_0$ is  the tree-level
cross section, $\beta$  is  given by 
\begin{equation}
\beta = 2 {{\alpha} \over {\pi}} {\cal L}  (Q^2)
\end{equation}
and $\varepsilon  = \varepsilon_i \varepsilon_f$ is the product of two, in principle
different,  cutoffs for ISR and FSR. 

The series expansion of eq.~(\ref{eq:sigll}) up to third order in $\alpha$ reads:
\begin{equation}
\sigma \simeq \sigma_0  \left( 1  + \beta \ln \varepsilon + {{1} \over {2!}} \beta^2 \ln^2
\varepsilon +  {{1} \over {3!}} \beta^3 \ln^3 \varepsilon \right)  + {\cal  O } (\beta^4) . 
\label{eq:series}
\end{equation}
It is worth noting that in  the LL approximation the coefficients $a_n $ are proportional to
the $n$-th power of the IR logarithm $l$. 

\begin{table}[!ht]
\newcommand{\perm}{{\times 10^{-3}}}
\def\alf1{ {\alpha \over \pi} }
\def\half{ {1 \over 2} }
\caption{
The canonical coefficients indicating the generic magnitude of various
leading and subleading contributions up to third-order, for a non-calorimetric ES. 
The collinear logarithm  $ L_{t} = L = \ln(|t|/m_e^2) $ is calculated for 
$\vartheta_{min}=30$~mrad and $\vartheta_{min}=60$~mrad and for two values 
of the c.m. energy: at LEP1 ($\protect\sqrt{s}=M_Z$), where
the corresponding $|t|=(s/4)\vartheta_{min}^2$ are 1.86 and 7.53~GeV$^2$, and
at LEP2 energy ($\protect\sqrt{s}=200$~GeV), where the corresponding $|t|$ 
are 9 and 36~GeV$^2$, respectively (from ref.~\cite{bharep96}). 
}
\label{tab:canonical-coeff}
\begin{tabular}{llllll}
\hline
    \multicolumn{2}{c}{ }
  & \multicolumn{2}{c}{  $\vartheta_{min}=30$~mrad  }
  & \multicolumn{2}{c}{ $\vartheta_{min}=60$~mrad   } \\  
\hline
    \multicolumn{2}{c}{ } & LEP1 &  LEP2 & LEP1 &  LEP2 \\  
\hline 
${\cal O}(\alpha L    )$     &    $  4 \alf1 L$         
       &  $137\perm$   &  $152\perm$   &  $150\perm$   &  $165\perm$  \\  
\hline 
${\cal O}(\alpha      )$     &    $  \alf1 $    
       &  $2.3\perm$  &  $2.3\perm$    &  $2.3\perm$   &  $2.3\perm$  \\ 
\hline  
${\cal O}(\alpha^2L^2 )$     &    $  \half \left(4 \alf1 L\right)^2 $ 
       &  $9.4\perm$   & $11\perm$   & $11\perm$       & $14\perm$  \\ 
\hline  
${\cal O}(\alpha^2L )$       &    $ \alf1 \left(4 \alf1 L\right) $ 
       &  $0.31\perm$  & $0.35\perm$  & $0.35\perm$    & $0.38\perm$ \\ 
\hline  
${\cal O}(\alpha^3L^3)$      &    $ {{1} \over {3 \fact} } \left( 4 \alf1 L \right)^3 $ 
       &  $0.42\perm$  & $0.58\perm$  & $0.57\perm$    & $0.74\perm$  \\ 
\hline
\end{tabular}
\end{table}

In order to exploit the information contained in eq.~(\ref{eq:series}), it is necessary to
specify the main features of the ES one is considering. Let us begin with a non-calorimetric
ES. It is not a realistic case; nonetheless, it can be considered as a very useful benching
situation. Such an ES is characterized by the fact that in principle it is possible to
separate final-state fermions from the photon(s) radiated by them, in such a way that a
maximum photon energy cutoff, irrespective of the fact that the photon comes from initial- or
final-state particles, is meaningful. For such an ES, one can take in eq.~(\ref{eq:series})
\begin{equation}
\varepsilon \simeq \varepsilon_i^2 \simeq \varepsilon_f^2 \simeq \varepsilon_{nc}^2 , 
\end{equation} 
in such a way that eq.~(\ref{eq:series}) becomes
\begin{equation}
\sigma \simeq \sigma_0  \left( 1  + 2 \beta \ln \varepsilon_{nc} 
+ {{1} \over {2!}} ( 2 \beta)^2 \ln^2 \varepsilon_{nc} 
+  {{1} \over {3!}} (2 \beta)^3 \ln^3 \varepsilon_{nc} \right)  + {\cal  O } (\beta^4) .
\label{eq:seriesnc}
\end{equation}
By defining now as {\it canonical coefficients} the coefficients appearing in front of the IR
sensitive terms, the canonical coefficients of the ${\cal O} (\alpha^n L^n)$ corrections, with
$1 \le n \le 3$, can be directly read off eq.~(\ref{eq:seriesnc}), and are $2 \beta$,
$(2  \beta)^2 / 2$ and  $(2 \beta)^3 / 3!$, respectively. The algorithms described in
Appendix~\ref{sect:upc} allow to take these corrections into account naturally.  
Going beyond the LL approximation,
eq.~(\ref{eq:seriesnc}) does not provide information any more. Actually, the non-leading corrections
are typically process dependent, and can be computed only by means of a full diagrammatic 
calculation.
On the other hand, a non-collinear  photon is
known to produce a correction whose typical size is $\alpha / \pi$, which sets the size of
the non-leading  ${\cal O} (\alpha)$ correction. 
Moreover, the ${\cal O} (\alpha^2 L)$ corrections come
from configurations for which there is one collinear and one non-collinear photon, so that
their typical size is $2 \beta \alpha / \pi$. The situation is summarized in 
Tab.~\ref{tab:canonical-coeff}, where also a numerical estimate of the canonical coefficients
for a non-calorimetric measurement is given. It is worth noticing  that  both these non-leading
corrections are relevant for obtaining  a theoretical error of the order of 0.1\%. 

Let us now consider the more realistic case of a calorimetric ES 
(see Sect.~\ref{sect:lummon}).  For such an ES, it is not possible in principle to separate a
final-state fermion from its accompanying radiation. This means that one is effectively
almost inclusive on FSR, \idest\ that $\varepsilon_f \simeq 1$ and
$\varepsilon \simeq \varepsilon_i = \varepsilon_c$. Equation
(\ref{eq:seriesnc}) is then modified as follows:  
\begin{equation}
\sigma \simeq \sigma_0  \left( 1  +  \beta \ln \varepsilon_{c} 
+ {{1} \over {2!}} \beta^2 \ln^2 \varepsilon_{c} 
+  {{1} \over {3!}} \beta^3 \ln^3 \varepsilon_{c} \right)  + {\cal  O } (\beta^4) .
\label{eq:seriesc}
\end{equation}
By comparing eq.~(\ref{eq:seriesc}) with eq.~(\ref{eq:seriesnc}), one obtains the relation
between the canonical coefficients for a non-calorimetric ES and those of a calorimetric ES. 
In particular, the ${\cal O} (\alpha^n L^n)$ coefficients are reduced by factor of 2, 4 and 8
for $n=1,2$, 3, respectively. One can not expect {\it a priori} a reduction in the coefficient
of the ${\cal O}(\alpha)$ correction. Hence, there is a reduction of a factor of 2 in the
coefficient of the ${\cal O} (\alpha^2 L)$ correction. The situation is summarized in
Tab.~\ref{tab:nc-c}. 

\begin{table}[!ht]
\newcommand{\perm}{{\times 10^{-3}}}
\def\alf1{ {\alpha \over \pi} }
\def\half{ {1 \over 2} }
\caption{
The relation between the canonical coefficients for a non-calorimetric ES (see
Tab.~\ref{tab:canonical-coeff}) and a calorimetric ES. }
\label{tab:nc-c}
\begin{tabular}{llll}
\hline
   & non-cal. ES & reduction factor & cal. ES \\
\hline
${\cal O}(\alpha L    )$     &    $  4 \alf1 L$  & 1/2  & $  2 \alf1 L $ \\      
\hline 
${\cal O}(\alpha      )$     &    $  \alf1 $   & 1      & $ \alf1 $     \\ 
\hline  
${\cal O}(\alpha^2L^2 )$     &    $  \half \left(4 \alf1 L\right)^2 $ & 1/4 & 
$ \half \left(2 \alf1 L\right)^2 $ \\
\hline  
${\cal O}(\alpha^2L )$       &    $ \alf1 \left(4 \alf1 L\right) $ & 1/2 &
$ \alf1 \left(2 \alf1 L\right) $ \\
\hline  
${\cal O}(\alpha^3L^3)$      &    $ {{1} \over {3 \fact} } \left( 4 \alf1 L \right)^3 $ &
1/8 & $ {{1} \over {3 \fact} } \left( 2 \alf1 L \right)^3 $ \\
\hline
\end{tabular}
\end{table}

\begin{figure}[hbt]
\begin{center}
\epsfig{file=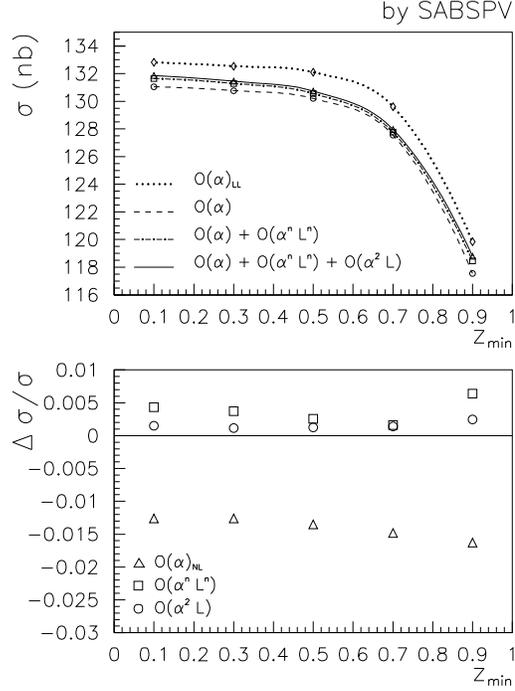, width=8truecm}
\end{center}
\caption{The effect of photonic corrections to the SABH process at different perturbative
order/accuracy as a function of $z_{min} = E_+ E_- / E_{beam}^2$, at $\sqrt{s} = 92.3$~GeV. 
$E_{+,-}$ are the energies
deposited in the positron and  electron clusters, respectively. The ES adopted is CALO2 (see
ref.~\cite{bharep96}). The tree-level cross section for this ES is  $\sigma_0 = 140.02$~nb. The
first plot shows the cross section including LL ${\cal O} (\alpha)$, exact ${\cal O} (\alpha)$, 
exact ${\cal O} (\alpha)$ plus higher-order LL  and the same  plus non-leading 
${\cal O} (\alpha^2)$ corrections. In the second plot, $\Delta \sigma / \sigma$ means 
the effect of non-leading ${\cal O} (\alpha)$ corrections (triangles), of higher-order LL  
corrections (squares), and an estimate af the effect of non-leading ${\cal O} (\alpha^2 L)$
corrections (circles). The numerical results  have been obtained by {\tt
SABSPV}~\cite{sabsyr95,sabscpc95,a2l96}. 
} 
\label{fig:sabrc}
\end{figure}

At present, there is complete control of the ${\cal O} (\alpha L, \alpha, \alpha^2 L^2,
\alpha^3 L^3)$ corrections, for an arbitrary ES.~\footnote{Actually, at present a theoretical error of
technical origin is attributed to the ${\cal O} (\alpha^3 L^3)$ corrections, 
that however is negligible with
respect to the one due to missing ${\cal O} (\alpha^2 L)$ corrections (see Sect.~\ref{sect:sabtoterr}).} 
The ${\cal O} (\alpha^2 L)$ corrections are
not fully under control, in the sense that are either analytically known for non-realistic
(non-calorimetric) ES's or  only approximated for realistic (calorimetric) ES's (see
Sect.~\ref{sect:sabct} for more details). As it will be seen in the following, the approximate
knowledge  of the ${\cal O} (\alpha^2 L)$ corrections is the main source of theoretical error on the
SABH cross section. An illustrative plot concerning photonic  radiative corrections to the SABH
process can be found in Fig.~\ref{fig:sabrc}. The tree-level  cross  section corresponding to the
ES adopted is $\sigma_0 = 140.02$~nb. As can be seen in the first plot, the photonic corrections
reduce the cross section by an amount ranging from around 6\% to around 16\%, depending on the
$z_{min}$ cut (see the Figure for the definition of $z_{min}$). In the second plot, one can see 
that the non-leading ${\cal O} (\alpha)$ corrections reduce the cross section by  around 
$1.5$\%, the higher-order LL  corrections  increase  it by about 0.5\%, and  the non-leading 
${\cal O} (\alpha^2 L)$ corrections introduce a further increase of about $0.1 \div 0.2$\%. All 
these  corrections are relevant for a theoretical prediction  at the 0.1\% level. 

There are also other radiative corrections to the SABH process, which however are much smaller
than the photonic ones. They are the following: 
\begin{itemize}
\item {\it light pairs}: these corrections arise from photons converting into $f  \bar f$ pairs,
and are dominated by $e^+  e^- $ pairs; they give typically a contribution of the order of few
$10^{-4}$ with  respect to the tree-level cross section (see~\cite{jsw97_prd55} and
references therein);  
\item {\it QED radiative corrections to $Z$-$\gamma $ interference}:  they  alter sizably the
tree-level $Z$-$\gamma $ interference contribution, since it changes sign when crossing the
resonance; they are under control (see~\cite{bmp95}, \cite{jpw_z95} and references therein). 
\end{itemize}
Non-QED corrections other than vacuum polarization are absolutely negligible. 

\subsection{Computational Tools}
\label{sect:sabct}
In the present section the basic features of the computer codes available for the calculation
of the SABH process cross section are briefly summarized. The aim of this discussion is to
make an inventory of the theoretical approaches to the problem, and of their realizations in
the form of {\tt FORTRAN} codes, rather than to give an exhaustive description of the programs,
which can be found in the literature.  

\noindent
\underline{\tt BHAGEN95}\cite{bhagen95} --- It is a MC  integrator for both small- and 
 large-angle
 Bhabha scattering. It is a structure function based program (see Appendix~\ref{sect:sfm}) for
 all-orders resummation, including complete photonic ${\cal O} (\alpha)$ and leading
 logarithmic ${\cal O} (\alpha^2 L^2)$ corrections in all the channels (long writeup 
 in~\cite{bharep96}). 

\noindent
\underline{\tt BHLUMI}\cite{bhlumi95,bhlumi_404} --- It is a MC event generator for small-angle
 Bhabha scattering. It includes multi-photon radiation in the framework of YFS exclusive
 exponentiation (see Appendix~\ref{sect:yfs}). Its matrix element includes complete  
 ${\cal O} (\alpha)$ and leading logarithmic ${\cal O} (\alpha^2 L^2)$ corrections. Some
 non-leading  ${\cal O} (\alpha^2 L)$ corrections are also taken into  account. The program
 provides the full event in terms of particle flavors and four-momenta with an arbitrary number
 of additional  radiative photons. It is the standard package used by the LEP Collaborations  for
 the calculation  of the theoretical  SABH cross section. 

\noindent
\underbar{\tt LUMLOG} --- It is a MC event generator for the SABH process (part of
{\tt BHLUMI}, see~\cite{bhlumi95}). Photonic corrections are treated at the leading logarithmic
level, in the strictly collinear approximation, and in inclusive way. Structure functions
exponentiated up to ${\cal O} (\alpha^3 L^3)$ are included. At the ${\cal O} (\alpha^2)$ it
includes the leading corrections of the kind  ${\cal O} (\alpha^2 L^2)$. It is used to improve
{\tt OLDBIS} (more on this later) in the sector of higher-order photonic corrections. 

\noindent
\underbar{\tt NLLBHA}\cite{nllbha95} --- It is the {\tt FORTRAN} translation of the only available fully
analytical non-leading second-order calculation. At the ${\cal O} (\alpha^2)$ it includes all the
next-to-leading corrections ${\cal O} (\alpha^2 L)$. It is also able to provide 
${\cal O} (\alpha^3 L^3)$ photonic corrections and light pair corrections, including the
simultaneous emission of photon and light pair. It is a semi-analytical result, at present
available for a bare ES. 

\noindent
\underbar{\tt OLDBIS}\cite{bk83} --- It is a classical MC event generator for the Bhabha
process (the modernized version is part of {\tt BHLUMI}, see~\cite{bhlumi95}). It includes exact ${\cal
O} (\alpha)$ photonic corrections. It is used to improve {\tt LUMLOG} in the sector of ${\cal  O}
(\alpha)$ photonic corrections. 

\noindent
\underbar{\tt OLDBIS+LUMLOG} --- It is the ``tandem'' developed in parallel with {\tt BHLUMI} in order
to take into account higher-order corrections ({\tt LUMLOG}) on top of the exact ${\cal O} (\alpha)$
result ({\tt OLDBIS}). The matching between ${\cal  O} (\alpha )$ and higher-order corrections is
performed in {\it additive} form. This means that no ${\cal  O} (\alpha^2 L)$ corrections are
present. 

\noindent 
\underbar{\tt SABSPV}\cite{sabsyr95,sabscpc95} --- It is a MC integrator, designed  for
small-angle Bhabha scattering. It is based on a proper matching of the exact 
${\cal O} (\alpha)$  cross section for $t$-channel photon exchange~\cite{cr89,gmnp93} and of the leading
logarithmic results for the full Bhabha scattering cross section in the structure function
approach~\cite{40thieves}. 
The matching is performed both in {\it additive} and {\it factorized} forms, the
first form being used for comparisons only. In its default mode (factorized cross section) it
includes the bulk of the ${\cal O} (\alpha^2 L)$ corrections. 

\begin{figure}[hbtp]
\begin{center}
\epsfig{file=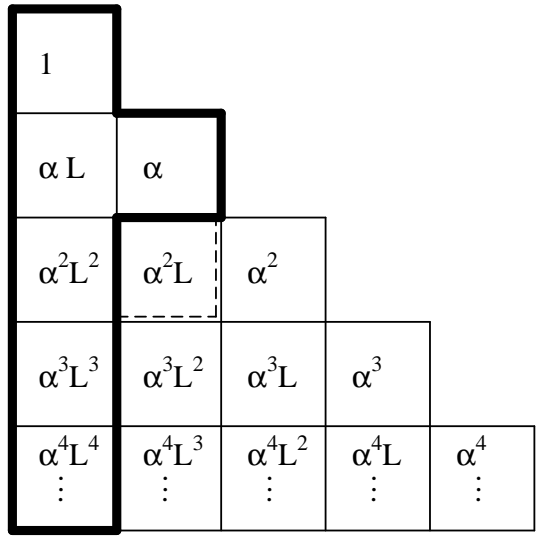, width=7truecm}
\end{center}
\caption{The general structure and the present situation concerning QED corrections 
to the SABH process. The thick line isolates the corrections exactly known for every ES. 
The dashed line points out the ${\cal O} (\alpha^2 L)$ corrections. They are at present 
the main source of theoretical error, since they are either approximately known for realistic ES's or
exactly known for unrealistic ES's.}
\label{fig:corrections}
\end{figure}

The situation of the computer codes available for the SABH process can be summarized as
follows: 

\begin{itemize}
 
 \item with the exception of {\tt LUMLOG}, all the other codes implement exact  ${\cal O} (\alpha)$
 photonic corrections for the SABH process; with the exception of {\tt OLDBIS}, all the other codes
 implement some  form of higher-order photonic corrections; 
 
 \item {\tt BHAGEN95}, {\tt OLDBIS+LUMLOG} and {\tt SABSPV} in the additive form include higher-order
 corrections in the leading logarithmic approximation;  
 
 \item {\tt BHLUMI} and {\tt SABSPV} in its default form include also 
 the bulk of the ${\cal  O} (\alpha^2 L)$
 corrections, for every ES; 
 
 \item {\tt NLLBHA} includes the full set of the ${\cal  O} (\alpha^2 L)$ corrections, but limited
 to a bare ES. 

\end{itemize}

The situation  concerning the QED corrections at present under control for every ES is 
described in Fig.~\ref{fig:corrections}. 
 
\subsection{The Total Theoretical Error}
\label{sect:sabtoterr}

Any theoretical prediction performed by means of a perturbative theory is affected by an 
intrinsic uncertainty because the perturbative series must be truncated at some finite order
(with the possible exception of those corrections that can be resummed to all the perturbative
orders, such as, for instance, the universal photonic corrections).
Hence the theoretical error of a perturbative prediction is dominated by the largest unknown
corrections, \idest\ by the lowest order corrections not under control. When applying the
above definition to the SABH process, it turns out that there are several sources of
theoretical error, namely the ones quoted in Tab.~\ref{tab:toterr}. 
Up to date, the most
important one is due to the missing photonic ${\cal O} (\alpha^2  L)$ corrections, which at
present are not fully under control as shown in Fig.~\ref{fig:corrections}. 

\begin{table}
 \caption{
 Summary of the total (physical+technical) theoretical uncertainty
 for a typical calorimetric detector.
 For LEP1, the above estimate is valid for the angular range
 within   $1^{\circ}-3^{\circ}$, and
 for  LEP2  it covers energies up to 176~GeV, and
 angular ranges within $1^{\circ}-3^{\circ}$ and $3^{\circ}-6^{\circ}$
 (from ref.~\cite{comdoc96}).}
 \label{tab:toterr}
 \begin{tabular}{llll}
   \hline
           & LEP1 & LEP1 & LEP2 \\
   \hline
   Type of correction/error & Past & Present  & Present \\
   \hline 
   (a) Missing photonic ${\cal O}(\alpha^2 L)$ & 0.15\%      & 0.10\%    & 0.20\% \\
   (b) Missing photonic ${\cal O}(\alpha^3 L^3)$ & 0.008\%   & 0.015\%   & 0.03\% \\
   (c) Vacuum polarization &    0.05\%      & 0.04\%    & 0.10\% \\
   (d) Light pairs &    0.01\%      & 0.03\%    & 0.05\% \\
   (e) $Z$-boson exchange  &    0.03\%      & 0.015\%   &  0.0\% \\
   \hline
   Total  & 0.16\%      & 0.11\%    & 0.25\% \\
   \hline
 \end{tabular}
\end{table}

Moreover, when considering the calculation of a cross section for a realistic ES, one  has to
implement the theoretical formulation of the problem in a computational tool, typically a
{\tt FORTRAN} code performing all the numerical integrations. 
\begin{figure}[hbt]
\begin{center}
\epsfig{file=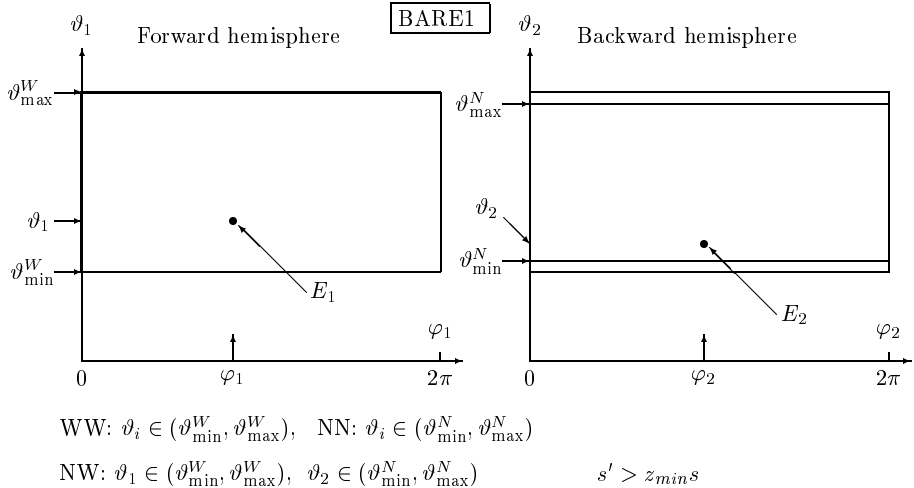,width=12truecm}
\end{center}
\caption{  Geometry and acceptance of the simple (non-calorimetric) ES BARE1.
  This ES restricts the polar angles $\vartheta_i$ 
  in the forward/backward hemispheres and requires
  a certain minimum energy to be detected simultaneously in 
  both hemispheres.
  Photon momentum is not constrained at all.
  The entire ``fiducial'' $\vartheta$-range, 
  \idest\  the wide (W) range, is
      $(\vartheta_{\min}^W, \vartheta_{\max}^W )=(0.024,0.058)$~rad 
  and the narrow (N) range is
      $(\vartheta_{\min}^N, \vartheta_{\max}^N ),$
  where $\vartheta_{\min}^N=\vartheta_{\min}^W+\delta_\vartheta$,
        $\vartheta_{\max}^N=\vartheta_{\max}^W-\delta_\vartheta$ and 
        $\delta_\vartheta=(\vartheta_{\max}^W - \vartheta_{\min}^W)/16$.
  This ES can be symmetric WW or NN,
  or asymmetric NW (see the description in the figure).
  The energy cut $s' > z_{min}s$ involves the momenta of the outgoing $e^\pm$
  ($s'=(q^+ + q^-)^2$) only  (from ref.~\cite{bharep96}).
}
\label{fig:bare1sel}
\end{figure}
In so doing, other sources of
uncertainty are added, that are of technical origin, such as, for  instance, approximations in
the formulae, numerical algorithms performing the phase  space integrations and so on. On
the whole, also the actual implementation in itself of a given theoretical formulation gives
rise to a finite contribution to the total theoretical error. 

\begin{figure}[hbt]
\begin{center}
\epsfig{file=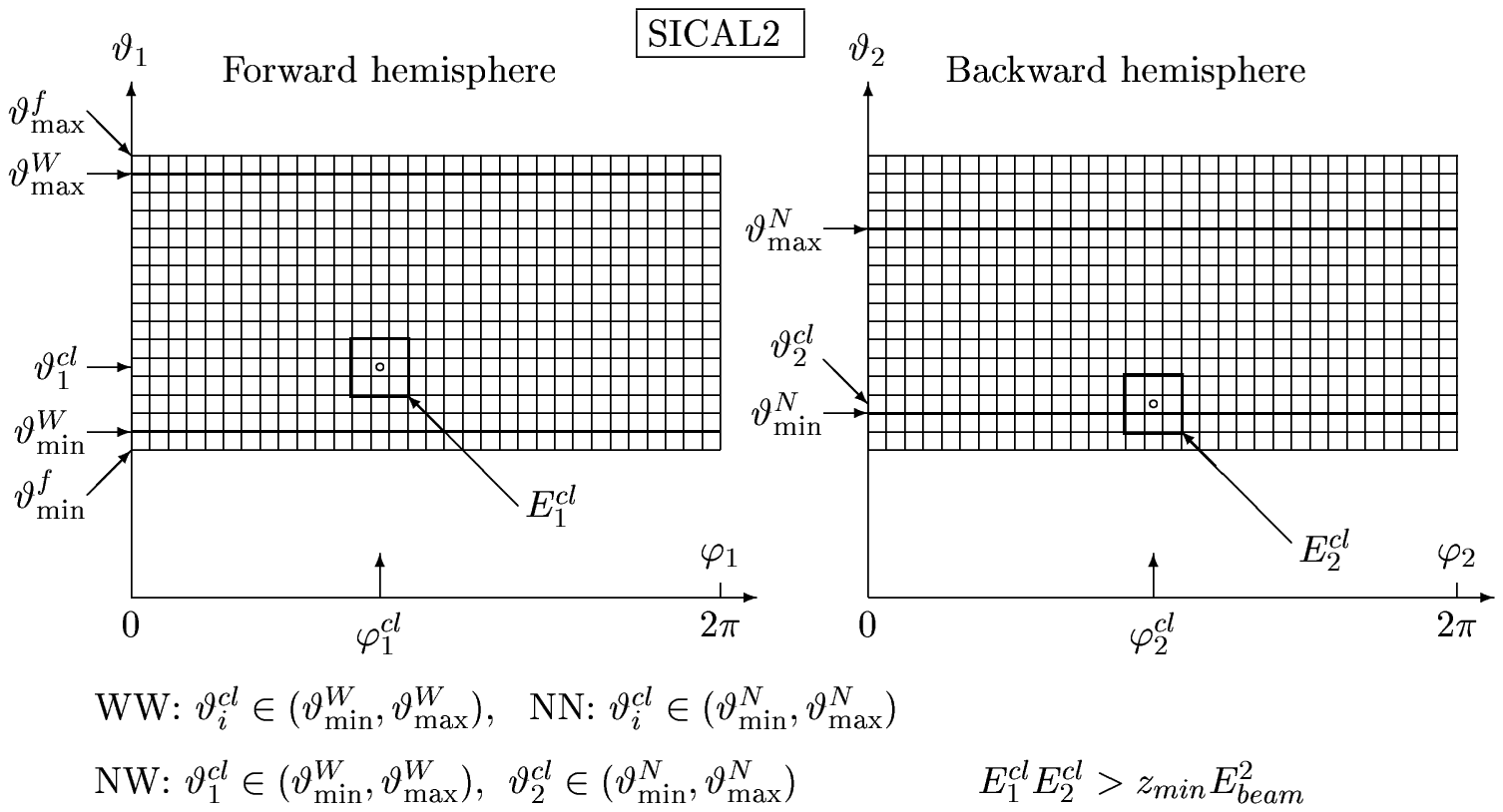,width=12truecm}
\end{center}
\caption{  Geometry and acceptance of the calorimetric ES SICAL2.
  This ES restricts the polar angles $\vartheta_i$ 
  in the forward/backward hemispheres and requires
  a certain minimum energy to be detected simultaneously in 
  both hemispheres.
  No restrictions on azimuthal angles $\varphi_i$ are there.
  The entire ``fiducial'' $\vartheta$-range, 
  $(\vartheta_{\min}^f,\vartheta_{\max}^f )=(0.024,0.058)$~rad,
  includes the wide (W) range
      $(\vartheta_{\min}^W, \vartheta_{\max}^W )$
  and the narrow (N) range
      $(\vartheta_{\min}^N, \vartheta_{\max}^N )$
  exactly as depicted in the figure.
  This ES can be symmetric WW or NN,
  or asymmetric NW.
  The energy cut and the $\vartheta$-cuts involve the definition 
of the {\it cluster}. 
Each side detector consists of $16\times 32$ equal {\it plaquettes}.
  A single plaquette registers the total energy of electrons and photons.
  The plaquette with the maximum energy, together with its $3\times 3$ 
  neighborhood, is called cluster.
  The total energy registered in the cluster is $E_{i}^{cl}$
  and its angular position is $(\vartheta_i^{cl},\varphi_i^{cl})$, $i=1,2$.
  More precisely the angular position of a cluster is the average 
  position of the {\it centers } of all $3\times 3$ plaquettes, weighted 
  by their energies (the definitions of $\varphi$'s are 
adjusted in such a way that
  $\varphi_1=\varphi_2$ for back-to-back configuration). 
  The plaquettes of the cluster which spill over the angular range 
  (outside thick lines) are also used to determine the total energy and 
  the average position of the cluster,  as in the  backward hemisphere (from ref.~\cite{bharep96}).
}
\label{fig:sical2sel}
\end{figure}

From now on, the precision reached in principle by means of a given theoretical 
formulation will be referred to as the ``physical precision'' of the approach, 
as distinct from the precision reached  in
its actual implementation, which will  be referred to as the ``technical precision''.   

As a  matter of fact, the LEP  Collaborations use the MC {\tt BHLUMI} to 
compute the SABH theoretical cross section. Hence, a key issue is  determining the 
theoretical error of the prediction by {\tt BHLUMI}. Since the codes described in the previous Section
differ from one another in the treatment of higher-order next-to-leading corrections, a careful
comparison of their predictions together with a deep understanding of their differences 
can be used to infer an estimate of the total theoretical error on the
SABH cross section. 

The most  extensive work in this  directions has  been performed in the context  of the 
Workshop  {\it Physics at LEP2}, held at CERN, Geneva, during 1995,  and in  particular within  the
Working Group  ``Event Generators for Bhabha scattering''~\cite{bharep96}, whose main tasks were
\begin{itemize}
 \item to make an inventory of all the available MC event
  generators, developed by independent collaborations,
  for  Bhabha processes at LEP1 and
  LEP2, both at  small and large scattering angles;
 \item to improve the understanding of their theoretical uncertainties by 
  means of systematic comparisons of MC's
  between themselves and with non-MC approaches.
\end{itemize}

The main emphasis was put on SABH processes, because of the pressing
need to match the theoretical precision of the calculations with the
much improved experimental accuracy ($\leq 0.1\%$) of the luminosity 
measurement. In particular, the main achievement of the Working Group, 
which is the result of a combined effort by several collaborations addressing several 
theoretical and  experimental issues, was the reduction of the theoretical error on the SABH 
cross section from 0.16\% to 0.11\%  for typical  ES's at LEP1, and a first 
estimate of the  theoretical  error on the SABH cross section at LEP2~\cite{comdoc96}.

The various components of the theoretical error on the SABH cross section 
are quoted in Tab.~\ref{tab:toterr} (see ref.~\cite{comdoc96}), where a summary of the past and
present situation at LEP1 together with the present estimate valid for LEP2 is given. 
The errors in the Table are understood to be attributed to the cross section for any 
typical (asymmetric) ES, for a LEP1 experiment in the angular range $1^{\circ}-3^{\circ}$,
calculated by {\tt BHLUMI~4.03}~\cite{bhlumi95}. In the case of LEP2, the estimate extends 
to the angular range $3^{\circ}-6^{\circ}$, and also to  a possible narrower angular range 
(say $4^{\circ}-6^{\circ}$) that may be necessary due to the effect of synchrotron radiation 
masks in the experiments. The entries include combined technical and physical precision.

\begin{figure}[htb]
\begin{center}
\epsfig{file=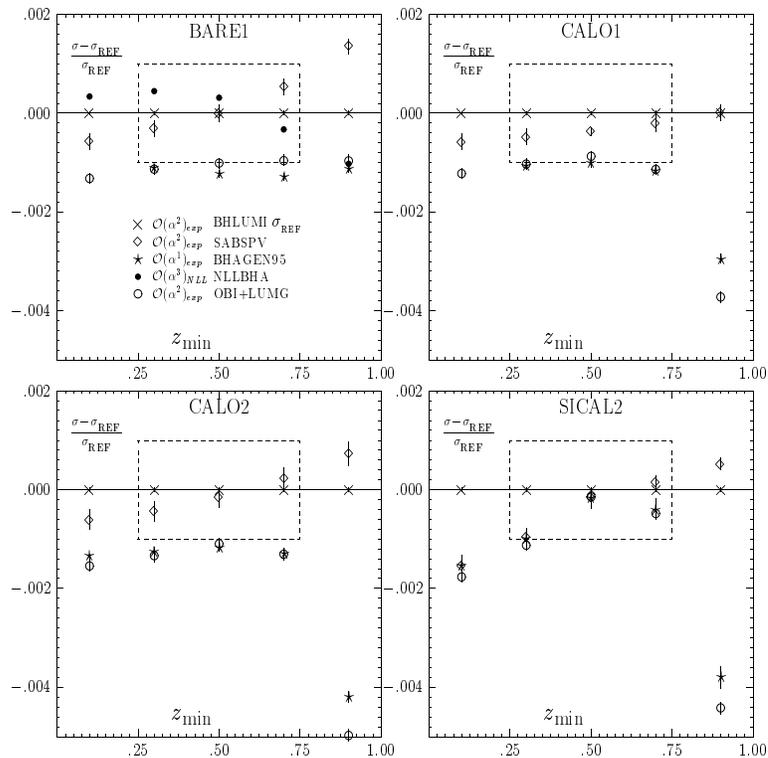,width=10truecm}
\end{center}
\caption{ Monte Carlo results for the symmetric Wide-Wide ES's 
 BARE1, CALO1, CALO2 and SICAL2, 
 for matrix elements beyond first order.
 $Z$-boson exchange, up-down interference and vacuum polarization
 are switched off.
 The c.m. energy is $\protect\sqrt{s}=92.3$~GeV. 
 $z_{min}$ is the cut
 condition on the final-state energies, defined as $ E_+ E_- / E^2  \geq 
 z_{min}$, $E_{-,+}$ being the final-state energy of the bare electron and
 positron, respectively. The {\it fiducial } angular range is 0.024-0.058~rad
 (for more details on the
 ES, the reader is referred to~\cite{bharep96}).
 In the plot, the  ${\cal O}(\alpha^2)^{YFS}_{exp}$ cross section
from {\tt BHLUMI~4.03} is used as a reference cross section (from refs.~\cite{bharep96,comdoc96}).
}
\label{fig:comdoc1}
\end{figure}

As can be seen in the Table, at the present stage the theoretical error is still dominated 
by the error on photonic corrections, quoted in entries (a) and (b), and in particular by 
the one due to missing ${\cal O} (\alpha^2 L)$ corrections of entry (a). Since the error of 
entry (a) is by far dominant with respect to all  the other ones, it is worth devoting some 
space to describe how it has been estimated, namely by adopting  the following procedure:
\begin{itemize}
 \item only the photonic corrections to the dominant
  part of the SABH cross section, namely the one due to $t$-channel photon
  exchange, have been considered as a first step; 
 \item four families of ES's have been defined (for the details concerning the definitions of the
  ES's the  reader is referred to~\cite{bharep96}); the simplest one, BARE1 (see
  Fig.~\ref{fig:bare1sel}), is an ES 
  in which cuts are applied only to the ``bare'' final-state fermions; 
  the other ones, CALO1, CALO2 and
  SICAL2, are calorimetric ES's, implementing more and more complex clustering
  algorithms;  in  particular, SICAL2 (see Fig.~\ref{fig:sical2sel}) 
  is very similar to a ``real'' experimental ES;
  since photonic corrections are very sensitive to the details of the ES, defining these four 
  ES's allows to span in detail the photonic phase space; even if the ES BARE1 is far from 
  realistic, the presently available analytical calculation  including the complete set of
  ${\cal O} (\alpha^2 L)$ corrections refers to such an ES, and so provides a very important 
  cross-check of the MC programs;
 \item  all the available codes have been run for all the ES's, varying inside any 
  ES the threshold requirements for the final-state fermions/clusters;
 \item a test concerning the technical precision  has been performed, namely 
  comparing the exact up to ${\cal O} (\alpha)$ cross sections provided by the various codes; 
  agreement at the level of a few $10^{-4}$ relative deviation has been achieved;
 \item finally, the results of the codes including the full higher-order
  photonic corrections have been compared, 
  for all the situations explored (see Fig.~\ref{fig:comdoc1}).
\end{itemize}

The result of this procedure allowed the definition of ``one-per-mill regions'',
referring to realistic threshold cuts, within which most of the predictions lie.
Moreover, for those cases for which the predictions do not lie within the
``one-per-mill regions'', the reasons for the deviations involved have been
carefully investigated and eventually understood.

\begin{figure}[hbt]
\begin{center}
\epsfig{file=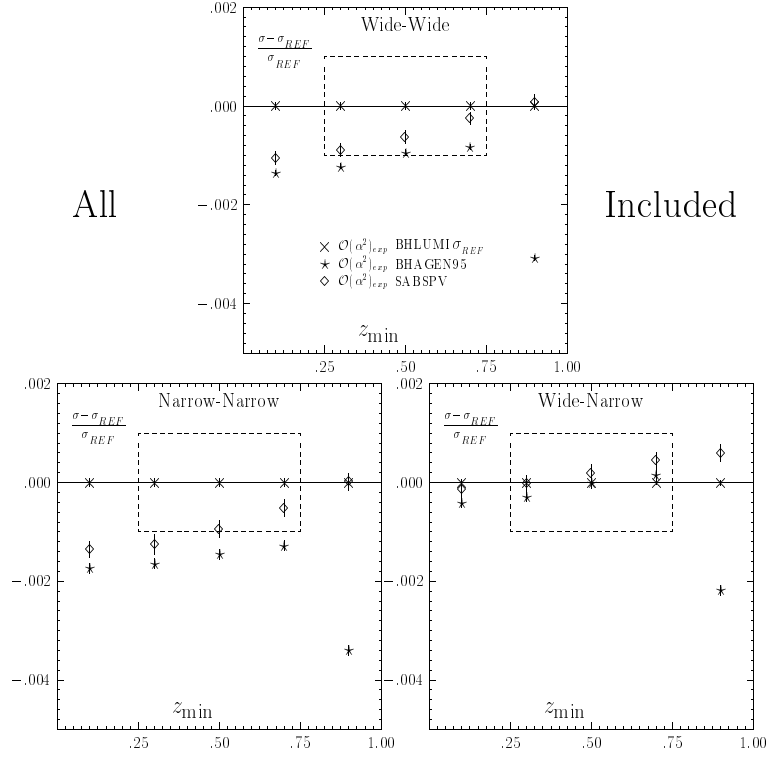,width=10truecm}
\end{center}
\caption{ Monte Carlo results for various symmetric/asymmetric           
 versions of the CALO2 ES,
 for matrix elements beyond first order.                    
 $Z$-boson exchange, up-down interference and vacuum polarization       
 are switched ON.                                               
 The c.m. energy is $\protect\sqrt{s}=92.3$~GeV.
 $z_{min}$ is the cut
 condition on the final-state energies, defined as $ E_+ E_- / E^2 \geq 
 z_{min}$, $E_{-,+}$ being the final-state energy of the electron and
 positron clusters, respectively. 
 The {\it fiducial } angular range is 0.024-0.058~rad (for more details on the
 ES, the reader is referred to~\cite{bharep96}). 
 In the plot, the ${\cal O}(\alpha^2)^{YFS}_{exp}$ cross section     
from {\tt BHLUMI~4.03} is used as a reference cross section (from refs.~\cite{bharep96,comdoc96}).                          
}
\label{fig:comdoc2}
\end{figure}

An analogous procedure has been followed after the inclusion of all the
relevant radiative corrections (vacuum polarization, $Z$-boson exchange contributions 
and so on), and extending the comparisons also to asymmetric ES's, leading to the 
results shown as an example in  Fig.~\ref{fig:comdoc2}.
A similar analysis has also been performed for the first time in situations which 
are typical at the LEP2 experiments. The conclusion drawn at the end of all these 
comparisons is that now the theoretical uncertainty due to uncontrolled ${\cal O} (\alpha^2 L)$ 
corrections is reduced from 0.15\% to 0.10\% for the LEP1 situation, and estimated to be
0.20\% at LEP2. As far as entry (b), the ``missing photonic ${\cal O} (\alpha^3 L^3) $''
uncertainty, is concerned, new estimates of the effect have resulted in a more
conservative theoretical error, namely 0.015\% to be compared with the old
estimate of 0.008\%.

As far as all the other entries in Tab.~\ref{tab:toterr} are
concerned, namely entries (c), the ``vacuum polarization'' uncertainty, 
(d), the ``light pairs'' uncertainty, and (e), the ``$Z$-boson exchange'' uncertainty,
two of them, (c) and (e),  are reduced with respect to the previous situation
thanks to several new fits of the hadronic contribution to the vacuum polarization,
and some additional original work on the $Z$-boson exchange contribution done during the 
workshop. New estimates, both MC and analytical, of
the light pairs contribution, (d), featuring more complete calculations
done during the workshop, have resulted in a more conservative estimate
of the pairs effect uncertainty of 0.03\%.

In conclusion, the total (physical + technical) theoretical uncertainty 
on the SABH cross section for a typical calorimetric detector is at present 
0.11\% at LEP1 and 0.25\% at LEP2. While the theoretical precision reached in the LEP2 case is
sufficient, further improvements in  the LEP1 case are desirable. 

\subsection{Recent Developments and Perspectives}
\label{sect:sabdp}

After the completion of  the  Working Group  ``Event Generators for 
Bhabha scattering''~\cite{bharep96}, some additional work, relevant for
a further reduction of  the theoretical uncertainty on the SABH cross 
section, has been done by  the 
{\tt BHLUMI}~\cite{jmwy_npps,jw_plb389,jmwy96,jetal97_tcbp, wetal97_pcbp} and 
{\tt SABSPV}~\cite{a2l96} groups.   

\begin{figure}[hbt]
\begin{center}
\epsfig{file=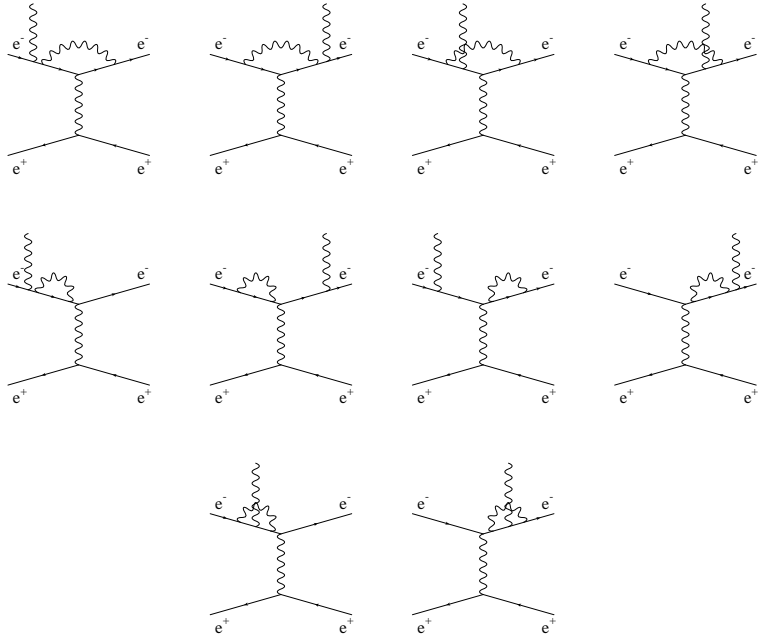,width=10truecm}
\end{center}
\caption{Examples of virtual  corrections to single bremsstrahlung in the SABH process 
(see ref.~\cite{jmwy_npps,  jmwy96}). Only  graphs concerning the real emission by 
the electron line are shown. }
\label{fig:a2bremm}
\end{figure}

Concerning the {\tt BHLUMI} group, the exact virtual one-loop  corrections to the 
hard bremsstrahlung process in the SABH scattering  (see Fig.~\ref{fig:a2bremm}) 
have been computed~\cite{jmwy_npps,jmwy96}. 
These results are needed to complete
the exact treatment of the ${\cal O} (\alpha^2 L)$ photonic corrections, since the contributions 
from double bremsstrahlung~\cite{jwy93} and the two-loop electron form factor~\cite{bmr72} are known. 
None of these corrections is at present implemented in {\tt BHLUMI}~\cite{jmwy_npps}.   
It has  to be  noticed
that this is the first exact, completely differential, result for the virtual one-loop  corrections 
to the hard bremsstrahlung process in the SABH scattering. 
Thanks to  this result, the authors are able to estimate the size of the missing 
${\cal O} (\alpha^2)  L$ part  in {\tt BHLUMI}, for a realistic NW ALEPH SICAL luminometer,  
finding it below  $2 \times 10^{-4}$ of  the Born cross  section in the massless limit.
Work is in progress in order to generalize the result to the massive case, and in order 
to compare its effect with the semi-analytical calculation  of the {\tt NLLBHA} 
group~\cite{nllbha95}.

Concerning the {\tt SABSPV} group, the theoretical formulation has been refined  in
order to  eliminate a  phase-space approximation in the ${\cal O} (\alpha^2 L)$
sector and  the result has been analytically checked against the available 
complete ${\cal O} (\alpha^2 L)$ results already present in the literature for  
an academic ES (namely a BARE ES) in the soft-photon approximation, both for
the annihilation and scattering channels~\cite{a2l96}. For the annihilation channel, 
the results have been compared in particular with the ones shown in ref.~\cite{kf}, 
finding that 
\begin{itemize}
\item the ${\cal O} (\alpha )$ perturbative result is exactly recovered, by
construction; 
\item \underline{all} the IR-singular terms,
namely the ones containing $\ln^2  \varepsilon  $ and 
$\ln  \varepsilon  $, where $\varepsilon = \Delta E / E$,  
are exactly recovered at the level of ${\cal O}
(\alpha^2 L_s^2)$,  ${\cal O} (\alpha^2 L_s)$ and ${\cal O} (\alpha^2 )$, 
where $L_s = \ln (s / m^2 )$; 
\item the difference between the two
results starts at the level of $ (\alpha / \pi )^2 L_s $ times a constant;    
\end{itemize} 
in particular, such a difference reads
\begin{equation}
{{\delta \sigma} \over {\sigma_0}} \Bigg\vert_{(\alpha^2 L_s)}= 
\left( {\alpha \over \pi} \right)^2 L_s \left[ 
3 \zeta (3) - {3 \over 2} \zeta (2) + {3 \over 16} \right],
\label{eq:schdiff} 
\end{equation}
where $\delta \sigma $ is the difference between the cross section
of~\cite{kf} and the cross section of ref.~\cite{a2l96}. 
The difference  numerically amounts to a relative deviation of about 
 $1.7 \times 10^{-4}$. The residual difference is 
 at ${\cal O} (\alpha^2) $ times a constant and is numerically irrelevant. 
For   the scattering channel,  the results  have been compared with  the ones of  
ref.~\cite{nllbha95}.  The results  of  the comparison are  the same as  in  the  
annihilation case up to the ${\cal O} (\alpha^2 L_t )$ corrections,  
namely the difference appears at the level of $ (\alpha / \pi )^2 L_t $ 
times a constant and reads
\begin{equation}
{{\delta \sigma} \over {\sigma_0}} \Bigg \vert_{(\alpha^2 L_t)}= 
2 \left( {\alpha \over \pi} \right)^2 L_t \left[ 
3 \zeta (3) - {3 \over 2} \zeta (2) + {3 \over 16} \right],
\label{eq:tchdiff} 
\end{equation}
where $\delta \sigma $ is the difference between the cross section
of~\cite{nllbha95} in soft approximation 
and the cross section of ref.~\cite{a2l96}.
This difference numerically amounts  to a relative deviation of about 
$2.2 \times 10^{-4}$, since the overall factor of two is compensated by the
fact that $L_t \simeq 2/3 L_s$. In this case, also an additional difference
appears, namely at the level of the IR-sensitive truly 
${\cal O} (\alpha^2)$ terms, which reads
\begin{equation}
{{\delta \sigma} \over {\sigma_0}} \Bigg \vert_{(\alpha^2 )}= 
- \left( {\alpha \over \pi} \right)^2  \left[ 4 \ln^2 \varepsilon + 8 \ln
\varepsilon \right], 
\label{eq:irdiff}
\end{equation} 
and is numerically irrelevant for a   realistic  situation. 
Therefore, the ${\cal O} (\alpha^2 L )$ corrections
taken into account by the method of ref.~\cite{a2l96} 
represent the bulk of the complete set. By taking  now into  account that  
 the canonical  coefficient of the ${\cal O} (\alpha^2 L )$ corrections 
 for a calorimetric (realistic) ES is one half  of the corresponding one
for the BARE ES (see Tab.~\ref{tab:nc-c}), and considering a safety factor of  
three, the authors
of~\cite{a2l96} estimate that  the overall theoretical error of the approach, 
as  far  as QED corrections are concerned, is $\delta  \sigma  /  \sigma  \simeq 
3 \times 10^{-4}$. 
\begin{figure}[hbt]
\begin{center}
\epsfig{file=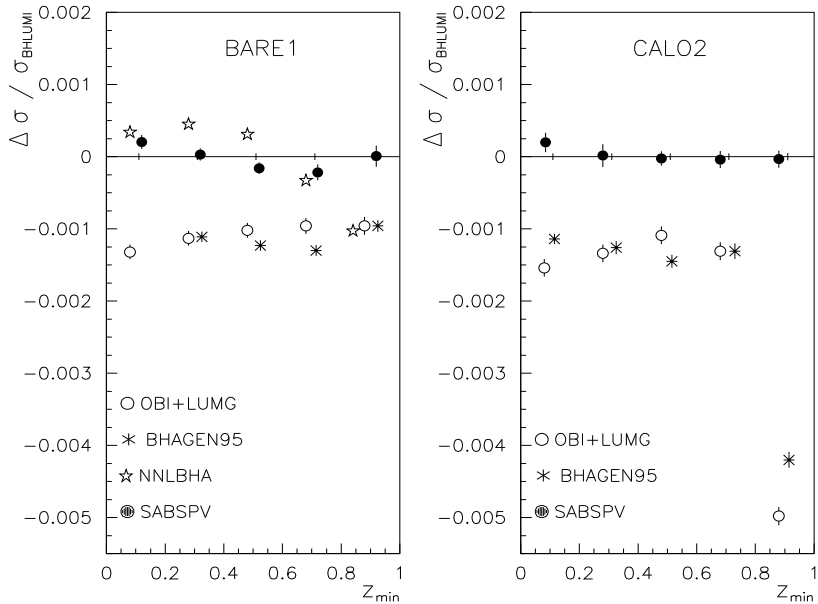,height=9truecm,width=12truecm}
\end{center}
\caption{Comparison of Monte Carlo's. 
The relative differences between the codes 
involved in the comparison 
and the cross section by {\tt BHLUMI 4.03} taken as a reference cross section are
shown as functions of the cut $ z_{min} = E_+ E_- / E_{beam}^2 $. 
$ E_{+,-} $ are the energies deposited in the positron and electron clusters, 
respectively. The details of the clustering algorithms (BARE1  and CALO2) are given 
in~\protect\cite{bharep96}. The c.m. energy is 
$ \protect\sqrt{s} = 92.3 $~GeV. For more details see ref.~\cite{a2l96}.}
\label{fig:barecalo}
\end{figure}

The situation is described in Fig.~\ref{fig:barecalo}, where two ES's   have been 
considered, namely a bare and a calorimetric ES. As can be seen, {\tt BHAGEN95}  and  
{\tt OLDBIS+LUMLOG}, which do not take into account ${\cal O} (\alpha^2 L )$ corrections, 
differ from {\tt BHLUMI}, which is taken as the reference cross section,  by about 0.1\%. 
{\tt NLLBHA} shows  differences which are  contained  within 0.1\%; since the {\tt NLLBHA} 
formulae
coincide with the {\tt  SABSPV} ones up to the ${\cal O} (\alpha^2 L )$ corrections in the  
soft-photon region,  
such differences in the soft-photon region are presumably due to  lack of exponentiation  
in {\tt NLLBHA}. 
{\tt BHLUMI} and {\tt SABSPV} differ at most  by 0.025\%, consistently with  the independent 
estimates of accuracies provided by  their  authors,  respectively. 

The results shown pave the way to a  definite  reduction of the theoretical  luminosity 
error at  the level  of 0.05\%. This  is anyway not far from the smallest possible theoretical  error, 
since the vacuum polarization uncertainty is 0.04\% in its  own (see Tab.~\ref{tab:toterr}).

%% file: z0.txi
\section{$Z^0$ Physics}
\label{sect:z0phys}
\begin{figure}[hbt]
\begin{center}
\epsfig{file=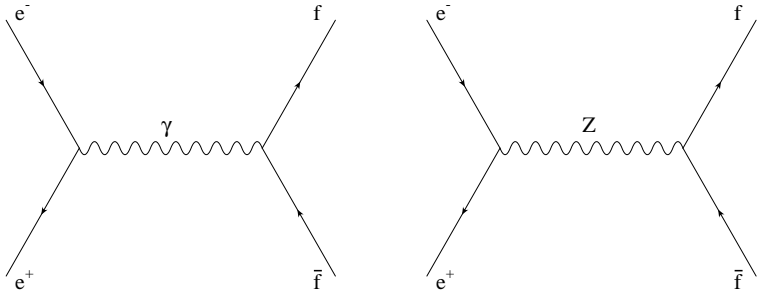,width=10truecm}
\end{center}
\caption{The tree-level Feynman diagrams for the process $e^+ e^- \to  
\gamma, Z \to  f \bar f  \, \, \, (f \neq e)$. }
\label{fig:bornz}
\end{figure}

The most important processes at the $Z^0$ resonance are represented  by
two-fermion production. In the case in which there are no electrons in the
final state, the tree-level Feynman  diagrams for the $s$-channel annihilation 
$e^+ e^- \to  \gamma, Z \to  f \bar f  \, \, \,(f \neq e)$ are depicted in
Fig.~\ref{fig:bornz}. In the presence of  electrons  in the final
state, there are additional $t$-channel amplitudes, which at large scattering angles
are essentially  backgrounds to the
dominant $s$-channel annihilation (see Fig.~\ref{fig:bhabhatree} in 
Sect.~\ref{sect:sablm}). 
The high degree of precision (at the level of 0.1\% or better) 
reached in the measurement of the electroweak parameters at the $Z^0$ pole 
requires the inclusion in the theoretical predictions of radiative 
corrections beyond the Born approximation, in order to perform a 
meaningful comparison between theory and experiment. Radiative 
corrections to $e^+ e^- \to \gamma, Z \to f \bar f$ processes are due 
to electroweak and strong forces as described by the Standard 
Model (SM) of the fundamental interactions, based on the gauge group 
$SU(2) \otimes U(1) \otimes SU(3)$. At a difference from the small-angle Bhabha 
process, where the calculation of QED radiative corrections only 
is enough to reach the aimed 0.1\% theoretical precision, 
the two-fermion production in $e^+ e^-$ neutral current processes 
receives important contributions also from the electroweak 
and strong sector of the theory. Therefore, the fine structure of the SM can be 
probed by precision physics at LEP/SLC in a highly non trivial way at 
the level of quantum loops. The issue of radiative corrections has been studied
 at length in the past.  For  instance, the situation before the starting of 
 LEP operations  is reviewed 
in refs.~\cite{yrlep88}, \cite{yrlep189}  and~\cite{greco88}. More  recently, in view of
the final analysis of LEP1 data, the theoretical studies have been updated 
in ref.~\cite{yrwg95}.  

In the present section, the basic aspects concerning the theoretical treatment 
of radiative corrections in precision calculations for the $Z^0$ resonance 
are reviewed, through the analysis of the main ingredients of 
pure weak, QCD and QED corrections to two-fermion production 
amplitudes. It is also discussed how the theoretical formulae are realized 
in the form of computational tools used for data analysis. The uncertainties 
associated to the theoretical predictions 
and their impact on the precision calculation of the $Z$-boson observables are 
finally analyzed.

\subsection{Realistic Observables and $Z^0$ Parameters}
\label{sect:z0par}

The data published by the LEP Collaborations and used to extract information on the 
SM and its possible extensions refer to two main classes of observables:
\begin{itemize}

\item realistic observables;

\item the $Z^0$ parameters or {\it pseudo-observables}.

\end{itemize}
 
The cross sections $\sigma(s)$ 
and the forward-backward asymmetries $A_{FB}(s)$ of the two-fermion production processes
\begin{eqnarray}
e^+ e^- \rightarrow (\gamma, Z) \rightarrow f {\bar f} (n\gamma),
\label{basicffng}
\end{eqnarray}
quoted as  functions of the centre of mass (c.m.) energy and  including 
real and virtual photonic 
corrections (\idest\   ``dressed by QED'') are collectively referred to as realistic 
observables.  
They  are  defined  as  
\begin{eqnarray}
&&\sigma = \sigma_F + \sigma_B , \\
&&A_{FB} = { {\sigma_F - \sigma_B} \over {\sigma_F + \sigma_B} } ,
\end{eqnarray}
where $\sigma_{F,B}$ are the cross sections in the forward and backward hemisphere, 
respectively, \idest\     
\begin{eqnarray}
\sigma_F = 2 \pi \int_0^1 d \cos\vartheta \, {{d \sigma} \over {d \Omega}}, \quad 
\quad \sigma_B = 2 \pi \int_{-1}^0 d \cos\vartheta \, {{d \sigma} \over {d \Omega}}. 
\label{eq:sfb}
\end{eqnarray}

In the tree-level approximation, for unpolarized initial- and final-state fermions, the differential
cross section $d \sigma_0 / d \Omega$, as obtained from the calculation of the Feynman diagrams
shown in Fig.~\ref{fig:bornz}, is given by 
\begin{equation}
{{d \sigma_0} \over {d \Omega}} = {{d \sigma_0^\gamma} \over {d \Omega}} 
+ {{d \sigma_0^{\gamma Z}} \over {d \Omega}} +
{{d \sigma_0^Z} \over {d \Omega}} , 
\label{eq:sborn}
\end{equation}
where  the $\gamma$, $\gamma Z$ and $Z$ contributions read
\begin{eqnarray}
& &{{d \sigma_0^\gamma} \over {d \Omega}} = {{\alpha^2 Q_f^2 N_c} \over {4 s}} \left( 1 +  
\cos^2 \vartheta \right) , \nonumber \\
& &{{d \sigma_0^{\gamma Z}} \over {d \Omega}} = 
-{{\alpha  G_\mu M_Z^2 Q_f N_c} \over {4 \sqrt{2} \pi s}} \hbox{\rm  Re} \big( \chi(s) \big) 
\left[ g_v^e g_v^f \left( 1 +  
\cos^2 \vartheta \right) + 2 g_a^e g_a^f \cos \vartheta \right] , \nonumber \\
& &{{d \sigma_0^Z} \over {d \Omega}} = {{G_\mu^2 M_Z^4 N_c} \over {32 \pi^2 s}} \vert 
\chi(s) \vert^2 \left[ \big( (g_v^e)^2  + (g_a^e)^2 \big)
\big( (g_v^f)^2  + (g_a^f)^2 \big) \left( 1 +  \cos^2 \vartheta \right) \right. \nonumber \\
& &\qquad \,\,\, \left. + 8 g_v^e g_a^e g_v^f g_a^f \cos \vartheta  \right] .
\label{eq:sborncontr}
\end{eqnarray}
In  eq.~(\ref{eq:sborncontr}) the symbols are defined as follows. $\vartheta$ is the fermion
scattering angle. $G_\mu$ is the muon decay constant,
$\alpha$ is the QED coupling constant and $M_Z$ is the $Z$-boson mass. $Q_f$ is the final-state
fermion  charge, in units of the positron charge.  
$N_c$ is  the colour factor, $N_c = 1,3$ for leptons
and quarks respectively. $g_v^i$  and $g_a^i$ are the tree-level 
vector and axial-vector couplings of the $i$th 
fermion to the $Z^0$  boson, given by 
\begin{equation}
g_v^i = I_3^i - 2 Q_i \sin^2 \vartheta_W , \qquad \qquad \qquad g_a^i = 
I_3^i ,  
\end{equation}
$I^3_i$ being the third weak-isospin component of the $i$th fermion and $\sin^2 \vartheta_W$ the
squared sine of the weak mixing angle. 
At last, $\chi(s)$ is the resonating factor
\begin{equation}
\chi(s) = {{s} \over {(s - M_Z^2) + i \Gamma_Z M_Z}} , 
\end{equation} 
$\Gamma_Z$ being the total $Z$-boson width. 

From the explicit expression of  the differential cross section given in 
eqs.~(\ref{eq:sborn}) and
(\ref{eq:sborncontr}) it is straightforward to  derive the expressions for the total cross section
and forward-backward asymmetry. Concerning the total cross section, at $\sqrt{s} = M_Z$ the 
$Z$-boson contribution can be written in terms of total  and partial  $Z$-boson widths (see
eq.~(\ref{eq:defpar}) below), while the $\gamma  Z$ term is exactly vanishing and the $\gamma $
contribution gives an effect contained below 1\%. As far  as the forward-backward asymmetry is
concerned, still at  $\sqrt{s} = M_Z$, it takes the form shown  in 
eq.~(\ref{eq:defasymfb})  below.   

The data presented by the LEP Collaborations 
for the realistic observables concern typically 
two different experimental configurations:
\begin{itemize}   

\item extrapolated set-up: a cut on the invariant mass of the final-state fermions or 
on the invariant mass of the event after initial-state radiation (ISR) 
alone is imposed; the 
data corresponding to this inclusive situation are also said {\it perfect data}; 

\item ``realistic" set-up: simple kinematical cuts, such as cuts on the energies or the 
invariant mass of the final-state products, angular acceptance and acollinearity angle of 
the outgoing fermions, are imposed. 

\end{itemize}

Owing to the critical dependence of the QED corrections on the applied cuts, the comparison of 
the theory with the data for both the above set-up necessarily requires the availability 
of formulations (and relative computational tools) as complete as possible in the treatment 
of QED effects. Moreover, when aiming to fit ``realistic" observables, a special effort must be 
devoted to the development of compact analytical and semi-analytical formulae for QED corrections.   

The $Z^0$ parameters, or {\it pseudo-observables}, 
are extracted from the measured cross sections and
asymmetries after some deconvolution or unfolding procedure. Basically, these quantities  
are determined by the LEP experiments by means of combined fits to the hadronic and 
leptonic cross sections and leptonic asymmetries, after corrections for the effect of 
ISR.  
Besides ISR, other radiative corrections or specific uninteresting effects can be depurated from 
the measurements in the unfolding procedure. For example, the deconvoluted forward-backward 
asymmetry includes the $Z$-boson exchange only (\idest\    after subtraction of the 
$\gamma$ and $\gamma Z$ contributions) and also final-state QED and eventually QCD corrections 
are subtracted from the experimental data. 
For the quantities relative to the  $e^+ e^-$ final state,
the $t$ and $s$-$t$ interference contributions are subtracted as well. 
More details  about  the fitting  procedure adopted by the LEP Collaborations for the extraction of
the $Z^0$ parameters are given in  Sect.~\ref{sect:fits}. 
Therefore, the 
$Z^0$ parameters are secondary quantities or, in this sense, {\it pseudo-observables}. They 
are anyway of utmost importance for the precision tests of the SM, since they depend 
on the details of the internal structure of the electroweak theory. The  $Z^0$ 
parameters considered in the literature are given in Tab.~\ref{tab:obs}.
\begin{table}[htb]
\caption{The most relevant $Z^0$ parameters.}
\label{tab:obs}
\begin{tabular}{ll}
\hline
Observable & Symbol \\
\hline
hadronic peak cross-section  & $\sigma_h $ \\
partial leptonic and hadronic widths & $\Gamma_l$ $(l=e,\mu,\tau)$, 
$\Gamma_{c}$, $\Gamma_{b}$ \\
total width &    $\Gamma_Z $ \\  
hadronic width &  $\Gamma_h $  \\
invisible width & $\Gamma_{inv} $  \\
ratios & $R_l $, $R_b$, $R_c $ \\
forward-backward  asymmetries & $A_{FB}^{l}$, $A_{FB}^b$, $A_{FB}^c$    \\ 
polarization asymmetries & $P^{\tau}$, $P^{b}$ \\
left-right  asymmetry (SLC)  & $A_{LR}^e$ \\
effective sine & $\sin^2 \vartheta^{l}_{eff}$,  $\sin^2 \vartheta^{b}_{eff}$ \\
\hline
\end{tabular}
\end{table}
The {\it effective sine} is defined as
\begin{equation}
4\,|Q_f| \sin^2 \vartheta^{f}_{eff} \, = \, 1 - {{g_V^f} \over {g_A^f}},
\end{equation}
where $Q_f$ is the electric charge of the fermion $f$ in units of the
positron charge, $g_V^f$ and $g_A^f$ are the {\it effective} neutral current vector and 
axial-vector couplings of the $Z^0$ to a fermion pair $f \bar f$, respectively.
By definition, the total and partial widths of the $Z^0$  include final-state 
QED and QCD radiation. The invisible $Z$-boson width $\Gamma_{inv}$, 
the ratios $R$ and the hadronic peak cross section are defined as
\begin{eqnarray}
&&\Gamma_{inv} \,  =  \, \Gamma_Z - \Gamma_e - \Gamma_{\mu} - \Gamma_{\tau}
 - \Gamma_h , \\
&&R_l \,  =  \, {\Gamma_h \over \Gamma_l }  , \\
&&R_{b,c} \,  =  \, {\Gamma_{b,c} \over \Gamma_h }  , \\
&&\sigma_{had}^0 \,  =  \, 12 \pi \, { {\Gamma_e \Gamma_h} \over {M_Z^2 \Gamma_Z^2} } .  
\label{eq:defpar}  
\end{eqnarray} 
By definition, $\sigma_{had}^0$ includes only the $Z$-boson exchange. Forward-backward, left-right
and polarization asymmetries, 
unlike the widths, are depurated, as said above, of the effects of QED and QCD corrections and,
like $\sigma_{had}^0$, refer to pure $Z$-boson exchange. 
This allows to express them as simple combinations 
of the effective $Z^0$ couplings as follows:
\begin{eqnarray}
&&A_{FB}^{f}  =  {3 \over 4} { A}_e { A}_f,   
\label{eq:defasymfb} \\
&&A_{LR}^e    =  { A}_e,  
\label{eq:defasymlr} \\
&&P^{f}     =  - { A}_f,
\label{eq:defasym}
\end{eqnarray}
where
\begin{equation}
{ A}_f = { {2 g_V^f g_A^f} \over {(g_V^f)^2+(g_A^f)^2} } .
\label{eq:afdef}
\end{equation}
Analogously, the decay width of the   
$Z^0$  boson into a $f \bar f$ pair  is given by the   
following expression: 
\begin{eqnarray}  
\Gamma_f = 4 N_c \Gamma_0 [(g_V^f)^2 R_V^f + (g_A^f)^2 R_A^f] ,  
\label{eq:widthpar}  
\end{eqnarray}  
where $\Gamma_0$ is given by $\Gamma_0 = G_{\mu} M_Z^3 / 24 \sqrt{2} \pi $, and   
$R_V^f$ and $R_A^f$ are factors taking into account QED and QCD   
final-state radiation (FSR)  and mass effects.   
 
\subsection{Electroweak Corrections}
\label{sect:z0ew}
 The high precision reached in the   
measurements at LEP and SLC allows to test the SM   
at the level of its radiative corrections.   
This can be done by virtue of the renormalizability of the   
theory~\cite{thooft71}. The one-loop radiative corrections  
to $e^+ e^- \to f \bar f$ can be divided   
into two classes,   separately gauge invariant: the pure QED corrections   
and the electroweak ones. The former are obtained by   
adding a virtual or real photon line to the lowest order   
Feynman   
diagrams, while the latter consist of all the remaining   
diagrams at one-loop order.   
As will be seen in the following the two   
classes of corrections are very different in many aspects   
and therefore different methods of calculation are required   
in order to reach the necessary theoretical accuracy.   
Contrary to the pure QED corrections, the electroweak   
corrections amount numerically to a few percent so that   
a one-loop approximation is already almost satisfactory. 
On the other hand, they   
depend on the fundamental parameters of the theory, so that   
particular attention must be devoted to the inclusion of   
potentially large higher-order effects in order   
to fully exploit the experimental accuracy to get   
information on the still unknown parameters.  
In the   
present Section the fundamental ingredients of electroweak   
corrections are illustrated and discussed.  

\subsubsection{One-loop Feynman diagrams}
\label{sect:olfd}
The calculation of the electroweak corrections involve the   
evaluation of diagrams with closed loops, where an   
integration over the loop momentum is present which diverges   
for large momenta (ultraviolet divergences). For this reason a   
regularization procedure is needed in order to deal with finite   
and mathematically well defined integrals, before implementing a renormalization  program. 
The most used    method   
in the case of gauge theories is the dimensional   
regularization~\cite{bg72,ashm72,hv72}   
because it allows to maintain   
Lorentz and gauge invariance at any step of the calculation.   
Some details about this procedure are given in Appendix~\ref{sect:scalint}.   
  
The electroweak one-loop corrections to $e^+ e^- \to f \bar f$  
can be classified in vector boson self-energies,   
fermion self-energies, vertex and box corrections. 
The fermion self-energies are commonly accounted as part of the   
vertex corrections.  
The vector boson self-energies, also referred  to as {\it oblique corrections}, 
include the following transitions:   
$\gamma $-$ \gamma$, $\gamma$-$ Z$, $Z $-$ \gamma$, $Z $-$ Z$ and   
$W $-$ W$.  

\begin{figure}[hbt]
\begin{center}
\epsfig{file=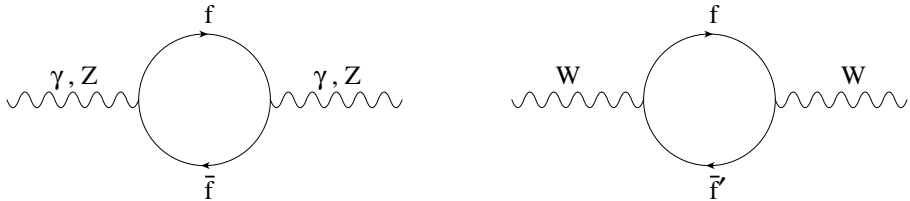, width=12.truecm}
\end{center}
\caption{One-loop fermionic contributions to vector-boson self-energies. }
\label{fig:fselfenergies}
\end{figure}
  
All the above classes of corrections can be written   
in terms of general   
two-, three- and four-point functions introduced in   
ref.~\cite{pv79}, which in turn can be reduced to certain   
combinations of the basic scalar one-, two-, three- and four-point   
integrals introduced in ref.~\cite{hv79}. Some simple examples   
of the procedure are given in Appendix~\ref{sect:scalint}.   
 The general expressions   
of the scalar form factors for arbitrary momentum and masses   
are not simple functions of their arguments and have to be   
computed numerically by means of the relations given in   
refs.~\cite{pv79,gv80,hv79}.  

\begin{figure}[hbt]
\begin{center}
\epsfig{file=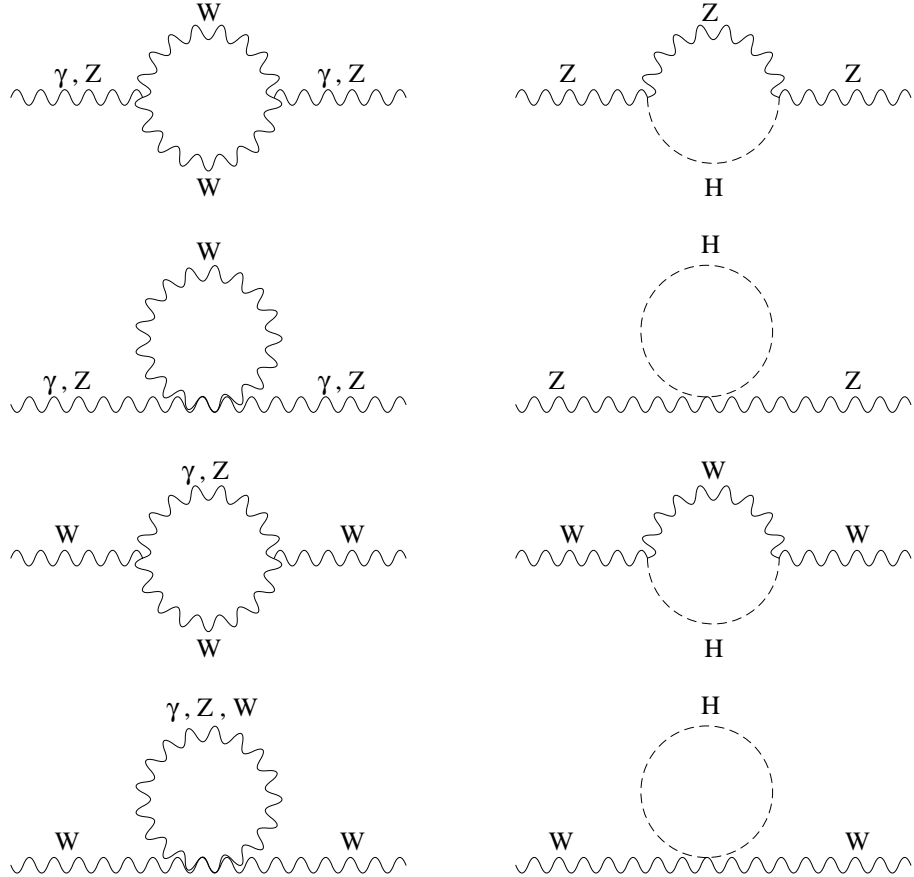, width=12.truecm}
\end{center}
\caption{One-loop bosonic contributions to vector-boson self-energies. }
\label{fig:bselfenergies}
\end{figure}
  
Within the set of the vector boson self-energies two different   
kinds of corrections can be distinguished for their   
different numerical relevance:   
the fermion-loop contribution which is the leading one 
(see  Fig.~\ref{fig:fselfenergies})  
and the bosonic contribution (see  Fig.~\ref{fig:bselfenergies}). 
Being gauge invariant by   
themselves, the fermion-loop corrections can be resummed   
with some procedure in order to take into account of reducible   
higher-order effects.  
The complete expressions for the vector boson self-energies   
in terms of two-point functions can be found in   
refs.~\cite{pv79,gv80,giam91}. Here only the fermion   
contributions are recalled because they deserve some important   
comments. The $\gamma $-$ \gamma$ transition is given   
by~\cite{gv80}  
\begin{eqnarray}  
S_{\gamma}(p^2) = {{e^2}\over {16 \pi^2}} p^2   
\sum_f Q_f^2 N_c   
\{ 8 B_{21}(p^2, m, m) - 4 B_0(p^2, m, m) \} ,  
\end{eqnarray}  
where  $B$ are two-point functions 
and the sum is extended to all charged   
fermions.   
This expression is the same as obtained in   
QED, since the SM coupling between photon and   
fermions recovers the QED case.   
The quantity   
\begin{eqnarray}  
\Pi^{\gamma \gamma} (p^2) = {{S_{\gamma}(p^2)}\over{p^2}}  
\end{eqnarray}  
is known as the photon vacuum polarization. It is interesting   
to work out the two-point functions for the following   
asymptotic cases, for a fermion of given flavour:  \\
$\bullet$ light fermions $ ( \vert p^2 \vert >> m^2 ) $ \\ 
\begin{eqnarray}  
\Pi^{\gamma \gamma} (p^2) = {{e^2}\over {12 \pi^2}} Q_f^2   
\left( -\Delta + \ln m^2 - {5\over 3}  
+ \ln {{\vert p^2 \vert }\over {m^2}}   
- i \pi \right)  , 
\end{eqnarray}  
$\bullet$ heavy fermions $ ( \vert p^2 \vert << m^2) $  \\
\begin{eqnarray}  
\Pi^{\gamma \gamma} (p^2) = {{e^2}\over {12 \pi^2}} Q_f^2   
\left( -\Delta + \ln m^2 - {{p^2}\over {5 m^2}}  \right) .  
\end{eqnarray}  
Apart from the terms $\Delta - \ln m^2$, which are related   
to the ultraviolet divergences and are removed by the   
renormalization procedure, it is worth noticing the term   
proportional to $ \ln ({{\vert p^2 \vert} / {m^2})} $ in   
the light fermions limit. At energies of the order of $100$~GeV,   
as is the case at LEP,    
this terms is the origin of large corrections coming from the   
light fermion spectrum of the theory. While the leptonic   
contribution can be unambiguously calculated, the hadronic one is affected   
by a large uncertainty, because the   
light-quark  masses can not be unambiguously  defined. For this   
reason the hadronic contribution to the photon vacuum   
polarization is calculated by means of a dispersion   
relation as described in Appendix~\ref{sect:vacpol}.   
As far as the {\it top}-quark contribution to   
the vacuum polarization is  concerned, taking into account the 
{\it top}-quark mass value   
of about $170 \div  180$~GeV,   
the heavy fermion limit of   
$ \Pi^{\gamma \gamma} $ can be taken, resulting in a   
small effect. This happens because the $U(1)_{e.m.}$ group   
of the SM is not spontaneously broken so that   
the decoupling theorem~\cite{appcar} can be applied.  
  
The combination of two-point form factors present in the   
fermionic contribution to the $ \gamma $-$ Z $ transition is the   
same as the $ \gamma \gamma $ case apart from the couplings.   
It is at this point worth noticing that, contrary to the case of   
the photon self-energy, the bosonic contribution to the mixed   
$ \gamma $-$ Z $ transition is not proportional to $ p^2 $ so   
that it is different from zero for $ p^2 = 0 $.    
  
The massive vector boson self-energies have the following   
expressions:  
\begin{eqnarray}  
S_+(p^2) = {{g^2}\over {16 \pi^2}} N_c \sum_d \Big[ &2& p^2   
\{ B_{21}(p^2, m_d, m_u) + B_1(p^2, m_d, m_u) \} \nonumber\\  
&+& (m_u^2 - m_d^2) B_1(p^2, m_d, m_u) - m_d^2   
B_0(p^2, m_d, m_u) \Big] \, ,   
\end{eqnarray}  
where $g$ is the $ SU(2) $ coupling constant and the sum is   
extended over the fermionic doublets;  
\begin{eqnarray}  
S_0(p^2) = {{g^2}\over {16\pi^2 }} N_c \sum_f \Big[ &p^2&   
{{(g_v^f)^2 + (g_a^f)^2}\over {4 c_{\vartheta}^2}}  
\{ 8 B_{21}(p^2, m, m)  - 4 B_0(p^2, m, m) \} \nonumber \\  
&-& {{m^2} \over {2 c_{\vartheta}^2}}  
B_0(p^2, m, m) \Big] ,  
\end{eqnarray}  
where the   
sum is extended to all fermions. As before it is interesting to   
consider the two limits of light and heavy fermions:  \\
$\bullet$ light fermions  
\begin{eqnarray}  
S_+(p^2) &=&  {{g^2}\over {12 \pi^2}}   
{{p^2}\over {4}} N_c (-\Delta + \ln \vert p^2 \vert - i \pi) , \\  
S_0(p^2) &=& {{g^2}\over {12 \pi^2}}   
{{(g_v^f)^2 + (g_a^f)^2}\over {4 c_{\vartheta}^2}} p^2   
(-\Delta + \ln \vert p^2 \vert - i \pi) ;  
\end{eqnarray}  
$\bullet$ heavy fermions   
\begin{eqnarray}  
S_+(p^2) &=& {{g^2}\over {12 \pi^2}} N_c \Big[ -{{p^2}\over {4}}  
(\Delta - \ln m_u^2) - {{3 m_u^2}\over {8}} (\Delta -   
\ln m_u^2 + {1\over 2}) \Big] , \\  
S_0(p^2) &=& {{g^2}\over {12 \pi^2}} N_c \Big[   
-{{(g_v^f)^2 + (g_a^f)^2}\over {4 c_{\vartheta}^2}} p^2   
- {{3 m_u^2}\over {8 c^2_{\vartheta}}} \Big] (\Delta - \ln m_u^2) ,  
\end{eqnarray}  
where the contribution of a single doublet in the charged vector boson case 
has been written.   
Some comments are in order here. The ultraviolet divergences   
are associated only with the real parts of the self-energies,   
as the imaginary parts are directly related to the widths of the   
gauge bosons. Furthermore it appears in the heavy fermions   
limit a term proportional to $m_u^2$, that becomes numerically   
relevant in the case of the {\it top} quark.  

\begin{figure}[hbt]
\begin{center}
\epsfig{file=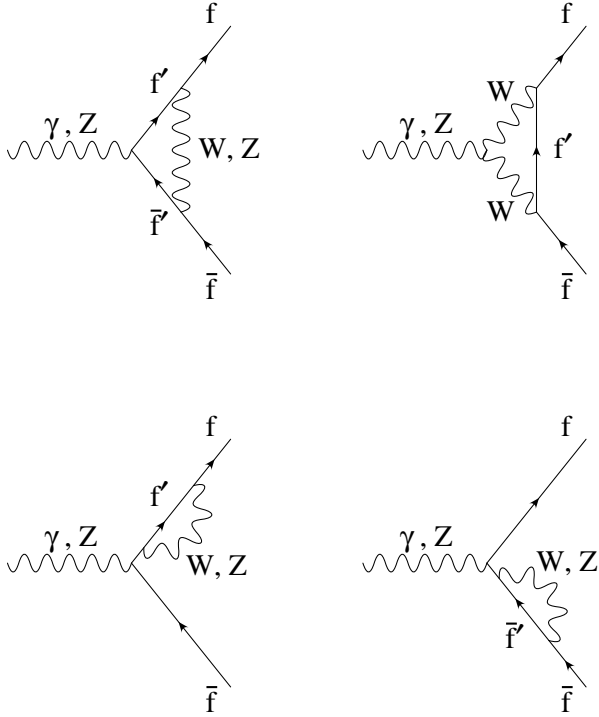, width=8truecm}
\end{center}
\caption{One-loop vertex corrections. }
\label{fig:vertex}
\end{figure}
  
As far as the vertex corrections are concerned (see Fig.~\ref{fig:vertex}), 
neglecting the   
external fermion masses only two combinations of three-point   
functions $C (p^2, m_f, M_V, m_f)$ are present in the 
calculations~\cite{pv79,giam91}:  
\begin{eqnarray}  
C_{ff}^V (p^2) = &-& 2 C_{24}(p^2, m_f, M_V, m_f) - \big[  
C_{11}(p^2, m_f, M_V, m_f) \nonumber \\  
&+& C_{23}(p^2, m_f, M_v, m_f)\big] p^2 + 1 ,   
\end{eqnarray}  
when only one gauge boson is present in the loop, and   
\begin{eqnarray}  
C_f^{VV}(p^2) = &6&C_{24}(p^2, M_V, m_f, M_V) + \big[  
C_0(p^2, M_V, m_f, M_V) \nonumber \\  
&+& C_{11}(p^2, M_V, m_f, M_V) + C_{23}(p^2, M_V, m_f, M_V) \big]   
p^2 - 1 ,   
\end{eqnarray}  
when two gauge bosons circulate in the loop.   
In the case of the $ b \bar b $ final state,  
due to the effect of   
internal {\it  top}-quark lines, five different combinations of three-point   
functions are needed, whose explicit expression can be found   
in~\cite{giam91}. As for the massive vector boson   
self-energies, terms of order $m_t^2$ originate from   
the vertex corrections to the $b \bar b $ final state~\cite{zbb}.      
In order to complete the vertex corrections, the wave function   
factors of the external fermions have to be added, which can   
be obtained from the fermion self-energy diagrams~\cite{gv80}.   

\begin{figure}[hbt]
\begin{center}
\epsfig{file=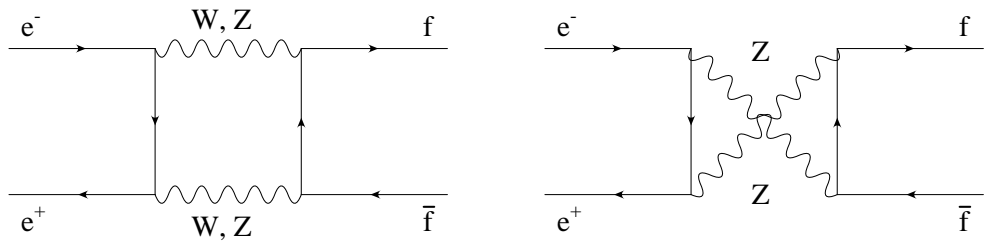, width=13.truecm}
\end{center}
\caption{One-loop box corrections. }
\label{fig:box}
\end{figure}

    
Last, the contribution of box diagrams for the process   
$e^+ e^- \to f \bar f $ (see Fig.~\ref{fig:box}) 
can be written in terms of two   
combinations of four-point functions. They are ultraviolet   
finite, apart from the unitary gauge, and since they are not   
resonating at the $Z^0$ peak, their effect is numerically very   
small, so that generally they are neglected for LEP1 physics,   
although their inclusion is necessary for the gauge invariance   
of the calculation. As will be discussed in Sect.~\ref{sect:tfewqcd}, the   
box corrections become relevant far away from the $Z^0$ peak.  

It is worth noticing that, from the point of view  of  the determination 
of unknown  SM parameters, the bosonic corrections involving the insertion 
of internal Higgs-boson lines  are particularly interesting, since they 
introduce a  dependence of the theoretical predictions on the unknown Higgs-boson 
mass $m_H$.  

\subsubsection{Calculational  schemes}  
\label{sect:rsch} 
The tree-level Lagrangian of the SM   
 contains in its electroweak part a certain number   
of free parameters which need to be fixed by comparison with   
experimental data.    
The choice of the lagrangian parameters and their relations   
with a set of experimental data is the content of a   
renormalization scheme.   
  
The gauge sector of the minimal SM   
is characterized by   
three free parameters fixed by three input data   
points, which have to be known with high experimental   
accuracy, as they are the input for precision calculations.     
At present the three most precise experimental quantities are   
the fine structure constant $\alpha$, measured  by means of one   
the methods quoted in Appendix~\ref{sect:vacpol}, the muon decay constant $G_{\mu}$,   
measured through the muon lifetime, and the   
$Z$-boson mass measured at LEP1. Their precision makes them   
the most widely used experimental input data for the  
calculation of radiative corrections.   
Having defined the parameters, the calculation of    
any other physical quantity can be used as a test of   
the theory by direct comparison of the theoretical prediction for   
the observable with its experimental value.  
   
The relations between the lagrangian parameters   
and the measurable quantities depend on the order at which   
perturbation theory is carried on.   
In particular, going beyond the tree-level   
approximation, the situation is complicated by the   
appearance of ultraviolet divergences which make the   
differences between the tree-level parameters and the   
radiatively corrected ones infinitely large. Because of this,    
a commonly adopted procedure is the introduction of   
counter-terms: each bare parameter is split into the   
renormalized parameter and a counter-term which absorbs   
the ultraviolet divergences plus finite terms.   
The values of this finite parts depend on the renormalization   
prescriptions. Due to the renormalization group invariance, all   
renormalization schemes are equivalent. However, since the   
radiative corrections are obtained from a finite order   
perturbative expansion (eventually supplemented by resummation   
to all orders of some class of gauge invariant contributions),   
the predictions for physical quantities are scheme dependent.   
The differences between two calculations,  performed at  
a fixed perturbative  order in two   
different schemes,  appear in  higher-order terms.   
This fact allows to estimate   
the size of missing higher-order effects by comparing   
the results obtained with calculations based on different   
renormalization schemes (see Sect.~\ref{sect:z0thu}). Given the   
high precision of the experiments at the $Z^0$ peak, it is very   
important to keep under control these higher-order effects.   
In the literature several schemes   
have been adopted for the calculation of radiative corrections.   
With the exception  of  the low energy   
scheme~\cite{gv80,v80,acc80, accp81},   
which has been used before the advent of LEP, it is worth   
mentioning the following calculational schemes adopted for   
various formulations of radiative corrections to observables   
at the $Z^0$ resonance:   
\begin{itemize}   
\item the on-shell scheme~\cite{onshsch};  
\item the ${\overline{MS}}$ scheme~\cite{msbarsch};   
\item the $G_{\mu}$ scheme~\cite{gmusch};    
\item the * scheme~\cite{starsch}.  
\end{itemize}   
In the following a brief review of the various schemes is   
given.   
  
The basic idea of the on-shell scheme is the   
observation that Thomson scattering and the particle   
masses set natural scales where the parameters $e$, $M_Z$, $M_W$,   
$m_H$ and $m_f$ can be defined. The finite parts of   
the mass counter-terms are fixed by the renormalization   
conditions that the particle propagators have poles at   
their physical masses. In the case of unstable particles like   
the $W^\pm$ and $Z^0$ bosons the mass is not uniquely defined.   
For LEP1 physics the vector boson mass is commonly defined   
as the zero of the real part of the inverse propagator.   
At one-loop order, after a Dyson resummation of the self-energy   
diagrams and  neglecting external fermion masses, 
the propagator is given by the expression  
\begin{eqnarray}  
{1 \over {p^2 + {M_V^0}^2 - S_V(p^2)}},   
\label{eq:invprop}  
\end{eqnarray}  
where the Lorentz structure has been neglected   
for simplicity and $M_V^0$ is the bare mass.  In the case  of the $Z$-boson propagator, due to
$\gamma Z$ mixing eq.~(\ref{eq:invprop}) is modified by reducible higher-order terms, which for
simplicity are not considered here.  
Introducing the mass counter-term means   
the following relation between the bare mass   
$M_V^0$ and the physical mass $M_V$:  
\begin{eqnarray}  
{M_V^0}^2 = M_V^2 + \delta M_V^2 .   
\label{eq:baremass}  
\end{eqnarray}  
Inserting the above   
expression for the bare mass in the    
propagator of eq.~(\ref{eq:invprop}), the expression of the counter-term   
can be easily identified as  
\begin{eqnarray}  
\delta M_V^2 = {\rm Re}S_V(M_V^2).  
\end{eqnarray}  
  
The charge counter-term is fixed by the condition that   
$e$ equals the $e e \gamma $ coupling constant in the Thomson   
limit of Compton scattering.  At one-loop order, the   
$e e \gamma $ coupling receives contributions only from the photon   
self-energy and the mixed $\gamma $-$ Z$ transition, because the   
corrections related to the external particles cancel each other   
as a consequence of a generalization of the QED Ward identity.   
The expression for the charge counter-term reads   
\begin{eqnarray}  
{{\delta e}\over {e}} = {1\over 2}\Pi_{\gamma \gamma}(0) -   
{{s_W}\over {c_W}}{{\Sigma_{\gamma Z}(0)}\over {M_Z^2}} ,   
\end{eqnarray}  
where $s_W$ and $c_W$ are the sine and cosine of the weak mixing   
angle defined in the SM by   
\begin{eqnarray}  
\sin^2 \vartheta_W = 1 - {{M^2_W}\over {M^2_Z}} .  
\label{eq:swons}  
\end{eqnarray}  
According to ref.~\cite{sirlin80}, the above definition of   
$\sin^2 \vartheta_W$ is assumed to be valid   
to all orders of perturbation theory. 
  
As a matter of fact, the precision of the $W$-boson mass measurement   
is not adequate for precision physics at the $Z^0$ resonance.   
For this reason the input $W$-boson mass is replaced by the more   
accurate value of $G_{\mu}$ obtained from the $\mu$-lifetime    
$\tau_{\mu}$:    
\begin{eqnarray}  
{{1}\over {\tau_{\mu}}} = {{G_{\mu}^2 m_{\mu}^5}\over   
{192 \pi^3}} \left( 1 - {{8 m_e^2}\over {m_{\mu}^2}}\right)   
\left[ 1 + {{\alpha }\over {2 \pi }}   
\left( 1 + {{2 \alpha}\over {3 \pi}}   
\ln {{m_{\mu}}\over {m_e}} \right)  
\left( {{25}\over {4}} - \pi^2 \right) \right] .  
\label{eq:taulifetime}  
\end{eqnarray}  
This equation, obtained within the effective four-fermion   
Fermi interaction with the inclusion of QED radiative  corrections  
up to ${\cal O} (\alpha^2)$, is used as the definition of $G_{\mu}$ in terms   
of the experimental $\mu$ lifetime. Calculating the last one   
 within the SM at one-loop order,   
the following relation can be   
established between $G_{\mu}$ and $M_W$:   
\begin{eqnarray}  
{{G_{\mu}}\over {\sqrt{2}}} = {{e^2}\over {8 s_W^2 M_W^2}}   
(1 + \Delta r) ,   
\label{eq:gmw}  
\end{eqnarray}  
where $\Delta r$ is a ultraviolet finite combination of   
counter-terms and loop diagrams.  
Since $\Delta r$ is a function of $e$, $M_W$, $M_Z$, $m_H$ and   
$m_t$, eq.~(\ref{eq:gmw}) can be solved iteratively for $M_W$.   
By inspection of the various contributions,   
$\Delta r$ can be written in the following   
form:   
\begin{eqnarray}  
\Delta r = \Delta \alpha - {{c_W^2}\over {s_W^2}} \Delta \rho   
 + (\Delta r)_{rem} ,  
\label{eq:deltar}   
\end{eqnarray}  
where $\Delta \alpha $ is the fermionic contribution to the   
photonic vacuum polarization, and $\Delta \rho $ is the   
following ultraviolet finite   
combination~\cite{rv75,v77}:  
\begin{eqnarray}  
\Delta \rho = {{\Sigma_Z (0)}\over {M_Z^2}} -   
{{\Sigma_W (0)} \over {M_W^2}} ,   
\label{eq:deltarho}  
\end{eqnarray}  
which is quadratic in the {\it top}-quark mass   
\begin{eqnarray}  
\Delta \rho = N_c {{G_\mu m_t^2} \over {8 \pi^2 \sqrt{2}}}.  
\end{eqnarray}
$\Delta \rho$ is the corrections to the $\rho$  parameter, defined as $\rho = M_W^2 / M_Z^2 \cos^2
\vartheta_W$, which is equal to one in the minimal  SM at the tree level~\cite{rv75,v77,cfh78}.  
$(\Delta r)_{rem}$ takes into account non-leading corrections, among which the most interesting ones,
from the phenomenological point of view, are due to Higgs-boson loops, yielding as leading 
contribution the following asymptotic logarithmic term ($m_H \gg
M_W$)~\cite{vscreen77,yrew89,yrdeltar89}
\begin{equation}
(\Delta r)_{rem}^{Higgs} \simeq {{\sqrt{2} G_\mu M_W^2} \over {16 \pi^2}}
\left\{ {11 \over 3} \left( \ln {{m_H^2} \over {M_W^2}} - {5 \over 6}
\right) \right\}, 
\label{eq:drhiggs}
\end{equation}
and to  {\it  top}-quark loops, yielding a logarithmic term  given by
\begin{equation}
(\Delta r)_{rem}^{top} = {{\sqrt{2} G_\mu M_W^2} \over {16 \pi^2}} 2
\left( {{c_W^2} \over {s_W^2}} - {1 \over 3} \right) \ln{{m_t^2} \over {M_W^2}} + \ldots
\end{equation}
At the one-loop order, no quadratic Higgs-boson mass corrections appear as a consequence of the
so-called {\it custodial} symmetry of the Higgs-boson sector of the minimal SM. 

Due to the presence of large contributions to  $\Delta r$ of the kind  
$\alpha \ln ({{M_Z} / {m_f}})$,   
these terms are resummed to all orders by writing eq.~(\ref{eq:gmw})   
in the following form~\cite{sirlin84}:   
\begin{eqnarray}  
{{G_{\mu}}\over {\sqrt{2}}} =   
{{e^2}\over {8 s_W^2 M_W^2 (1 - \Delta r)}} .  
\label{eq:gmwr}  
\end{eqnarray}  
This form takes into account to a good approximation also the   
terms of  the order of  
$\alpha^2 \ln ({{m_f} / {M_Z}})$~\cite{sirlin84}.   
More details on higher-order terms are given in Sect.~\ref{sect:hoew}.  
  
The method outlined above of parameter renormalization   
is sufficient to obtain finite $S$-matrix elements.   
However propagators and vertices by themselves are not finite.   
To this aim also field renormalization has to be carried out.   
This allows to fulfill further renormalization conditions,   
in particular the vanishing of the mixed $\gamma$-$ Z$ transition   
for real photons. The different ways of   
implementing field renormalization and the different gauges   
in which the calculations are performed are the main differences   
among the various realizations of the on-shell scheme present   
in the literature.   
  
In the ${\overline{MS}}$ scheme, the counter-terms are defined only   
through the divergent parts proportional to $\Delta - \ln m^2$  
of the radiative corrections to the bare parameters,   
as these are fixed only by the bare Lagrangian, and do not depend   
on particular renormalization conditions as is the case   
for the on-shell scheme. 
The ${\overline{MS}}$   
expressions of the parameters can be directly obtained   
replacing the bare parameters with the ${\overline{MS}}$ ones,    
and replacing at the same time the quantity $\Delta $ in the   
radiative corrections to the bare parameters with $\ln \mu^2 $, 
$\mu$ being the renormalization  scale.   
In so doing, the ${\overline{MS}}$ parameters depend on the   
arbitrary scale $\mu$, which can be naturally chosen to be $M_Z$ for   
electroweak calculations.   
The ${\overline{MS}}$ electroweak mixing angle can be defined as   
\begin{eqnarray}  
{\hat s}^2 = 1 - {{{\hat M}_W^2}\over {{\hat M}_Z^2}} ,   
\label{eq:shat}  
\end{eqnarray}  
where ${\hat M}_{W,Z}$ are the ${\overline{MS}}$ vector boson masses.  
  
According to refs.~\cite{dfsb351} and~\cite{dsb352}, a way to link directly   
${\hat s}^2 $ to the experimental inputs $\alpha $, $G_{\mu}$,   
and $M_Z$ is to calculate the radiative corrections to $\mu $   
decay within the ${\overline{MS}}$ framework introducing the   
correction $\Delta {\hat r}_W $ together with a generalized   
relation between the $W$- and $Z$-boson physical masses:  
\begin{eqnarray}  
&&{\hat s}^2 (1 - \Delta {\hat r}_W) =   
{{\pi \alpha}\over {\sqrt{2} G_{\mu} M_Z^2}} ,  \\  
&&M_W^2 = {\hat c}^2 {\hat \rho} M_Z^2 ,  
\end{eqnarray}  
where ${\hat \rho}^{-1} = 1 - \Delta {\hat \rho}$ .  
 
Within the ${\overline{MS}}$ scheme the relation giving the  
$M_W$-$M_Z$ interdependence can be expressed as  
\begin{eqnarray} 
{{M_W^2}\over {M_Z^2}} = {{{\hat \rho}}\over 2}  
\left\{ 1 + \left[ 1 -  
{{4 A^2}\over {M_Z^2 {\hat \rho} (1 - \Delta {\hat r}_W)}}  
\right]^{{1\over 2}} \right\} ,  
\end{eqnarray} 
with $A = \big( \pi \alpha /  
\sqrt{2} G_{\mu} \big)^{{1\over 2}}$. 
 
The advantage of adopting the ${\overline{MS}}$ scheme 
is twofold~\cite{sirlin89}. First, this 
calculational framework leads to effective couplings which absorb 
the largest part of the {\it top}-quark and Higgs-boson mass dependence of the 
radiative corrections. Secondly, the knowledge of the gauge coupling constants   
at the $Z$-boson mass can be naturally extrapolated to higher energies in order to 
test scenarios of Grand Unification.
 
The idea of the $G_{\mu}$ scheme is to directly relate the   
bare parameters to the input data $\alpha$, $G_{\mu}$   
and $M_Z$. This can be realized according to different methods, for example 
by assuming on-shell parameters~\cite{timme86} or $\overline{MS}$ parameters~\cite{pp89,giam91}. 
Here a brief account of the second realization is given, where the bare parameters
are chosen to be the $SU(2)$ coupling constant   
$g$, the $W$-boson mass $M_W$ and the squared sine of the weak mixing   
angle $s^2_{\vartheta}$.  
Starting from the observation that there is no   
one to one correspondence between lagrangian parameters and   
experimental data, in this scheme no attempt is made to define   
renormalized parameters (in particular $s_{\vartheta}^2$)   
beyond lowest order.   
A set of three fitting equations~\cite{v77,giam91} for the   
bare parameters is introduced,    
corresponding to the definition of $\alpha $ from Thomson   
scattering, $G_{\mu}$ from the $\mu $ lifetime and $M_Z$   
from the zero of the real part of the inverse $Z^0$ propagator:   
\begin{eqnarray}  
d_i^{exp} = d_i^{th}(g, M_W, s_{\vartheta}, \Delta),   
\quad \quad (i = 1,2,3). 
\label{eq:fittingeq}  
\end{eqnarray}
In  eq.~(\ref{eq:fittingeq}), $d_i^{th}$  are the theoretical expressions  of 
$\alpha$, $G_{\mu}$ and $M_Z$ in   terms of  $g$,  $M_W$, $s^2_{\vartheta}$ 
and $\Delta$, whereas  $d_i^{exp}$  are   the corresponding experimental values.  
The equations can be solved to first order in perturbation   
theory with respect to the bare parameters which contain   
ultraviolet divergences represented by the quantity $\Delta $.   
They  can be properly modified to account for   
resummation of relevant gauge invariant   
higher-order effects~\cite{giam91}.   
In the calculation of any physical quantity by means of one-loop 
diagrams calculated with tree-level parameters   
and tree diagrams calculated with up to one-loop order   
parameters the ultraviolet divergences cancel. This procedure   
is suitable for an order by order numerical renormalization.   
In the numerical calculation, the quantity $\Delta $ enters as   
an arbitrary parameter which can assume any numerical value   
without changing the final result.   
  
The condition $\Sigma_{Z \gamma} = 0 $ is fulfilled through   
a proper redefinition of the bare coupling $g^0$, which   
automatically guaranties that the sum of all $Z f \bar f$   
vertices is ultraviolet finite~\cite{giam91}.   
  
It is worth noticing that in the $G_{\mu}$ scheme the $W$-boson mass   
plays the same r$\hat{ \hbox{\rm o}}$le as any other physical observable,   
contrary to the on-shell scheme where at the same time it   
appears as a prediction and as an intermediate parameter for   
the calculation of other observables.   
  
In the $*$ scheme, an effective lagrangian   
to one loop   
is introduced, and the universal effects of oblique corrections   
resummed to all orders are absorbed in the following   
three running parameters (free of ultraviolet divergences):   
$e_*^2(q^2)$, $s_*^2(q^2)$ and $G_{\mu *}^2(q^2)$. The values   
of these three starred functions have to be fixed at a certain   
set of $q^2$'s and then can be calculated at any other scale   
by means of a set of evolution equation. The choice of  
ref.~\cite{kennlynn89} is to fix $e_*^2$ and $G_{\mu *}^2$ at   
$q^2 = 0$ by means of $\alpha $ and $G_{\mu}$, and $s_*^2$ at the   
$Z$-boson mass scale by means of the pole of the $Z^0$ propagator.  
In order to cure the problem $\Sigma_{\gamma Z}(0) \neq 0$ in the   
bosonic sector, a rediagonalization of the neutral current   
sector in the one-loop lagrangian is performed including in the   
self-energies the contribution of the universal and gauge   
dependent parts of the vertex corrections. 

Besides the calculational  schemes described above, one should also  mention 
the {\it $Z$-peak subtracted} representation of $e^+ e^- \to f \bar f$ processes at one loop,
recently suggested in~\cite{zpsub}. In this approach, the input parameter $G_\mu$ is replaced  
by  quantities measured on top of the $Z^0$ resonance. For instance, for leptonic final states the
new input  parameters are the leptonic $Z$-boson width $\Gamma_l$ and the effective sine 
$\sin^2 \vartheta_{eff}^l$, that reabsorb the bulk of loop effects, while the residual one-loop
corrections are contained  in quantities subtracted at $Q^2 = M_Z^2$. This approach has been proved
to be particularly powerful for the description of $e^+ e^- $ annihilations above the $Z^0$ peak,
especially if an investigation of models of new physics is the final goal. 

\subsubsection{Transition amplitudes and effective couplings}  
\label{sect:taaec}   
The basic ingredients of loop calculations described in the   
previous subsections can be used to evaluate transition   
amplitudes for the physical process $e^+ e^- \to f \bar f $.   
Before discussing the radiative corrected amplitude, it   
is worth recalling the general structure of the tree-level   
amplitude which, excluding for simplicity the case of Bhabha   
scattering, reads~\cite{ewwg95}:  
\begin{eqnarray}  
{\cal M} \sim {1\over s} \left[ Q_e Q_f \gamma_{\alpha}   
\otimes \gamma^{\alpha} + \chi \gamma_{\alpha}   
(g_v^e - g_a^e \gamma_5 ) \otimes  \gamma^{\alpha} 
(g_v^f - g_a^f \gamma_5 )) \right] ,  
\label{eq:born}   
\end{eqnarray}  
where $\chi $ is the propagator ratio   
\begin{eqnarray}  
\chi = {s\over {s - M_Z^2 + i \Gamma_Z (s)  M_Z}} , 
\label{eq:proprat}  
\end{eqnarray}  
and the symbol $\Gamma_\alpha \otimes \Gamma^\alpha$ means product of  spinorial  currents. 
The $s$-dependent width $\Gamma_Z (s) \simeq s \Gamma_Z / M^2_Z$ appearing in eq.~(\ref{eq:proprat})
is due to the imaginary part of the $Z$ self energy.  
In eq.~(\ref{eq:born}),   the photon and the $Z$-boson exchange diagrams are   
well separated and moreover the $Z^0$ contribution is written   
in a factorized form. By virtue of these two properties,     
simple relations between the pseudo-observables defined in 
Sect.~\ref{sect:z0par}  and the  measured quantities can be established.   
  
However the two   
features are lost when the electroweak non-photonic corrections   
(self-energy diagrams, $Z f \bar f$ vertex insertions and   
weak boxes) are considered.   
In this case the matrix element can be written in the following   
way:  
\begin{eqnarray}  
{\cal M} \sim {1\over s} \left[ \alpha (s) \gamma_{\alpha}   
\otimes \gamma^{\alpha} + \chi \left( F_{vv}^{ef}(s,t)   
\gamma_{\alpha} \otimes \gamma^{\alpha}  -   
F_{va}^{ef}(s,t) \gamma_{\alpha} \otimes   
\gamma^{\alpha}  \gamma_5  \right. \right. \nonumber \\  
- \left.  \left. F_{av}^{ef}(s,t) \gamma_{\alpha} \gamma_5 \otimes   
\gamma^{\alpha} +   
F_{aa}^{ef}(s,t) \gamma_{\alpha} \gamma_5 \otimes   
\gamma^{\alpha} \gamma_5   
\right) \right] ,   
\label{eq:corramp}  
\end{eqnarray}  
where $\alpha (s)$ is the QED running coupling constant with   
only fermionic contributions (in order to have a gauge   
invariant term) and the form factors $F_{ij}^{ef}$ are complex   
valued functions of the input parameters and of the kinematical   
variables $s$ and $t$. Usually the self-energies contributions   
are resummed in their fermionic component to take into account   
large higher-order reducible diagrams. The bosonic parts are   
instead expanded at ${\cal O}(\alpha)$, in  order to preserve 
gauge invariance.   
The dependence of the form factors on   
$t$ is only due to the presence of box diagrams. Being non-resonant 
at the $Z^0$ peak, their numerical effects are   
very small and can be neglected, even if from a theoretical   
point of view the gauge invariance is no more respected.   
Other non-resonating   
contributions, such as the bosonic insertions to the photon   
propagator and photon-fermion vertex corrections, can be   
safely neglected. Another commonly used approximation is the   
$Z^0$ pole approximation, which means to fix the scale in   
the form factors at the value $M_Z^2$, since the weak corrections   
depend very mildly on the scale around the $Z^0$ peak. After these approximations   
the factorization of the $Z^0$ contribution is re-established   
and the form factors take on the form:   
\begin{eqnarray}  
F_{ij}^{ef}(M_Z^2) = G_i^e(M_Z^2) G_j^f(M_Z^2) .  
\label{eq:corrcoupl}  
\end{eqnarray}  
This means that the $Z$-boson part of the corrected amplitude can be   
obtained from the tree-level expression by replacing the vector   
and axial vector couplings with the corrected versions defined   
above by eq.~(\ref{eq:corrcoupl}).   
The imaginary parts of the couplings $G_i^f(M_Z^2)$   
are generally small with respect to the real parts, so that   
the following effective couplings are commonly introduced:  
\begin{eqnarray}  
g_{V,A}^f = {\rm Re} G_{v,a}^f(M_Z^2) .  
\label{eq:effectcoupl}  
\end{eqnarray}  
  
As  already noticed in Sect.~\ref{sect:z0par},  the decay width of the   
$Z^0$  boson into a $f \bar f$ pair  is given by the   
following expression in terms of the effective 
couplings:  
\begin{eqnarray}  
\Gamma_f = 4 N_c \Gamma_0 [(g_V^f)^2 R_V^f + (g_A^f)^2 R_A^f] .  
\label{eq:zwidth}  
\end{eqnarray}  
  
In the literature other sets of parameters replacing   
the effective couplings are present. For example according to   
ref.~\cite{yrew89} the parameters $\rho_f$ and $k_f$ are introduced   
related to $g_V^f$ and $g_A^f$ by the following relations:  
\begin{eqnarray}  
\rho_f &=& 4 (g_A^f)^2 \\  
{{g_V^f}\over {g_A^f}} &=& 1 - 4 \vert Q_f \vert s_W^2 k_f , 
\label{eq:rhofkfdef}  
\end{eqnarray}  
with  
\begin{eqnarray} 
\rho_f &=& 1 + \delta \rho_f , \\  
k_f &=& 1 + \delta k_f. 
\label{eq:deltafdeltakf}  
\end{eqnarray} 
By means of these parameters,  the $Z^0$ partial widths  
can be written in the following way   
(neglecting for simplicity final-state QED and QCD   
interactions and   
mass effects):  
\begin{eqnarray}  
\Gamma_f = \Gamma_0 N_c^f \rho_f   
[4 (I_f^3 - 2 Q_f s_W^2 k_f)^2 + 1] . 
\label{eq:widthrhofkf}  
\end{eqnarray}  
The above described procedure of step-by-step approximations, leading from  the exact one-loop
amplitude  to an approximate amplitude formally identical,  or very similar, to the 
Born-approximation one, but written in
terms of form factors evaluated at the $Z^0$ peak, is known as  Improved Born Approximation
(IBA). Some different realizations of IBA's are available in the literature, and can be found for
instance in~\cite{yrew89,ibapv,topaz0np}.    

\subsubsection{Higher-order electroweak corrections}  
\label{sect:hoew}  
Given the high precision reached in the experimental measurements,  
particular attention must be paid to include in  
the electroweak corrections potentially large higher-order effects.  
These generally originate from terms containing logarithms of the  
type $\ln (M_Z/m_f)$, where $m_f$ stand for a generic fermion mass,  
or from contributions proportional to the {\it top}-quark mass.  
In Sect.~\ref{sect:rsch} the prescription based on renormalization  
group arguments~\cite{am81,sirlin84} 
\begin{eqnarray} 
1 + \Delta r \to {1\over {1 - \Delta r}}  
\end{eqnarray} 
has been introduced to include to all orders the leading  
logarithms of ${\cal O}(\alpha^n \ln^n (M_Z/m_f))$, contained  
in $\Delta \alpha$ of eq.~(\ref{eq:gmw}),   
giving the interdependence  
between $M_W$, $M_Z$ and $G_{\mu}$. However, with this  
prescription the powers $(\Delta \rho)^n$ are not correctly  
resummed. In ref.~\cite{chj89} it  has been shown how to  
take into account in the on-shell scheme the leading  
two-loop contributions proportional to $m_t^4$ by means  
of the replacement  
\begin{eqnarray} 
{1\over {1 - \Delta r}} \to {1\over {1 - \Delta \alpha}} 
\cdot {1\over {1 + {{c_W^2}\over {s_W^2}}\Delta {\bar \rho}}}  
+ \Delta r_{rem} , 
\end{eqnarray} 
where $\Delta {\bar \rho}$ includes the contribution of  
two-loop one particle irreducible diagrams: 
\begin{eqnarray} 
\Delta {\bar \rho} = 3 x_t \left[ 1 +  
x_t \rho^{(2)} \left( {{m_H}\over {m_t}} \right) \right],  
\quad x_t = {{G_{\mu}m_t^2}\over {8 \pi^2 \sqrt{2}}}. 
\label{eq:deltabarrho}  
\end{eqnarray} 
The two-loop correction $\rho^{(2)}$ was first calculated  
in ref.~\cite{hoogbij} in the limit of light Higgs-boson mass  
$m_H << m_t$, yielding $\rho^{(2)} = 19 - 2\pi^2$. More  
recently the calculation has been carried out by several  
groups~\cite{barbtwo,fjttwo,dhltwo} for arbitrary values  
of the ratio $m_H / m_t$ in the limit of vanishing gauge  
coupling constants.  
In ref.~\cite{sirlin84} it has been shown that the fermionic  
mass singularities of ${\cal O}(\alpha^2 \ln (M_Z/m_f))$ are  
correctly taken into account by keeping $\Delta r_{rem}$ in  
the denominator, \idest\    
\begin{eqnarray} 
{1\over {1 - \Delta r}} \to {1\over {\left( 1 - \Delta \alpha  
\right) \left( 1 + {{c_W^2}\over {s_W^2}}\Delta {\bar \rho}  
\right) - \Delta r_{rem}}}. 
\end{eqnarray} 

The effects of two-loop heavy Higgs-boson contributions have been  
investigated in refs.~\cite{bijveltman84}  and~\cite{bcs93}. 

Very recently the two-loop sub-leading corrections of  
${\cal O}(\alpha^2 m_t^2/M_W^2)$ have been calculated for the vector-boson masses  
correlation~\cite{dtwol95,dgvtwol96,dgstwol97}, and  
found to be potentially of the order of the leading $m_t^4$  
contributions, depending on the Higgs-boson mass. Moreover, the exact two-loop
Higgs-boson mass dependence has also become available~\cite{deltarh2}, introducing 
a shift in the $W$-boson mass contained within 4~MeV~\cite{deltarwm2} with respect to the
calculation of  ref.~\cite{dgvtwol96,dgstwol97}.

Another place where higher-order electroweak effects play  
an important r$\hat{ \hbox{\rm o}}$le is the partial width $Z \to f {\bar f}$.  
The effective couplings contain leading universal parts,  
arising from self-energies and counter-terms, and  
flavour-dependent parts coming from vertex corrections.  
The leading universal contributions are given by~\cite{ewwg95}  
\begin{eqnarray} 
(\delta \rho_f)_{univ} &=& \Delta \rho + \ldots , \\ 
(\delta k_f)_{univ} &=&  
{{c_W^2}\over {s_W^2}} \Delta \rho + \ldots . 
\end{eqnarray} 
The higher-order terms can be incorporated by means of  
the following replacements: 
\begin{eqnarray} 
\rho_f \to {1\over {1 - \Delta {\bar \rho}}} + \ldots \\ 
k_f \to 1 + {{c_W^2}\over {s_W^2}} \Delta {\bar \rho}  
+ \ldots , 
\end{eqnarray}  
with $\Delta {\bar \rho}$ given by eq.~(\ref{eq:deltabarrho}).  
 
The partial decay width $Z \to b {\bar b}$ contains an additional  
$m_t$ dependence due to vertex diagrams with virtual {\it  top}-quark lines,  
resulting in additional leading terms of the order  
$G_{\mu} m_t^2$ for the effective couplings $\rho_b$ and $k_b$.  
The additional leading two-loop electroweak effects of the order  
$G_{\mu}^2 m_t^4$ have been computed in  
refs.~\cite{barbtwo} and~\cite{fjttwo}, and amount to the following  
redefinition of the effective couplings: 
\begin{eqnarray} 
\rho_b &\to& \rho_b (1 + \tau_b)^2 , \\ 
k_b &\to& {{k_b}\over {1 + \tau_b}},  
\end{eqnarray} 
where $\tau_b$ is given by  
\begin{eqnarray} 
\tau_b = -2 x_t \left[ 1 + x_t \tau^2 \left(  
{{m_t^2}\over {m_H^2}} \right) \right] ,  
\end{eqnarray} 
where $\tau^2 (m_t^2 / m_H^2)$ can be found in~\cite{barbtwo,fjttwo}. 

At the level of two-loop or higher-order corrections, 
other phenomenologically relevant  
contributions come from the mixed electroweak-QCD corrections,   
discussed in the next Section. 
\subsection{QCD Corrections}
\label{sect:qcdcorr}
QCD corrections play a significant  r$\hat{ \hbox{\rm o}}$le 
as higher-order corrections 
due to  virtual  gluon  insertions in electroweak loops (mixed electroweak-QCD 
corrections) and as final-state  corrections in the hadronic decay channels. With the exception of
very recent improvements, an account of perturbative QCD calculations for the $Z$-boson observables
can be found in~\cite{yrwg95} and~\cite{qcd96}.  

\subsubsection{Mixed electroweak-QCD corrections}
\label{sect:z0mix}

The need of very accurate predictions demanded by the high precision of the
measurements on the $Z^0$ peak, motivated various authors to perform calculations 
of higher-order effects beyond the one-loop   and higher-order  
electroweak corrections  already discussed in Sect.~\ref{sect:z0ew}. Great effort was spent in the 
calculation of corrections due to the insertion of virtual gluons in the electroweak 
quark loops, known as {\it mixed electroweak-QCD corrections}, giving rise to contributions 
simultaneously depending on the weak coupling $G_\mu$ and the coupling of the strong
interaction $\alpha_s$. Higher-order corrections of this kind have been calculated 
both for $W$- and $Z$-boson self-energies, thus affecting all $Z \to f \bar f$ decay 
channels, and specifically for the peculiar $Z \to b \bar b$ vertex.   

\begin{figure}[hbt]
\begin{center}
\epsfig{file=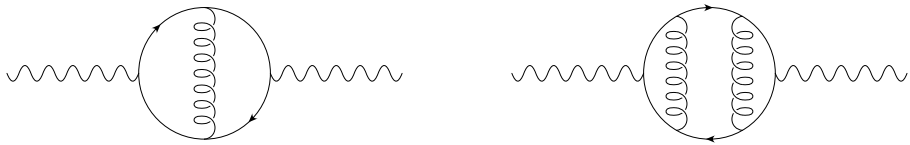, width=12truecm}
\end{center}
\caption{Example of Feynman diagrams for QCD corrections to gauge-bosons 
self-energies. }
\label{fig:mix}
\end{figure}

QCD corrections to the gauge-bosons self-energies have been computed in the SM considering 
the insertion of one and two gluons into the internal quark loops. Typical 
Feynman diagrams contributing to these corrections are depicted in 
Fig.~\ref{fig:mix}.

The diagrams involving the exchange of a single gluon, described for instance by the 
first graph in Fig.~\ref{fig:mix}, give rise to two-loop corrections of the order of 
$\alpha \alpha_s$, that are now exactly known~\cite{aas} and are 
dominated by terms of order $G_\mu \alpha_s m_t^2$. The insertion of two 
virtual gluons, of the kind depicted in the second graph of Fig.~\ref{fig:mix}, 
generates a three-loop contribution of the order of $\alpha \alpha_s^2$, 
that is partially under control and introduces effects of the form 
$G_\mu \alpha_s^2 m_t^2$~\cite{aas2}. In formulae, the effect of such corrections to 
the $\Delta\rho$ factor introduced in Sect.~\ref{sect:z0ew} can be cast, in the 
heavy {\it top}-quark limit, as follows
\begin{eqnarray}
\Delta\rho \to \Delta\rho \cdot \left(1 + \delta\rho_{QCD} \right)
\end{eqnarray} 
where $\delta\rho_{QCD}$ is the QCD correction to the leading $G_\mu m_t^2$ term
\begin{eqnarray}
\delta\rho_{QCD} \, = \, -2.86 a_s - 14.6 a_s^2,    \quad \quad 
a_s = {\alpha_s(m_t) \over \pi} \, ,
\end{eqnarray}    
the running coupling constant $\alpha_s$ being evaluated at the energy scale 
given by the {\it top}-quark mass. It is  worth noticing that the QCD correction to the $\rho$
parameter  is negative, and tends to screen the electroweak contribution.\footnote{The SUSY-QCD
corrections to $\Delta\rho$ have been recently computed; they could introduce a shift of the
order of 10~MeV on the $W$-boson mass~\cite{susyqcd}.} 

A second source of higher-order mixed effects comes from QCD corrections to the 
$Z \to b \bar b$ electroweak vertex once corrected by the one-loop contributions 
depicted in Fig.~\ref{fig:vertex} (see Sect.~\ref{sect:z0ew}). 
Due to the presence of virtual 
{\it top}-quark lines in the one-loop $b \bar b$ vertex, significant non-universal 
mixed corrections, depending on $\alpha_s$ and $m_t$, are additionally present for 
the $b \bar b$ final state. The effects induced by ${\cal O}(\alpha_s)$ 
gluon radiation introduce two-loop corrections to the leading electroweak term of 
the order of $G_\mu \alpha_s m_t^2$~\cite{aasbb} 
and to the $\ln(m_t/M_W)$ term of the order of $G_\mu \alpha_s \ln(m_t/M_W)$, the 
latter with a very small numerical coefficient~\cite{aasbb2}. The missing next-to-leading
corrections of ${\cal O}(\alpha \alpha_s)$  to  the partial width $\Gamma_b$ have been very recently
calculated~\cite{hss97}, thus improving the SM prediction for the $Z \to b \bar b$ vertex.   

Finally, final-state mixed corrections to the $Z \to q \bar q$ decay channels, 
due to the interplay between QED and QCD radiation off final-state quarks, 
are exactly available~\cite{aask92} and taken into account in the theoretical predictions. 
For example, the $Z^0$ decay width into $q \bar q$ final states receives 
a contribution from these mixed corrections given by the factor
\begin{eqnarray}
\delta^{fs}_{\alpha \alpha_s} \, = \, Q_f^2 {3 \over 4} \, {\alpha(s) \over \pi} \, 
\left[1 - {1 \over 3} {\alpha_s(s) \over \pi} \right] \, \quad 
{\rm with} \, \, s = M_Z^2 \, ,
\end{eqnarray}
that can be easily deduced from the QCD results for the $Z \to q \bar q$ decay width, 
discussed in the next Section. 

\subsubsection{Final-state QCD corrections}
\label{sect:z0qcd}

\begin{figure}[hbt]
\begin{center}
\epsfig{file=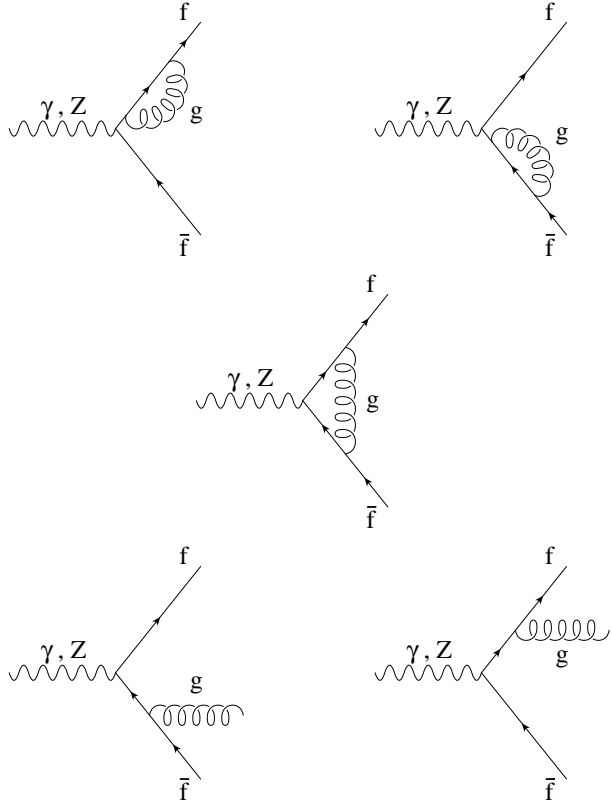, width=8truecm}
\end{center}
\caption{Feynman diagrams for final-state ${\cal O}(\alpha_s)$ QCD corrections. }
\label{fig:qcdfs}
\end{figure}

In addition to the classes of radiative corrections discussed 
in the two previous Sections, an adequate theoretical description of the
processes explored at LEP requires, as a further basic ingredient, 
the inclusion of the QCD corrections to the $Z \to q \bar q$ decay channels, 
arising from the emission of real and virtual gluons off the final-state 
quarks (see Fig.~\ref{fig:qcdfs}). 
Among the observables measured at the $Z^0$ peak, 
these corrections affect the total hadronic cross section 
$\sigma_{\rm had}$, the FB asymmetry of heavy-quark ($q=c,b$) 
production $A_{FB}^{c,b}$, as well as the total hadronic 
width $\Gamma_h$ and the partial widths into $c$ and $b$ quarks $\Gamma_{c,b}$. 
When discussing final-state QCD effects in hadronic $Z$-boson decays, it is necessary to 
distinguish between corrections to the production of light, or massless, 
quarks and corrections to the heavy-quark production.   

In the case of light quarks ($q=u,d,s$) with $m_q \simeq 0$, the QCD 
final-state correction to the cross section (or, equivalently, to the 
partial decay width) is completely known up to ${\cal O}(\alpha_s^3)$. This high 
perturbative accuracy is of course of the utmost importance in view 
of a precise determination of the strong coupling constant in $Z$-boson 
decays, as it  will be shown in  the following. 
The result for the $Z^0$ partial widths into massless quarks can be written as 
\begin{eqnarray}
\Gamma = \Gamma_0 \, \left[ (g_V^f)^2+(g_A^f)^2 \right]  K_{QCD}
\label{eq:gamqcd}
\end{eqnarray} 
with 
\begin{eqnarray}
\Gamma_0 = N_C^f {{G_\mu M_Z^3 \sqrt{2}} \over {12 \pi}} \quad (N_C^f = \hbox{\rm colour factor})  \, , 
\end{eqnarray}
$g_V^f$ and $g_A^f$ being  the {\it effective} electroweak couplings. 
$K_{QCD}$ is the up to three-loop 
correction factor for massless quarks given by~\cite{as2,as3}
\begin{eqnarray}
K_{QCD} \, = 1 \, + \, {\alpha_s \over \pi} \, 
+ \, 1.41 \, \left( {\alpha_s \over \pi} \right)^2 \,
-12.8 \, \left( {\alpha_s \over \pi} \right)^3 \,.
\label{eq:wqcd}
\end{eqnarray}
Actually, leading and next-to-leading second-order 
{\it top}-quark mass corrections to the $q \bar q$ decay 
width in the massless limit are also available in the literature~\cite{as2mt} 
and are taken into account in 
the more refined (but too lengthy to be shown here) QCD radiation factors 
usually implemented in standard computational tools for LEP1 physics. The complete 
expressions, together with a discussion of the numerical effects of the various 
contributions, can be found in the recent review paper of ref.~\cite{qcd96}.

For massive quarks the situation is different, due to the presence of finite mass terms. 
In such a case, the vector and axial-vector contribution to the $Z$-boson decay width receive 
different corrections, known at different perturbative orders. Actually, the 
calculation of the axial-vector part of the hadronic $Z$-boson decay rate is more 
involved than that of the  vector part. This is because the heavy quark does not 
decouple in the axial-vector part and hence one cannot avoid calculating massive 
diagrams. In particular, the $b \bar b$ final state receives peculiar contributions from 
{\it top}-quark dependent two-loop diagrams in the axial-vector part.
Consequently, the  formula of eq.~(\ref{eq:gamqcd}) in the massless limit 
needs to be integrated with additional correction factors for massive quarks, of the form
\begin{eqnarray}
\Gamma = \Gamma_0 \, \left[ (g_V^{c,b})^2 \, R_V +(g_A^{c,b})^2 
R_A \right]  . 
\end{eqnarray} 
The coefficients in the perturbative expansions
\begin{eqnarray}
R_V \, &=& \, c^1_{V} {\alpha_s \over \pi} + c^2_{V} 
\left( {\alpha_s \over \pi} \right)^2
+ c^3_{V} \left( {\alpha_s \over \pi} \right)^3 + \dots ,  \\
R_A \, &=& \, c^1_{A} {\alpha_s \over \pi} + c^2_{A} 
\left( {\alpha_s \over \pi} \right)^2 + \dots , 
\end{eqnarray}
depending on the heavy-quark masses and $m_t$, 
are calculated up to third order in the vector part and up to second order 
in the axial-vector part~\cite{asmass,qcd96,qcd96b,qcdwg95}.  
Given the above formulae for the 
quark partial widths in the massless limit and massive case, the total hadronic width and 
cross section can be obtained as the sum over the partial contributions of 
each $q \bar q$ channel. 

Recently,  the non-factorizable part of the two-loop ${\cal O} (\alpha  \alpha_s)$ correction
to the hadronic $Z$-boson 
width has  been calculated, introducing an extra negative contribution  
  of about  $-0.6$~MeV~\cite{aasnew}. 

Concerning the forward-backward asymmetry for heavy quarks, the QCD 
corrections to $A_{FB}$ are available in the literature~\cite{afbqcd}. At the 
order $\alpha_s$, they yield a correction of the kind 
\begin{eqnarray}
A_{FB}  \to A_{FB} \, \left(1 - k {\alpha_s \over \pi} \right)
\end{eqnarray}
where $k$ depends on the mass of the heavy quark, and its  explicit  expression can be found
in~\cite{hfyr}. 

Last, a few comments are in order. First, since the above QCD factors depend on the 
perturbative expansion parameter $\alpha_s$ and the latter depends on the energy scale 
according to the renormalization group  equation for the $\beta$ function, a formula for 
the {\it running} of $\alpha_s$ 
needs to be specified. Indicating with $\mu$ the renormalization scale and with 
$n_f$ the number of active quark flavours of mass $m_q \ll \mu$, the solution for 
$\alpha_s^{(n_f)}(\mu)$ typically used  reads~\cite{qcd96,qcd96b} 
\begin{eqnarray}
\alpha_s(\mu) = {\pi \over {\beta_0 L} } \, \left\{1 \, - \, 
{1 \over {\beta_0 L}} { {\beta_1 \ln L} \over {\beta_0}} \, + 
{1 \over {\beta_0^2 L^2} } \left[{\beta_1^2 \over \beta_0^2} 
\left(\ln^2 L -\ln L -1 \right) + {\beta_2 \over \beta_0} \right] \right\} \, ,
\label{eq:alphas}
\end{eqnarray}  
with
\begin{eqnarray}
\beta_0 \, &=& \, \left( 11 - {2 \over 3} n_f \right) / 4 \, ,   \quad \quad 
\beta_1 \,  = \, \left( 102 - {38 \over 3} n_f \right) / 16   \, , \nonumber \\
\beta_2 \, &=& \, \left({2857 \over 2} - {5033 \over 18} n_f + 
{325 \over 54} n_f^2 \right) / 64 \, .
\label{eq:betai}
\end{eqnarray}
In the above formulae $L = \ln (\mu^2/\Lambda^2_{\overline{MS}} )$. Actually
eq.~(\ref{eq:alphas}), together with the coefficients of eq.~(\ref{eq:betai}), is valid 
in the modified minimal subtraction ($\overline{MS}$) scheme and corresponds to the 
three-loop expansion of the QCD $\beta$ function. It is used with $n_f = 5$ for the number 
of active flavours 
and it is technically called next-next-to-leading solution of $\alpha_s(\mu)$. At present, 
 also the four-loop coefficient in the $\overline{MS}$ scheme is  
available~\cite{as4r}, thanks to the recent calculation of the coefficient $\beta_3$ of the QCD
$\beta $  function~\cite{rvl97}. A second comment concerns the mass correction terms entering the 
QCD factors for heavy-quarks. Analogously to the coupling constant, the quark masses 
are running parameters in QCD and obey the renormalization group 
equation controlled by the anomalous 
dimension $\gamma_m$, that is known up to three-loop accuracy. The masses appearing in the 
formulae for $c$-  and $b$-quark production have thus to 
be understood as running ($\mu$ dependent) quark masses. The relations used to account for 
such effect read as follows~\cite{yrwg95}:
\begin{eqnarray}
&&\overline{m}(\mu) \, = \, \overline{m}(m^2)
\exp \left\{-\int_{\alpha_s(m^2)}^{\alpha_s(\mu)} \, 
dx {\gamma_m(x) \over \beta(x)} \right\} ,  \nonumber \\ 
&&m_q \, = \, \overline{m}_q(m_q^2) \, \left\{ 1 + {4 \over 3} 
{ {\alpha_s(m_q)} \over {\pi} }  + 
K_q  { {\alpha_s^2(m_q)} \over {\pi} } + {\cal O}(\alpha_s^3) \right\}, 
\end{eqnarray}
where $m_q$ is the so-called {\it pole} mass (defined in quantum field theory 
as the position of pole of a renormalized quark propagator), $\overline{m_q}$ is the 
$\overline{MS}$ running mass and $K_{c,b} = 13.3, 12.4$. Notice that the running mass  
depends on $\alpha_s$. Using, for instance, for the $b$-quark pole mass 
$m_b = (4.7 \pm 0.2)$~GeV and $\alpha_s(M_Z) = 0.12$, one obtains a running  mass 
at the $Z^0$ peak of $\overline{m}_b(M_Z) \simeq 2.8$~GeV, leaving non-negligible 
mass effects. The running $c$-quark mass $\overline{m}_c$ is 
about a factor of five smaller than $\overline{m}_b$, thus making $c$-quark mass 
corrections almost invisible in the $Z$-boson decays~\cite{qcd96}.  

\begin{figure}[hbt]
\begin{center}
\epsfig{file=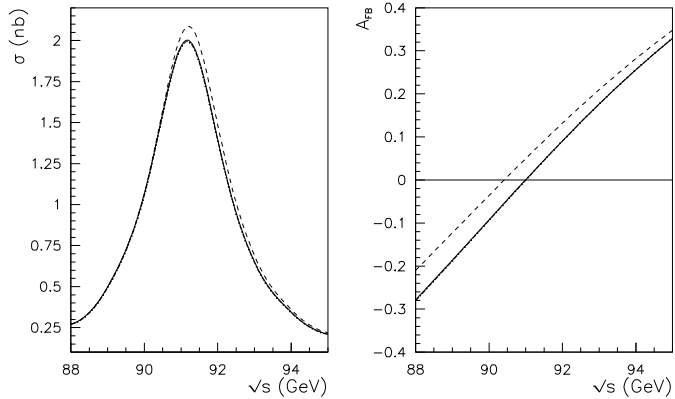, width=10truecm}
\end{center}
\caption{Comparison between the cross section and the forward-backward
asymmetry for $e^+ e^- \to \mu^+ \mu^-$, as computed in the Born  approximation
(dashed line) and including all the one-loop and
the relevant  higher-order electroweak and QCD corrections (solid line).   
Numerical results for this last case  by
{\tt TOPAZ0}~\cite{topaz0}.}
\label{fig:iba0}
\end{figure}
\begin{figure}[hbt]
\begin{center}
\epsfig{file=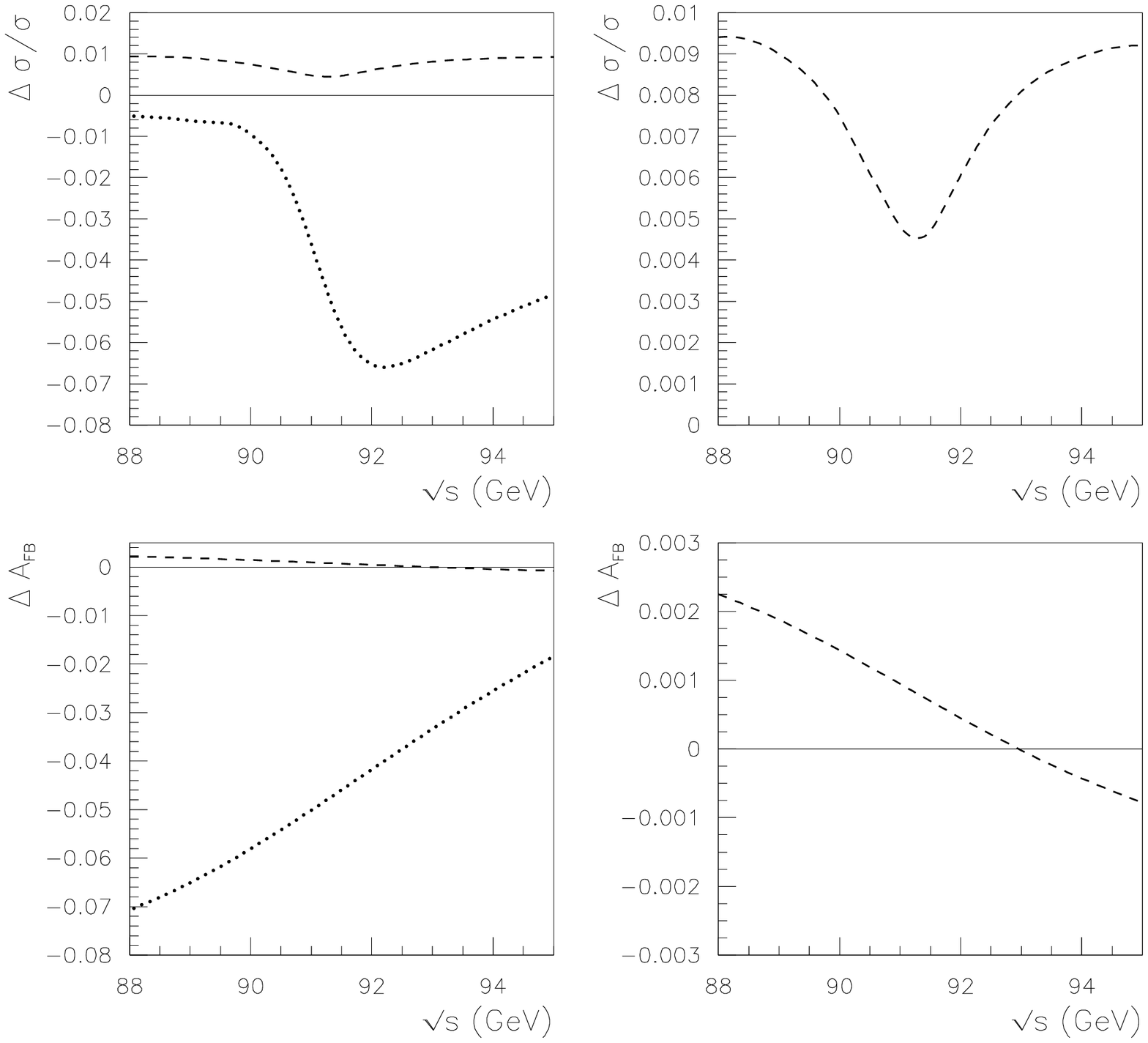, width=10truecm}
\end{center}
\caption{The left windows show the bulk of electroweak/QCD corrections as 
obtained by means of a
``tree-level like'' calculation (dotted line) and the residual effect  of a
complete electroweak/QCD calculation (dashed line). The right windows show a
blow-up of this last difference.  
Numerical results for the full electroweak/QCD prediction  by
{\tt TOPAZ0}~\cite{topaz0}.}
\label{fig:iba}
\end{figure}

Before turning  to QED corrections, it is  worth quantifying the effect of the 
electroweak and QCD corrections discussed up  to  now. To this aim,
Fig.~\ref{fig:iba0} shows the comparison between the total cross section and
the forward-backward  asymmetry for the process $e^+ e^- \to \mu^+ \mu^-$  at 
the tree level (dashed line) and the full prediction including all 
the one-loop and
the relevant  higher-order electroweak and QCD corrections  (solid line).  
In Fig.~\ref{fig:iba}  relative  deviations for the cross section and absolute 
deviations for the asymmetry are shown. The dotted  lines represent the 
difference between strictly  tree-level predictions and ``tree-level
like'' results, obtained by running the QED coupling constant (and hence $\sin^2 
\vartheta_W$) and including final-state QED and QCD corrections in the total 
$Z$-boson width in the propagator. The dashed lines represent the deviations of
the last  prediction from a full electroweak/QCD calculation. 
The full electroweak/QCD results have been produced by means of {\tt  TOPAZ0}
in its default mode, neglecting theoretical uncertainties (see
Sect.~\ref{sect:z0thu}). As can be seen from the figures, the bulk of the
corrections is due to the running of the QED coupling constant and to
final-state QCD corrections to the $Z$-boson width. However, there is an 
effect of $0.5 \div 1$\% on the cross section and up to 0.002 on the asymmetry
due to non-trivial  corrections as computed by means of a full  
electroweak/QCD calculation.   

\subsection{QED Corrections} 
\label{sect:z0qed}
As already  said in Sect.~\ref{sect:rcsabs}, QED corrections originate from those 
diagrams with extra real and/or virtual photons added to the tree-level
graphs (see Figs.~\ref{fig:isr}, \ref{fig:fsr} and \ref{fig:embox} below). 
Although these corrections are not particularly interesting with respect to the 
underlying theory, they need a special attention at LEP, as at any 
other $e^+ e^-$ collider, because the size of their effects strongly depends on 
the experimental cuts,  
and therefore their proper treatment constitutes the unavoidable 
link between the data taking and the physics analysis~\cite{greco88}. In fact, they are large 
corrections at high energies because, as already explained in Sect.~\ref{sect:rcsabs}, 
they are dominated by logarithmic contributions of the kind 
$(\alpha / \pi) L$,  $L = \ln (s/m_e^2) $  being the so-called  
 {\it collinear logarithm}. At LEP1, where $s \simeq M_Z^2$, $L$ is of the 
order of 25, and the {\it effective } expansion
parameter in perturbation theory is $\beta = 2 \alpha (L-1)/ \pi  \simeq 0.1$ rather 
than $\alpha$. The Breit-Wigner line-shape of the $Z^0$ resonance 
in Born approximation is  sensibly modified by the QED corrections for the following 
typical effects~\cite{nt87,yrls89}:
\begin{itemize}

\item the peak height is lowered by around $25\%$;

\item the peak position is shifted towards higher energies by around 100~MeV; 

\item a hard radiative tail, that increases the lowest-order cross section,
      appears above, say, 93~GeV.        

\end{itemize}

\begin{figure}[hbt]
\begin{center}
\epsfig{file=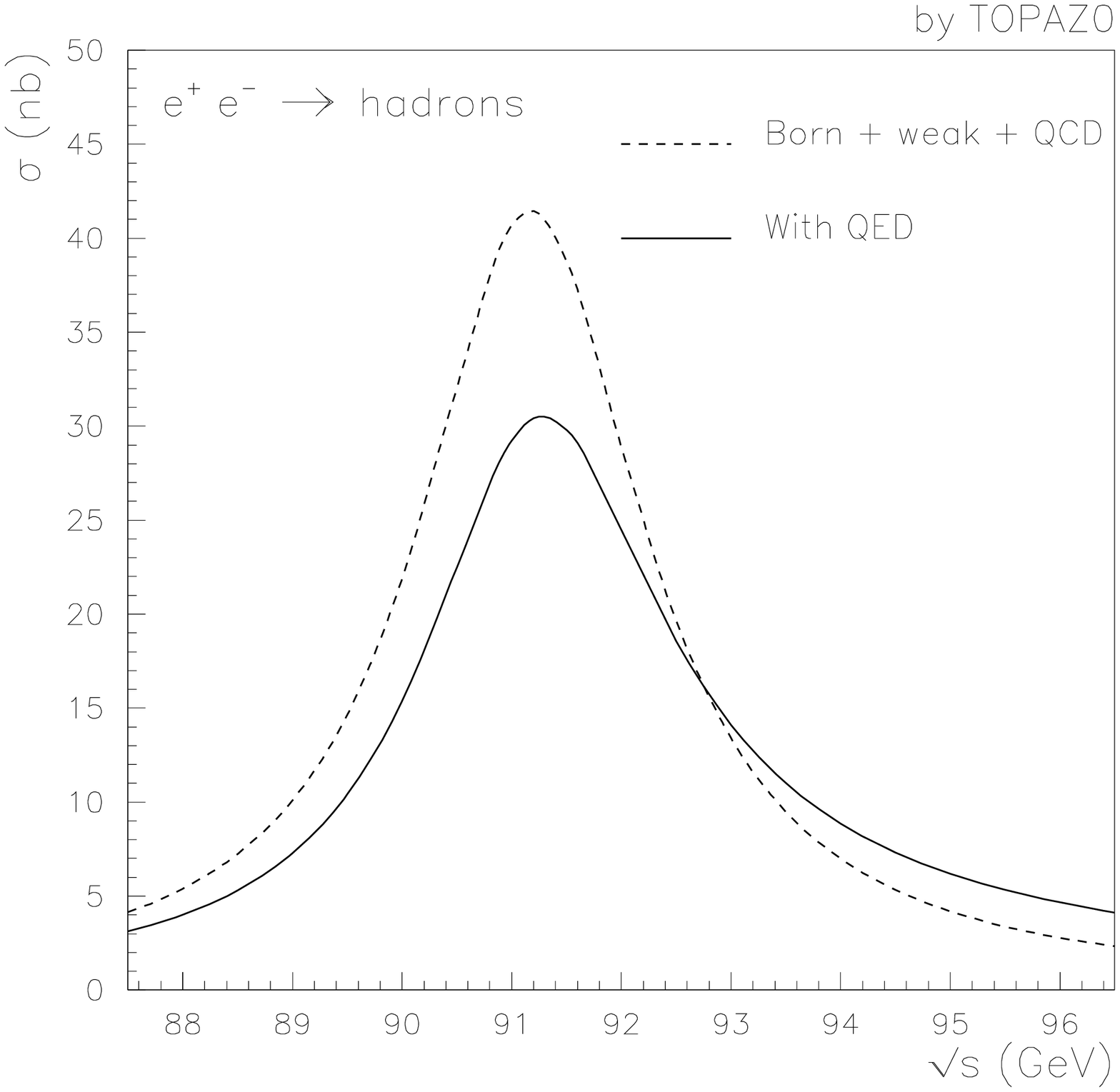, width=10truecm}
\end{center}
\caption{The effect of QED corrections on the $Z^0$ line-shape of
$e^+ e^- \to$~hadrons. The dashed line is the QED-deconvoluted 
cross section, the solid line is the QED corrected one. Numerical results by
{\tt TOPAZ0}~\cite{topaz0}.}
\label{fig:zpeak}
\end{figure}

The significant distortion introduced by the QED corrections on the $Z^0$ line-shape of
the process $e^+ e^- \to$~hadrons can be clearly seen in Fig.~\ref{fig:zpeak}.  

\begin{figure}[hbt]
\begin{center}
\epsfig{file=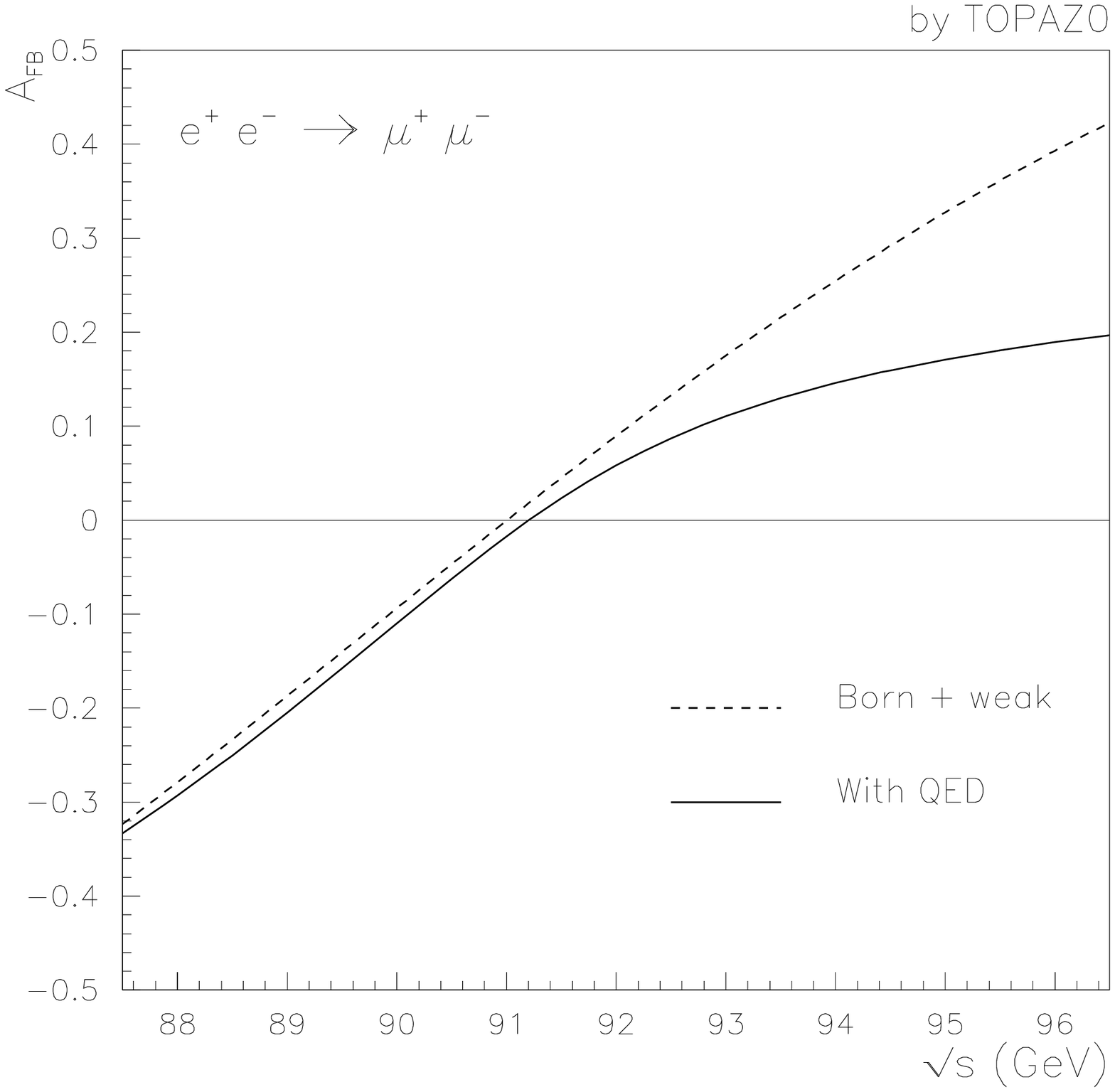, width=10truecm}
\end{center}
\caption{The effect of QED corrections on the forward-backward asymmetry 
$A_{FB}$ of $e^+ e^- \to \mu^+ \mu^-$ around the $Z^0$ peak. The dashed line is the 
QED-deconvoluted 
asymmetry, the solid line is the QED corrected one. Numerical results by
{\tt TOPAZ0}~\cite{topaz0}.}
\label{fig:zafb}
\end{figure}

Also the forward-backward  asymmetry, $A_{FB}$, which crosses the zero in the proximity 
of $M_Z$, is significantly affected by the QED corrections,   
that shift the position of its zero and globally change its shape~\cite{mnt89,yrafb89}. 
The QED  effects on
$A_{FB}$ for the reaction 
$e^+ e^- \to \mu^+ \mu^-$ are illustrated in Fig.~\ref{fig:zafb}.

The whole set of QED corrections to $s$-channel $e^+ e^-$ annihilations can be 
divided into three subsets, each separately gauge-invariant:

\begin{itemize}

\item initial-state corrections;

\item final-state corrections;

\item initial-final state interference.

\end{itemize}

The initial-state corrections are by far the dominant ones, because, at a
difference  from final-state 
and initial-final state interference contributions, they are responsible for a reduction of 
the c.m. energy available for the hard-scattering reaction. 

\subsubsection{Initial-state radiation} 
\begin{figure}[hbt]
\begin{center}
\epsfig{file=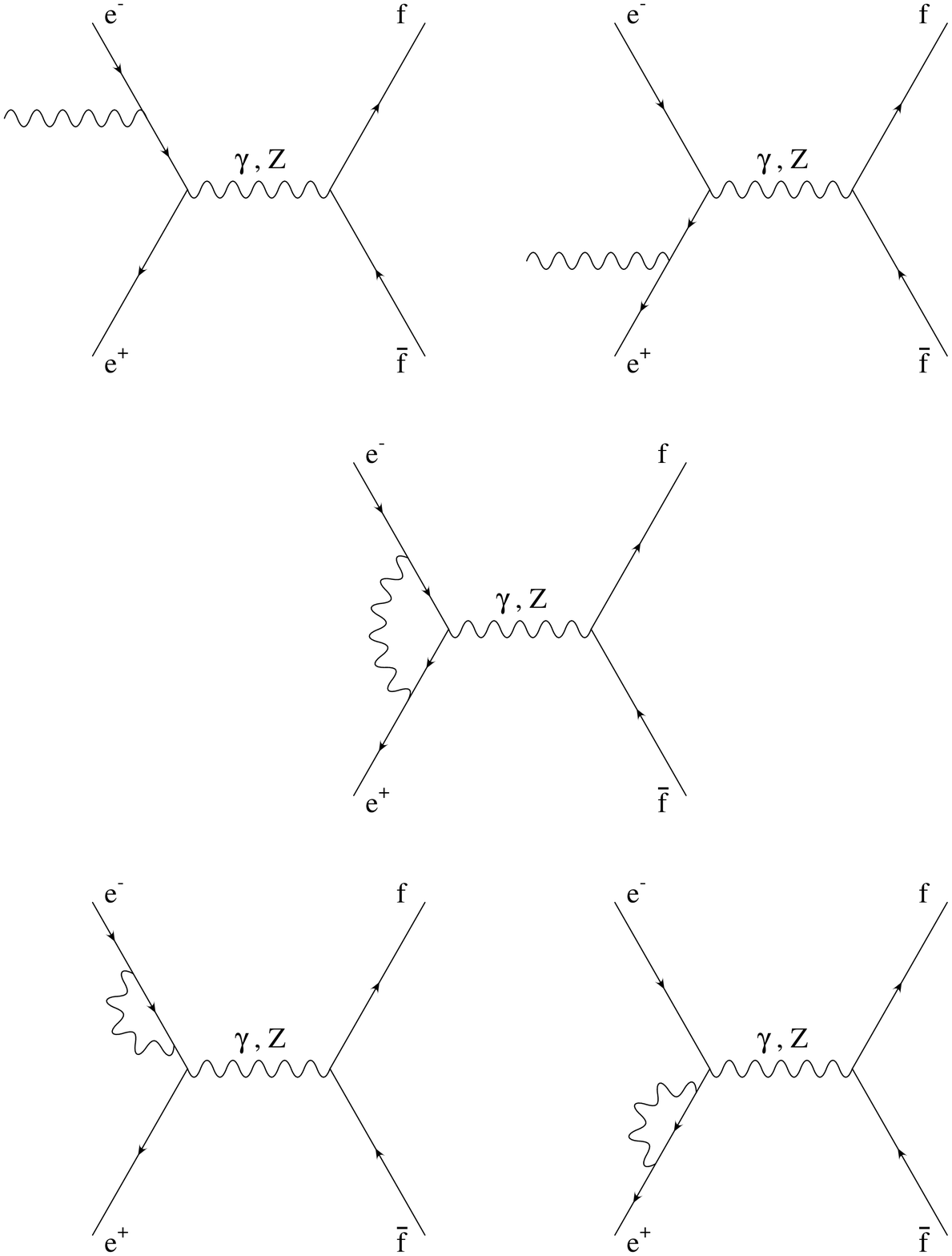, width=9truecm}
\end{center}
\caption{The Feynman diagrams for ${\cal O}(\alpha)$ initial-state
QED corrections to $s$-channel process $e^+ e^- \to \gamma, Z^0 \to f \bar f$. }
\label{fig:isr}
\end{figure}

The Feynman diagrams for ${\cal O}(\alpha)$ initial-state QED corrections to a 
generic $s$-channel process $e^+ e^- \to \gamma, Z^0 \to f \bar f$ 
are depicted in Fig.~\ref{fig:isr}.

Since at LEP1 energies the electron-positron collision takes place at energies very close 
to the  $Z$-boson mass, $M_Z$, the radiation emitted by the initial-state (ISR)
is essentially radiation of soft photons. This remarkable property is a direct 
consequence of the fact that the maximum energy of the photons emitted by the initial-state 
can not exceed the ratio $\varepsilon_{max} = \Gamma_Z / M_Z$,  with
$\varepsilon_{max} \simeq 0.03$. In fact, the finite $Z$-boson width, $\Gamma_Z$,  
acts as a natural cut-off that strongly inhibits the emission of those hard photons, of 
energy fraction larger than $\varepsilon_{max}$, that would prevent the formation of the  
$Z^0$ resonance. This soft-photon dominance at the $Z^0$ peak is a very useful guideline when 
facing the problem of including finite order 
perturbative corrections, beyond the leading logarithmic (LL) approximation,  that are actually 
 demanded by the high-precision level of the LEP1 measurements.  

As recalled in Sect.~\ref{sect:rcsabs}, the $n$-th order contribution to a QED corrected cross section
$\sigma^{(n)}$ can be cast in the form
\begin{equation}
\sigma^{(n)} = \left( {{\alpha} \over {\pi}} \right)^n \sum_{k=0}^n  
a_k^{(n)} {\cal L}^k , 
\label{eq:signz0}
\end{equation}
where ${\cal L} = {\cal L}(s) = L-1$.  
The coefficients $a_k^{(n)}$ in eq.~(\ref{eq:signz0}) are given by the following expression
\begin{equation}
a_k^{(n)} = \sum_{j=0}^k b_{kj} l^j , 
\label{eq:akeqz0}
\end{equation}
where $l$ is the so called IR logarithm  given by $l  =  \ln (E / \Delta E)$, $\Delta E$  being 
the maximum energy of the emitted photons. 
This general expansion obviously applies to the 
IS QED corrected cross section. For example, the up to 
${\cal O}(\alpha)$ cross section in 
the  soft-photon approximation, as obtained by standard diagrammatic techniques, reads
\begin{eqnarray}
\sigma^{(\alpha)}(s) = 
\sigma_0(s) \Bigg\{1 + {\alpha \over \pi } \Bigg[- 2 l (L-1) + {3 \over 2}L 
+ {\pi^2 \over 3} - 2 \Bigg] \Bigg\} ,   
\label{eq:alfa}
\end{eqnarray}
which shows a large ${\cal O}(\alpha)$ negative correction,   
clearly indicating that higher-order QED corrections are 
required for a high-precision reconstruction of the $Z^0$ line-shape. These higher-order effects 
are kept under control at all orders by employing one of the approaches described 
in Appendix~\ref{sect:upc}. For  instance, in the  structure  function (SF)  approach the
LL-corrected cross section, in the extrapolated set-up, can be written  as 
\begin{equation}
\sigma (s)  = \int_0^{\Delta  E   /E}  dx  H(x,s)  \sigma_0 \left( (1-x) s\right),
\label{eq:sigrad} 
\end{equation}
$H  (x,s)$ being the QED {\it  radiator}  of  eq.~(\ref{eq:radh}), and  $\sigma_0 (s)$ being the  
lowest order cross section, possibly including all the
short-distance process dependent corrections, as the electroweak and QCD ones previously discussed.  
This allows to exactly reproduce the LL content of eq.~(\ref{eq:signz0}) at any order and, in
particular, that of eq.~(\ref{eq:alfa}) at ${\cal O}(\alpha)$ (see the series expansion in the LL 
approximation given by eq.~(\ref{eq:series}) in Sect.~\ref{sect:rcsabs} ). The 
sub-leading terms 
present in eq.~(\ref{eq:alfa}) are not naturally reproduced by the perturbative expansion 
of the cross section obtained via the algorithms for the calculation of universal photonic corrections.
It is worth noting that they are numerically important for a description of the $Z^0$ 
line-shape with a theoretical accuracy at the 0.1\% level. For this reason, they are 
usually incorporated in the algorithms for the resummation of LL corrections, by means of a 
proper matching with the perturbative diagrammatic results. For instance, in the  
SF approach, 
a standard procedure consists in replacing the Gribov-Lipatov form factor 
in front of the exponentiated term of the SF's
$(\beta / 2) \, (1 - x)^{{\beta \over 2} - 1}$ by means of 
a soft+virtual $K$-factor $\Delta'_{S+V}$ given,  
up to the first order in $\alpha$, by
\begin{eqnarray}
\Delta_{S+V}'(s) = 1 + {\alpha \over {2 \pi} } \Bigg[{3 \over 2}L 
+ {\pi^2 \over 3} - 2 \Bigg] , 
\label{eq:alfak}
\end{eqnarray} 
in such a way that the up to ${\cal O}(\alpha)$ 
cross section of eq.~(\ref{eq:alfa}) is exactly reproduced.
By virtue of the soft-photon dominance 
in the ISR mechanism previously discussed, the hard-photon non-leading contributions, 
left-over in the 
above derivation, turn out to be numerically unimportant at LEP1. 
For the applications to precision physics at the $Z^0$ peak, 
the inclusion of sub-leading terms in the SF approach is pushed up to ${\cal O}
(\alpha^2)$~\cite{nt87}, 
relying upon the exact second-order calculation of initial-state QED corrections to 
$s$-channel processes available in the literature~\cite{bmr72,burg85,bbvn88}. 
By  exploiting again the soft-photon dominance, 
one can extract from the second-order complete calculation 
the soft+virtual contributions only, so that the up to ${\cal O}(\alpha^2)$ $K$-factor, to be 
placed in front of the electron SF, can be cast in the form
\begin{eqnarray}
\Delta'_{S+V} (s) = 1 + \Delta_{S+V}^{' \, (\alpha)}(s) + \Delta_{S+V}^{' \, (\alpha^2)}(s)
\end{eqnarray}
and is calculable through the relation
\begin{eqnarray}
\Delta_{S+V} (s) = \left( \Delta'_{S+V}(s) \right)^2 - {\pi^2 \over 24} \beta^2  .
\label{eq:ddprime}
\end{eqnarray}
The quantity  $\Delta_{S+V} (s)$ contains the non-leading non-IR sensitive diagrammatic corrections to 
the cross section corrected by the inclusion of  soft+virtual  photons up to  ${\cal  O}   (\alpha^2)$,
analogous to the one shown   in eq.~(\ref{eq:alfa}). Equation~(\ref{eq:ddprime}), that relates 
$\Delta_{S+V} (s)$ and $\Delta'_{S+V} (s)$, can  be derived from the explicit integration of 
eq.~(\ref{eq:masterd}) in Appendix~\ref{sect:upc}   in the soft-photon approximation. 
By inspection,  $\Delta_{S+V} (s)$ is given by 
\begin{eqnarray}
\Delta_{S+V} (s) = 1 + \left( {{\alpha} \over {\pi}} \right) \Delta_{S+V}^{(\alpha)}(s) 
+ \left( {{\alpha} \over {\pi}} \right)^2  \Delta_{S+V}^{(\alpha^2)}(s),  
\label{eq:delta2}
\end{eqnarray}
where the perturbative corrections $\Delta_{S+V}^{(\alpha)}(s)$ and 
$\Delta_{S+V}^{(\alpha^2)}(s)$ are explicitly given by 
\begin{eqnarray}
&&\Delta_{S+V}^{(\alpha)}(s) = {3 \over  2} L + {{\pi^2} \over {3}} - 2 , \nonumber \\
&&\Delta_{S+V}^{(\alpha^2)}(s)  =  \left( {9 \over 8} - {\pi^2 \over 3} \right) \, L^2  
+ \Big[- {45 \over 16} +  {11 \over 12} \, \pi^2 +  3\,\zeta(3)\Big]\,L 
+ \hbox{\rm constant terms} . 
\label{eq:alfak2}
\end{eqnarray}
In eq.~(\ref{eq:alfak2}) $\zeta$ is the Riemann function, with $\zeta(3) \simeq 1.202$.
With respect to the exact second-order calculation, 
${\cal O} (\alpha^2)$ sub-leading non-soft corrections are neglected,   but they are 
negligible in the resonance region, their contribution to the cross section 
being at the level of 0.01\%~\cite{yrls89,jsw91,s92,mnp97}. The above described procedure for
matching LL results with exact finite-order calculations is valid in the soft-photon
approximation. It is worth noticing, however, that a general prescription is also known, valid
for any ES, \idest\  taking into account also hard-photon contributions, as can be found
in~\cite{a2l96}. 

A complete treatment of IS ${\cal O} (\alpha^2)$ QED effects to the $Z^0$ line-shape 
requires, as a last ingredient, the inclusion of the corrections arising 
from the conversion of a photon into leptonic pairs and/or hadrons (pair corrections).
These contributions, which are dominated by $e^+ e^-$ pairs, can be included 
using the ${\cal O} (\alpha^2)$ formulae available in the literature~\cite{pairs}. Their 
main effect is a reduction of the peak cross section of around 0.3\%, hence significant
in the light of the remarkable experimental precision of the LEP1 data. 

Equation~(\ref{eq:sigrad}) allows to understand the main effects of ISR on the $Z^0$ line
shape. Actually, since $\Delta E / E$ is of the order of  $\Gamma_Z /  M_Z $ around the
resonance, the explicit calculation of the integral appearing in eq.~(\ref{eq:sigrad}) in the
soft-photon limit leads to the following reduction of the peak height:  
\begin{equation}
{{\sigma} \over {\sigma_0}} \simeq \left( {{\Gamma_Z} \over {M_Z}} \right)^\beta \simeq 0.75 . 
\end{equation}
On the other hand, eq.~(\ref{eq:sigrad}) is nothing but a weighted average of the lowest-order
cross section for c.m. energies smaller than the nominal one, the weight being the
radiator $H$ itself. This leads to a shift of the corrected peak position towards higher
energies by about 100~MeV, and to  the raising of a radiative tail above the resonance (see
Fig.~\ref{fig:zpeak}). It  is worth noticing, in  particular,  that the contribution of
soft-photon exponentiation to the shift of the peak  position  is of the order of 15~MeV, and
thus phenomenologically relevant in  view of the experimental  precision of the determination
of the $Z$-boson mass. 

Before concluding the discussion  on  ISR, it is worth noticing that the analogue of 
eq.~(\ref{eq:sigrad}) for the corrected forward-backward asymmetry in principle does not
apply. Indeed, since $A_{FB}$  is a less inclusive quantity  than the total  cross section,
kinematical  effects can  show up, thus introducing the need for improving 
eq.~(\ref{eq:sigrad}). However, at  the resonance these effects are negligible due to
soft-photon dominance, and anyway  they  can be naturally taken into account in a more
appropriate SF formulation, as described in~\cite{mnpi93}. 

\subsubsection{Final-state radiation}
\begin{figure}[hbt]
\begin{center}
\epsfig{file=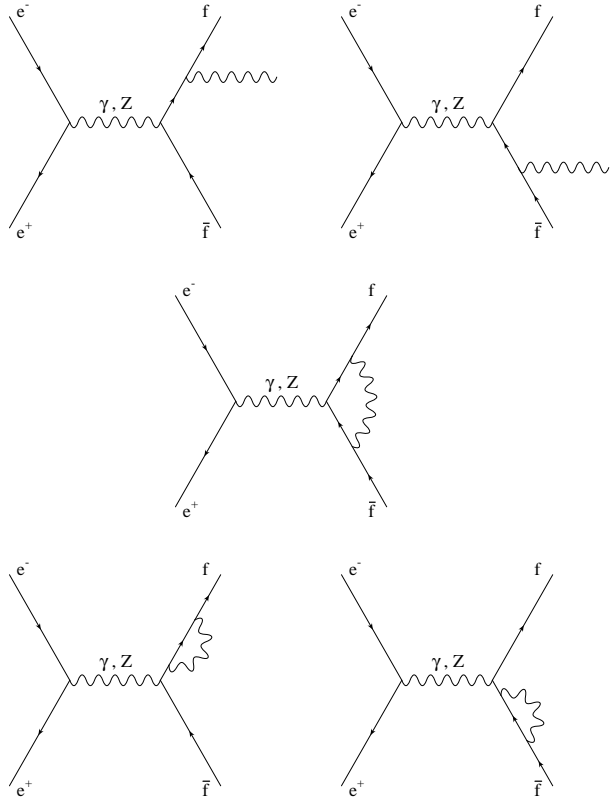, width=8truecm}
\end{center}
\caption{The Feynman diagrams for ${\cal O}(\alpha)$ final-state 
QED corrections to $s$-channel process $e^+ e^- \to \gamma, Z^0 \to f \bar f$. }
\label{fig:fsr}
\end{figure}

The final-state  QED corrections, although numerically smaller than the IS ones 
under many typical experimental conditions, are an essential ingredient for a high-precision 
phenomenology of the LEP1 processes. They must be included in any formulation 
of $Z^0$ physics that aims at a theoretical precision at the level of 0.1\%. 
The Feynman diagrams for ${\cal O}(\alpha)$ final-state QED corrections to a 
generic $s$-channel process $e^+ e^- \to \gamma, Z^0 \to f \bar f$ 
are depicted in Fig.~\ref{fig:fsr}.

With respect to the emission of photons by the initial  state, the mechanism of 
bremsstrahlung by the final-state particles is no more characterized by 
the soft-photon dominance discussed above, simply because the radiation process occurs after the 
formation of the $Z^0$ resonance. This implies that the control on final-state QED corrections at the 
level of sub-leading and constant terms necessarily requires a full perturbative 
calculation, including hard photon bremsstrahlung. A treatment of FSR 
completely flexible with respect to {\it any} kind of ES requires the usage of numerical   
(Monte Carlo) techniques in order to perform the phase space integrations in the presence 
of arbitrary cuts. However, when considering  particular experimental selection criteria 
actually adopted by the LEP Collaborations, it is possible to obtain the final-state 
${\cal O} (\alpha)$ QED correction according to a completely analytical procedure~\cite{abl92,mnpa93}.  
In this case, the final-state corrected cross section can be obtained by calculating the 
exact matrix element associated to the gauge-invariant set of Feynman diagrams describing 
final-state real photon radiation in $e^+ e^-$ annihilation  (see Fig.~\ref{fig:fsr}), 
integrating it over the phase space allowed 
by cuts and matching the obtained result with the soft+virtual correction. 
For example, if one assumes that only an invariant mass cut of the type $M^2(f \bar f) 
\geq  s_0$ 
is present, then the above strategy leads to the following analytical correction 
factors~\cite{topaz0np,kleiss92,mnpa93,mnpi93}
\begin{eqnarray}
\delta_{_{F+B}}^{fs} &=& {\alpha \over \pi} \, Q_f^2 \, \Big\{  \left[ x + {1 \over 2}
\, x^2 + 2 \,  \ln(1-x) \right] \, L_f \nonumber\\
&+& x \, \left( 1 + {1 \over 2} \,x \right) \,\ln\,x + 2\,\ln(1-x)\left(
\ln\,x - 1 \right) \nonumber \\
&+& 2\, {\rm Li_2} (x) \Big\}   , \label{eq:fscfpb} \\
\delta_{_{F-B}}^{fs} &=&  {\alpha \over \pi} \, Q_f^2 \,\Big\{  \left[ x + {1 \over 2}
\, x^2 + 2 \, \ln(1-x) \right] \,  L_f \nonumber\\
&+& 2\, \ln(1-x) \left( \ln\,x - 1 \right) -2 \, x  \nonumber \\
 &+& 2 \, {\rm Li_2}(x) \Big \} ,
\label{eq:fscfmb} 
\end{eqnarray}
where $L_f  = \ln(s / m_f^2)$,  
 $x = s_0/s$ and $(F \pm B)$ denotes the forward $\pm$ backward cross sections 
defined in eq.~(\ref{eq:sfb}). In eq.~(\ref{eq:fscfpb}) and (\ref{eq:fscfmb}),   
${\rm Li_2}(x)$ 
is the dilogarithm function. It is worth observing that when no cuts are applied 
(\idest\   in the limit $x \to 0$ in the latter equations) the final-state correction factor 
for the total cross section reduces to $3  \alpha Q_f^2  /   4  \pi$, that is the 
well-known inclusive final-state correction, while the asymmetry does not get any effect. 
The very small value (0.17\% for fermions with unit charge) of the 
final-state correction to the fully inclusive cross section is a consequence of the cancellation 
of mass and collinear singularities established by the Kinoshita-Lee-Nauenberg 
theorem~\cite{kln6264}. 
Besides an invariant mass cut, also the effect of acollinearity and/or energy threshold cuts can be 
treated analytically, and the expressions for the corresponding ${\cal O}(\alpha)$ correction 
factors can be found in the literature~\cite{mnpa93}. 
The availability of such results in analytical form is 
of utility for $Z^0$ physics at LEP1 in order to test Monte Carlo programs and to develop 
fast computational tools for fitting realistic observables.    

Whenever particularly tight cuts are imposed, the treatment of FSR at the ${\cal O}(\alpha)$ can 
become inadequate and a procedure of resummation of the LL contributions 
needs to be advocated~\cite{nt88}. The leading terms can be quite easily identified either analyzing 
the explicit perturbative results or invoking one of the algorithms for the 
universal photonic corrections described in Appendix~\ref{sect:upc}. This allows to see that 
final-state leading contributions are of the form
\begin{eqnarray}
2 \, {{\alpha} \over {\pi}} \, Q_f^2 \, \ln(1-s_0/s) \left( 
L_f - 1 \right),
\end{eqnarray}
so that they can be extracted from the finite order correction and summed up to all orders
according to the preferred resummation technique. By using, for instance, a ``naive" exponentiation 
procedure, the final-state correction factor, including higher-order leading 
contributions and exact sub-leading terms at ${\cal O}(\alpha)$, can be cast in the form
\begin{eqnarray}
\delta^{fs} \, = \, \Delta_{V+S}^{' \, (\alpha)} (s) \, \exp 
\Big\{ 2 \, {{\alpha} \over {\pi}} \, Q_f^2 \, \ln(1-s_0/s) \left( L_f - 1 \right) \Big\}
\, + \, \delta^{fs,r},
\end{eqnarray}  
where $\Delta_{V+S}^{' \, (\alpha)} (s)$ is the soft+virtual $K$-factor of eq.~(\ref{eq:alfak}) 
(provided that the substitution $m_e \to m_f$ is performed in the collinear logarithm) and 
$\delta^{fs,r}$ is the residual correction factor including sub-leading terms 
as obtained after depuration of leading logarithms from the complete ${\cal O} (\alpha)$ 
 results.

All the above discussion and formulae con\-cer\-ning FSR are valid for a 
non-ca\-lo\-ri\-me\-tric 
ES (see Sect.~\ref{sect:rcsabs} for more details) according to which the 
energy of the final-state fermions measured by the experimental apparatus coincides 
with the energy of the ``bare" particles, regardless of the emission of 
collinear final-state photons. However, when considering the process of large-angle Bhabha scattering, 
this assumption turns out to be unrealistic.  
Actually, what is detected in the real environment is an 
{\it electromagnetic jet} of half-opening angle $\delta_c$, where $\delta_c$ is an 
experimental parameter describing the resolution power of the calorimeter. 
This electromagnetic effect can be analytically accounted for by adding to the 
${\cal O}(\alpha)$ final-state correction for ``bare" electrons the contribution due to 
a hard photon of energy fraction greater than $1-x$, where $x$ is $s_0/s$ 
for an invariant mass cut and $2 E_0 / \sqrt s$ for an energy threshold cut, and collinear 
with the final fermion within an angle $0 \le \vartheta_\gamma \le \delta_c$.
For electrons in the energy regime of LEP the contribution reads~\cite{calo}
\begin{eqnarray}
F_{coll} \, &=& \, 2 \, {{\alpha} \over {\pi}} \, C \nonumber , \\
C &=& - \ln (1-x) \Big[ \ln \Big ( 1+r^2 x^2 \Big) -1 \Big] 
\nonumber \\
&+& \Big[ {1 \over 4} - \Big( 1 - {1 \over 2} (1-x) \Big)^2 \Big] 
\ln \Big( 1+r^2 x^2 \Big) \nonumber \\
&+& {{\pi^2} \over {3}} + {9 \over 4} - {5 \over 2} (1-x) 
+ {1 \over 4} (1-x)^2 \nonumber \\
&+& 2 \ln x \ln (1-x) + \, 2 {\rm Li_2} (1-x) , 
\label{eq:coll}
\end{eqnarray}
where 
\begin{equation}
r = {{\delta_c \sqrt s} \over {2 m_e}} , 
\end{equation}
$m_e$ being the electron mass. Equation~(\ref{eq:coll}) 
holds under the conditions $\delta_c \ll 1$~rad
 and $r \gg 1$, that are both very well satisfied at LEP, where 
 $\delta_c$ is of the order of a few degrees. It is worth stressing that 
the effect of the calorimetric measurement,  superimposed  over the correction  for ``bare'' 
final-state  particles, depends very critically on the energy or 
invariant mass threshold, being at the level of 0.1\% for low energy thresholds (around 
1~GeV) but raising to order 1 per cent at, say, 10~GeV. 

\subsubsection{Initial-final state interference}
\begin{figure}[hbt]
\begin{center}
\epsfig{file=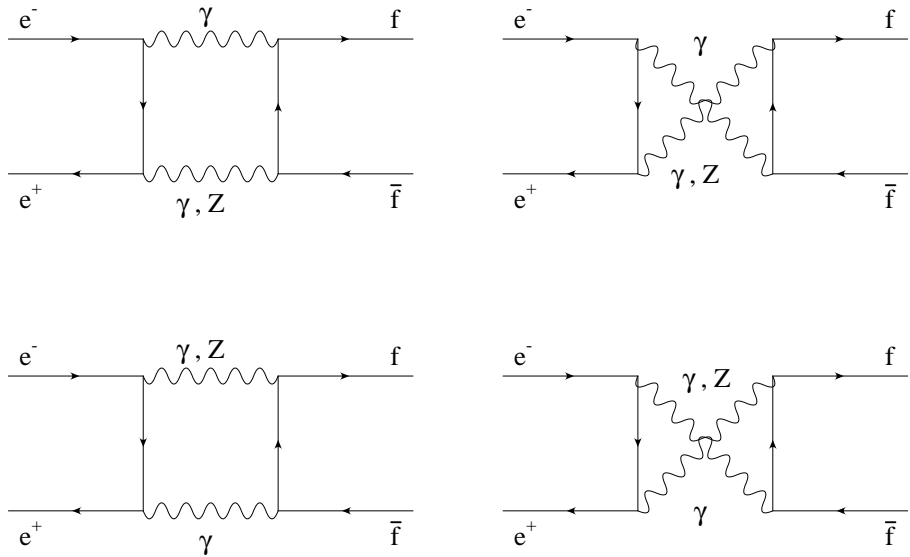, width=12truecm}
\end{center}
\caption{The Feynman diagrams for ${\cal O}(\alpha)$  
QED box  corrections to $s$-channel process $e^+ e^- \to \gamma, Z^0 \to f \bar f$. }
\label{fig:embox}
\end{figure}

While the initial- and final-state QED corrections are functions of $s$ only, the 
initial-final  state interference correction depends also on the fermion scattering angle 
$\vartheta$ and therefore induces a modification of the $\vartheta$ dependence of the 
Born differential cross section. It receives contributions from the interferences  between 
the real radiation  diagrams of  Figs.~\ref{fig:isr}  and  \ref{fig:fsr} 
and from the interference  between 
the tree-level amplitudes  and the QED box diagrams shown in  Fig.~\ref{fig:embox}.  
The leading angular dependent terms that enter the result 
are of the form 
$\ln (t / u ) =  \ln \left( \tan (\vartheta / 2) \right)$~\cite{hollik90,bkj82}. 
This explains why the initial-final state 
interference 
is numerically small under many realistic experimental conditions, even if its actual magnitude 
crucially depends on the applied cuts. A physical argument can also be advocated to 
understand the r$\hat{ \hbox{\rm o}}$le 
of the initial-final state interference in the region of  the $Z^0$ resonance. In fact, 
when the $Z^0$  is produced close to its mass shell, the wave functions for
initial- and 
 final-state radiation are separated in space-time due to the finite $Z^0$ lifetime, 
so that their overlap is small.  This implies that for loose cuts the interference effect is 
typically at the level of $10^{-3}$ or even smaller. Only when particularly tight cuts are imposed 
or one moves away from the peak, the above argument is invalidated and the interference can 
become more sizeable. In order to keep under control all the situations, the correction due to 
the initial-final state interference is included in the theoretical predictions, generally using the 
existing exact ${\cal O} (\alpha)$ calculations (including hard bremsstrahlung), that 
are sufficient for the realistic ES. 
For very tight,  unrealistic,  photon energy cuts, a proper exponentiation  procedure can  be
 introduced.

\subsection{Computational Tools}
\label{sect:z0ct}

In order to perform actual precision tests of the electroweak theory in 
$e^+ e^-$ collisions at the $Z^0$ pole, very precise measurements of cross sections and 
asymmetries are required, that in turn imply, together with a high 
statistics, a deep knowledge of systematics effects, such as the acceptance, 
selection efficiency and backgrounds for a given reaction. The tools used by LEP 
Collaborations to determine experimental acceptances and efficiencies are unavoidably 
Monte Carlo event generators since they allow to simulate experimental cuts with the 
maximum flexibility. Just to give some examples, the LEP experiments commonly use the 
generators  
{\tt HERWIG}~\cite{herwig} and {\tt PITHYA/JETSET}~\cite{jetset} 
to simulate the process $e^+ e^- \to$~hadrons, 
{\tt KORALZ}~\cite{koralz} for the production of $\mu$ and $\tau$ pairs, 
{\tt BHAGENE3}~\cite{bhagene3}, {\tt BHWIDE}~\cite{bhwide} and {\tt UNIBAB}~\cite{unibab} 
for the large-angle Bhabha scattering.  {\tt HERWIG}  and {\tt PITHYA/JETSET} are generators
for the study of  hadronic final  states in $e^+  e^-$, $e p $ and $p p$ collisions,
describing the phase of parton cascade in  the framework  of perturbative QCD and the
hadronization  mechanism by using independent theoretical  models.  {\tt KORALZ} is a Monte
Carlo program developed  for  $s$-channel $e^+ e^- \to f \bar f$ processes, including
electroweak loops and using YFS exponentiation (see Appendix~\ref{sect:upc}) for the treatment
of photonic radiation.  Large-angle Bhabha generators will be described in the following.

However, in addition to Monte Carlo Generators, 
fast analytical and semi-analytical programs are of utmost importance for 
$Z^0$ precision physics, because, although they can account for simple kinematical cuts only, 
they are necessary tools in order to extract the electroweak parameters from the 
measurements according to an iterative fitting procedure. These fitting programs 
are also  said {\it electroweak libraries}. As a matter of fact, 
the fitting programs mostly used by the LEP Collaborations are {\tt BHM}~\cite{bhm94},  
{\tt MIZA}~\cite{miza}, 
{\tt TOPAZ0}~\cite{topaz0} and {\tt ZFITTER}~\cite{zfitter92}.  
However, the extreme complexity in the calculation of
 radiative corrections and the relevance of improvements and cross-checks for 
  precision physics at LEP, motivated a number of theoretical groups to 
develop completely independent and original electroweak libraries. 
The availability of various independent electroweak libraries 
turned out to be particularly useful in the context of the job of the  
Electroweak Working Group at CERN~\cite{ewwg95}. The aim of such a collaboration work was the   
estimate of the theoretical error inherent to 
the SM predictions for $e^+ e^- \to f \bar f$ processes, in view 
of the final analysis of LEP precision data (see  Sect.~\ref{sect:z0thu} for 
more details). The basic features of the codes that contributed to reach the above 
task are shortly described in the following, with particular emphasis on their 
main physics input as well as their numerical output. However, for more details on  
technical aspects of these programs and their underlying formulation, the reader is referred 
to the original literature and to ref.~\cite{ewwg95}.

\noindent
\underline{\tt BHM}~\cite{bhm94} --- It is a semi-analytical program for the 
calculations of the $Z^0$ parameters and realistic observables  
for an extrapolated set-up only. It relies upon the on-shell renormalization scheme 
for the formulation of weak loops and the QED radiator for the treatment of ISR  (see
Appendix~\ref{sect:upc}).   

\noindent
\underline{\tt LEPTOP}~\cite{leptop94} --- It is an analytical code, developed 
by the ITEP Moscow group,  giving predictions for the 
the $Z^0$ parameters in the so-called $\overline{\alpha}$-Born approximation. It consists in taking as
``tree-level approximation'' the one obtained by using $\alpha (M_Z)$ overall,   
and computing $\sin^2 \vartheta_W$ from $\alpha (M_Z)$, $G_\mu$ and $M_Z$, with the residual
electroweak  corrections on top of this. 
 
\noindent
\underbar{\tt TOPAZ0}~\cite{topaz0} --- It is a semi-analytical program, developed  by 
the Pavia and Torino  
groups, that can be used to calculate both {\it pseudo-observables} and realistic observables, the 
 latter over both an extrapolated and a more realistic set-up.  It is additionally able to 
calculate the full Bhabha scattering cross section at large angles. It employes the 
$\overline{MS}$  scheme for the treatment of the 
electroweak corrections. QED corrections are exactly included 
at ${\cal O}(\alpha)$ for $s$-channel processes, at the LL level for $t$ and $s$-$t$ contributions 
to the large-angle Bhabha scattering. On top of that, higher-order QED corrections are 
implemented using the SF approach (see Appendix~\ref{sect:upc}). 

\noindent
\underbar{\tt WOH}~\cite{woh} --- It is an analytical code for the calculation of the 
{\it pseudo-observables} only. It is based on the on-shell renormalization scheme and, 
basically, it leads back to the same approach of {\tt BHM}, 
so that {\tt BHM/WOH} are not completely independent of one another.

\noindent
\underbar{\tt ZFITTER}~\cite{zfitter92} --- It is a semi-analytical program, of the Dubna-Zeuthen group, 
that, as {\tt TOPAZ0}, allows to obtain predictions for both the {\it pseudo-observables} and 
the realistic ones, but excluding large-angle Bhabha scattering. 
The on-shell renormalization scheme is used for electroweak loops; QED corrections 
are exactly treated at ${\cal O}(\alpha)$, together with soft-photon exponentiation for 
ISR and FSR.

To summarize, all the above  codes can provide predictions for the $Z^0$ parameters 
and therefore can all be used as fitting tools to these de-convoluted quantities. {\tt BHM}, 
{\tt TOPAZ0} and {\tt ZFITTER} can also calculate realistic observables and fit data 
for such observables. {\tt TOPAZ0} is also additionally able to calculate the full Bhabha scattering 
observables. Concerning the treatment of electroweak loops, the codes employ 
completely independent  calculational schemes, with the exception of
{\tt BHM/WOH} that essentially are based on the same approach. QED corrections in the three 
QED dressers are treated according to different theoretical methods, whereas QCD corrections 
are common to a large extent in all the codes. 
How these codes were used as {\it tools} to 
estimate the intrinsic theoretical uncertainties in precision calculations for the $Z^0$ resonance  
is explained in Sect.~\ref{sect:z0thu}. 

The special r$\hat{ \hbox{\rm o}}$le 
played by large-angle Bhabha (LABH) scattering in $Z^0$ precision physics 
led the to development of dedicated semi-analytical and Monte Carlo programs for such a 
process. Two of them, {\tt BHAGEN95} and {\tt TOPAZ0}, have been previously described, the former 
in Sect.~\ref{sect:sabct} and the latter here in the present Section. The other programs used by the LEP
experiments for the study of $e^+ e^- \to e^+ e^-$ at large angles are briefly described 
in the following. It is worth pointing out that the Bhabha process, when considering 
large scattering angles and the energy region around the $Z^0$ peak, is basically 
a $s$-channel resonant process with ``small'' non-resonant contributions, so that its 
dynamics is completely different from that of the same process at small scattering angles, 
that has been already discussed in details in Sect.~\ref{sect:sablm}. In particular, 
whereas non-QED effects other than vacuum polarization are absolutely negligible 
in the SABH case, an accurate treatment of the electroweak loops is a necessary 
ingredient for a precision calculation of the LABH reaction~\cite{bhr91,alibaba91,topaz0np}. 
As a consequence, 
the computational tools (and their underlying formulations) for the LABH process are very 
different with respect to the ones used for the luminosity monitoring and already described 
in Sect.~\ref{sect:sabct}. 

\noindent
\underline{\tt ALIBABA}~\cite{alibaba91} --- It is a semi-analytical program, implementing exact 
${\cal O}(\alpha)$ weak and QED corrections to all Bhabha channels. The higher-order QED effects 
consist of LL ${\cal O}(\alpha^2)$ plus soft-photon exponentiation.    

\noindent
\underline{\tt BHAGENE3}\cite{bhagene3} --- It is a Monte Carlo event generator, including one-loop and the
most relevant two-loop electroweak corrections. The ${\cal O}(\alpha)$ QED corrections uses 
the exact matrix element and are supplemented with higher-order soft-photon effects. 
 
\noindent
\underbar{\tt BHWIDE}\cite{bhwide} --- It is a  recent Monte Carlo event generator, based 
on the electroweak library of {\tt ALIBABA} for the treatment of 
${\cal O}(\alpha)$ weak and QED corrections. It includes 
multiphoton radiation in the framework of  Yennie-Frautschi-Suura (YFS) exponentiation (see
Appendix~\ref{sect:upc}) and can be 
considered as the extension of the code {\tt BHLUMI} (see Sect.~\ref{sect:sabct}) 
to large angles. 

\noindent
\underbar{\tt UNIBAB}\cite{unibab} --- It is a Monte Carlo event generator, relying upon 
an updated version of the {\tt ALIBABA} electroweak library. The QED corrections are implemented 
assuming $s$-channel dominance and using photon shower algorithms for ISR and
FSR (see Appendix~\ref{sect:upc}).
 
A more extensive description of the LABH programs, together with global comparisons between the
semi-analytical and Monte Carlo programs, can be found in the recent work of 
ref.~\cite{bharep96}. At LEP1, when analyzing LABH data, the common procedure employed
 is the 
so-called $t$-channel subtraction, where $t$ and $s$-$t$ contributions are subtracted from the 
data. As a matter of fact, the two programs commonly used to perform this unfolding are 
{\tt ALIBABA} and {\tt TOPAZ0}. Furthermore, besides the programs  described above, one should 
also mention the codes {\tt ALISTAR} and {\tt MIZA}, that are basically rearrangements of 
{\tt ALIBABA} developed for fitting purposes and specific experimental needs.

\subsection{Theoretical Uncertainties}
\label{sect:z0thu}

The very high experimental precision reached by the LEP Collaborations in the 
measurement of the $Z^0$ parameters and realistic observables (of the order of 0.1\% or
even below), necessarily requires a careful estimate of the intrinsic uncertainties 
associated to the SM theoretical predictions for these observables. In fact, as 
already discussed in Sect.~\ref{sect:sabtoterr} when addressing the problem of the total 
theoretical error in the luminosity measurement, any prediction obtained {\it via} 
a perturbative expansion is intrinsically affected by an uncertainty that is mainly 
due to neglecting  higher-order contributions. The evaluation of such an 
uncertainty turns out to be a particularly severe problem whenever considering    
precision calculations for the $Z^0$ resonance, since the higher-order  contributions 
depend in a highly
non-trivial way on the whole stuff of radiative corrections in the SM (weak, QCD, 
mixed ew/QCD, QED). For this reason, a combined effort by different groups of 
theorists is in principle the best strategy to address this delicate subject. Such an 
effort was pursued by the collaboration work of the Electroweak Working Group,  
held at CERN, Geneva, during 1994 and concluded at the beginning of 1995. The
composition of the Electroweak  Working Group consisted of the authors of those formulations 
and relative computational programs containing at that time the state-of-the-art 
of the radiative corrections to the electroweak $Z^0$ observables: {\tt BHM}, 
{\tt LEPTOP}, {\tt TOPAZ0}, {\tt WOH} and {\tt ZFITTER}. They have been already
described in some detail in the previous Section. Because all these codes basically 
include the same content in the perturbative expansion 
( \idest\ exact ${\cal O}(\alpha)$ electroweak corrections plus higher-order leading 
contributions), but differ in the choice of the renormalization scheme and in the 
treatment of higher-order sub-leading effects, they are ideal {\it tools} 
to provide an {\it estimate} of the theoretical uncertainties. The latter can be 
classified as follows:
\begin{itemize}

\item {\it Parametric uncertainties:} are  related to the ex\-pe\-ri\-men\-tal 
precision of the input pa\-ra\-me\-ters. Among them, the largest un\-cer\-tain\-ty comes from 
$\alpha (M_Z)$: $|\Delta \alpha^{-1} (M_Z) |$ = 0.09; other sources are the errors on 
$\alpha_s (M_Z)$ and on the heavy-quark masses. These uncertainties can be estimated comparing the 
effects of variations in the input parameters within their errors~\cite{ewwg95}.  

\item {\it Scheme-dependence uncertainties:} are  associated to the choice of the 
calculational scheme. In fact, in a given renormalization framework the 
truncation of the perturbative series is realized in some specific way. An estimate of 
these uncertainties can be obtained by comparing the predictions for a given 
observable obtained with two codes based on different renormalization schemes. 
This is possible since, as discussed in Sect.~\ref{sect:z0ct}, {\tt BHM/WOH} and 
{\tt ZFITTER} employ the on-shell scheme, {\tt TOPAZ0} makes 
use of the $\overline{MS}$ scheme and an original approach is implemented in 
{\tt LEPTOP} .  

\item {\it Intrinsic uncertainties:} are  inherent in a given  
renormalization framework. They are a consequence of the still missing 
higher-order terms, reflecting in a certain degree of arbitrariness on how 
to combine the various theoretical ingredients in order to get  the final expressions
 for the observables. This arbitrariness can be quantified using the concept 
 of {\it option}, \idest\ a set of alternative but equally plausible implementations 
of the building blocks of the radiative corrections within a given 
calculational scheme~\cite{gp94}.  

\end{itemize}

To better illustrate the meaning of {\it intrinsic uncertainties}, it is worth
presenting and discussing a few examples of {\it option}. A first one~\cite{dbt96}
refers to different possible implementations of QCD and QED final-state corrections to a 
given $Z^0$ partial width $\Gamma_f \, = \, \Gamma (Z \to f \bar f)$. Let us 
suppose that ${\cal O}(\alpha)$ QED and ${\cal O}(\alpha^2_s)$ QCD corrections 
to the above width are known but that the mixed ${\cal O}(\alpha \alpha_s)$ 
corrections have not yet been calculated. Therefore, one can combine the two above 
final-state corrections either according to a {\it factorized} representation
\begin{eqnarray}
\Gamma_f \, = \, \Gamma_f^{EW} \, \left( 1 + {3 \over 4} Q_f^2 {\alpha \over \pi} 
\right) \left[1 + {\alpha_s \over \pi} + c_2 \, {\alpha_s^2 \over \pi^2} \right],
\end{eqnarray}
or an {\it expanded} one
\begin{eqnarray}
\Gamma_f \, = \, \Gamma_f^{EW} \, \left[ 1 + {3 \over 4} Q_f^2 {\alpha \over \pi} 
+ {\alpha_s \over \pi}  + c_2 \, {\alpha_s^2 \over \pi^2} \right].
\end{eqnarray}
In the latter equations,   $\Gamma_f^{EW}$ denotes the electroweakly-corrected 
partial width. Owing to the lack of knowledge of the exact ${\cal O}(\alpha \alpha_s)$ 
corrections, the two realizations are of course equally correct and could be 
both implemented into a program as two different {\it options}. They differ by a 
term of the order of $3 \alpha \alpha_s Q_f^2  / 4 \pi^2$, that 
can be seen as a naive estimate of the unknown mixed corrections. The explicit 
calculation of such corrections should reduce the uncertainty moving it 
to ${\cal O}(\alpha \alpha_s^2)$. Indeed, this correction is today available, 
giving the result  
\begin{eqnarray}
\Gamma_f \, = \, \Gamma_f^{EW} \, \left[ 1 + {3 \over 4} Q_f^2 {\alpha \over \pi} 
- {1 \over 4} Q_f^2 {\alpha \over \pi} {\alpha_s \over \pi} + {\alpha_s \over \pi} 
+ c_2 \, {\alpha_s^2 \over \pi^2} \right],
\end{eqnarray} 
so that the two above realizations can not be longer seen as options. 

Another theoretical uncertainty, which is related to the pure weak  
corrections to the pseudo-observables,  
is the leading-remainder splitting in the effective couplings  
discussed in Sect.~\ref{sect:taaec}. These generally contain a leading  
universal part, which is usually resummed, and a non leading  
(remainder) part. The way of performing the separation between  
leading and remainder part and the way of treating the last are  
not uniquely defined, so that they are a source of theoretical  
uncertainty. A typical example is provided by $\Delta r$  
(the same reasoning, as the one shown in the following, also holds  
for the quantities $\delta \rho_f$ and $\delta k_f$) introduced  
in eq.~(\ref{eq:gmw}) and split into a leading and a remainder part  
according to eq.~(\ref{eq:deltar}), where the leading term contains the  
light fermion mass singularities and the terms proportional  
to $m_t^2$.  
In this separation the remainder contains, among others, logarithmic  
contributions in the {\it top}-quark and Higgs-boson masses, which can be numerically  
important.   
In resumming $\Delta r$ as in eq.~(\ref{eq:gmwr}), to take into account  
of the leading terms to all orders, different ways of treating  
$\Delta r_{rem}$ are in principle possible, \idest\   they are different  
options which give an estimate of the associated theoretical  
uncertainty~\cite{ewwg95}:  
\begin{eqnarray} 
& & {1\over {1 - \Delta r_L - \Delta r_{rem}}}, \nonumber \\ 
& & {1\over {1 - \Delta r_L}}\left( 1 + {{\Delta r_{rem}}\over  
{1 - \Delta r_L}} \right) , \nonumber \\ 
& & {{1 + \Delta r_{rem}}\over {1 - \Delta r_L}} , \nonumber \\  
& & {1\over {1 - \Delta r_L}} + \Delta r_{rem}. \nonumber 
\end{eqnarray} 
In the specific case of $\Delta r$, the last two expansions are not  
valid according to the arguments given in ref.~\cite{sirlin84}.
  
The two examples given above should clarify the meaning of intrinsic  
theoretical error and how to estimate it by considering proper options. Actually,  
apart from the factorization of QCD and EW corrections and the leading-remainder 
splitting, other options have been carefully examined  
by the Electroweak  Working  Group, such as the choice of the scale of $\alpha $ in the  
non-leading corrections (in particular the vertex corrections),  
the linearization of the radiatively corrected quantities like  
$\sin^2\vartheta_{eff}^l$, different choices of implementing  
the resummation of the vector-bosons self-energies, in particular  
the terms related to the Higgs-boson contribution, and the choice of the  
scale of $\alpha_s $ in the mixed electroweak-QCD corrections.  
The results obtained by the work 
of the Electroweak Working Group, based on a careful analysis of 
both the pseudo-observables and realistic ones, allowed to draw at 
that time (beginning of 1995) the following conclusions on the theoretical accuracy of 
the precision calculations for the $Z^0$ resonance~\cite{ewwg95}:
\begin{itemize}

\item the differences between the predictions of different computational tools are 
small compared to the experimental errors;

\item the parametric uncertainty due to $\Delta \alpha_{had}$ is the dominant 
source of error and the real ``bottleneck" to improve the theoretical accuracy; 
only new accurate measurements of the cross section of $e^+ e^- \to$~hadrons at low energy
could reduce it;

\item complete one-loop calculations, supplemented with higher-order leading 
effects are adequate  for $Z^0$ precision physics, but the control on 
two-loop electroweak sub-leading corrections would significantly reduce 
the still remaining uncertainty.

\end{itemize} 

The first two conclusions still remain valid today. Concerning the third point, 
progress in the calculation of important two-loop electroweak sub-leading corrections 
has been achieved after the conclusion of the work of the Electroweak  Working  Group. 
In fact, the two-loop next-to-leading heavy {\it top}-quark contributions of the 
order of $G_{\mu} m_t^2 M_Z^2$ to the $M_W$-$M_Z$ interdependence and to 
$\sin^2 \vartheta^{lept}_{eff}$ have been completely calculated 
in the $\overline{MS}$ framework and for two different realizations of the on-shell
scheme~\cite{dgvtwol96,dgstwol97} (see also~\cite{deltarh2,deltarwm2}). 
 These newly calculated corrections give a nice reduction of the theoretical error of  
 electroweak origin, especially for $M_W$ and 
$\sin^2 \vartheta^{lept}_{eff}$~\cite{dgstwol97,dgsp97}, even if it should be noticed that 
they have not yet been  implemented, at the time of writing, in standard computational tools 
for LEP1 precision physics.\footnote{After the completion of this work, the two-loop electroweak sub-leading
corrections of refs.~\cite{dgvtwol96,dgstwol97}, together with the QCD calculation of ref.~\cite{aasnew} and
the QED effects  studied in ref.~\cite{mnp97}, have been implemented in a new version of the code {\tt
TOPAZ0}~\cite{topaz040}. } 
\begin{figure}[hbt]
\begin{center}
\epsfig{file=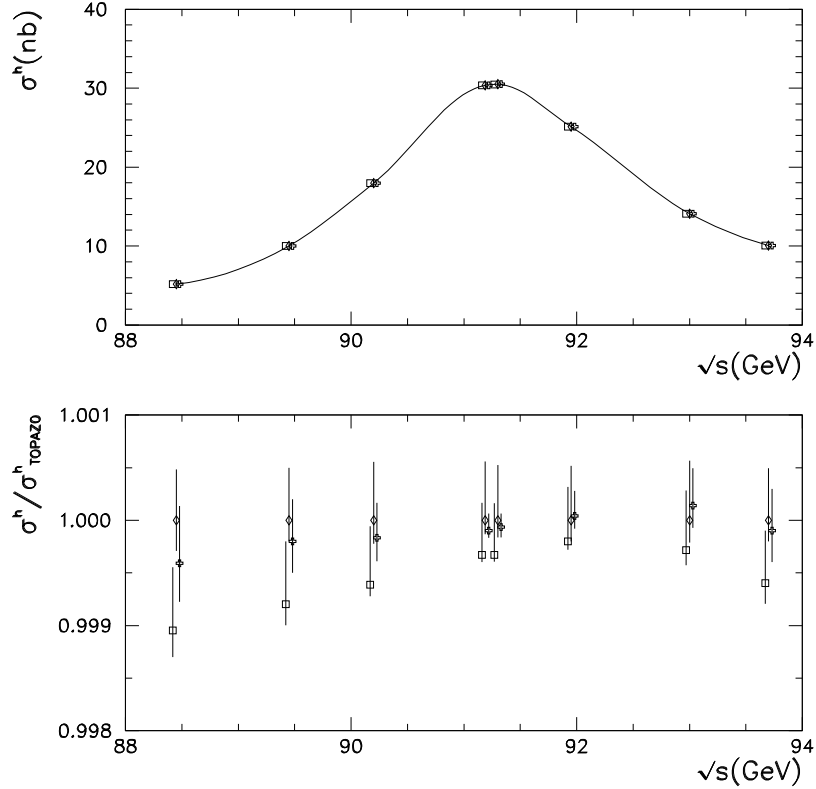,width=12truecm}
\end{center}
\caption{A comparison between theoretical predictions for the hadronic cross section. The squares, diamonds and
crosses represent {\tt BHM}~\cite{bhm94}, {\tt TOPAZ0}~\cite{topaz0} and {\tt ZFITTER}~\cite{zfitter92},
respectively (from ref.~\cite{ewwg95}). The lower window shows the predictions of the codes together
with their estimate of the theoretical error. }
\label{fig:hadcomp}
\end{figure} 

\begin{figure}[hbt]
\begin{center}
\epsfig{file=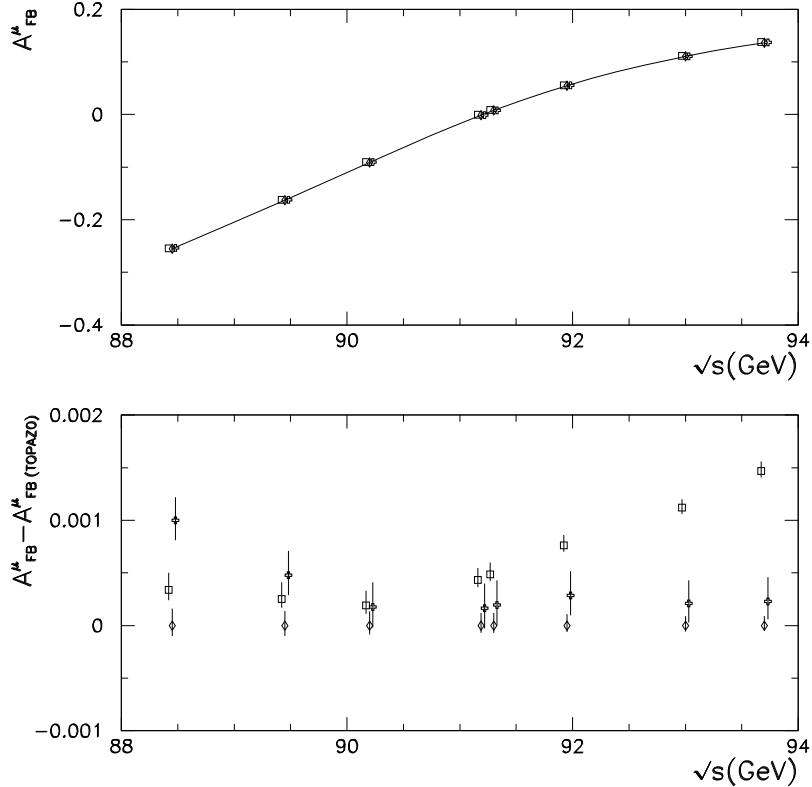,width=12truecm}
\end{center}
\caption{The same as in Fig.~\ref{fig:hadcomp} for the $\mu$ forward-backward 
asymmetry (from ref.~\cite{ewwg95}).}
\label{fig:mufbcomp}
\end{figure}

Concerning the theoretical uncertainty associated to the treatment of QED corrections, of 
utmost importance for the study of the realistic distributions, the conclusions of the 
analysis performed by the Electroweak Working Group, as obtained through a critical 
comparison of QED dressers 
over both an extrapolated and realistic set-up, can be summarized as follows:
\begin{itemize}

\item for $s$-channel annihilation processes, the theoretical error is of the order of 
0.1\%, almost independently of the considered c.m. energy (see 
Figs.~\ref{fig:hadcomp} and \ref{fig:mufbcomp});

\item for full Bhabha scattering, 
the theoretical accuracy can be estimated, with due caution, 
to be at the level of 0.1-0.2\% before and 
strictly around the peak, growing to about 1\% (depending on the imposed experimental cuts) 
on the hard radiative tail.

\end{itemize}  
 
It is worth pointing out that the above uncertainties associated to the photonic corrections 
actually match the statistics and systematics of the LEP1 data for the realistic distributions. 
In fact, whenever a somehow large theoretical error is present for full Bhabha scattering, 
this shows up in an energy regime (the hard radiative tail above the peak) 
where the experimental precision is much lower (\idest\ around an 
order of magnitude larger) than the corresponding theoretical accuracy. 

As far as $s$-channel processes are considered, the estimate of the QED theoretical accuracy 
provided by the Electroweak Working Group has been recently reinforced by a novel 
investigation of the 
effects of LL ${\cal O}(\beta^3)$ photonic corrections to the QED radiator~\cite{mnp97}, 
usually neglected 
in standard computational tools and not taken into account in the analysis 
of the Electroweak Working
Group. 
In fact, it has been shown that the ${\cal O}(\beta^3)$ corrections introduce, on an  
extrapolated cross section computed with a standard QED radiator with up to 
${\cal O}(\beta^2)$
non-soft terms, a systematic shift of around -0.07\%, that confirm {\it a posteriori} 
the estimate of the Electroweak Working Group. However, the size of the effect also indicates that these LL 
third-order QED corrections should be carefully taken into account in the theoretical predictions 
for the realistic observables of two-fermion production, in the light of the latest experimental 
data for some realistic observables, such as the cross section of $e^+ e^- \to$~hadrons.

Concerning the estimate 
of the theoretical error in the large-angle Bhabha scattering, it substantially traces 
back to a certain disagreement between the predictions of {\tt ALIBABA} and {\tt TOPAZ0} 
far from the peak and due, as discussed in the literature~\cite{topaz0np}, 
to a different implementation of 
QED final-state corrections, yielding a difference in the treatment of ${\cal O}(\alpha^2)$ 
sub-leading contributions. In fact, in {\tt ALIBABA} final-state corrections are 
implemented factorizing the leading terms only and summing up at ${\cal O}(\alpha)$ non-leading 
effects, whereas in {\tt TOPAZ0} a fully factorized prescription is followed, 
in order to preserve the so-called classical limit according to which the cross section of 
a real scattering process must vanish in the absence of electromagnetic radiation. 
This fundamental property is not guaranteed by an additive formulation. Furthermore, as shown and 
extensively discussed in \cite{a2l96}, a fully factorized form allows to keep under control in an 
effective way the bulk of ${\cal O}(\alpha^2)$ next-to-leading QED corrections of the order of   
${\cal O}(\alpha^2 L) $, that are on the contrary completely absent in the additive prescription 
(see also the discussion given in Sects.~\ref{sect:rcsabs}, \ref{sect:sabtoterr}  and   
\ref{sect:sabdp} on 
the r$\hat{ \hbox{\rm o}}$le 
of the ${\cal O}(\alpha^2 L)$ corrections in the present theoretical error 
to the SABH process). Therefore, with respect to an additive formulation, 
a factorized formulation is more accurate from the point of view of the perturbative content as 
well as more theoretically founded. However, after the completion  of the work  performed
in  ref.~\cite{ewwg95}, a much more extensive analysis of the large-angle Bhabha process has 
been performed
in ref.~\cite{bharep96}. In this last analysis,  several codes other than {\tt ALIBABA}  and 
{\tt  TOPAZ0},
noticeably {\tt  BHWIDE}, have been used for the comparisons. Moreover, also  more realistic ES's have
been adopted, namely calorimetric ES's in addition to the unrealistic BARE ES adopted in  
ref.~\cite{ewwg95}, and this is  a key  point in order to obtain a reliable error  estimate (see the
discussion in Sect.\ref{sect:sablm} for more details). 
Thanks to these results, it may  be concluded that the theoretical error of  the
large-angle Bhabha scattering  is  0.3\% on peak, and 0.5\% off  peak~\cite{bharep96}.

%% file: fits.txi
\section{Fits to Precision Data}
\label{sect:fits}
The application of the theoretical results discussed in 
Sect.~\ref{sect:z0phys} to the analysis of precision data 
collected at LEP1/SLC leads to the indirect determination of 
the fundamental parameters of the Standard Model (SM). Actually, 
the high experimental accuracy makes the electroweak 
measurements on the $Z^0$ peak sensitive to the mass of the particles 
circulating in the loops, although they are not energetically 
accessible. Hence the precision data can be used to infer valuable 
constraints on the 
{\it top}-quark and the Higgs-boson masses, as well as on the 
value of the strong coupling constant $\alpha_s$. The level of agreement 
between theory and experiment allows in addition to derive hints on possible 
new physics scenarios beyond the standard description of fundamental 
particles and interactions. 

The present section is devoted to illustrate the procedure adopted 
by the experiments to extract the electroweak $Z^0$ parameters from the directly 
measured production cross sections and forward-backward asymmetries. The
interpretation of the experimental results in terms of the SM parameters 
is discussed. The most recent indirect limits obtained for the {\it top}-quark 
and Higgs-boson masses {\it via} the virtual effects due to radiative corrections 
to the precision observables are presented, and 
compared with present information from direct searches. 
The determination of $\alpha_s$ from precision data at the
$Z^0$ pole is compared with other measurements in different processes and at 
different energy scales. The issue of a 
more model-independent attitude towards the precision measurements 
is in conclusion addressed, discussing 
the $S$-matrix approach as a framework for a model-independent determination of 
the $Z$-boson parameters, and the $\varepsilon$ parameterization of pure weak 
loops as a strategy for a model-independent analysis of the electroweak 
quantities. The implications of precision data for new physics beyond the SM 
are briefly examined.
\subsection{Determination of the Electroweak  Parameters}
\label{sect:detpar}
The ``primary'' observables measured by the LEP experiments are the cross sections and 
forward-backward asymmetries of the two-fermion processes 
$e^+ e^- \to \gamma Z \to f \bar f$ quoted as a function of the centre of  mass (c.m.) energy (see
Sect.~\ref{sect:z0par}). 
The channels considered  are the 
hadronic ($q \bar q$) and leptonic ($l^+ l^-$, $l=e,\mu,\tau$) final states.
The cross section data are determined by making use of the relation of 
eq.~(\ref{eq:sigexp}) 
in Sect.~\ref{sect:lummon}, once the number of events for 
a given channel has been determined  
and corrected for the trigger efficiency, the geometrical acceptance and 
the efficiency of the selection cuts. Analogously, 
the data for the forward-backward (FB) 
asymmetry $A_{FB}$ are obtained exploiting the definition 
\begin{eqnarray}
A_{FB} \, = \, { {N_F - N_B } \over {N_F + N_B} } ,  
\end{eqnarray}
where $N_F (N_B)$ is the number of events collected with a forward (backward) scattered 
fermion. This determination of the FB asymmetry is known as  the  {\it counting} 
method. Alternatively, the FB asymmetry is obtained from a maximum 
likelihood fit of an asymmetric differential cross section of the form  
\begin{eqnarray}
{{d\sigma} \over {d\cos\vartheta} } \, \propto \, 
1 + \cos^2\vartheta + {8 \over 3} A_{FB} \cos\vartheta 
\label{eq:dcs}
\end{eqnarray}
to the experimental angular distribution. The expression given by eq.~(\ref{eq:dcs}), where 
$\vartheta$ is the angle between the incoming electron and the outgoing fermion, is the 
form predicted by the electroweak theory in the lowest-order approximation 
and applies to $s$-channel annihilation only. For $e^+ e^- \to e^+ e^-$, due to the 
presence of $t$-channel contributions, a similar but more complicated expression is 
used. This second determination of the FB asymmetry is known as the {\it fitting} or {\it
likelihood} method.

From the measured hadronic and lep\-to\-nic cross sections and lep\-to\-nic 
for\-ward-back\-ward 
a\-sym\-\-metries, the pa\-ra\-me\-ters of the $Z^0$ 
resonance are extracted by the LEP Collaborations 
by means of a combined fit to the data~\cite{expew}. 
The $Z^0$ parameters are obtained using a $\chi^2$ 
minimization procedure with the correlations among the data (common experimental errors, 
theoretical luminosity error, uncertainties from the LEP energy calibration, etc.) 
taken into account using a covariance matrix. Several sets of parameters can be (and actually 
are) introduced to parameterize the measurements. However, in order to make possible a
combination of the results obtained by each of the four LEP Collaborations as 
well as for comparison purposes, 
the LEP experiments use a standard set containing {\it nine} free $Z^0$ parameters:
\begin{itemize}

\item the  $Z$-boson mass, $M_Z$, and its total width $\Gamma_Z$;

\item the hadronic pole cross section $\sigma_{h}^0$, defined by eq.~(\ref{eq:defpar}) in 
Sect.~\ref{sect:z0par};

\item the ratios $R_{e,\mu,\tau} = \Gamma_h / \Gamma_{e,\mu,\tau}$;

\item the pole asymmetries $A_{FB}^{e}, A_{FB}^{\mu}, A_{FB}^{\tau}$, that can be 
expressed in terms of the effective vector and axial-vector neutral current couplings of the 
fermions according to the relations given by eqs.~(\ref{eq:defasymfb})  
and (\ref{eq:afdef}) in 
Sect.~\ref{sect:z0par}.

\end{itemize} 
These parameters are chosen because they are most directly related to the experimental 
quantities and are weakly correlated. If the assumption of lepton universality 
is additionally introduced, then the numbers of parameters reduces from nine to five, \idest\
\begin{eqnarray}
M_Z, \quad \Gamma_Z, \quad \sigma_{h}^0, \quad R_l = \Gamma_{h}/\Gamma_l, 
\quad  A_{FB}^{l} , 
\end{eqnarray}
that actually is another standard set of parameters commonly used by the LEP Collaborations. 

The extraction of the $Z^0$ parameters from a combined fit to the hadronic and leptonic 
cross sections and leptonic forward-backward asymmetries requires that an appropriate 
parameterization for these last realistic  observables is introduced. 
By exploiting the relation  
between the effective couplings and the $Z^0$ partial width,  
the cross section can be written in terms of mass, total and    
partial widths of the $Z$ boson, without relying upon  
any particular assumption on the validity of  
the Standard  Model (SM)~\cite{bcms90,bbvm90,yrls89}:   
\begin{eqnarray}   
\sigma(s) = {{12 \pi \Gamma_e \Gamma_f}\over {\vert    
s - M_Z^2 + i M_Z \Gamma_Z(s)\vert^2}} \left(    
{{s}\over {M_Z^2}} + R_f {{s - M_Z^2}\over {M_Z^2}}    
+ I_f {{\Gamma_Z}\over {M_Z}} + \ldots \right) +  
\sigma_{\gamma} ,  
\end{eqnarray}   
where  
\begin{eqnarray} 
\Gamma_Z(s) = \Gamma_Z \left( {s\over {M_Z^2}} +  
\varepsilon {{s - M_Z^2}\over {M_Z^2}} + \ldots \right) ,  
\end{eqnarray}
and the partial widths are understood expressed in terms of the $Z^0$ parameters.   
The term $\sigma_{\gamma}$ identifies the $\gamma $-exchange  
contribution known from QED, $R_f$ and $I_f$ describe the  
$\gamma$-$Z$ interference and $\varepsilon $ is the correction  
due to finite final-state fermion mass effects.  
This procedure justifies the typical Breit-Wigner ansatz for  
the $Z^0$ contribution adopted by the LEP Collaborations,  
\begin{eqnarray} 
\sigma_Z^f = {{12 \pi }\over {M_Z^2}} \Gamma_e \Gamma_f  
{s\over {(s-M_Z^2)^2 + s^2 \Gamma_Z^2 / M_Z^2}} , 
\end{eqnarray} 
for a given $f \bar f$ channel, although the forms actually used by each  
experiment may differ in some detail.  
The photon-exchange contribution is calculated from QED and the 
$\gamma$-$Z$ electroweak interference is taken into account as well, usually computing it 
within the SM. Although the interference contribution is small around the $Z^0$ peak, 
it is important for the precise measurement of $M_Z$ and therefore some more general fit, 
where the SM dependence of the $\gamma Z$ effect is somehow relaxed, are actually 
performed by the experiments (see Sect.~\ref{sect:smatrappr} below). 

\begin{figure}[hbt]
\begin{center}
\epsfig{file=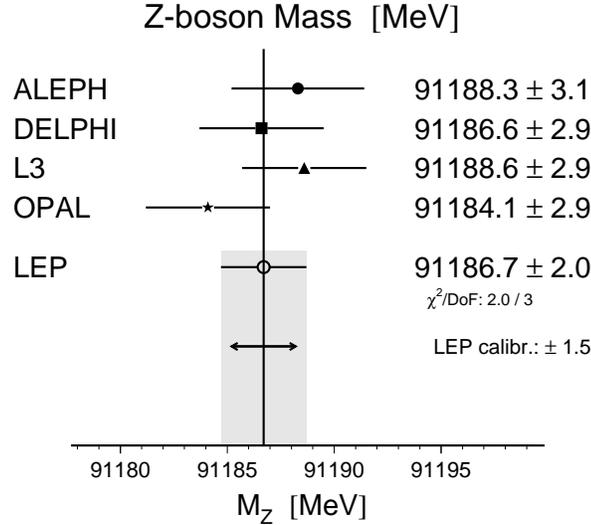, width=8truecm}
\end{center}
\caption{The $Z$-boson mass according to  the most recent data 
analyses~\cite{quast97,dward97}. }
\label{fig:mz}
\end{figure}

\begin{figure}[hbt]
\begin{center}
\epsfig{file=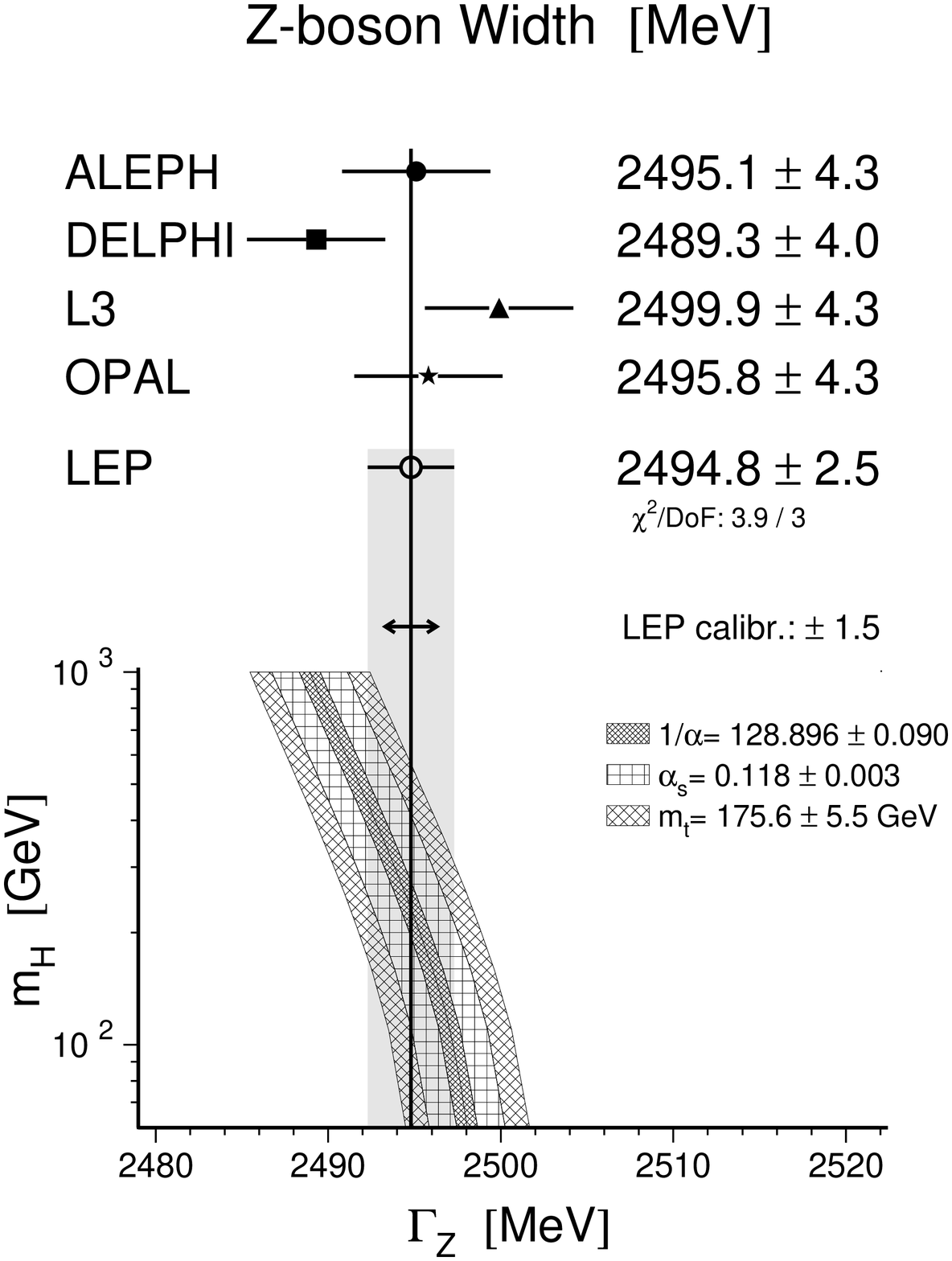, width=8truecm}
\end{center}
\caption{The $Z$-boson width according to  the most recent data 
analyses~\cite{quast97,dward97}. }
\label{fig:gz}
\end{figure}

It is worth pointing out that the above parameterization 
is introduced in order to describe just the short-distance behaviour of the cross section. 
Indeed, in the fit of the $Z^0$ parameters to the measured data, 
the QED radiative corrections enter as an unavoidable theoretical 
ingredient, and they are taken into account according to the algorithms described in
Sect.~\ref{sect:z0qed} and in Appendix~\ref{sect:upc}.   
A complication in the fitting procedure arises when considering the Bhabha channel 
$e^+ e^- \to e^+ e^-$, because of the presence of $t$-channel contributions. To minimize 
this contamination the experiments adopt a selection with large polar angle 
acceptance (typically $45^\circ < \vartheta < 135^\circ$) where the $s$-channel $Z$-boson  
exchange largely dominates. The remaining $t$ and $s$-$t$ contributions are generally calculated 
by  using the programs {\tt ALIBABA} and {\tt TOPAZ0}, in order to correct the 
cross section. The most recent LEP results for the $Z$-boson mass and width are shown in
Figs.~\ref{fig:mz} and \ref{fig:gz}, respectively. 

In addition to the above $Z^0$ parameters determined by means of a combined 
nine (five) parameters fit to the cross sections and 
asymmetries, other important electroweak observables are measured by the 
LEP experiments. Each of them is obtained according to a well-defined 
experimental strategy applied to a specific $f \bar f$ final-state. A 
first example is given by the $\tau$ polarization asymmetry $P^\tau$ that is 
defined as~\cite{jwyr89}
\begin{eqnarray}
P^\tau \, = \,  { {\sigma_R - \sigma_L} \over {\sigma_R + \sigma_L} },
\end{eqnarray} 
where $\sigma_{R(L)}$ denotes the $\tau$-pair cross section for the production of 
a right-handed (left-handed) $\tau^-$. This quantity is experimentally 
determined by measuring the longitudinal polarization of $\tau$ 
pairs produced in $Z^0$ decays. The left-right asymmetry $A_{LR}$ is another 
important manifestation of parity violation in weak interactions, that, however, 
requires longitudinal polarization of the initial-state electrons. Owing to the 
difficulties in achieving a high degree of polarization at LEP collider 
because of the presence of depolarizing effects, the left-right asymmetry has not 
been measured by the LEP experiments but by the SLD Collaboration at SLC, where 
the conditions for polarization measurements are much more favourable. In fact, 
an average degree of polarization of the incoming electrons $<P_e>$ reaching 
77\% has been obtained at SLC. The left-right asymmetry is determined by the 
SLD experiment exploiting the relation~\cite{frey97}
\begin{eqnarray}
A_{LR} \, = \, { {1} \over {<P_e>} } \, 
{ {N_{e_l} - N_{e_r}} \over {N_{e_l} + N_{e_r}} },
\end{eqnarray} 
where $N_{e_l}$ and $N_{e_r}$ are the observed numbers of hadronic events using 
left-handed and right-handed electron beams, respectively.

Further interesting electroweak observables in $Z^0$ decays 
are derived from the analysis of the final states containing 
$c$ and $b$ quarks ({\it heavy flavours}). 
These measurements require sophisticated tagging techniques of heavy flavours 
(such as mass or lifetime tagging), whose description is beyond the aim of the 
present review (see for instance ref.~\cite{btag} for an overview 
of the tagging methods used at LEP). The $Z^0$ parameters that are determined from these 
studies are
\begin{eqnarray}
R_b = \Gamma_b / \Gamma_{h}, \quad R_c = \Gamma_c / \Gamma_h, \quad
A_{FB}^{b}, \quad A_{FB}^{c},
\end{eqnarray} 
where $\Gamma_{b(c)}$ is the $Z^0$ width for  the decay into $b(c)$ quarks, 
$A_{FB}^{b}, A_{FB}^{c}$ are the $b$- and $c$-quark pole forward-backward 
asymmetry. As already remarked in Sect.~\ref{sect:z0par}, the results for 
$A_{FB}^{b}$ and $A_{FB}^{c}$ are quoted after having subtracted from the 
data, besides the effect of initial-state  radiation (ISR), 
also the effects of QCD corrections, in order to deal with 
a pure electroweak observable.   

\begin{figure}[hbt]
\begin{center}
\epsfig{file=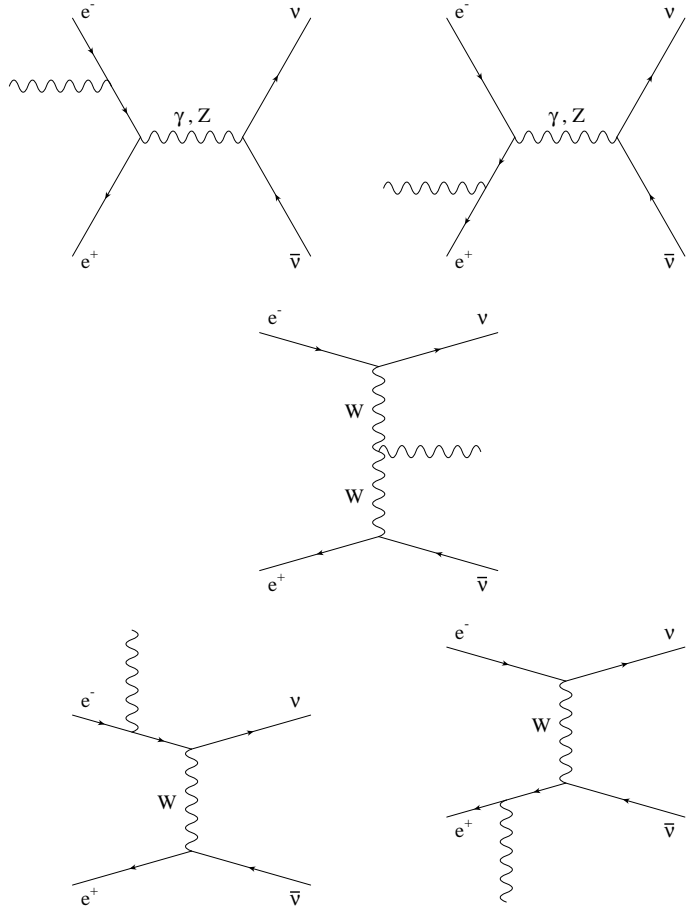, width=9truecm}
\end{center}
\caption{The tree-level Feynman diagrams for the process $e^+ e^- \to \nu \bar\nu \gamma$. }
\label{fig:ncr}
\end{figure}

Once the above ``primary'' $Z^0$ parameters are extracted from the data, 
other important additional quantities, such as the $Z^0$ partial widths 
$\Gamma_{h}$, $\Gamma_{l}$ and $\Gamma_{inv}$, are 
derived. $\Gamma_{inv}$ denotes the $Z$-boson invisible width that can be obtained  
{\it via} the relation (assuming lepton universality)
\begin{eqnarray}
\Gamma_{inv} \, = \, \Gamma_Z - \Gamma_{h} - 3 \Gamma_{l} .
\end{eqnarray}
Its present experimental value  is~\cite{quast97,dward97}
\begin{eqnarray}
\Gamma_{inv} = 500.1 \pm 1.8~{\rm MeV}.
\end{eqnarray}
This quantity is of particular interest for the extraction of the 
number of light neutrino species $N_\nu$. In order to 
get rid of the bulk of the non-negligible {\it top}-quark mass dependence of the 
partial widths, $N_\nu$ can be conveniently derived from the comparison 
of the experimental ratio $\Gamma_{inv}/\Gamma_{l}$ with the 
corresponding SM prediction, yielding the result~\cite{quast97,dward97}
\begin{eqnarray}
N_\nu \, = \, 2.993 \pm 0.011 . 
\end{eqnarray}
This determination, that firmly proves the existence of three standard 
light neutrino families, agrees well with the independent measurement 
obtained by means of the process $e^+ e^- \to \nu \bar\nu \gamma$, 
known as {\it radiative neutrino counting} reaction. It gives rise 
to a signature where only a single bremsstrahlung photon and nothing 
else is seen in the detector. The Feynman diagrams contributing in the SM to 
this process are depicted in Fig.~\ref{fig:ncr}.

The study of the {\it radiative neutrino counting} events offers the 
possibility to determine in a rather direct way the partial width of the 
$Z^0$ into invisible particles, since the $Z$-boson exchange contribution, largely 
dominating the $e^+ e^- \to \nu \bar\nu \gamma$ cross section around the $Z^0$ peak, is 
proportional to the number of neutrinos. For that reason, this procedure is known as
{\it direct determination of the invisible width}. In practice, the measured 
cross section as a function of the c.m. energy is compared with the SM calculation 
for different numbers of neutrino generations. A meaningful comparison between
theory and experiments requires that electroweak radiative corrections, 
especially the large effects introduced by ISR, to $e^+ e^- \to \nu \bar\nu \gamma$ 
are taken into account, although the statistical precision of the  
direct determination of the invisible width is smaller than the one achieved 
in the measurements of the observables of two-fermion 
production processes (see ref.~\cite{yrnc89} and references therein). 
Yet, the study of {\it radiative neutrino counting} events 
clearly rules out the existence of a fourth family with  light neutrinos in the SM.

In addition to the number of light neutrinos, there are other 
derived electroweak parameters that deserve mention for their r$\hat{ \hbox{\rm o}}$le in 
precision tests of the electroweak theory. From the partial widths 
of the $Z^0$ into charged leptons, $b$ and $c$ quarks, the $\tau$ polarization
asymmetry, the left-right asymmetry as well as the forward-backward asymmetry 
in leptonic and heavy-flavour channels,  the LEP/SLC experiments 
determine the vector and axial-vector neutral current couplings of fermions. 
These results for the effective $Z$-boson couplings $g_V^f$ and $g_A^f$  make use of the 
relations between the $Z^0$ parameters and the couplings themselves 
given by eqs.~(\ref{eq:defasymfb})-(\ref{eq:defasym}) and  (\ref{eq:afdef})  
for the asymmetries and 
polarizations,  
and eq.~(\ref{eq:widthpar})  for the partial widths given in Sect.~\ref{sect:z0par}.
In particular, the comparison between electron, muon and tau couplings is in 
good agreement with lepton universality, with a precision of about 0.2\% for $g_A$ and
$5 \div 10$\% for $g_V$~\cite{dward97}.  Furthermore, since the 
forward-backward and polarization asymmetries depend only on the 
ratio $g_V^f/g_A^f$, it is possible to express the asymmetry measurements 
in terms of a single parameter given by the effective weak mixing angle, \idest\  
\begin{eqnarray}
4\,|Q_f| \sin^2 \vartheta^{f}_{eff} \, = \, 1 - {{g_V^f} \over {g_A^
f}}.
\end{eqnarray}
The value of $\sin^2 \vartheta^{f}_{eff}$ is flavor-dependent 
due to the effects of weak vertex corrections that are non-universal. However, 
since the most precise results are obtained for the charged leptons, the 
mostly quoted value for the effective weak mixing angle coincides with 
$\sin^2 \vartheta^{l}_{eff}$, that allows to combine into 
a single electroweak parameter all the leptonic asymmetry measurements.

It is worth pointing out in conclusion that, as a result of the procedure 
described in the present Section, the $Z^0$ parameters that are extracted 
by the LEP experiments carry dependence on the whole stuff of
electro-weak, mixed electroweak/QCD as well as, for the quantities 
referring to  $q \bar q$ final states, QCD radiative corrections,  
reviewed in Sects.~\ref{sect:z0ew}-\ref{sect:qcdcorr}. Therefore, these 
observables can be conveniently used to measure quantum effects of the electroweak 
interactions, \idest\ to derive interesting constraints on the SM parameters that enter 
the predictions {\it via} radiative corrections only, noticeably the {\it top}-quark mass 
$m_t$, the Higgs-boson mass $m_H$ and the value of the strong coupling 
constant at the $Z^0$ pole $\alpha_s(M_Z)$. Such an analysis assumes the 
validity of the SM {\it ab initio}. However, if an appropriate 
strategy is organized, a model-independent analysis of the $Z^0$ parameters can be also performed, 
exploiting the high level of precision of the data in order to constrain 
new physics predicted by possible extensions of the SM. 
The physical and theoretical implications of the precision measurements of the 
electroweak $Z^0$ parameters are discussed in more detail in the following.   

\subsection{Standard Model Fits}
\label{sect:smfits}
A summary of the most recent measurements of the $Z^0$ parameters obtained 
by the LEP and SLC experiments is given in Tab.~\ref{tab:lepdata}. According 
to a standard presentation, the data are compared
 with the SM predictions (third column) corresponding to a fit of the 
electroweak data in terms of $m_t$, $m_H$ and $\alpha_s(M_Z)$, that will be
described in more detail in the following. In the fourth column one can also see 
the {\it pulls} derived from the fit, \idest\ the difference between each 
theoretical result and the corresponding experimental measurement, in units 
of the measurement error. A few but important comments about Tab.~\ref{tab:lepdata} 
are in order here.

The first remark concerns the precision level reached in the experimental 
measurements at LEP and SLC. The  $Z$-boson mass $M_Z$ 
is now known with a relative error of $2 \times 10^{-5}$ (absolute 
error of $\pm 2$~MeV). The total $Z^0$ width and the hadronic peak cross section 
are determined with a relative experimental uncertainty of the order of 
$1 \times 10^{-3}$ (with an absolute error for $\Gamma_Z$ of around $\pm 3$~MeV). 
All the asymmetry measurements have an absolute error smaller than $0.01$, 
generally at the level of a few $0.001$, and the effective weak mixing angle 
is derived with a precision of around $\pm 0.0003$. Therefore, in the light of 
these numbers, the LEP1/SLC experimental program can be considered as a 
great success, being the precision reached better than that expected at 
the beginning of operation.    

\begin{table}[hbt]
\caption{
Electroweak data according to the most recent determination 
(see refs.~\cite{quast97} and \cite{dward97}). 
}
\label{tab:z0par}
\begin{tabular}{llll}
\hline 
Quantity  & Data (Jerusalem '97) & Standard Model & Pull  \\
\hline
$M_Z$ [GeV] & 91.1867(20)	& 91.1866 & $+0.04$ \\
$\Gamma_Z$ [GeV] & 2.4948(25) & 2.4966 & $-0.73$ \\
$\sigma^0_h$ [nb]	& 41.486(53)  & 41.467 & $+0.36$ \\
$R_l$	& 20.775(27) & 20.756 & $+0.71$ \\ 
$R_b$ (LEP+SLC) & 0.2170(9) & 0.2158 & $+1.38$ \\ 
$R_c$ (LEP+SLC) & 0.1734(48) & 0.1722 & $+0.24$ \\
$A^l_{FB}$ &  0.0171(10) & 0.0162 & $ +0.89$ \\
$A_\tau$ & 0.1411(64) & 0.1471 & $-0.93$ \\
$A_e$ & 0.1399(73) & 0.1471 & $-0.98$ \\ 
$A^b_{FB}$ & 0.0984(24) & 0.1031 & $-1.95$ \\
$A^c_{FB}$ & 0.0741(48)	& 0.0737 & $+0.09$ \\ 
$A_b$  (SLC direct) & 0.900(50) & 0.935 & $-0.69$  \\
$A_c$  (SLC direct) & 0.650(58) & 0.668 & $-0.31$ \\
$\sin^2\theta_{eff}({\rm\hbox{LEP-combined}})$ & 0.23199(28) & 0.23152 & $+1.68$ \\
$A_{LR}\rightarrow  \sin^2\theta_{eff}$ (SLC) & 0.23055(41) &	0.23152	& $-2.37$ \\
$M_W$ [GeV] (CDF/D0+LEP2) & 80.43(8) & 80.375 & $+0.69$ \\
$1 - {{M_W^2} \over {M_Z^2}} \, (\nu  N)$ & 0.2254(37) & 0.2231 & $+0.63$ \\ 
$m_t$ [GeV] (CDF/D0) & 175.6(5.5) & 173.1 & $+0.45$ \\
\hline
\end{tabular}
\label{tab:lepdata}
\end{table}

A second comment regards the level of agreement between theory and experiment. 
No significant evidence for departures from the SM predictions is 
present. Close to the end of the LEP1/SLC experimental program, the data support 
the SM in a remarkable way. Looking  at the data in more detail, one could point out 
that there is an around $2\sigma$ deviation of the measured value of $A^b_{FB}$ 
with respect to the SM prediction, that $R_b$ is $1.4\sigma$ away from 
the SM expectation, that the determination of the effective weak mixing 
angle from $A_{LR}$ at SLC differs significantly from the LEP average. 
However, one should also emphasize that the observed pulls just follow the 
pattern that can be expected from a normal distribution of experimental 
measurements~\cite{dward97,gual97}. 
Moreover, the disagreement between the measured values of  $R_b$ and $R_c$ and their SM expectation,
that  received much attention in the past (see for instance ref.~\cite{mlmz0}  and references
therein), has now substantially disappeared. 
Presumably, future improved measurements of the $Z^0$ parameters, 
especially in the $b$-quark sector at LEP and $A_{LR}$ at SLC, will 
clarify the above issues.
\begin{figure}[hbt]
\begin{center}
\epsfig{file=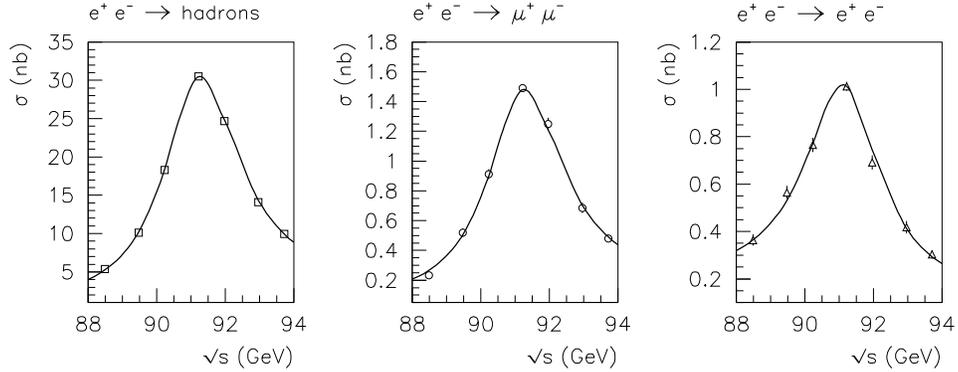, width=14.5truecm}
\end{center}
\caption{Data-theory comparison for realistic cross section data. Experimental data from
ref.~\cite{opal1}. Theoretical predictions by
{\tt TOPAZ0}~\cite{topaz0}.}
\label{fig:sigfit}
\end{figure}

Last but not least, the electroweak data contained in Tab.~\ref{tab:lepdata}, by virtue of 
their high precision, can be used to derive constraints on the  {\it top}-quark 
and  the Higgs-boson masses, as well as on the value of $\alpha_s(M_Z)$, fully exploiting the 
predictive power of the SM, as a renormalizable quantum field theory, beyond 
the tree-level approximation (for instance, several fits performed by various 
authors, using older
data sets,   can be found in 
refs.~\cite{hhm97,ds97,hollik96,holetal96,gurtu96,passarino96,efl96,dsw96,cp96}).  
Hence, the sensitivity of the electroweak data to quantum loops 
elevates the {\it precision physics} to the level of {\it discovery physics}. Moreover, 
the analysis can be generalized in order to constrain, disfavor or even rule out various 
extensions of the SM.
\begin{figure}[hbt]
\begin{center}
\epsfig{file=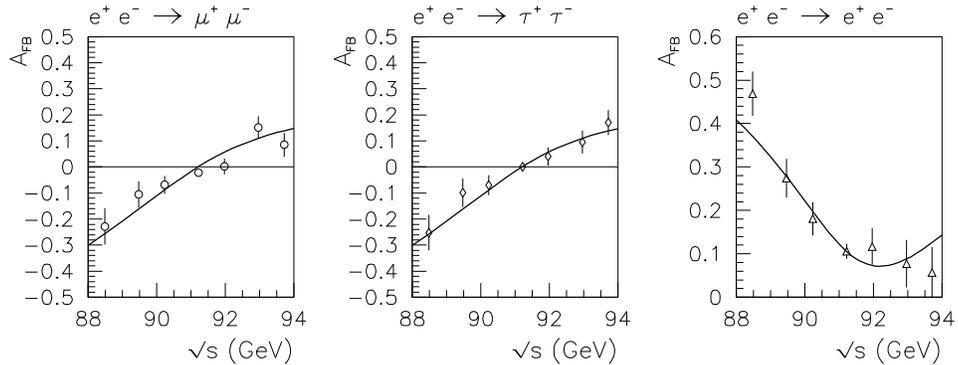, width=14.5truecm}
\end{center}
\caption{Data-theory comparison for realistic forward-backward asymmetry  data. 
Experimental data from
ref.~\cite{opal1}. Theoretical predictions by
{\tt TOPAZ0}~\cite{topaz0}.}
\label{fig:afbfit}
\end{figure}

The procedure described above is the one that is commonly adopted by the LEP experiments. Its main
advantage consists in the fact that it allows the extraction of parameters (the $Z^0$ parameters)
independent of the experimental details and hence easily comparable from experiment to experiment,
both at different machines and energy scales. Moreover, and as a corollary, the procedure offers the
possibility of fitting the unknown SM parameters by making use of a theoretical machinery which is
completely independent of the experimental selection criteria. On the other hand, it is intrinsically
a two-step procedure, requiring firstly the determination of the $Z^0$ parameters and secondly the
fitting  of the unknown  SM  parameters. It is worth noticing that an alternative but not antithetic    
procedure  is possible, namely considering as fundamental experimental quantities the data for the
realistic observables (see Sect.~\ref{sect:z0par}), and fitting the unknown SM parameters  directly
on them. This last procedure requires the use of theoretical tools that are able to provide
predictions for the realistic observables, taking into account all the details of the experimental
set-up. As a consequence, the required theoretical framework is sensibly more involved, but on the
other hand the procedure is a single-step one. 
This last strategy has been used in the past by the LEP
Collaborations  and also by other authors~\cite{mnpr93,topaz0np}, yielding results compatible with
the outcomes of the standard procedure. Figures~\ref{fig:sigfit} and~\ref{fig:afbfit} show the comparison
between the experimental data for realistic cross sections and forward-backward asymmetries of 
ref.~\cite{opal1} and the theoretical prediction  at best fit performed by {\tt TOPAZ0}~\cite{topaz0}. 

\subsubsection{Determination of the {\it top}-quark mass}
\label{sect:tqm}
\begin{figure}[hbt]
\begin{center}
\epsfig{file=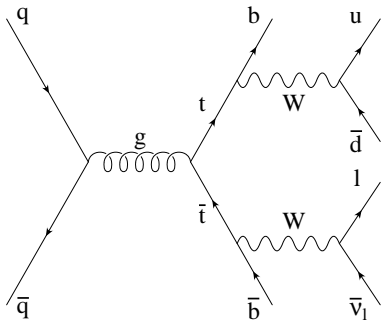, width=5truecm}
\end{center}
\caption{The main production mechanism for the {\it top} quark at the TEVATRON, 
yielding a six-fermion final state. }
\label{fig:toph}
\end{figure}

One the most important recent achievements in particle physics has been the 
discovery of the {\it top} quark by the experiments CDF and D0 at the $p \bar p$ 
collider TEVATRON at FERMILAB in Chicago~\cite{topdisc}. The present average of the mass values 
reported by the CDF and D0 Collaborations is 
$m_t = 175.6 \pm 5.5$~GeV\cite{topexp97}, with a measured 
{\it top}-quark production cross section in fairly good agreement 
with the QCD prediction~\cite{qcdtop}. 
The main production mechanism for the {\it top} quark at the TEVATRON is 
the quark-antiquark annihilation into a gluon followed by the creation of a 
$t \bar t$ pair. Due to its short lifetime, the {\it top} quark can not be detected directly 
but only {\it via} its decay products, \idest\  the weak decay $t \to W^+ b$, followed 
by the subsequent leptonic or hadronic $W$-boson decays $W \to l \bar\nu_l, u \bar d$. 
Hence, the production of the {\it top} quark at the hadron colliders gives rise to  
six-fermion final states ($b \bar b u \bar d l \bar \nu_l, b \bar b u \bar d d \bar u$), 
as depicted in Fig.~\ref{fig:toph}.

The {\it top}-quark mass is reconstructed by CDF and D0 measuring the energy and momenta 
of the final-state products in the dilepton (where both $W$ decay leptonically) and 
lepton-plus-jets (where one $W$ decays leptonically and the other one hadronically)
events.\footnote{Recently CDF employed also the fully hadronic sample, that however gives a less
precise determination of the {\it top}-quark mass. }  
The large value obtained for $m_t$ (the {\it top} quark is as heavy as a heavy nucleus) has as 
a direct consequence that the total width of this particle depends on  the third 
power of the mass, reaching the value of about 1.5~GeV for $m_t \simeq 175$~GeV. As said above, this 
corresponds to an extremely short lifetime of about $10^{-24}$~s, so that the {\it top} quark
decays immediately after its formation and $t \bar t$ bound states cannot be formed.

The direct evidence for the {\it top} quark in proton collisions and the measurement of its 
mass in a range around 170-180~GeV, in agreement with the indirect determination performed at
LEP/SLC,  constitutes a striking success for the SM at the 
quantum level and a remarkable confirmation of the precision tests of the electroweak 
interaction at LEP/SLC. Indeed, although the LEP energy is not sufficient to 
produce real {\it top} quarks, the effects of this particle and, in particular, of its 
high mass, can be felt in the precision measurements, {\it via} the radiative corrections induced 
by the loops containing the {\it top} quark as a virtual particle. In particular, thanks to the 
high statistics collected at the $Z^0$ resonance and the leading quadratic dependence of 
the theoretical predictions on $m_t$, the mass value of the {\it top} quark can be 
deduced from the virtual effects in electroweak observables with a precision 
comparable to the accuracy of the direct measurement.\footnote{The values quoted for $m_t$  refer to
the pole mass.}

Since the mass of the {\it top} quark is at present known from CDF and D0 with a
rather good precision, it is reasonable to adopt two different strategies in fitting the 
{\it top}-quark mass from the precision data, namely
\begin{itemize}

\item consider all the available electroweak data with the exclusion of the CDF/D0 
measurement of $m_t$; 

\item consider the whole set of data, including $m_t$ from CDF/D0.

\end{itemize}
The former procedure allows to establish whether the estimate of $m_t$ from radiative 
corrections is in agreement with the direct measurement at the TEVATRON. The latter, 
being a more general type of fit, allows to test the overall consistency of the SM 
in a more complete way and to obtain the present best estimate of derived quantities 
such as the $W$-boson mass and $\sin^2\vartheta_{eff}$. 

The results obtained according to the above  strategies, when fitting the data 
with $m_t$, $m_H$ and $\alpha_s(M_Z)$ as free parameters, are summarized in 
Tab.~\ref{tab:asmtmh}. 
\begin{table}[hbt]
\caption{
Fits to $\alpha_s$, $m_t$ and $m_H$ (from ref.~\cite{dward97}). 
}
\label{tab:asmtmh}
\begin{tabular}{llll}
\hline 
Parameter & LEP (incl. $M_W$) & All but $M_W$ and $m_t$ & All data \\
\hline
$m_t$ [GeV] & 158$^{+14}_{-11}$ & 157$^{+10}_{-9}$ & $173.1 \pm 5.4 $ \\
$m_H$ [GeV] & 83$^{+168}_{-49}$ & 41$^{+64}_{-21}$ & 115$^{+116}_{-66}$ \\
$\alpha_s(M_Z)$ & $0.121 \pm 0.003$ & $0.120 \pm 0.003$ & $0.120 \pm 0.003$ \\
$\chi^2/dof$ & 8/9 & 14/12 & 17/15 \\
\hline
\end{tabular}
\end{table}
\begin{figure}[hbt]
\begin{center}
\epsfig{file=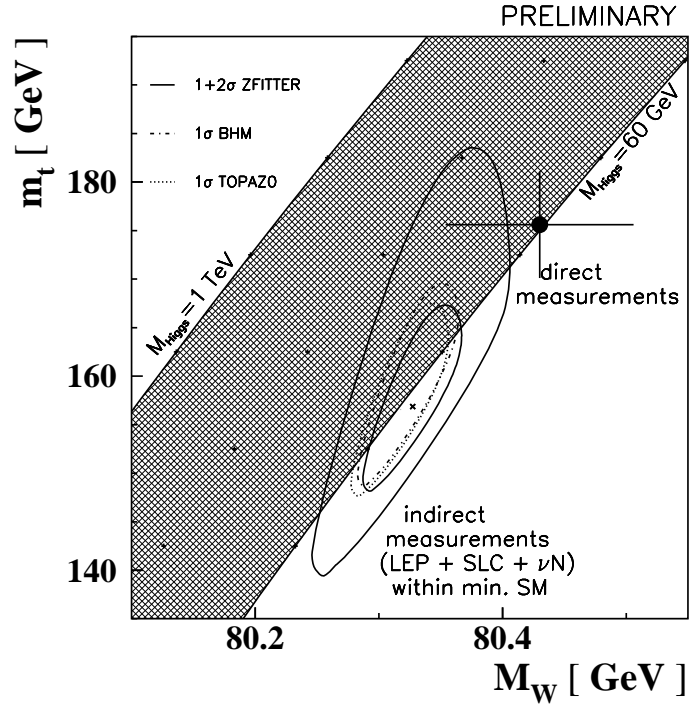,width=10truecm}
\end{center}
\caption{Contour plot showing the $top$-quark  mass  $m_t$ 
versus the $W$-boson mass. The shaded area shows the SM prediction for the
Higgs-boson mass  ($M_{Higgs}$ in the figure) between 60~GeV and 1~TeV. The
theoretical predictions are performed by means of the codes {\tt BHM}~\cite{bhm94}, {\tt
TOPAZ0}~\cite{topaz0} and {\tt ZFITTER}~\cite{zfitter92}
(see ref.~\cite{quast97}).}
\label{fig:mwmt}
\end{figure}
As can be seen from the  table, the fitted mass of the {\it top} quark is 
somewhat lower than 
the direct measurement, even if in well agreement with it. This compatibility, as already 
remarked, is an impressive confirmation of the SM at the level of quantum effects 
and justifies combining the direct and indirect derivation of $m_t$.
The $1\sigma$ 
errors returned by the fit include the uncertainties due to the 
error on $\alpha(M_Z)$ (inducing a $m_t$ variation of about $\pm 4$~GeV) and to 
missing higher-order electroweak corrections (affecting $m_t$ by about $\pm 1$~GeV).  

By adopting the procedure of including the direct measurements of $m_t$ among the input data, 
then typical results obtained are~\cite{quast97,dward97}
\begin{eqnarray}
&&m_t = 173.1 \pm 5.4~{\rm GeV}~, \nonumber \\
&&m_H = 115^{+116}_{-66}~{\rm (or}~m_H < 420~{\rm GeV~at~95\%~CL}),  \nonumber\\
&&\alpha_s(M_Z) = 0.120 \pm 0.003.
\label{eq:fit2}
\end{eqnarray}
From this fit, an indirect measurement of the $W$-boson mass $M_W$ can be derived (see
Fig.~\ref{fig:mwmt}), \idest\
\begin{eqnarray}
M_W \, = \, 80.375 \pm 0.030~{\rm GeV}.
\end{eqnarray}
This small error of 30~MeV on $M_W$ from radiative corrections is clearly a 
challenge for future direct precision measurements of $M_W$ at LEP2 and at the TEVATRON.

\subsubsection{Determination of the Higgs-boson mass}
\label{sect:hm}
As already shown in Sect.~\ref{sect:tqm}, the electroweak precision data can be 
used to infer indirect limits on the mass of the yet elusive Higgs boson. As 
discussed  in Sect.~\ref{sect:z0ew}, the leading effect of the Higgs-boson mass on the observables 
is only logarithmic and correlated with $m_t$. These difficulties, associated with other 
problematic aspects that will be discussed later, naturally set the question how 
reliably the Higgs-boson mass $m_H$ can be predicted from electroweak data. Clearly, the 
answer to this question is of utmost importance for planning the search for the Higgs boson 
at LEP2 and future accelerators. 

\begin{figure}[hbt]
\begin{center}
\epsfig{file=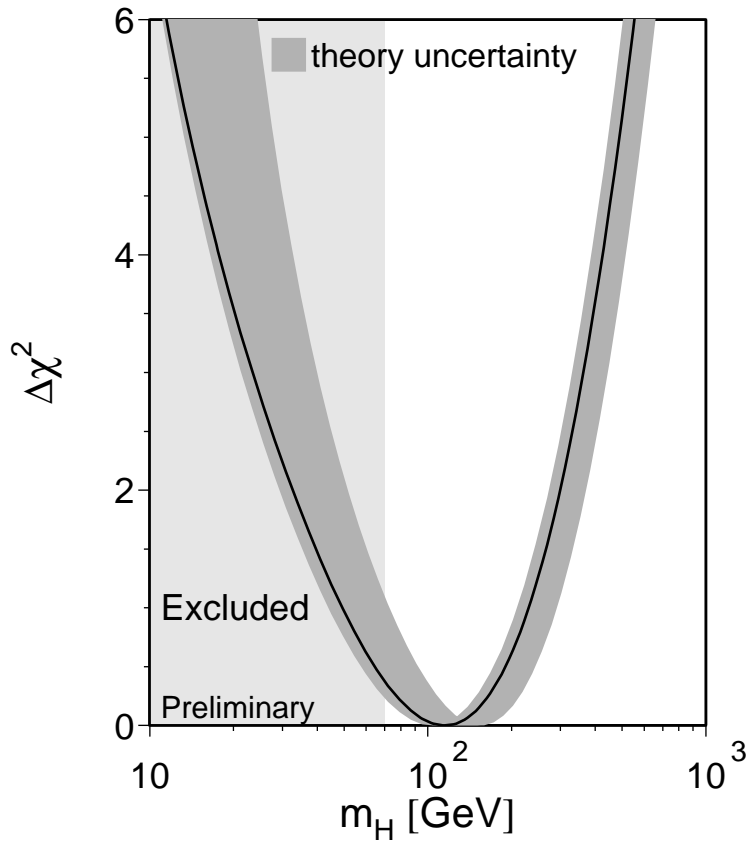,width=10truecm}
\end{center}
\caption{$\Delta \chi^2$ versus the Higgs-boson mass  (see  ref.~\cite{dward97}).}
\label{fig:mhsyst}
\end{figure}

The indirect bounds that can be put on $m_H$ from radiative corrections effects and their 
validity are matter of debate in the community of particle physicists and an unambiguous, 
unique answer to the above question can not be given, especially whenever one takes into 
account the large amount of investigations and detailed analyses 
present in the literature, many of which appeared after the discovery of the {\it top} quark 
at the TEVATRON~\cite{chan97,rosner97,dsw96b,hioki96,dsk95,dgs95,ch95,
dskk94,efl94,mnpp94,norvy94,norv94}. 
Nonetheless, there are aspects concerning the constraints on $m_H$ that 
are common to most of the analyses of precision data and therefore deserve special mention. 
Before entering the details of the discussion, it is worth recalling that 
negative searches at LEP put the present lower limit on the Higgs-boson mass 
$m_H > 77$~GeV at 95\%~CL, derived assuming the validity of the minimal SM~\cite{murray97,janot97}. 

The first common feature concerns the dependence of the fitted central value for $m_H$ and 
of the $1\sigma$ errors on the set of input data considered. This rather strong 
dependence can be clearly seen, for example, from Tab.~\ref{tab:asmtmh}, looking at the 
variations of several tens of GeV on $m_H$, when considering the LEP data alone, 
all the available the  data with the exception  of $M_W$ and $m_t$, 
and the whole set of data. 
It has been in particular emphasized by several authors that the exclusion of the SLD 
data for $A_{LR}$ moves up $m_H$ considerably, as a symptom of the clash between the 
$\sin^2\vartheta_{eff}^{l}$ determination from LEP and SLC. Actually, the SLD data 
alone leads to an unnatural bound on $m_H$ at the $1\sigma$ level, in conflict 
with the limit from direct searches at LEP~\cite{gurtu96,chan97,s97}.

\begin{figure}[hbt]
\begin{center}
\epsfig{file=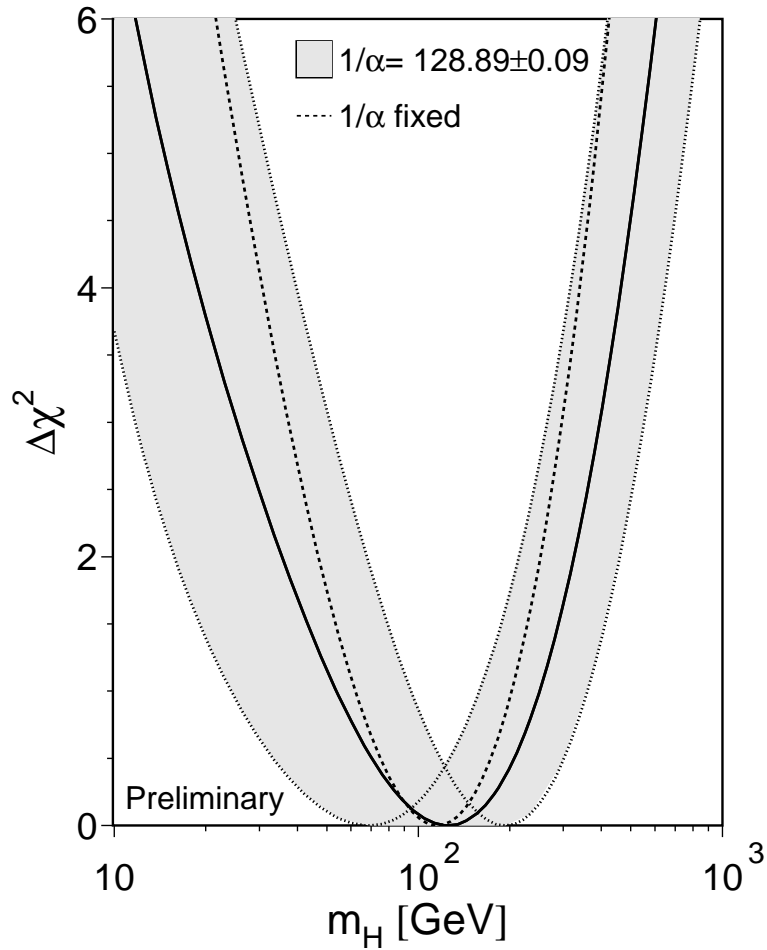,width=10truecm}
\end{center}
\caption{$\Delta \chi^2$ versus the Higgs-boson mass, taking into account the parametric
uncertainty on $\alpha (M_Z)$  (from~\cite{bolek96}).}
\label{fig:chi2mh_aem}
\end{figure}

A second common feature concerns the effect of including  in the fit the theoretical uncertainties, 
both {\it intrinsic} and {\it parametric}, of the SM calculations 
discussed in Sect.~\ref{sect:z0thu}. Contrary to 
$m_t$, that is marginally influenced by missing higher-order electroweak corrections and 
parametric uncertainties, the upper bounds on $m_H$ are sensibly affected by 
parametric and intrinsic theoretical errors~\cite{mnpp94}. For example, the upper bound at 95\%~CL 
can vary by around 100~GeV as a result  
of different implementations of radiative corrections beyond the one-loop level, so that 
the $\chi^2$ as a function of $m_H$ appears as a band, representing the {\it intrinsic}
theoretical error of the SM predictions, rather than a single curve 
(see Fig.~\ref{fig:mhsyst}).\footnote{The recently calculated two-loop next-to-leading electroweak
corrections of the order of $G_\mu^2 M_Z^2 m_t^2$ are expected to reduce this theoretical
uncertainty, provided they are included in the standard electroweak libraries. Actually, 
after the completion of
this work, the program {\tt TOPAZ0} has been upgraded to include these corrections, together with the
results of refs.~\cite{aasnew} and \cite{mnp97} concerning QCD and QED corrections,
respectively~\cite{topaz040}.} Similar 
considerations apply to the $\chi^2$ distribution as a function of $m_H$  
when the input parameter of the fit $\alpha(M_Z)$ changes within its error. This 
parametric variation is actually larger than the one induced 
by different treatments of higher-order effects, pointing out the limitation 
imposed by the uncertainty on $\alpha(M_Z)$ in a sensible derivation of $m_H$ 
from precision data (see Figs.~\ref{fig:chi2mh_aem}  and \ref{fig:mhas}). 

\begin{figure}[hbt]
\begin{center}
\epsfig{file=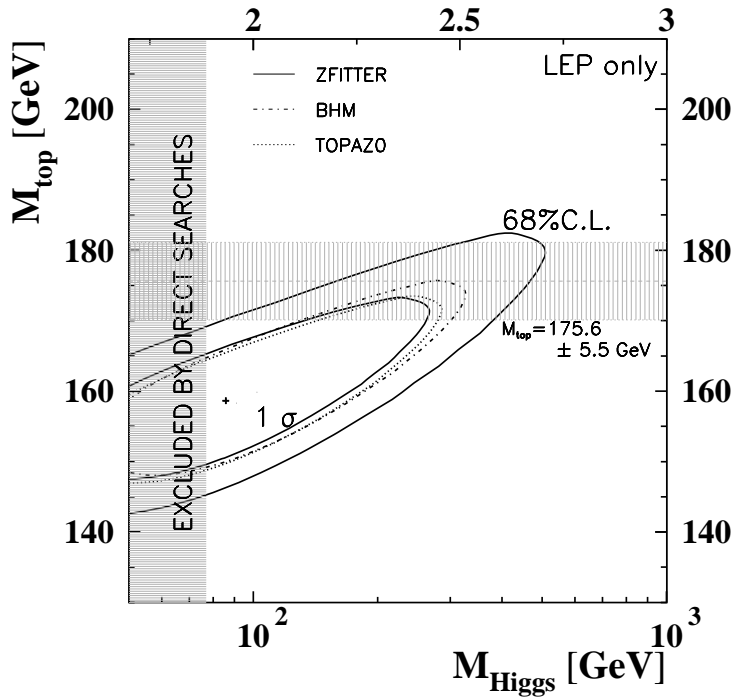,width=10truecm}
\end{center}
\caption{Contour plot showing the $top$-quark  mass  ($M_{top}$ in the figure) 
versus the Higgs-boson mass ($M_{Higgs}$ in the figure). The
theoretical predictions are performed by means of the codes {\tt BHM}~\cite{bhm94}, {\tt
TOPAZ0}~\cite{topaz0} and {\tt ZFITTER}~\cite{zfitter92}
(see ref.~\cite{quast97}).}
\label{fig:mhmt}
\end{figure}

In the light of the above {\it caveat} in establishing a quite precise upper bound on 
$m_H$, it turns out to be difficult going beyond the conclusion that the electroweak 
precision data, even when accounting for the direct measurement of $m_t$, imply 
an indirect upper bound at 95\%~CL of around 400-500~GeV, with a preference for 
a ``light'' Higgs boson of mass around $100  \div 150$~GeV (see Figs.~\ref{fig:mhmt} and 
\ref{fig:mhas}).

\subsubsection{Determination of $\alpha_s$}
\label{sect:alphas}
Besides the mass of the {\it top} quark and of the Higgs particle, the precision 
measurements of the electroweak parameters at the $Z^0$ pole provide the opportunity 
of an accurate determination of the coupling constant of the strong interactions 
$\alpha_s$. Actually, $e^+ e^-$ machines are an ideal laboratory for QCD studies. Being the
hadronic activity in $e^+ e^-$ collisions restricted to the final state, the 
measurement of $\alpha_s$, as well as further tests of QCD, can be carried out in 
a particularly clean environment, where the experimental signatures of 
hadronic events are largely free of backgrounds (see ref.~\cite{revqcdexp} 
for reviews on the subject). 

\begin{figure}[hbt]
\begin{center}
\epsfig{file=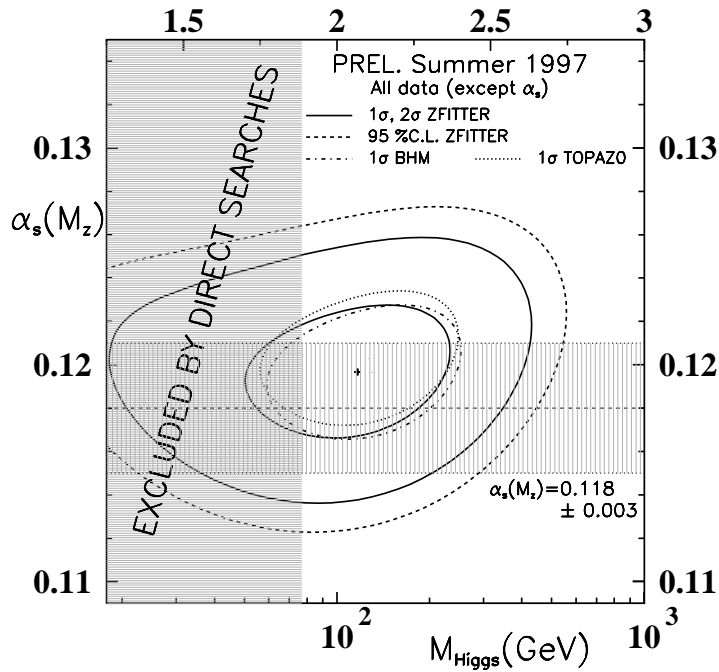,width=10truecm}
\end{center}
\caption{Contour plot showing $\alpha_s (M_Z)$ versus the Higgs-boson mass
($M_{Higgs}$ in the figure). 
The theoretical predictions are performed by 
means of the codes {\tt BHM}~\cite{bhm94}, {\tt
TOPAZ0}~\cite{topaz0} and {\tt ZFITTER}~\cite{zfitter92}
(see ref.~\cite{quast97}).}
\label{fig:mhas}
\end{figure}

Owing to the renormalization group dependence of the strong coupling constant 
on the hard energy scale $Q$ (see the discussion in Sect.~\ref{sect:z0qcd}), a 
reference energy scale needs to be specified when quoting a value for $\alpha_s$. 
Because of the large amount of data collected at LEP1 and SLC, it has become 
conventional to use $Q^2 = M_Z^2$,  $M_Z$ being the $Z$-boson mass, provided that the measured value
of $\alpha_s$ is run from the scale at which the measurement takes place to $M_Z$ (from  now on,
$\alpha_s$ will be used as a short-hand  notation for $\alpha_s(M_Z)$). 
This is in turn 
a great advantage since it allows to compare independent determinations of $\alpha_s$ 
in different reactions ($e^+ e^-$ annihilation, hadron-hadron collisions and 
deep-inelastic lepton-hadron scattering) at different energy scales, covering a 
range of $Q^2$ from roughly a few to $10^5$~${\rm GeV}^2$. In particular, 
``low-$Q^2$'' and ``high-$Q^2$'' results can be compared, thus providing a non-trivial 
check of the running of $\alpha_s$ as predicted by QCD, and a world-average value of 
$\alpha_s$, as obtained by averaging all the available measurements once extrapolated to 
the $Z$-boson mass, can be derived. 
 
\begin{figure}[hbt]
\begin{center}
\epsfig{file=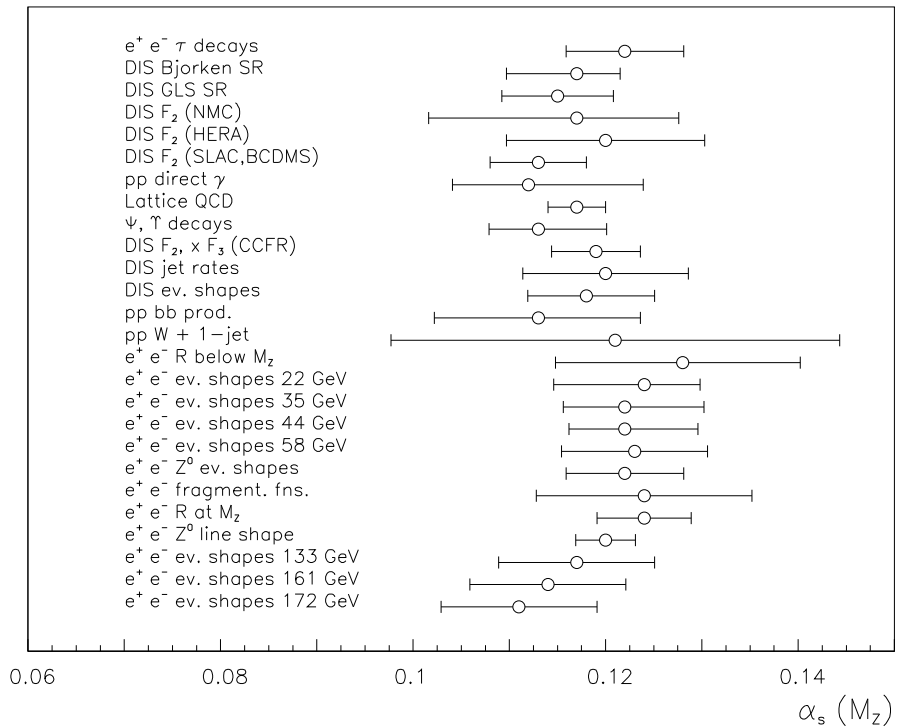,width=15truecm}
\end{center}
\caption{The $\alpha_s(M_Z)$ values and total errors from  measurements at $e^+ e^-$, 
lepton-hadron and hadron-hadron machines, and calculations from lattice  QCD. 
The results are ordered vertically in $\protect{\sqrt{Q^2}}$.}
\label{fig:alphas}
\end{figure}

The data for the electroweak parameters can be used 
to measure $\alpha_s$ according to two different procedures, \idest\
\begin{itemize}

\item performing a fit, with $\alpha_s$ as a free parameter, 
to the single observable given by the ratio of hadronic to leptonic width $R_l$;

\item performing a global fit to the whole set of the electroweak data (usually 
including the direct measurement of $m_t$), by allowing the three parameters $m_t$, $m_H$ and 
$\alpha_s$ or the two parameters $m_H$ and $\alpha_s$ to vary.

\end{itemize} 

\begin{table}[hbt]
\caption{World averages of $\alpha_s(M_Z)$}
\label{tab:alphasav}
\begin{tabular}{lll}
\hline 
Value & Error &  Reference \\
\hline
0.118 & 0.003 & ref.~\cite{schmelling96} \\
0.118 & 0.005 & ref.~\cite{burrows97} \\
0.118 & 0.004 & ref.~\cite{stirling97} \\
0.119 & 0.006 & ref.~\cite{bethke97} \\
0.119 & 0.004 & ref.~\cite{gual97} \\
0.119 & 0.005 & ref.~\cite{catani97} \\
0.119 & 0.004 & ref.~\cite{dward97} \\
\hline
\end{tabular}
\end{table}

For the inclusive ratio $R_l$, the QCD corrections are known up to ${\cal O}(\alpha_s^3)$ for 
massless quarks, as discussed in Sect.~\ref{sect:z0qcd}, and the electroweak contributions are well 
under control since the main part of the weak radiative corrections, being common to the leptonic and 
hadronic $Z^0$ decay channels, cancel in the ratio of partial widths. From the experimental point of view, 
$R_l$, being a ratio, is known with a precision higher than that of single partial widths and is 
slightly affected by the luminosity uncertainty.
Using the combined LEP result for $R_l$ the fitted value of $\alpha_s$ is~\cite{quast97}
\begin{equation}
\alpha_s \, = \, 0.124 \pm 0.004 ({\rm exp.}) \pm 0.002 (m_H) ,  
\end{equation}    
where the experimental error is essentially due to the limited data sample and 
the theoretical error ($m_H$) derives 
from the residual dependence on  $m_H$. 
It is worth noticing that 
the extraction of $\alpha_s$ from $R_l$ doesn't require any specific knowledge about 
the hadronization mechanism. This $\alpha_s$ determination at the $Z^0$ peak is exactly analogous 
to the procedure followed when considering the published measurements for the 
ratio $ R = \sigma(e^+ e^- \to {\rm hadrons}) / \sigma(e^+ e^- \to \mu^+ \mu^-)$ at energies 
below the $Z^0$ resonance. The value obtained in such a case when fitting $\alpha_s$ to $R$ 
as measured in the energy range $5 \leq \sqrt{s} \leq 65$~GeV is~\cite{burrows97,bes94}
\begin{equation}
\alpha_s \, = \, 0.128^{+0.012}_{-0.013} \pm 0.002 (m_H) ,
\end{equation} 
where the second theoretical error is due to the variation of $m_H$ in the range 60-1000~GeV.

An example of the second kind of determination of $\alpha_s$ from the $Z^0$ line-shape data is given by the 
three-parameter fit discussed at the end of Sect.~\ref{sect:tqm}, yielding the
result~\cite{quast97,dward97}
\begin{equation}
\alpha_s \, = \, 0.120 \pm 0.003 ({\rm exp.}) .
\end{equation} 
Detailed analyses of the theoretical uncertainties underlying the above determinations suggest that 
they can be estimated to contribute about $\pm 0.002$~\cite{ck95}. 
Actually, for all the input observables the 
non-perturbative effects are expected to be of the order of $1/M_Z$ and are 
hence usually neglected. 

It should be kept in mind that the results for $\alpha_s$ from  $Z^0$ data are 
obtained under the assumption that the hadronic width $\Gamma_{h}$ is given by the SM. Therefore,
a possible anomaly in $\Gamma_b$ is a potential source of bias in the $\alpha_s$ derivation.

Besides the above quoted determinations of $\alpha_s$, 
other measurements performed  at $e^+ e^-$, lepton-hadron and hadron-hadron machines 
need to be mentioned for their relevance, together  with the calculations from lattice QCD. 
The interested reader is referred to the  detailed 
compilations that can be found in refs.~\cite{burrows97},
\cite{stirling97}, \cite{bethke97} and \cite{dward97}, 
together with references to the original literature. 

A compilation of  $\alpha_s(M_Z)$ determinations  is given in Fig.~\ref{fig:alphas}, 
together with the corresponding total error. 
It is worth noticing that the most precise 
determination of $\alpha_s$ is obtained from a global fit to the whole set of precision 
data ($Z^0$ line-shape determination).
By inspection of Fig.~\ref{fig:alphas} it can be seen that there is agreement, within the errors, 
between ``low-$Q^2$'' and ``high-$Q^2$'' measurements. Contrary to a few years ago, 
where a discrepancy between low- and high-energy determinations was observed, the 
present status of $\alpha_s$ measurements is quite satisfactory, basically as a consequence 
of the fact that, while the $\alpha_s$ determination from deep-inelastic scattering and 
lattice QCD increased, the precise $\alpha_s$ measurement from fitting the 
electroweak data decreased. The problem of determining a world average of $\alpha_s$ has been
considered by several authors. In Tab.~\ref{tab:alphasav} the most recent central values and
corresponding errors are quoted, together with reference to the original literature. 

\subsection{Model Independent Approaches and Physics Beyond the Standard Model}
\label{sect:miappr}
The possibility of extracting predictions for the mass of the {\it top} quark and of the 
Higgs boson, as well as for $\alpha_s(M_Z)$, {\it via} the analysis of the electroweak 
precision data, as described in the previous sections, relies upon the assumption of the 
validity of the SM. In spite of the great success of such a strategy, it is of course  
important to exploit the high precision of LEP1/SLC measurements in order to 
test the validity of the SM in a way as independent as possible of the details of the 
underlying theory and, possibly, to derive also constraints on models of new physics 
beyond the SM. In order to achieve these goals, a model independent strategy for the 
analysis of precision data has to be organized. 
In the following, a  model independent way of  
extracting  $Z^0$  parameters (the $S$-matrix  approach) is shortly described. 
Next, the $\varepsilon$ parameterization as a tool for a model independent analysis of the 
$Z^0$ parameters is considered. 
At last, the implications of precision data for physics beyond the SM are briefly discussed. 

\subsubsection{$S$-matrix approach}
\label{sect:smatrappr}
The original idea behind the parameterization of the $Z^0$ line shape  
introduced in ref.~\cite{bcms90,bbvm90,yrls89} was  
to allow a model independent extraction of the fundamental  
parameters of the $Z^0$ resonance, such as its mass, width and  
decay rates. In the following years this procedure has been  
posed on a firmer ground by appealing to the $S$-matrix  
theory~\cite{chew,eden,bohm,martin85}.  
Within this framework, the scattering amplitude can be cast into  
the form of a Laurent expansion around the $Z^0$ resonance  
as follows~\cite{stuart91,stuart91ii,stuart97}:  
\begin{eqnarray} 
A(s) = {R \over {s - s_p}} + \sum_{n=0}^{\infty} B_n (s - s_p)^n =  
{R\over {s - s_p}} + B_0 + {\cal O}(\Gamma_Z^2/M_Z^2) , 
\label{eq:lspole}  
\end{eqnarray} 
where the quantities $R$, $s_p$ and $B_0$ are complex numbers.  
Taking into account that an overall phase in the amplitude is  
unobservable, the cross section depends on five real parameters,  
which are separately gauge invariant, because the amplitude is  
an analytical function of $s$. The position of the complex pole  
defines the mass $M$ and the width $\Gamma$ in a gauge invariant  
and process independent way~\cite{levy,peierls,cs86}  
through the relation  
\begin{eqnarray} 
s_p = M^2 - i M \Gamma . 
\end{eqnarray} 
In the expansion given by eq.~(\ref{eq:lspole}), the photonic  
contribution to the amplitude is part of the background  
denoted by $B_0$. Another approach followed in the  
literature~\cite{lrr91} is to introduce explicitly the photon  
exchange contribution according to the parameterization  
\begin{eqnarray} 
A(s) = {{R_Z}\over {s - s_p}} + {{R_{\gamma}}\over {s}} + B(s), 
\label{eq:ldpole}  
\end{eqnarray} 
which has been implemented in the computer code {\tt SMATASY}.  
In principle the parameterization of eq.~(\ref{eq:ldpole})  
may lead to some difficulties related to the fact that the  
coefficients $R_{\gamma}$ and $B_i$ are not independent  
quantities~\cite{lrr91,stuart97}, but they disappear if a truncated  
version is employed. Neglecting terms of the order of $\Gamma^2/M^2$,  
the cross section for a given final state can be written in the  
following form~\cite{riem97}: 
\begin{eqnarray} 
\sigma(s) = {4\over 3}\pi \alpha^2 \left[ {{r_T^{\gamma}}\over {s}}  
+ {{s r_T + (s - M^2) j_T}\over {(s - M^2)^2 + M^2 \Gamma^2}} \right] , 
\end{eqnarray} 
where $r_T^{\gamma}$ is the photon exchange term and the parameters  
$r_T$ and $j_T$ are related to the $Z$-boson exchange residuum and to the 
$\gamma$-$Z$ interference, respectively.  
On the same grounds as for the total cross sections, the asymmetries  
can be characterized around the $Z^0$ resonance by two  
parameters~\cite{riem92}: 
\begin{eqnarray} 
A(s) = A_0 + A_1\left( {s\over {M^2}} - 1 \right) +  
{\cal O}\left( {s\over {M^2}} - 1 \right)^2 ,  
\end{eqnarray} 
with  
\begin{eqnarray} 
& & A_0 = {{r_A}\over {r_T}} + {\cal O}(\Gamma^2 / M^2) , \\ 
& & A_1 = \left[ {{j_A}\over {r_A}} - {{j_T}\over {r_T}} \right] , 
\end{eqnarray} 
where $r_A$ and $j_A$ are the analogues of $r_T$ and $j_T$ for the numerators 
of the asymmetries. 
In order to allow for a realistic data analysis, the universal  
effects of ISR need also to be taken  
into account, following the procedures described in Sect.~\ref{sect:z0qed}  and 
in Appendix~\ref{sect:upc}. 
  
The $Z$-boson mass and width parameters extracted with the $S$-matrix  
approach described above differ from the commonly defined on-shell  
$Z$-boson mass and running width by two-loop and higher order corrections,  
resulting in a shift of 34~MeV and 1~Mev respectively, according  
to the following expressions~\cite{blrs,bbhvn88,wv91,stuart91}: 
\begin{eqnarray} 
& & M = M_Z - {{\Gamma_Z^2}\over {2 M_Z}} \simeq M_Z - 34 {\rm MeV} , \\ 
& & \Gamma = \Gamma_Z - {{\Gamma_Z^3}\over {2 M_Z^2}} \simeq  
\Gamma_Z - 1 {\rm MeV} . 
\end{eqnarray} 
 
Taking into account of the proper definitions, the $Z$-boson mass and  
width obtained by fitting the data with the $S$-matrix formalism  
are in agreement with those obtained within the SM,  
showing its internal consistency, even if the error on the  
$Z$-boson mass is larger in the $S$-matrix approach, because the  
$\gamma$-$Z$ interference is a free parameter. 
The fact that the $\gamma$-$Z$ interference  
is sizeable off peak can be used  to reduce  
the error on the parameter $j^{had}_T$. Actually, by adding to the peak results also 
data above and below the resonance, the values obtained  from such an analysis 
are $M_Z = 91.1882  \pm 0.0029$~GeV and $j^{had}_T = 0.14 \pm  0.12$~\cite{dward97}. 
These values are  to be compared with $M_Z  = 91.1867 \pm 0.0020$ from the  standard analysis, and    
$j^{had}_T = 0.22$ as the SM  expectation.
  
\subsubsection{Model-independent analysis of  precision data}
\label{sect:epsilon}
Since new physics 
effects can be more easily disentangled in the analysis of the $Z^0$ parameters 
if not obscured by our ignorance of SM parameters, 
it is convenient to introduce variables that are as free as possible from large $m_t$ effects 
and hence sensitive to new physics. This attitude is followed in the model independent 
approach proposed and further developed in ref.~\cite{eps}, where the validity of the SM is not 
assumed {\it ab initio} and appropriate variables (the $\varepsilon$ parameters) are defined 
in order to provide an efficient parameterization of the most important electroweak data 
with respect to the sensitivity to new physics. More specifically, four 
independent quantities, indicated as $\varepsilon_1, \varepsilon_2, \varepsilon_3$ and $\varepsilon_b$, 
are introduced in one to one correspondence with the observables $M_W / M_Z, \Gamma_l, 
A_{FB}^l$ (assuming lepton universality) and $\Gamma_b$, that are chosen as primary defining 
measurements of the $\varepsilon$ parameters. From these input data, the 
strategy consists in isolating the key quantities $\Delta\rho, \Delta r_W, \Delta k, \varepsilon_b$, 
that, as discussed in Sect.~\ref{sect:z0ew}, are the dominant effects in weak radiative corrections 
due to gauge bosons self-energies and vertex corrections to the $Z f \bar f$ coupling. 
Indeed, the four quantities $\Delta\rho, \Delta r_W, \Delta k, \varepsilon_b$, 
whenever calculated in the SM, are dominated, for sufficiently large 
{\it top}-quark mass values, by quadratic terms in $m_t$ of the order of $G_\mu m_t^2$.    
In particular, $\Delta r_W$ is connected to $\Delta r$ defined by eq.~(\ref{eq:deltar}) in
Sect.~\ref{sect:z0ew} according to the 
definition $(1 - \Delta r) = (1 - \Delta\alpha) (1 - \Delta r_W)$, in such a way that 
the running of $\alpha$ ($\alpha(M_Z) =\alpha /(1 - \Delta\alpha)$) due to known physics 
is extracted from $\Delta r$. Furthermore, $\varepsilon_b$ is introduced in order to take care of 
the important non-oblique corrections to the $Z \to b \bar b$ vertex discussed in 
Sect.~\ref{sect:z0ew}.
The explicit relations between the primary defining observables and the 
$\Delta\rho, \Delta r_W, \Delta k, \varepsilon_b$ factors can be found in the 
original literature~\cite{eps,epsrev}.
Since the aim, as already stressed, is to provide a parameterization unaffected as much as 
possible by the relative ignorance of $m_t$, it is convenient to keep 
$\Delta\rho$ ($\varepsilon_1 = \Delta\rho$) and $\varepsilon_b$ and introduce, in place 
of $\Delta r_W$ and $\Delta k$, the two following linear combinations
\begin{eqnarray}
\varepsilon_2 &=& c_0^2 \Delta\rho + {{s_0^2 \Delta r_W} \over {c_0^2-s_0^2} 
- 2 s_0^2 \Delta k} ,  \nonumber \\
\varepsilon_3 &=& c_0^2 \Delta\rho +  \left( {c_0^2-s_0^2} \right)^2 \Delta k ,
\end{eqnarray} 
where $s_0^2$ is an effective weak mixing angle containing only photon vacuum polarization 
effects, given by
\begin{eqnarray}
s_0^2 c_0^2 \, = \, { {\pi \alpha(M_Z)} \over {\sqrt{2} G_\mu M_Z^2} }, \quad 
\quad c_0^2 = 1 - s_0^2 .
\end{eqnarray} 
\begin{figure}[hbt]
\begin{center}
\epsfig{file=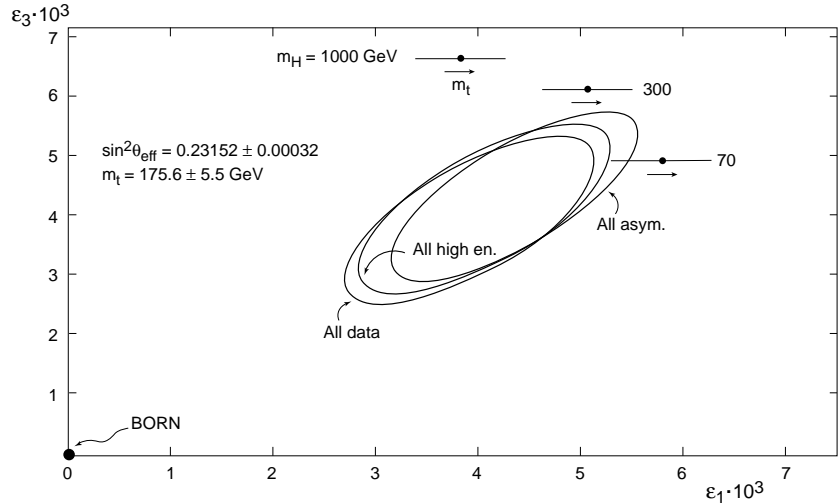,width=11truecm}
\end{center}
\caption{Electroweak precision data versus theory in the $\varepsilon_1$-$\varepsilon_3$ 
plane (from
ref.~\cite{gual97}; see also refs.~\cite{eps,epsrev}). }
\label{fig:fig6_alt96}
\end{figure} 
In this way, contributions of the order of $G_\mu m_t^2$ do not enter 
$\varepsilon_2$ and $\varepsilon_3$, that contain only logarithmic terms in $m_t$. Quadratic 
{\it top}-quark mass effects are confined in $\varepsilon_1$ and $\varepsilon_b$. Furthermore, the 
leading logarithmic terms for large Higgs-boson mass are contained in 
$\varepsilon_1$ and $\varepsilon_3$. For instance, the leading expressions for the $\varepsilon$ parameters
within the SM can be written as~\cite{eps,epsrev}
\begin{eqnarray}
&&\varepsilon_1 = {{3 G_\mu m_t^2} \over {8 \pi^2 \sqrt{2}}} - {{3  G_\mu M_W^2} \over  
{4 \pi^2 \sqrt{2}}}
\tan^2  \vartheta_W \ln \left( {{m_H} \over {M_Z}} \right) + \ldots \nonumber \\
&&\varepsilon_2 = -{{ G_\mu M_W^2} \over  {2 \pi^2 \sqrt{2}}} \ln \left( {{m_t} \over  {M_Z}} \right) 
+ \ldots \nonumber \\
&&\varepsilon_3 = {{ G_\mu M_W^2} \over  {12 \pi^2 \sqrt{2}}} \ln \left( {{m_H} \over  {M_Z}} \right) -
{{ G_\mu M_W^2} \over  {6 \pi^2 \sqrt{2}}} \ln \left( {{m_t} \over  {M_Z}} \right) + 
\ldots \nonumber \\
&&\varepsilon_b = - {{ G_\mu m_t^2} \over {4 \pi^2 \sqrt{2}}} + \ldots 
\end{eqnarray}
Finally, the relations between the defining observables    
 and the $\varepsilon_i$ can be inverted in order to get the formulae for the 
$\varepsilon_i$ in terms of the data. The explicit formulae read as follows
\begin{eqnarray}
&&\varepsilon_1 \, = \, -0.9882+0.011963 \, \Gamma_l/{\rm MeV} - 0.1511 \, x , \nonumber \\
&&\varepsilon_3 \, = \, -0.7146+0.009181 \, \Gamma_l/{\rm MeV} - 0.69735 \, x , \nonumber \\
&&\varepsilon_b \, = \, -0.62 \, \varepsilon_1 + 0.24 \, \varepsilon_3 + 0.436 \, 
\left( \Gamma_b / \Gamma_{b0} - 1 \right) , \nonumber \\
&&\varepsilon_2 \, = \, 1.43 \, \varepsilon_1 - 0.86 \, \varepsilon_3 + 0.43 \, \Delta r_W , 
\end{eqnarray}
where $x = g_V / g_A = 1 - 4 (1 + \Delta k) s_0^2$ and $\Gamma_{b0}$ is the value of $\Gamma_b$ 
in the limit when all the $\varepsilon$'s are neglected.
\begin{figure}[hbt]
\begin{center}
\epsfig{file=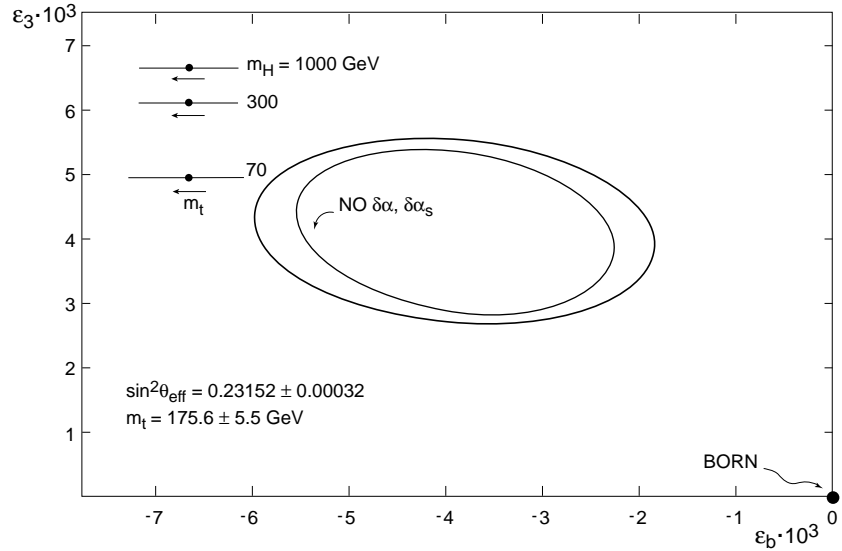,width=11truecm}
\end{center}
\caption{The same as in Fig.~\ref{fig:fig6_alt96} in the $\varepsilon_b$-$\varepsilon_3$ 
plane (from ref.~\cite{gual97}; see also refs.~\cite{eps,epsrev}). 
The effects of the errors on $\alpha(M_Z)$ and $\alpha_s(M_Z)$ are also shown. }
\label{fig:fig7_alt96}
\end{figure}

The parameters 
$\varepsilon_1$, $\varepsilon_2$ and $\varepsilon_3$ can be directly related to the variables 
$S$, $T$ and $U$, that are other quantities proposed in the literature 
in order to parameterize the oblique corrections~\cite{stu} 
and widely employed for model-independent studies of 
electroweak data~\cite{stu_bis}. However, while $S$, $T$ and $U$ are 
defined as deviations with respect to the SM predictions for specified values of 
$m_t$ and $m_H$, 
the $\varepsilon$ parameters are defined with respect to a reference approximation that 
is independent of $m_t$, in such a way that they are exactly zero in the SM in the 
limit of neglecting all pure weak loops~\cite{eps}. Moreover, it is worth noticing  
that the definitions for the $\varepsilon$ parameters are quite general because they do not 
refer to any particular model. Therefore, they are useful since they allow 
to perform a model-independent analysis of the electroweak precision data.

As a first step of such an analysis, it is possible to derive $1\sigma$ contours 
for the $\varepsilon$ parameters by using the minimal set of data given by 
$M_W / M_Z$, $\Gamma_l$, $A_{FB}^l$ and $\Gamma_b$.
To include additional observables in the $\varepsilon$'s analysis, further assumptions are 
required in order to maintain a consistent definition of the parameters. However, under 
appropriate hypotheses, the analysis can be generalized to include all the observables 
measured on the $Z^0$ peak, supplemented with low-energy electroweak measurements.
 The results obtained from the data 
can be then compared with the SM predictions, as a function of $m_t$ and $m_H$, 
obtained by using computational tools that contain the state-of-the-art of  
radiative corrections. Examples of such a 
procedure are given by Figs.~\ref{fig:fig6_alt96}-\ref{fig:fig7_alt96}.
From the comparison between theory and experiment in a given 
$\varepsilon_i - \varepsilon_j$ plane, it is possible with the present data to draw a number 
of significant conclusions.  
First, a preferred range for the {\it top}-quark and the Higgs-boson masses can be inferred, 
according to a strategy  that is different from what has 
been described in previous sections. Referring to the most recent data (with the 
TEVATRON result for $m_t$ excluded from the analysis), one can see a preference of the 
precision data for a ``light'' Higgs boson and a $m_t$ value somewhat lower than the 
CDF/D0 determination. These conclusions on  $m_t$ and $m_H$ corroborate 
the results obtained from SM fits discussed in Sects.~\ref{sect:tqm} and \ref{sect:hm}. 
Second, it is possible to observe a good agreement between the 
full SM predictions and the data. Noticeably, one can point out a strong evidence for 
pure weak radiative corrections, since the beautiful agreement between data and 
theory can not be simply explained in terms of an improved Born approximation based 
on tree-level SM plus QED and QCD corrections. This very important conclusion 
demonstrates by itself the constraining power of precision physics at LEP. The latter 
has been also emphasized by some authors that have studied the sensitivity of the data 
to a very peculiar subset of weak radiative corrections, namely the pure bosonic 
loops involving trilinear gauge-bosons and Higgs couplings~\cite{dsw96b}. As shown by these 
investigations, the LEP1/SLC data are so precise that they require the inclusion 
of such corrections in the SM calculation in order to precisely fit the data. Therefore, the present 
LEP1/SLC precision data feel the electroweak non-abelian couplings that are presently measured at 
LEP2~\cite{s97}.

\subsubsection{Physics beyond the Standard Model}
\label{sect:beyond}
As already stressed, the main virtue of the $\varepsilon$ parameterization is that they 
can be rather easily calculated in extensions of the SM, and therefore can be 
used to analyze models of new physics in the light of precision data. Typical examples 
studied in the literature are 
{\it technicolour models}~\cite{technicolor}, 
{\it models with an extended gauge group}~\cite{yrnp96,yrzp96}  and 
the {\it Minimal Supersymmetric Standard Model}~\cite{mssm}. Also specific analyses within specific
models of new physics beyond the SM have been considered in the literature, for instance concerning
models with an extended Higgs sector~\cite{giamplb89,giamplb90,lynnnardi92,blankhollik} or models
with extra fermionic generations~\cite{vnovikov,fourgen}. 
The main conclusions of such analyses are summarized in the following. 

Concerning {\it technicolour models}, 
it can be said that they tend to produce, in their typical realizations, corrections 
to the $\varepsilon$ parameters (large and positive to $\varepsilon_3$, large 
and negative to $\varepsilon_b$) that are disfavored 
by the data~\cite{tech_lep}. In short, simple technicolour  models 
are basically disfavored by LEP experiments~\cite{efl95}, even if they can not be ruled out 
completely since particularly sophisticated versions with non trivial 
behaviour under the electroweak group could avoid their typical 
bad consequences~\cite{tech_ns}.   

As far as {\it models with an extended gauge group} are concerned, the case of the 
simplest models with an extra $U(1)$ (and with an associated new 
neutral vector boson) has been addressed in the literature~\cite{u1}. For such a situation, 
two new parameters have to be introduced, the former being a mixing angle $\xi$ in order to 
define the mixture of the standard and the new neutral vector bosons, the latter being 
the shift $\delta\rho_M$ induced at the tree-level in the $\rho$ parameter by the 
above mixing. Considering the class of models based on E6 and for the left-right symmetric 
model, the main implication of precision data is that very strong constraints on 
$\xi$ and $\delta\rho_M$ can be deduced. In particular, the amount of mixing allowed 
is very small, less than, say, 1\%. It is worth observing that, in recent years, 
models containing an extra
$Z'$ boson with enhanced hadronic or almost vanishing leptonic couplings (hadrophilic or leptophobic
$Z'$) received particular attention as candidate models able to explain the simultaneous
anomaly of $R_b$ and $R_c$ data (see for instance ref.~\cite{mlmz0}). At present, the $R_b$-$R_c$
anomaly has  substantially disappeared, but anyway these model have not been  completely ruled  out
by the new data, and could be tested at LEP2~\cite{mplrv97}.    

\begin{figure}[hbt]
\begin{center}
\epsfig{file=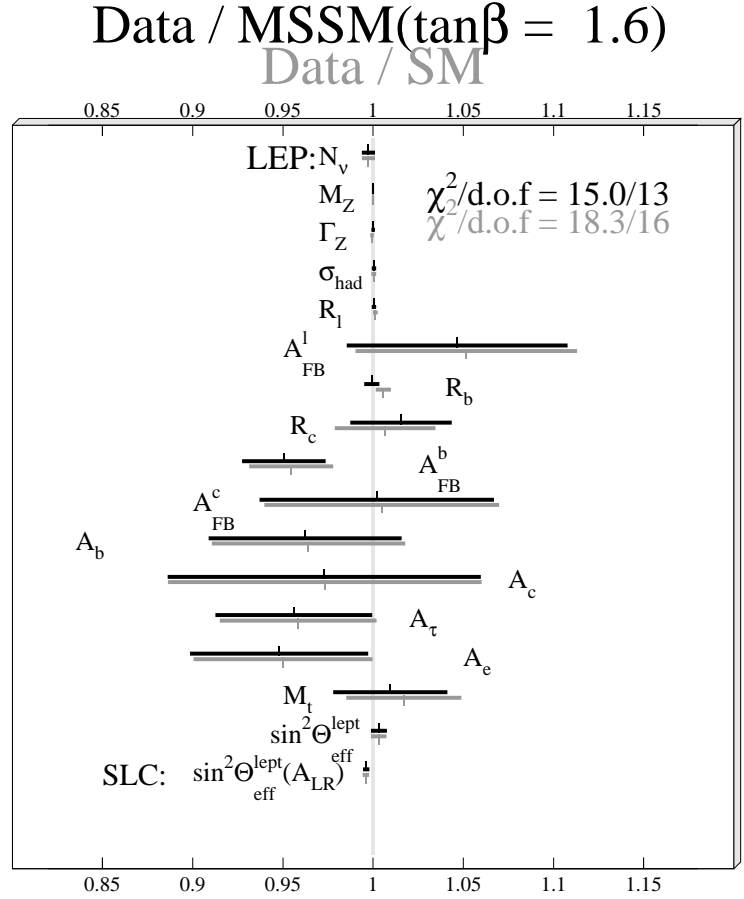, width=12truecm}
\end{center}
\caption{Experimental data normalized to the best fit results in the SM and MSSM 
(from ref.~\cite{holetal96}).}
\label{fig:mssm}
\end{figure}

The {\it Minimal Supersymmetric Standard Model}  (MSSM) deserves a special discussion 
since it is the most predictive framework beyond the SM. However, because of the 
very high number of free parameters, a convenient strategy when confronting theory and 
experiment is restricting to consider two limiting cases (the so-called 
``heavy'' and ``light'' MSSM) rather than attempting a direct fit to the data. 
The ``heavy'' MSSM implies all supersymmetric particles to be sufficiently massive. 
In this limit, the MSSM predictions for electroweak data 
essentially coincide with the results of the SM with a light Higgs boson, say, lighter 
than 100~GeV~\cite{bcf92}. Therefore, this particular realization of the MSSM can nicely 
accomodate the 
LEP data essentially in the same way as the SM does. In the ``light'' MSSM the masses 
of some of the superparticles are nearby their experimental lower limit. For such 
realization, the pattern of radiative corrections differs from that of the SM, with 
peculiar effects in vacuum polarization diagrams and $Z \to b \bar b$ vertex, that can 
be easily incorporated in the $\varepsilon$'s~\cite{zrcsusy}. These peculiar effects are reviewed in 
some detail
in~\cite{gualcrad96}. In particular, it is possible to find mechanisms able to explain 
a possible departure (if real) of $R_b$ from the SM~\cite{zbbsusy}. In conclusion, 
the MSSM agrees well with precision 
electroweak data, since it doesn't give rise, both in its ``heavy'' and ``light''
realizations, to effects that are inconsistent with the LEP measurements. This is due to the
particular nature of the MSSM as new physics model that doesn't alter significantly 
the structure of the SM. This positive outcome of precision physics at LEP pointing 
towards supersymmetric extensions of the SM is corroborated by independent 
investigations of the MSSM with respect to the electroweak data performed in \cite{holetal96}.
In these papers, instead of analyzing the MSSM implications on the data in terms 
of $\varepsilon$ parameters, a more orthodox approach has been followed, namely a 
complete calculation of electroweak precision observables has been carried out, 
using one value for the Higgs-boson mass and $\tan \beta$, together with the set of SUSY soft 
breaking parameters fixing the chargino/neutralino and scalar fermion sectors. 
In this kind of analysis, an optimal set of SUSY parameters can be extracted from a global 
fit to the data, showing how supersymmetry can consistently describe precision data 
without contradicting present limits from direct search on its mass spectrum (see
Fig.~\ref{fig:mssm}).  

Other specific models analyzed in  the light of the LEP/SLC data  
concern the {\it extension of the Higgs-boson  sector}. 
In the minimal SM the tree-level masses of the vector  
bosons are linked by means of the weak mixing angle through  
the relation $\rho = 1$, with  
\begin{eqnarray} 
\rho = {{M_W^2}\over {M_Z^2 \cos^2 \vartheta_W}}. 
\end{eqnarray} 
Since the Higgs-boson sector has not yet  
been tested with high accuracy, it is interesting to investigate a  
possible extension of the SM adopting an enlarged scalar   
sector. The simplest way to do this, by changing the tree-level expression of $\rho$,
is to introduce, in addition to  
the minimal doublet representation, a Higgs-boson triplet with a vacuum  
expectation value different from zero in the neutral sector.  
In this case the masses of the vector bosons are no more related by  
the weak mixing angle, and the $\rho$ parameter deviates from unity  
already at the tree level according to the following form: 
\begin{eqnarray} 
\rho_{tree} = {{\sum_i v_i^2 (I_i^2 - I_{3i}^2 + I_i)} \over  
{\sum_i 2 v_i^2 I_{3i}^2}},  
\end{eqnarray} 
where $I_i (I_{3i})$ is the weak isospin (third component) of the  
$i-$th Higgs-boson multiplet, and $v_i$ the respective vacuum expectation  
value. In addition to the standard Higgs boson, other Higgs  
particles belong to the physical spectrum. The scenarios  
arising in non minimal standard models with Higgs-boson triplets has  
been studied in refs.~\cite{giamplb89,giamplb90,lynnnardi92}.  
The calculation of the radiative corrections to precision  
observables within these  
non minimal models requires not only the evaluation of the extra  
loop diagrams involving the non standard Higgs bosons, but also  
an extension of the renormalization procedure. Due to the fact  
that the bare $W$- and $Z$-boson masses are not linked through the  
weak mixing angle, a fourth input data is necessary in order to  
perform the relative subtraction. In other words the $\rho$ parameter  
is not calculable as in the minimal standard model but need to be  
fixed by a renormalization condition. The procedure has been  
discussed in ref.~\cite{giam91} within the $\overline{MS}$ scheme  
using the physical $W$-boson mass as the fourth data point, in addition to  
$\alpha$, $G_{\mu}$ and $M_Z$.  
Recently, the radiative corrections to the full set of LEP1  
electroweak observables have been calculated in a triplet model with  
a neutral Higgs boson and a pair of charged Higgs particles in  
addition to the standard Higgs boson~\cite{blankhollik}.  
The adopted renormalization  
framework is an extension of the on-shell scheme, where the fourth  
input data has been chosen to be the effective leptonic mixing angle  
at the $Z^0$ resonance. The interesting result of this study is that  
for a variation  of the {\it top}-quark mass in the range $m_t = 175 \pm 6$~GeV,  
the predictions of the SM and of the triplet model are  
practically indistinguishable. As far as the Higgs-boson mass dependence is  
concerned, while the SM has a preference for a heavy  
Higgs boson from the observable $\Gamma_Z$ and for a light Higgs boson from the  
mixing angle measurement, in the triplet model a light Higgs particle is  
compatible with all precision observables.  
 
Models with {\it extra fermion generations}  have been  also investigated.  
As  previously discussed, one of the achievements of the  
LEP experiments is the determination of the number of light neutrinos  
($N_{\nu} = 2.993 \pm 0.011$) and hence of the number of fermionic generations.  
However, the theory does not exclude the presence of new sequential  
quark-lepton generations with heavy neutral leptons, whose production  
would be kinematically forbidden at LEP. The  experimental  
lower bounds on the masses of the hypothetical new fermions available in the literature 
(including LEP1.5 data) are of  
about $60$~GeV for the leptons and  $100$~GeV for the  
quarks~\cite{vnovikov}. In the literature, 
the indirect bounds obtained through the study of their effects on  
radiative corrections for LEP1 electroweak  
observables~\cite{vnovikov,fourgen} have been examined.   
A large mass splitting between the members of the doublets would  
spoil the agreement with the precision data, so the new heavy fermions  
should be almost degenerate in mass within a doublet. In order to reduce  
the number of parameters, a reasonable assumption is to work with  
quarks and leptons of equal masses. 
Within this scenario, the effect  
of one new generation can be compensated by increasing the  
fitted  value of the {\it top}-quark mass. 
An analysis based on older data sets shows that  only at most two new sequential  
generations are allowed~\cite{vnovikov}. 
However, as a general {\it caveat}, it should be noticed that 
new experimental data, both from LEP2 and the TEVATRON, could modify this scenario. 
Considering the effects of electroweak  
vacuum stability and the absence of a Landau pole in the Higgs-boson   
potential allows to put upper bounds on the masses of the new fermions,  
depending on the scale at which new physics will appear.  
In the case of absence of new physics up to the grand unification scale, the upper  
bounds on the masses of the fourth generation are of the order of  
$100$~GeV, a value viable for LEP2 direct searches~\cite{vnovikov}.  
 
As a last comment, it is worth pointing out that the present precision electroweak data coming from the
LEP analysis can be used to derive constraints on the structure of a grand unified theory (GUT). In
particular, within standard GUT's the evolution of the gauge couplings starting from the scale given by
the $Z$-boson mass $M_Z$ does not lead to the unambiguous
identification of a grand unification scale. Viceversa, such a grand unification scale can be
identified within the framework of supersymmetric GUT's~\cite{guts}. 
This is at present considered as an indication of new physics beyond the SM, and in particular of
supersymmetry~\cite{gual97}.  

%% file: lep2.txi
\section{Physics at LEP2}
\label{sect:lep2phys}
\begin{figure}[hbt]
\begin{center}
\epsfig{file=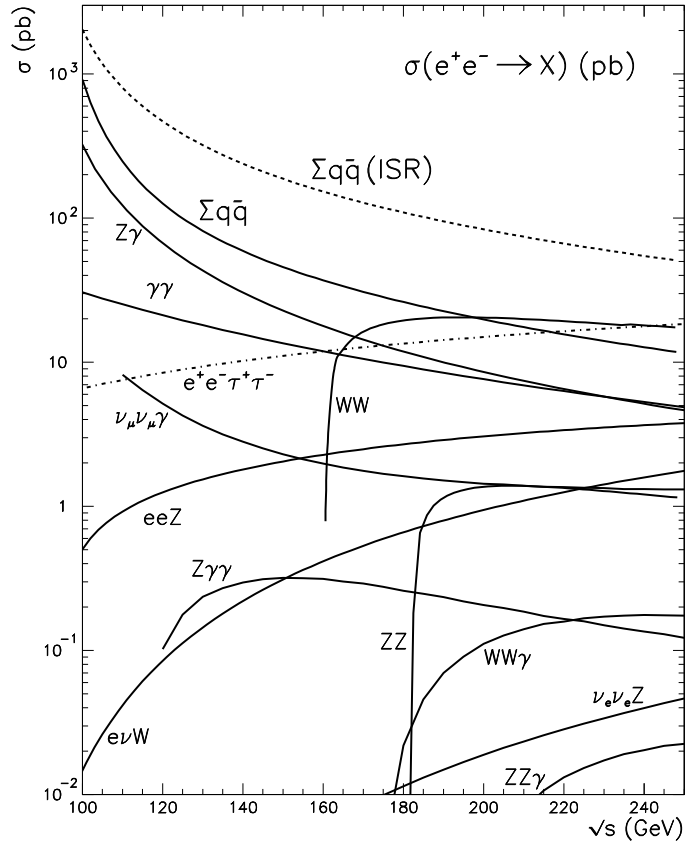, width=10truecm}
\end{center}
\caption{Standard Model cross sections at LEP  (from~\cite{yrsm96}).}
\label{fig:lep2overview}
\end{figure}

As discussed in the Introduction, the ALEPH, DELPHI, L3 and OPAL 
experiments at LEP terminated in the fall 1995 their data taking at 
energies around the mass of the $Z^0$ resonance. After a short run at 
intermediate energies (the so-called LEP1.5 phase), the collider enters a 
new phase of operation far away the 
$Z^0$ peak, namely in the range $\sqrt{s} \simeq 161 - 192$~GeV. The energy 
upgrade of the LEP machine offers the possibility of exploring a new 
energy regime for electron-positron collisions, beyond the threshold of 
$W$-pair production (see Fig.~\ref{fig:lep2overview}). The precision 
measurements 
of $Z$-boson properties 
at LEP1/SLC can be therefore completed by the experiments at LEP2 with 
the precision determination of the $W$-boson properties. In addition to 
the precision physics of the $W$ boson, experiments at LEP2 will continue to 
perform tests of the Standard Model (SM) in two-fermion 
production processes and in photonic 
final states, search for the Higgs boson, perform QCD studies as well as 
explore direct new physics signals~\cite{yrlep296}. Actually, as can be clearly 
seen from 
Fig.~\ref{fig:lep2overview}, there are at LEP2 
many interesting processes with measurable cross sections, 
such as $e^+ e^- \to f \bar f$,
$e^+ e^- \to \gamma\gamma$, $e^+ e^- \to \nu \bar\nu \gamma$, 
$e^+ e^- \to W^+ W^-, ZZ$, 
the latter yielding four-fermion ($4f$) final states.  

In the present section, the processes of main interest for precision studies of 
the
electroweak theory at LEP2 are examined. The two-fermion production processes 
and radiative events are discussed, pointing out new physics aspects with 
respect to LEP1. The physics of the $W$ boson in $4f$ production 
processes is then addressed, with particular emphasis on the status of 
theoretical 
predictions as well as first experimental results. The search for the SM Higgs 
boson in $4f$ final states is finally discussed.

\subsection{Two-Fermion Processes}
\label{sect:tfproc}
Although LEP2 was mainly conceived to measure precisely the properties of 
the $W$ boson through the production of $W$ pairs in $e^+ e^-$ 
collisions, the processes of quark and lepton pair production,  
still remaining among the most copious reactions at LEP2, is an important
physics item.  In this new 
energy regime, the two-fermion processes can be used to perform further 
tests of the SM, with enhanced sensitivity to the electroweak interference, 
as well as to obtain limits on possible extensions of the SM, such as  
the existence of new $Z'$ bosons or effective $4f$ contact
interactions~\cite{pward97}.
However, the de-convoluted cross sections for these processes are 
typically of the order of 10~pb in the LEP2 
energy regime, \idest\  three orders of magnitude 
smaller than the cross sections 
for the same processes at LEP1. Therefore, even when considering the highest 
final integrated luminosity expected at LEP2, \idest\  
$500~{\hbox {pb}}^{-1}$~\cite{mw96}, it will be   possible 
to measure the two-fermion observables with a statistical error of the 
order of 1\%, to be compared with the corresponding 0.1\% of LEP1. Nonetheless, 
in order to meaningfully compare theory and experiment, precise  
predictions are still demanded and, therefore, radiative corrections need to be 
known with a good precision and incorporated in the theoretical calculations. 
To this aim, it is worth noticing that at LEP2 new aspects in the sector of 
radiative 
corrections to two-fermion processes do appear with respect to LEP1, 
noticeably~\cite{yrsm96} 

\begin{itemize}

\item weak boxes (\idest\   diagrams containing 
$WW$ and $ZZ$ internal lines) are at the level of a few 
per-cent of lepton and hadron cross sections,  
and hence no longer negligible as at LEP1; 

\item initial-state radiation (ISR) of hard photons is not inhibited as around 
the 
$Z^0$ peak, where, as discussed in Sect.~\ref{sect:z0qed}, the $Z$-boson width 
$\Gamma_Z$ acts a natural cut-off; consequently, hard-photon bremsstrahlung 
corrections  become numerically relevant at LEP2.

\end{itemize}

The phenomenon of the so-called $Z^0$ {\it radiative return} constitutes the main 
manifestation of the relevance of hard-photon radiation in two-fermion 
production above the $Z^0$ peak. 
It corresponds to the emission of very energetic photons that reduce 
the two-fermion effective centre of mass (c.m.)  
energy back to the $Z$-boson mass, and considerably 
enhance the  cross section. Actually, one can observe an increase of 
about a factor of four in the total rate of two-fermion events at LEP2, between
the tree-level and the QED corrected predictions.  
This effect directly arises from the dependence of the two-fermion cross section 
on the c.m. energy, when going from LEP1 to LEP2. Actually, 
the Breit-Wigner behaviour 
for $\sqrt{s} \simeq M_Z$, followed by a fast $1 / s$ decrease after it, makes 
very likely the production of a $f \bar f$ pair with invariant mass 
$m_{f \bar f}$ clustered around $M_Z$,  accompanied by the emission of 
a real photon carrying away the residual energy. In fact, the photon energy 
spectrum 
shows a pronounced peak around the value $(1-M_Z^2/s)\,\sqrt{s}/2$,  
corresponding to those events (LEP1-like events) with an effective c.m. 
energy after ISR 
of $\sqrt{\hat s} \simeq M_Z$ ($Z^0$ {\it radiative return}) .

The above comments are referred to $s$-channel annihilation processes. 
As far as large-angle Bhabha scattering is concerned, it is worth noticing that
already some GeV off-resonance the cross sections is dominated by $t$-channel
photon exchange. Hence, large-angle Bhabha scattering at LEP2 is much more
similar to small-angle  than to large-angle Bhabha  scattering at LEP1 (see
ref.~\cite{bharep96} for   a detailed discussion). 

\subsubsection{Weak  corrections}
\label{sect:tfewqcd}
As already discussed in Sect.~\ref{sect:z0ew}, the one-loop pure weak corrections 
to $e^+ e^- \to \gamma Z^0 \to f \bar f$ arise from those diagrams 
that involve corrections to the vector boson propagators, from the 
set of vertex corrections (with the exclusion of the virtual photon 
contributions that are accounted among the QED effects) and from the box 
diagrams containing the exchange of two {\it massive} gauge bosons.  
In fact, the box diagrams where at least one boson is a photon are 
classified as belonging to the QED corrections simply because they are 
IR divergent and need to be combined with the photonic initial-final 
state interference in order to get a meaningful result. Therefore, 
the genuine weak boxes are those with two $W$- and $Z$-boson propagators.
Differently from the propagators and vertex corrections, they are 
ultra-violet 
finite in any gauge but the unitary one, where they actually 
give rise to UV infinities. Moreover, weak boxes introduce, in addition to $s$, 
a dependence on the scattering angle, that makes the parameterization of weak 
corrections in terms of ``running'' ($s$-dependent) effective couplings, 
successfully employed at LEP1, no longer viable at LEP2. In fact, since 
the box diagrams are negligibly small around the $Z^0$ pole, 
 any dependence on the scattering angle can be 
neglected in the calculation of pure weak effects around the peak. Off 
resonance, 
\idest\   in the LEP2 energy regime, the weak boxes become relevant corrections, 
due to the natural appearance of $WW$ and $ZZ$ thresholds. Therefore, they are 
actually included as relevant short-distance effects in the theoretical 
predictions 
for two-fermion physics above the $Z^0$ peak.     


In order to make some quantitative statements on the effects of weak 
boxes on two-fermion observables, it is necessary to remember that 
the contribution of weak boxes is not gauge invariant by itself. It follows that 
a sensible evaluation of these corrections, as well as a comparison of their 
size as obtained from different computational tools based on different 
theoretical 
framework, can not be carried out unless a gauge-invariant procedure of 
subtraction of weak boxes (the so-called {\it de-boxization}) is introduced. 
A possible procedure has been proposed in ref.~\cite{yrsm96}. Denoting the 
correction 
due to the $WW$ box diagrams by $B_{WW} (\xi)$ as computed in a general 
$R_{\xi}$ gauge, 
then $B_{WW} (\xi)$ can be split as
\begin{eqnarray}
B_{WW} (\xi) \, = \, B_{WW} (1) + \left(\xi^2 - 1 \right) \Delta(\xi) .
\end{eqnarray}
In ref.~\cite{yrsm96}, in order to get an estimate of the effect of weak boxes 
at LEP2, 
it was agreed to subtract from the full one-loop theoretical prediction
$B_{WW} (\xi = 1)$. This contribution at LEP1 
is of the order of $10^{-4}$ and   its evaluation far from the 
peak can give an idea of the numerical effects introduced by the weak boxes 
at LEP2. The effect of weak boxes, as defined above, has been evaluated for 
the most relevant two-fermion processes ($e^+ e^- \to \mu^+ \mu^-, 
u \bar u, d \bar d, b \bar b$, hadrons) and for several c.m. energies. The main 
conclusion of this analysis, based on the predictions of different tools such as 
{\tt TOPAZ0}, {\tt ZFITTER} and {\tt WOH} (see Sect.~\ref{sect:z0ct}), is that 
the weak boxes introduce a significant relative correction of the order of a few 
per cent, 
especially around the $WW$ threshold and at the highest LEP2 energies. In 
particular, 
for $\sigma_{\mu}$ the effect is positive, of the order of 1-2\% at 
$\sqrt{s} = 161-175$~GeV and negative, of the order 1\% at 205~GeV; for 
$\sigma_{had}$ 
the correction is positive at $\sqrt{s}=161$~GeV, of the order of 
1\%, negative and around 3\% and 4\% at 175 and 205 GeV, respectively. For the 
muon forward-backward asymmetry the effect of weak boxes is within 0.01.


\subsubsection{QED corrections}
\label{sect:tfqed}
Because of the overwhelming effect due to ISR in the LEP2 energy regime 
discussed above, 
some contributions associated to the emission of hard photons that are   
irrelevant at LEP1 play a new r$\hat{ \hbox{\rm o}}$le far above the $Z^0$ peak. 
Actually, the soft-photon dominance in the ISR mechanism emphasized in 
Sect.~\ref{sect:z0qed} for the energy region around the $Z^0$ resonance 
ceases to be  valid at LEP2,  
and previously negligible hard-photon 
bremsstrahlung effects, both at the leading and 
next-to-leading level, show up. This implies the need for 
an upgrade in the treatment 
of QED corrections with respect to LEP1, since the numerical impact of the new 
corrections is typically at the 1\% level and hence significant in view of the 
LEP2 experimental precision. 

\begin{figure}[hbt]
\begin{center}
\epsfig{file=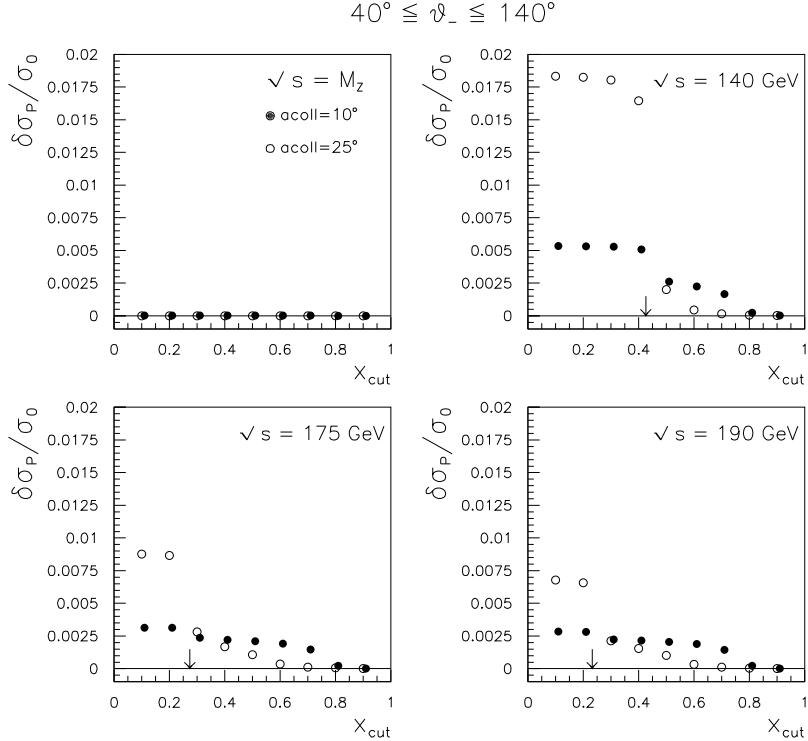, width=12truecm}
\end{center}
\caption{The theoretical error  of the ``$Z^0$-peak recipe'' for the $\mu$ 
pair production cross section (relative deviation), as a function of the invariant
mass cut $x_{cut}$ defined by $s'  > x_{cut} s$, $s'$ being the
invariant mass of the final-state fermion pair after ISR.  
The results correspond to two acollinearity cuts,  
for the same angular acceptance cut.
The arrows point at 
$x_{cut} = M_Z^2 / s$. }
\label{fig:fig3_ishp}
\end{figure}

The first example of such effects due to hard-photon bremsstrahlung is given by the 
${\cal O}(\alpha)$ NLO corrections arising from the emission of a 
hard {\it acollinear} photon by the initial state, because 
hard collinear photons
are already taken into account by the leading logarithmic (LL) 
prescriptions (see Appendix~\ref{sect:upc}).  
Since these corrections 
are strongly affected by the imposed experimental cuts and can be very 
difficultly cast in a fully analytic form, one can evaluate their size 
by performing an exact ${\cal O}(\alpha)$ perturbative calculation, 
integrating it numerically and depurating the final result of the universal LL 
as well as   
of the non-leading (process dependent) soft+virtual 
corrections~\cite{mnpishp97}. 
This procedure allows to isolate and quantify in a 
clean way the effect of hard-photon contributions in ${\cal O}(\alpha)$ ISR.    
The results of such an analysis can be found in ref.~\cite{mnpishp97}
for typical event selections 
(ES's) adopted by the LEP Collaborations, namely   

\begin{figure}[hbt]
\begin{center}
\epsfig{file=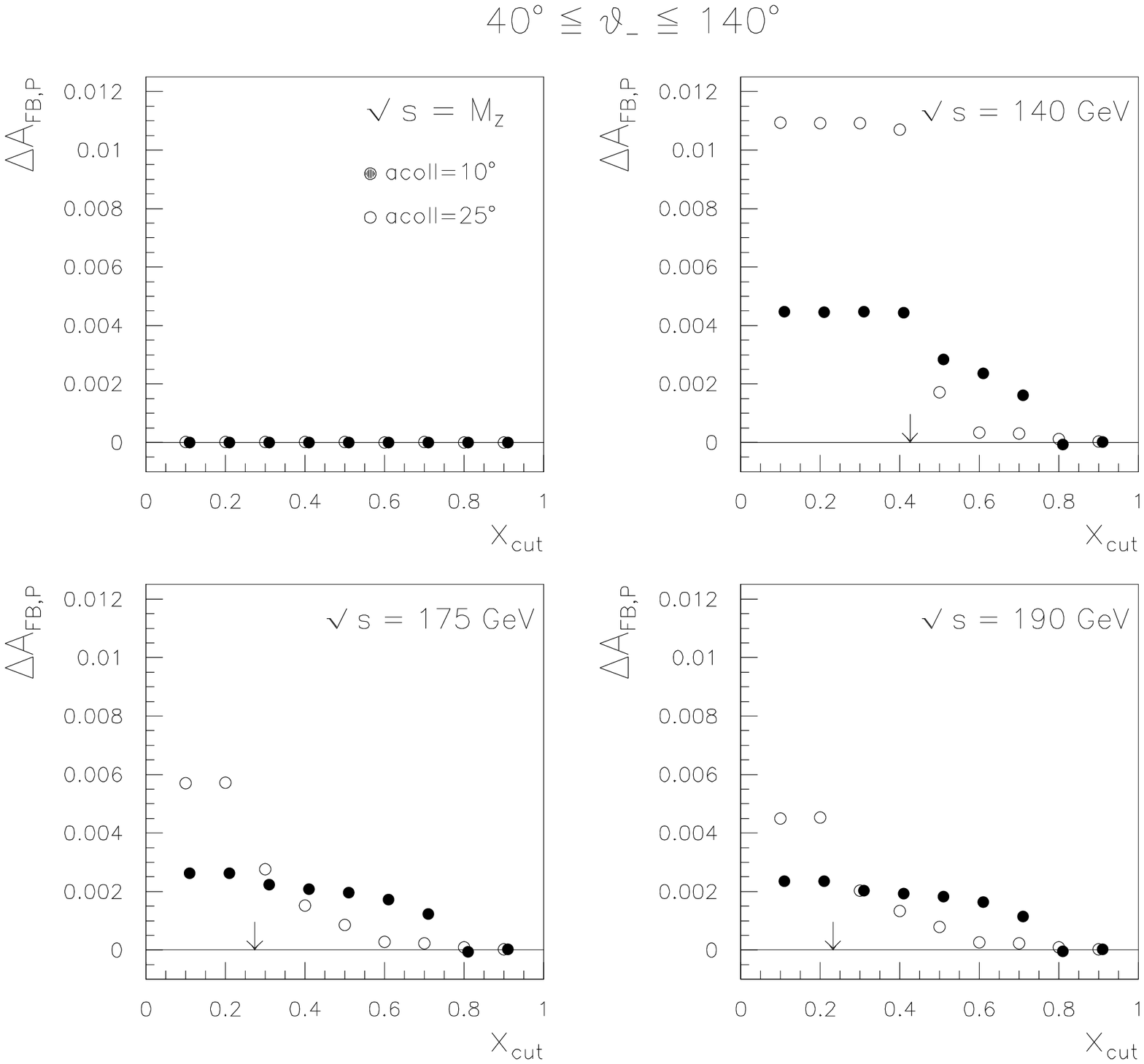,width=12truecm}
\end{center}
\caption{The same as in Fig.~\ref{fig:fig3_ishp}  for  the forward-backward
asymmetry (absolute deviation). }
\label{fig:fig4_ishp}
\end{figure}

\begin{itemize}

\item only a cut on the invariant mass of the event after ISR; 
this matches the ES adopted by recent analyses of hadron and lepton-pair 
production by the  LEP Collaborations at LEP1.5 and LEP2 energies;

\item fermion acceptance ($40^\circ \leq \vartheta_{-} \leq 140^\circ$) and 
acollinearity 
(acoll=$10^\circ,25^\circ $) cuts in association with an invariant mass one.

\end{itemize} 

For the second ES, the results are shown in 
Figs.~\ref{fig:fig3_ishp} and \ref{fig:fig4_ishp}. 
They refer to four c.m. energies: $\sqrt{s} = M_Z$ (as a 
test case), 
140, 175 and 190~GeV. In all the cases, the final
state considered is $\mu^+ \mu^-$. 

Whenever a $s'$ cut alone is imposed in the data analysis (that in practice 
coincides with what is done by the LEP Collaborations above the $Z^0$ peak),  
 ${\cal O}(\alpha)$ non-leading hard-photon corrections are compatible with zero 
for
hadronic and leptonic cross sections, and negligible at the $10^{-3}$ 
level for the lepton asymmetry. This applies both to loose and tight $s'$ cuts, 
\idest\  including or excluding the events with $Z^0$ radiative return. 
Therefore, 
for such typical ES's, those LEP1 calculations accounting for 
soft plus virtual NLO terms only ($Z^0$-peak recipe) 
can be safely extrapolated at higher 
energies without loss of accuracy. For more complicated ES's 
(see Figs.~\ref{fig:fig3_ishp} and \ref{fig:fig4_ishp}), including an 
acceptance and acollinearity cut, the theoretical error due to 
the neglect of the ${\cal O}(\alpha)$ non-leading hard-photon  terms 
can grow up to about $2.5\%$ for 
the cross section and to $1 \times 10^{-2}$ for the forward-backward 
asymmetry. However, if the $Z^0$ radiative return is cut away, the 
above theoretical error is limited to 0.3\% relative
deviation for the cross section and $3 \times 10^{-3}$ absolute deviation for 
the
forward-backward asymmetry, that are acceptable in the light of the 
LEP2 experimental precision~\cite{mnpishp97}. 

\begin{figure}[hbt]
\begin{center}
\epsfig{file=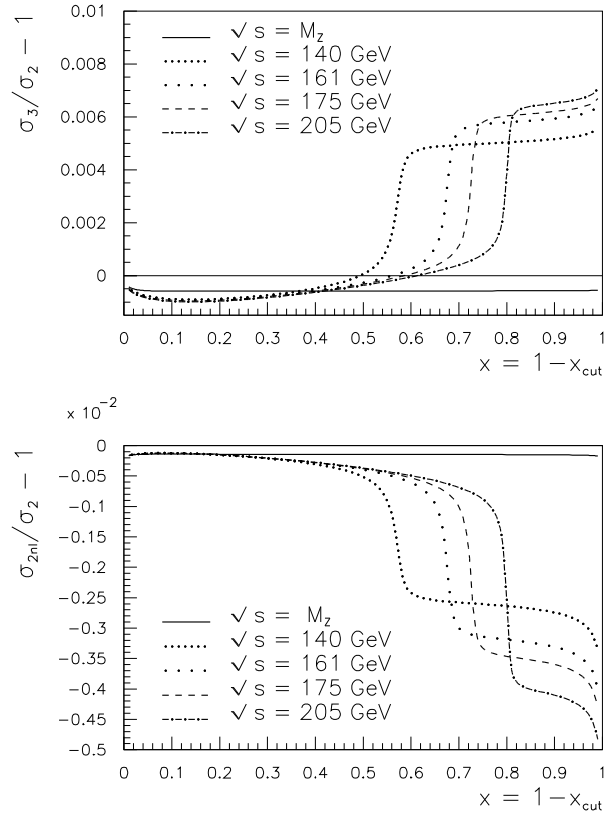,width=11truecm}
\end{center}
\caption{The effect of ${\cal O} (\beta^3)$ and ${\cal O}(\alpha^2 L)$ corrections on the $e^+ e^-
\to \mu^+ \mu^-$ total cross section as a function of the $s'$ cut (see ref.~\cite{mnp97} for more
details).} 
\label{fig:fig1_h3}
\end{figure}

The second example of hard-photon contributions damped at LEP1 but no longer 
negligible at LEP2 comes from the IS second-order non-leading corrections
 and third-order LL contributions. 
The former arise from the configurations where a hard photon is 
radiated off in the direction of incoming electron or positron in association 
with a
large-angle, acollinear additional hard photon, and provide 
corrections of the order of $\alpha^2 L$. The latter are due to the 
emission of three hard photons collinear to the colliding leptons, originating 
terms of the order of ${\cal O} (\beta^3)$. 
The ${\cal O}(\alpha^2 L)$ non-soft  
effects are exactly known from explicit perturbative calculation of the 
ISR spectrum~\cite{bbvn88}.
They can be easily included in the radiator or flux function (see
Appendix~\ref{sect:upc}) in such 
a way that the exact ${\cal O}(\alpha^2)$ calculation for an inclusive 
cross section is reproduced. The LL ${\cal O} (\beta^3)$ contributions, being 
universal photonic effects, can be kept under control employing one 
of the algorithms described in  Appendix~\ref{sect:upc}. Referring in 
particular to the structure function  
(SF) method, these LL ${\cal O} (\beta^3)$ corrections 
are known for the electron SF~\cite{sj91,cdmn92,s92} as well as the 
radiator~\cite{jsw91,mnp97}, both in factorized and 
additive form. It has to be noticed that the ${\cal O} (\alpha^2)$ 
non-leading non-soft corrections are in principle of the same order of magnitude 
as the third-order LL ${\cal O} (\beta^3)$ corrections. Therefore, 
a careful evaluation of the effects induced by higher-order 
hard-photon corrections in the LEP2 energy regime necessarily requires that 
both the contributions are simultaneously included in the theoretical 
predictions. 
The interplay between ${\cal O}(\alpha^2 L)$ and ${\cal O} (\beta^3)$
effects for the QED corrected muon-pair cross section, as a function of the $s'$ 
cut defined as $s' / s \ge
x_{cut}$, is shown in Fig.~\ref{fig:fig1_h3}. The relative deviations with 
respect
to the cross section computed by means of a standard additive radiator with up 
to ${\cal O}(\alpha^2)$ LL hard-photon corrections ($\sigma_2$)
 are shown for several c.m. energies. 
\begin{figure}[hbt]
\begin{center}
\epsfig{file=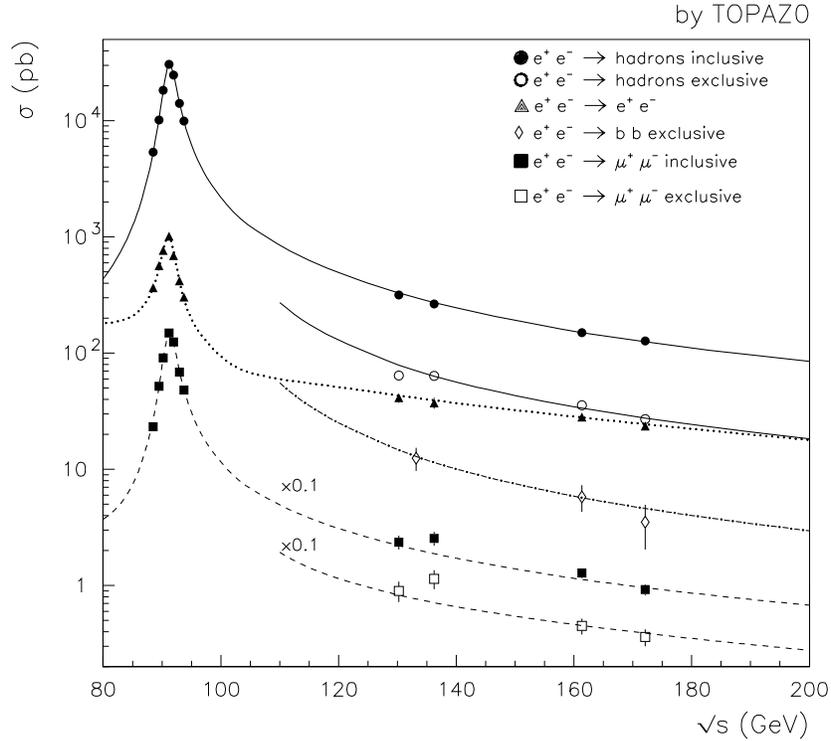, width=12truecm}
\end{center}
\caption{Data-theory comparison for two-fermion production cross sections at LEP. 
The experimental data
can be found in~\cite{opal1} 
(on peak) and~\cite{opal2} ($\sqrt{s} > 130$~GeV).
 The theoretical
prediction is performed by {\tt TOPAZ0}~\cite{topaz0}. }
\label{fig:2flep2}
\end{figure}
As can be seen, both the 
${\cal O} (\beta^3)$ (upper window) and  NL ${\cal O} (\alpha^2 )$ (lower
window)
corrections amount to a contribution of several 0.1\% when the $Z^0$ radiative
return is included, but they tend to compensate one another, being of the 
same order of magnitude but of opposite sign. When the $Z^0$
radiative return is excluded, or close the $Z^0$ resonance, the 
NL ${\cal O} (\alpha^2 )$ corrections are confined at the level of 0.01-0.02\%, 
whereas the ${\cal O} (\beta^3)$ ones remain at the level of 0.05-0.1\%. 
\begin{figure}[hbt]
\begin{center}
\epsfig{file=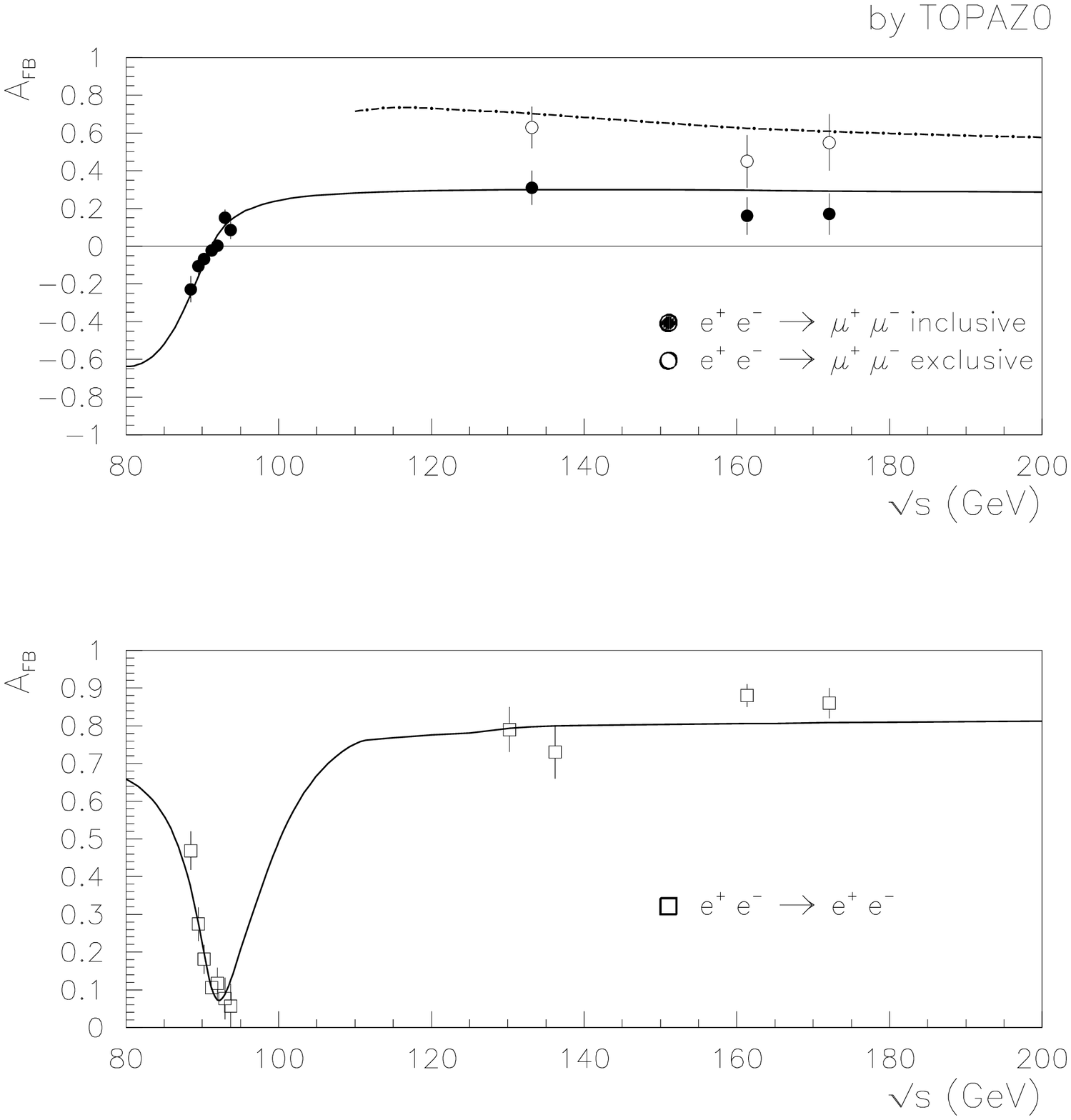, width=12truecm}
\end{center}
\caption{Data-theory comparison for the $e^+ e^- \to \mu^+ \mu^-$ and $e^+ e^-$ forward-backward
asymmetries at LEP. The experimental data
can be found in~\cite{opal1} 
(on peak) and~\cite{opal2} ($\sqrt{s} > 130$~GeV). 
The theoretical
prediction is performed by {\tt TOPAZ0}~\cite{topaz0}. }
\label{fig:2flep2afb}
\end{figure}
More precisely, the inclusion of the NL ${\cal O} (\alpha^2 )$ plus ${\cal O} 
(\beta^3)$ 
corrections above the $Z^0$ peak causes a reduction of the QED corrected 
muon-pair cross section of about $ - 0.1 $\% 
when the $Z^0$ radiative return is excluded and an enhancement of about 0.25\% 
when 
it is included. In the case of the hadronic cross section the effect of 
enhancement is around 0.4\%, \idest\   comparable with the expected experimental 
precision for such an observable.

The theoretical predictions for  cross sections and forward-backward asymmetries of two-fermion
processes, as obtained by means of the code {\tt TOPAZ0} including all the weak and QED corrections
discussed above, are compared with experimental data and shown as functions of the c.m. energy in
Figs.~\ref{fig:2flep2} and \ref{fig:2flep2afb}, respectively. 

\subsubsection{Radiative events}
\label{sect:singgamma}
As already discussed  in  Sect.~\ref{sect:fits}, the study of  single   
photon production in
the process  $e^+ e^- \to \nu \bar\nu \gamma $ has  been exploited at LEP1 as
an  alternative method for  the determination  of the invisible  $Z$-boson
width. At  LEP2, the radiative processes, both with one and more final-state
visible photons, still remain an important physics item, but within a 
different framework~\cite{yrsm96,mnpt95}. 
For instance, they are used  as QED or SM  
tests or, more generally, as tools to investigate physics beyond the SM. 
As a general feature, all these processes at LEP2 are affected by an
error which is dominated by statistics and is of the order of some per
cent. This in turn implies that a theoretical knowledge of the cross sections
with a theoretical error of the order of one per cent is mandatory, thus
requiring the inclusion of all the phenomenologically relevant radiative
corrections.  

The process $e^+ e^-  \to n \gamma$, $n \geq 2$, is well suited for testing
QED. Actually, it is marginally affected by weak or strong corrections, 
whose amount is below 1\%. Since it provides  a particularly clean signature,
it can also be used to  probe models predicting the existence of excited leptons
or contact interactions. At present, the data-theory comparison confirms pure
QED~\cite{choutko97}. 

\begin{figure}[hbt]
\begin{center}
\epsfig{file=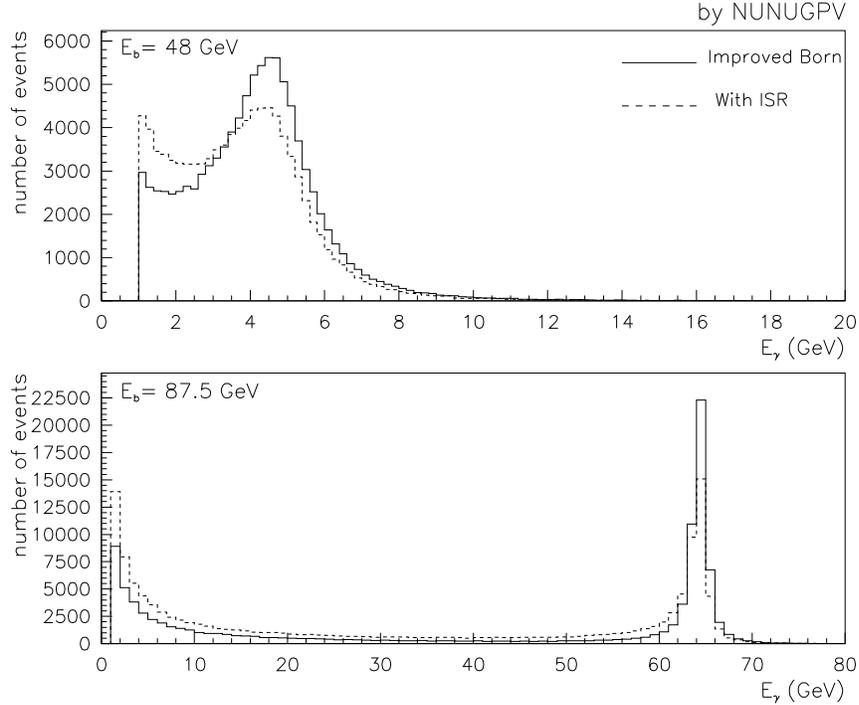, width=12truecm}
\end{center}
\caption{The energy spectrum of the observed photon at LEP1 and LEP2 energies.
Numerical  results by {\tt NUNUGPV}~\cite{nunug}. }
\label{fig:nc_eg}
\end{figure}

The production of single radiative events in the SM is dominated by
the process $e^+ e^- \to  \nu \bar\nu \gamma$, with a  sizeable background 
of radiative Bhabha events with the electrons lost. All  these events manifest
themselves as a single visible photon accompanied by missing transverse
momentum ($\rlap/{p_\perp}$). It is worth noticing that, at  a difference from
the LEP1 case  where the radiative photon is essentially soft, at LEP2, due to
the phenomenon of the $Z^0$ radiative return discussed above, the observable
photon is essentially a very energetic one, and thus easily taggable  (see
Fig.~\ref{fig:nc_eg}).  Standard
computational tools  employed by the LEP  Collaborations for the evaluation of
the SM cross sections are described  in refs.~\cite{koralz}, 
\cite{mmm}  and  \cite{nunug}. 
These events   can also be used to probe  processes beyond the
SM, such as production of neutralino-pairs  or excited neutrinos.
Moreover, they are in  principle sensitive to $WW\gamma $ anomalous couplings
(see Sect.~\ref{sect:ac} for more  details). Typical  examples of  new physics
are the following: 
\begin{itemize}
\item in the framework of the Minimal Supersymmetric Standard Model (MSSM), with
R-parity conservation and  the neutralino $\chi^0_1$ as the lightest 
supersymmetric particle  (LSP), the process of interest is $e^+ e^- \to
\chi^0_1   \chi^0_1 \gamma$~\cite{ammnp96}; 
\item in the framework  of the so-called ``no scale supergravity
model''~\cite{lnz96}, the
neutralino  $\chi^0_1$ should be visible through its radiative decay into a
photon and a gravitino $\tilde G$, $\chi^0_1 \to  \gamma \tilde G$, so that the
process  of interest is $e^+ e^- \to \gamma \tilde G \tilde G$. 
\end{itemize}  
Unfortunately, both the processes are hardly detectable because of their too 
low rate. Actually, at  present the data analyses do not point out any anomaly
with respect to  the SM predictions~\cite{chemarin97}. 

A radiative process that received particular attention during the last months 
is $e^+ e^- \to \gamma \gamma + \rlap/{p_\perp}$. Actually in the MSSM, 
such signatures could be produced by the process $e^+ e^- \to \chi^0_2 \chi^0_2
\to \gamma\gamma \chi_1^0 \chi_1^0$, where $\chi_2^0$ is the next-to-lightest 
neutralino decaying radiatively into $\chi_1^0\gamma$. In the framework of the 
``no scale supergravity model'', instead, the process of interest is 
$e^+ e^- \to \chi^0_1 \chi^0_1 \to \gamma\gamma \tilde G \tilde G$. At 
present there is no evidence for physics beyond the SM, and the data 
are used to put constraints on the masses of the supersymmetric
particles~\cite{chemarin97}.  

\subsection{Four-Fermion Processes}
\label{sect:ffproc}
As discussed in Sect.~\ref{sect:tfproc}, two-fermion production processes
certainly are an important physics item at LEP2, but the true novelty
is represented by the processes involving the production of $W$-boson pairs 
(and their backgrounds), that at LEP2 energies are kinematically accessible 
with cross sections comparable to the ones typical of two-fermion production. 
The study of $W$-boson pair production can add information concerning two
important items of the SM, namely the determination of the
$W$-boson mass~\cite{yrwmass96} and the structure of the 
triple gauge-boson couplings~\cite{yragc96}. 

At LEP1, given the set of input parameters $\alpha$, $G_\mu$ and $M_Z$ (see
Sects.~\ref{sect:z0phys} and \ref{sect:fits} for more details) all the other
observables can be derived within the SM. In particular, the
$W$-boson mass $M_W$ satisfies the relation (see eq.~(\ref{eq:gmwr}))
\begin{equation}
G_\mu = {{\alpha \pi} \over {\sqrt{2} M_W^2 (1 - M_W^2 / M_Z^2) }}
         {{1} \over {1 - \Delta r}} ,
\label{eq:mudec} 
\end{equation}
which is essentially the SM prediction for the muon decay. In
eq.~(\ref{eq:mudec}), $\Delta r$ represents the radiative corrections to the
tree-level matrix element, and hence depends, in particular, 
on the {\it top}-quark and
Higgs-boson masses, $m_t$  and $m_H$. At LEP2, the $W$-boson mass is measured
and therefore eq.~(\ref{eq:mudec}) becomes a constraint that, {\it via} $\Delta
r$, correlates $m_t$ and $m_H$ to one another. Of course, the more precise is
the knowledge of the $W$-boson mass, the stronger is the resulting constraint,
which can eventually shed  light on the Higgs-boson mass. 

Concerning the triple gauge-boson couplings, within the SM they
have a specific form determined by the underlying $SU(2) \otimes U(1)$
symmetry. Any departure from such a specific form (anomalous couplings) is in 
general responsible of a bad high-energy behaviour of the cross sections, 
spoiling the renormalizability of the theory. As a matter of fact, at  LEP2
energies an observable very sensitive to anomalous couplings is the angular
distribution of the $W$ bosons in $W$-pair production. 

In order to exploit the experimental information as a stringent test of the
SM, it is clear that one needs reliable theoretical  calculations
of the various $W$-pair production observables. It is also clear that the best
knowledge is mandatory, \idest\  the observables should be computed taking also 
into account all the relevant radiative corrections. 
\begin{figure}[hbtp]
\begin{center}
\epsfig{file=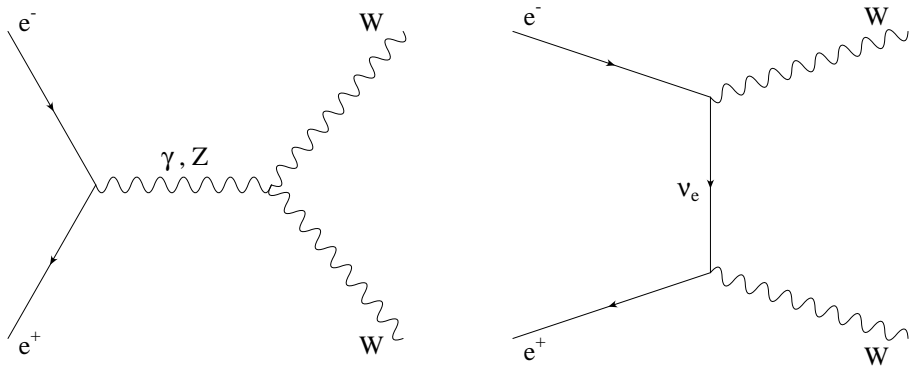, width=12truecm}
\end{center}
\caption{The tree-level Feynman diagrams  for  on-shell  $W$-pair production. }
\label{fig:wwonshell}
\end{figure}

A first step in this direction
consists in  calculating the cross section for $W$-pair production in  the so
called narrow-width (NW) approximation, \idest\   assuming that the 
$W$-boson width is zero and hence the $W$ bosons  are
stable particles (see Fig.~\ref{fig:wwonshell}). 
A huge amount of work along this direction has been performed in the
past~\cite{bd94,olww1,olww2}, and today the process
of on-shell $W$-pair  production  is known together  with  its full 
${\cal O} (\alpha)$ radiative corrections. 

\begin{figure}[hbt]
\begin{center}
\epsfig{file=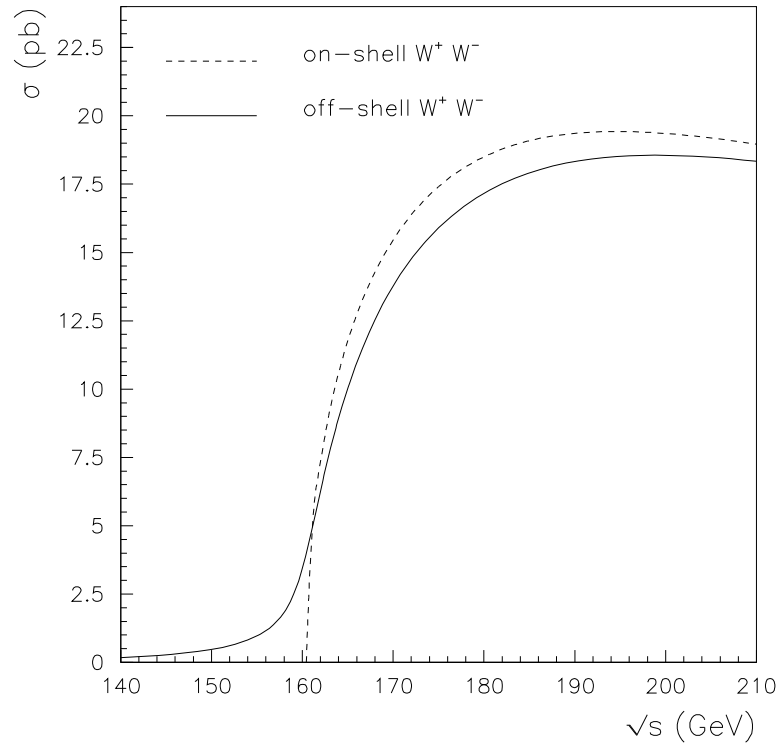, width=12truecm}
\end{center}
\caption{The effect of the finite $W$-boson width on the  inclusive cross
section. }
\label{fig:wwonoffshell}
\end{figure}

On the other hand, the $W$ bosons are, as well known, {\it unstable }
particles, so that 
the effect of their finite width is in principle important,
especially around the threshold region. Actually, whereas the cross section for
the production of two real $W$ bosons is exactly zero below $\sqrt{s} = 2 M_W$,
when the effect of the finite width is taken  into account the cross section
must become non-zero also below $\sqrt{s} = 2 M_W$, due to the possibility of
producing off-shell $W$'s which subsequently decay (see
Fig.~\ref{fig:wwonoffshell}). Moreover, the $W$
bosons, being unstable, can be only intermediate states of the reactions, the
true final  states being {\it four-fermion} $(4f)$ states. This  in turn  implies 
that,  for a  given  $4f$ final  state, besides the processes concerning the 
production and decay of a $W$-boson  pair, one  has to  consider  also  
those processes that  lead to  the same final state,  but {\it via} different
intermediate states (background processes). 

The above comments point out the need of going beyond the NW approximation,
both in order to obtain reliable cross sections and to study all the details
concerning the $W$-boson decay products. 
\clearpage
\subsubsection{Tree-level calculations}
\label{sect:tlcalc}
A first attempt in the direction of taking into account finite-width effects  
can  be found in the approach described in ref.~\cite{mnw86}.  If one is
interested in {\it  inclusive } $W$-boson decays, then the  total  cross 
section can be approximatively factorized into two stages,   
namely off-shell $W$-boson pair  
production times their  decay,  as   follows:   
\begin{equation}
\sigma (s) = B_{f_1 \bar  f_2} B_{f_3 \bar  f_4} \int_0^s d s_1 
             \rho (s_1) \int_0^{(\sqrt{s} - \sqrt{s_1})^2} d s_2
             \rho (s_2)  \sigma_0 (s, s_1, s_2) .  
\label{eq:mnw86}
\end{equation}
In  eq.~(\ref{eq:mnw86}) $\sigma_0 (s,s_1,s_2)$ represents  the total cross 
section for  the production of  two  $W$ bosons with squared  masses $s_1$ and 
$s_2$, respectively  (the Feynman diagrams of 
the production  process at the tree level are still given in Fig.~\ref{fig:wwonshell}).  
Its  explicit expression can  be
found  in ref.~\cite{mnw86}. $B_{f_a \bar f_b}$  is the
branching  ratio for  the decay channel  $W  \to f_a \bar f_b$.  $\rho(s_i)$
is the weight  factor describing the inclusive decay of  a $W$ boson of  
squared mass $s_i$. It   takes  into  account that  the $W$  bosons are 
produced  off-shell  and hence, rigorously speaking,  they  must   be  
described  by   a  propagator. The explicit   expression of $\rho  (s) $ is 
the following   
\begin{equation}
\rho (s) = {{1} \over {\pi}} {{\sqrt{s}  \Gamma(s)} \over 
           {(s - M_W^2)^2 + M_W^2 \Gamma(s)^2}}  ,  
\label{eq:rho}
\end{equation}
where $\Gamma  (s) $  is   the total $W$-boson  decay width. 
Equation~(\ref{eq:mnw86}) is a very simple and effective approximate description 
of the physics of $W$-pair production. Also, the main factorizable  QED
initial-state corrections
to  it have been considered in the literature and can be found in 
refs.~\cite{cdmn91,fkms95,ks95}. On the other hand, its main limit is
that it can  provide at most  double-differential spectra, namely the combined
invariant mass distribution of the $W$ bosons. If one  is interested in the
full description of the final-state products, a complete $4f$ calculation
is mandatory. 
 
\begin{figure}[hbt]
\begin{center}
\epsfig{file=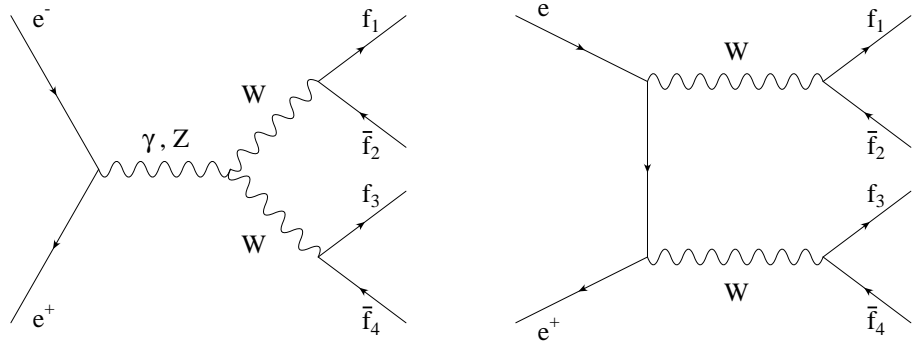, width=12truecm}
\end{center}
\caption{The tree-level Feynman diagrams for the CC03 class.  }
\label{fig:cc03}
\end{figure}

A  natural extension of eq.~(\ref{eq:mnw86}) to the description  of a full $4f$ 
final state can  be obtained by considering the diagrams of
Fig.~\ref{fig:cc03}, where the $W$ bosons are the intermediate states of the
reaction and hence the full  dynamical information on  the final-state 
fermions is contained. The diagrams considered constitute the so  called {\it CC03
class}, and  describe the full $4f$ process in the so called  double-resonant
approximation. The complete calculation of the matrix elements corresponding to
them, together with the appropriate  four-body phase space, allows one to go
beyond the ``off-shell  production times decay'' approximation of
eq.~(\ref{eq:mnw86}), and obtain the fully differential description of the
process.

\begin{figure}[hbt]
\begin{center}
\epsfig{file=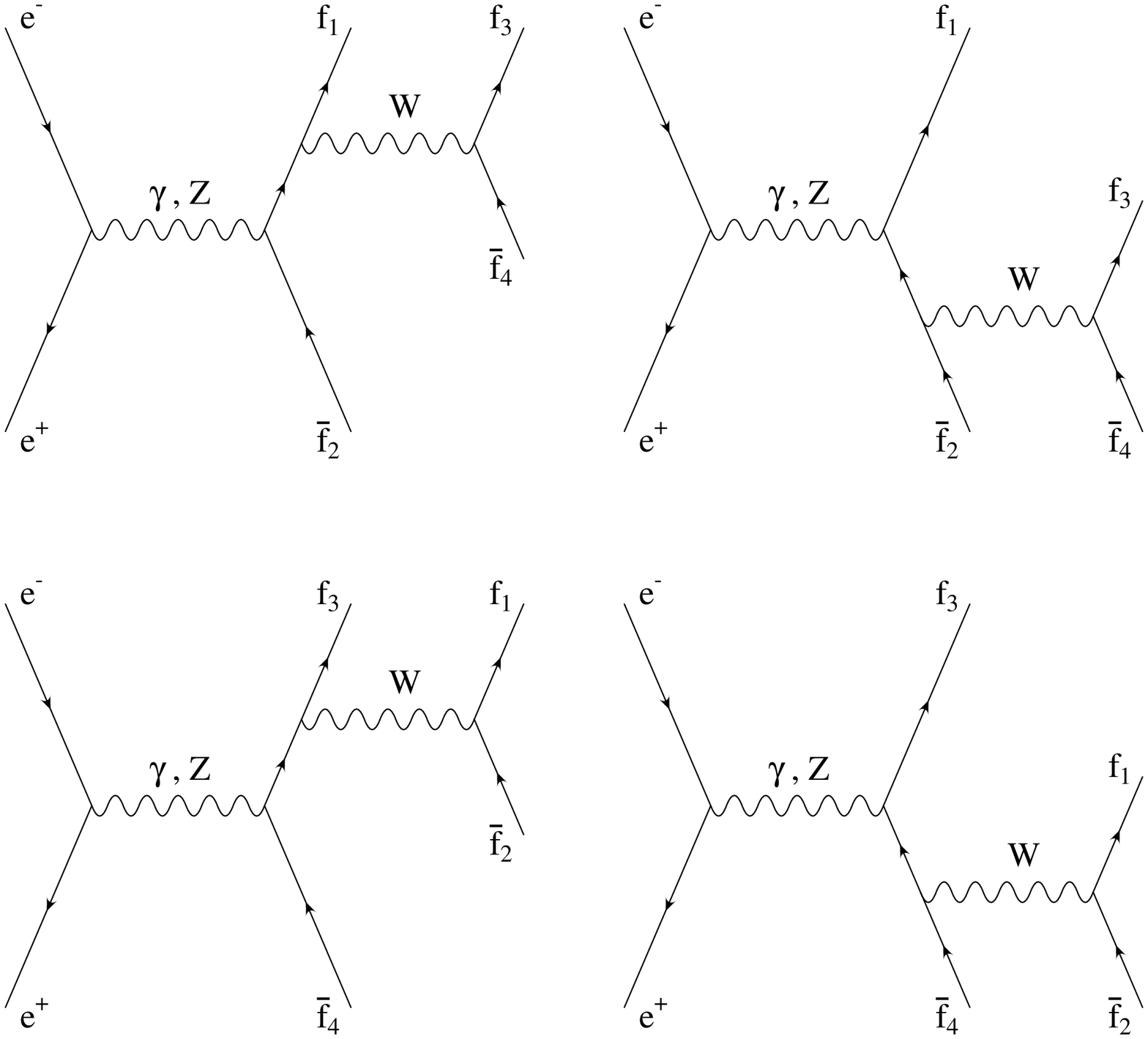, width=11truecm}
\end{center}
\caption{The additional eight tree-level Feynman diagrams for the CC11 class. Notice  
that in the semi-leptonic case ($e^+e^- \to  u \bar{d}  \mu^- \bar{\nu}_\mu$, 
for instance) the additional diagrams  are seven; in the fully-leptonic  case 
($e^+e^- \to  \tau^+ \nu_\tau  \mu^- \bar{\nu}_\mu$, for instance) the 
additional diagrams are  six.  }
\label{fig:cc11}
\end{figure}

On the other hand, as already observed in Sect.~\ref{sect:ffproc}, 
a given $4f$ final
state can be obtained also {\it via} other sub-processes. For instance, if one
considers a semi-leptonic channel, \idest\   a channel in which one $W$-boson
decays into hadrons and the other one into leptons, and restricts himself to
those channels that do not have electrons in the final  state, then the same
final state as in Fig.~\ref{fig:cc03} can also be generated by the diagrams of
Fig.~\ref{fig:cc11}. These diagrams (background diagrams), together with the
ones of Fig.~\ref{fig:cc03}, constitute the so called {\it CC11 class}, and are
characterized by having a single $W$ boson in the intermediate state, so that
they  are single-resonant. Of  course, the full matrix element for such a final
state is the sum of all the diagrams, and the squared matrix element includes
also  the interferences between double- and single-resonant diagrams.

In  the presence of electrons in the final state, the situation
is more involved. Actually, the full matrix element receives contributions also
from additional $t$-channel diagrams, so that the total number of diagrams
becomes 20 (the so called {\it CC20 class}). 

A full classification of the various
Feynman diagrams contributing to a give $4f$ final state goes beyond the aims of
the present paper. The interested reader is referred 
to refs.~\cite{yrsm96,yrwwcs96,yrwweg96}, and references 
therein, for a more
detailed account. Here it is worth noticing that for particular final states,
for instance fully hadronic final states, also neutral current (NC) and QCD 
backgrounds appear and become relevant.

The calculation of the scattering amplitudes for $2 \to 4$ processes is, 
already at the tree level, considerably more involved than the corresponding 
calculation for $2 \to 2$ processes, typical of LEP1/SLC physics. There are two
main reasons for this, namely the fact that a single Feynman amplitude for $2
\to 4$ is algebraically more involved and the fact that, typically, for a given
final state there are much more Feynman diagrams contributing. The
calculational techniques adopted in the literature can be classified as
follows: 

\begin{itemize}

\item helicity-amplitude techniques: in this approach, the scattering 
amplitude
for a given process, and for  a  given helicity pattern of the initial- and
final-state fermions, is computed analytically as a complex number by
exploiting the formal properties of the spin projection operators; the squared
modulus  of the amplitude is then computed numerically; the approach, in all
its actual implementations (see refs.~\cite{evform,wvdwform,covform,tmform} and references therein), 
is particularly powerful for massless fermions, albeit also mass effects can be
taken into account;  

\item automatic calculations: these approaches adopt both standard
techniques for the evaluation of the squared matrix element~\cite{comphep} and
the helicity amplitude formalism for the evaluation of  the scattering
amplitude~\cite{channellib}, properly interfaced  with software packages that
render the calculation of cross sections almost automatic; 

\item numerical evaluation of the generating functional for the connected
Green's functions: it  is  a  new method, presented in ref.~\cite{alpha},
in  which  the scattering  amplitude is computed numerically by means of  an
iterative algorithm starting from the effective  action of the theory and with 
no use of Feynman diagrams; it becomes strongly competitive with respect to 
standard techniques as the number of final-state particles becomes larger and
larger.   

\end{itemize}

\subsubsection{Gauge invariance}
\label{sect:ginv}
Being in the framework of a gauge theory, as is the case of  
the SM, means that the calculations of physical  
observables must be gauge invariant, or at least the gauge  
violations have to be confined well below the required  
theoretical accuracy. In the case of $4f$ final  
states there are two sources of problems connected with gauge  
invariance in theoretical calculations 
(for a review see for instance ref.~\cite{beencrad96}).  

The first one originates when the  
matrix element of the process is calculated considering only  
a subset of the Feynman diagrams contributing to the scattering  
amplitude. For example, in the case of $W$-boson production,  
the natural extension from the on-shell approximation to the  
more realistic off-shell production and subsequent decay  
would be the calculation of  
the three Feynman diagrams of the on-shell case with the two  
final-state fermionic currents attached to the virtual $W$-boson lines,  
\idest\   the double-resonant diagrams of Fig.~\ref{fig:cc03}.  
In ref.~\cite{bd94} it has been shown that from the evaluation  
of the double-resonant diagrams in an axial gauge,  
gauge dependent terms arise, which exhibit a single pole  
structure in either of the two invariant masses and are cancelled  
when also the single-resonant diagrams (see Fig.~\ref{fig:cc11}) 
contributing to the process  
are evaluated. The single- and the non-resonant contributions  
are generally suppressed by a factor $\Gamma_W / M_W \simeq 2.5\%$  
with respect to the double-resonant ones, but they need to be taken  
into account in view of the required theoretical accuracy. 
 
Even when the calculation of all the tree-level 
diagrams contributing to a given  
$4f$ final state is performed, the problem of gauge  
invariance is not yet solved because of the presence of   
singularities in the phase space due to the massive gauge boson  
propagators, which display poles at $p^2 = M^2$,  $p^2$ being the  
invariant mass of the decay products of the unstable bosons.  
These singularities are cured by the introduction of the finite  
widths of the gauge bosons, which shift the poles away from  
the real axis.  
However, in field theory the widths arise from the imaginary parts of  
higher-order diagrams describing the gauge boson self-energies,  
resummed to all orders. So the tree-level amplitude is supplemented  
with only a subset of higher-order contributions and this can destroy  
the gauge invariance of the calculation.  
 
A pragmatic approach, suitable for the construction of a  
lowest-order event generator, is  
the so called ``fixed-width scheme'', where the propagators  
$1 / (p^2 - M^2)$ are systematically replaced with  
$1 / (p^2 - M^2 + i\Gamma M)$ also for space-like momenta.  
This procedure has no physical motivation, because the propagator  
for space-like momenta does not develop any imaginary part. An  
improvement on this is the ``running-width scheme'', where the  
width is a function of $p^2$, equal to zero for space-like momenta,  
in agreement with the calculation of the imaginary part of the  
gauge boson self-energy.  
 
Since the resonant diagrams are not gauge invariant by themselves,  
both of the above mentioned schemes violate gauge invariance,  
and their reliability needs to be checked by a truly  
gauge invariant scheme, even if it cannot be uniquely defined.  
In the literature several gauge restoration schemes have been  
studied. The simplest one is the ``fudge-factor  
scheme''~\cite{bfz,kpgs}, which  
amounts to calculate the matrix element without any width  
(so that the calculation is manifestly gauge invariant) and  
introduce an overall factor of the form  
$(p^2 - M^2) / (p^2 - M^2 + i \Gamma M)$ for every singularity,  
in order to transform it in a resonance.  
The problem connected with  
this scheme is that when the double-resonant diagrams are not  
dominant, as is the case of charged current processes at energies  
below and at the $W$-pair production threshold, it can lead to  
large deviations~\cite{bz}.  
 
Another possibility is the ``pole  
scheme''~\cite{v63,stuart91,avow,s9603351},  
where the complete amplitude is decomposed according to the  
pole structure in gauge invariant subsets of double-,  
single-  and non-pole terms. Introducing the widths  
in the pole factors does respect gauge invariance. At present  
there is some debate about the correct way of implementing this  
scheme and about its validity in the vicinity of  
thresholds.  
Since the pole scheme is a gauge invariant decomposition of  
the amplitude according to its degree of resonance, it can be used  
as a starting point for the evaluation of higher-order corrections,  
as will be discussed in Sect.~\ref{sect:ffrc}.  
 
The most theoretically appealing solution to the gauge invariance  
problem is the ``fermion loop scheme''~\cite{bz,p95,bhp1,bhp2},  
where the minimal set of one-loop Feynman diagrams necessary for  
compensating the gauge violation caused by the self-energy graphs  
is included in the calculation. Since the lowest order decay widths  
of the gauge bosons are given by the imaginary parts of the  
fermion loops in the one-loop self-energies, according to this  
scheme the imaginary parts of all the other possible  
one-particle-irreducible fermionic one-loop corrections  
must be included. For the process $e^+ e^- \to 4f$,  
after the resummation of the vector boson self-energies,  
the only left out contributions are given by  
the imaginary parts of the fermionic corrections to the triple  
gauge-boson vertex, which have been calculated in ref.~\cite{bhp1}  
in the limit of massless fermions.  
The calculation has been extended in ref.~\cite{bhp2} to take  
into account the complete fermionic one-loop corrections,  
including real and imaginary parts, and all contributions of the  
massive {\it top} quark.  
 
Due to the complexity of the fermion loop scheme, the CPU time  
needed for the calculation of the cross section increases  
considerably with respect to the one performed within the  
fixed or running width scheme, based on a naive treatment  
of the bosonic widths. For this reason, from a phenomenological  
point of view, it is important to quantify the amount of  
gauge violating effects in $4f$ processes of interest  
at LEP2. As a result of the detailed investigations pursued in  
refs.~\cite{bhp1,bhp2}, the violations connected with the gauge  
group $SU(2)$ are not relevant in the LEP2 energy range.  
They become important for energies reached at the next generation  
of linear colliders. The case of $U(1)_{e.m.}$  
gauge invariance violation is different, since the effects are enhanced by  
a factor of ${\cal O}(s/m_e^2)$ in processes with an electron  
or positron in the final state, almost collinear to the incoming  
particle~\cite{kpgs,bz,bhp1,p95}. This happens, for instance, for the  
CC20 process $e^+ e^- \to e^- {\bar \nu}_e u {\bar d}$,  
which is particularly important for studying triple gauge-boson  
couplings. In ref.~\cite{bhp1} it has been shown that the  
running width scheme for this process gives a totally  
wrong result when the minimum scattering angle of the electron  
is set to zero, unless it is  
improved with the fermionic corrections to the trilinear gauge  
boson coupling. On the other hand, the fixed width scheme, which  
does not violate the electromagnetic current conservation, gives  
reliable results. It is worth noticing that the results for  
the total cross section, obtained by means of different methods respecting  
current conservation, agree to ${\cal O}(\Gamma_W^2 / M_W^2)$.  
In the limit $q^2 \to 0$, where $q^2$ represents the space-like  
momentum transfer of the electron through the photon, the  
fermion loop correction to the standard Yang-Mills vertex can  
be written in the following factorized form~\cite{bhp1}: 
\begin{eqnarray*} 
C_{fl} = 1 + i{{\gamma_W(p_+^2)}\over {p_+^2 - p_-^2}} , 
\end{eqnarray*} 
where  
\begin{eqnarray} 
\gamma_W(q^2) &=& {{\Gamma_W} \over {M_W}} q^2 , \quad q^2 \geq 0 ,  
\nonumber \\ 
\gamma_W(q^2) &=& 0 , \quad q^2 \leq 0 , \nonumber  
\end{eqnarray} 
and $p_{\pm}$ are the squared momenta flowing in the $W^{\pm}$-boson  
propagators. 
This approximate factorized prescription can be easily implemented   
in a Monte Carlo event generator. However, the validity of the above formula  
is  limited to the LEP2 energy range.  
 
As a last comment, it is worth noticing that the problems  
connected with $U(1)_{e.m.}$ gauge invariance in the CC20 processes  
are rendered much more mild when a cut on the scattering angle  
of the outgoing electron is imposed. For a realistic cut of $10^{\circ}$  
the results obtained with the fixed width scheme are compatible  
with the ones obtained with the fermion loop corrections~\cite{bhp1}. 
 
\subsubsection{Radiative corrections}
\label{sect:ffrc}
As already discussed in Sect.~\ref{sect:ffproc}, 
the main physical motivations of the LEP2 experiments are  
a precise determination of the $W$-boson mass and the study of  
the triple gauge-boson couplings.  
Since in the LEP2 energy range the integrated  
cross section is not very sensitive to the presence of anomalies  
in the Yang-Mill sector, the angular distributions are the  
best suited observables for the determination of the anomalous  
couplings. For this reasons the radiative corrections to both  
total and differential cross sections should be under control. 
Moreover, as will be illustrated in Sect.~\ref{sect:wmass},  
one of the methods used for the $W$-boson mass measurement  
(the direct reconstruction method) 
relies upon the knowledge of the energy loss due to  
ISR defined as  
\begin{eqnarray} 
<E_{\gamma}> = {1\over {\sigma_{tot}}}\int dE_{\gamma} 
 {{d\sigma }\over {dE_{\gamma}}} E_{\gamma} ,  
\label{eq:enloss}
\end{eqnarray} 
which requires the best possible control of photonic  
radiative corrections.
The other method (the threshold method) requires a good knowledge of the total
cross section in the threshold region, \idest\   a total cross  section
including all the relevant radiative corrections.  
However, it should also be kept in mind that, given the  
limited statistics of the LEP2 experiments with respect to LEP1,  
a theoretical relative accuracy of the order of $1\%$ at  
the $W$-pair production threshold and of $0.5\%$ at higher energy  
will be enough.  
 
Before discussing various aspects of radiative corrections to  
$4f$ processes, it is worth recalling the r$\hat{ \hbox{\rm o}}$le played  
by the $W$-boson mass at LEP2 with respect to the situation of LEP1.  
As already pointed out, 
in this last case it is an observable calculated by means of  
the input parameters $\alpha$, $G_{\mu}$ and $M_Z$, and,   
with the inclusion of the radiative corrections to the 
 $\mu$ decay, $M_W$ shows a parametric  
dependence on $m_{t}$, $m_{H}$ and $\alpha_s$, as results  
from eq.~(\ref{eq:gmwr}). In the case of LEP2 it is more natural  
to consider $M_W$ as a free parameter to be fitted by the  
experiments. As a consequence,    
eq.~(\ref{eq:gmwr}) can be solved with respect  
to $m_{top}$ as a parametric function of $\alpha_s$ and $M_{H}$,  
to be compared with the direct measurement from the
TEVATRON~\cite{dward97,taylor97},  
allowing for a stringent determination of the Higgs-boson mass.  
 
Turning now the attention to LEP2 processes, the ideal situation  
would be the knowledge of the complete set of ${\cal O} (\alpha)$ 
radiative corrections to  
processes with four fermions in the final state, but at present the calculation,  
due to its complexity,  is not available. In fact, such a
task would require the calculation of ${\cal  O} (10^3 \div 10^4)$ one-loop
diagrams. It is important to notice that, 
due to the presence of charged vector bosons in the Born amplitude,  
a separation between virtual photonic and pure  
weak corrections respecting the $SU(2)$ gauge invariance  
is not possible.  
Anyway a class of dominant corrections important for LEP2 physics  
can be singled out, and the effects of non-leading terms can be  
estimated relying upon the knowledge of the full one-loop  
electroweak corrections for the reaction $e^+ e^- \to W^+ W^-$, where  
the charged gauge bosons are considered in the on-shell  
limit~\cite{olww1,olww2}, including the bremsstrahlung  
process $e^+ e^- \to W^+ W^- \gamma$~\cite{wwg}.  
Since the exact ${\cal O}(\alpha )$ calculation, already for on-shell $W$'s, 
leads to complicated  
and lengthy expressions in terms of twelve $s$- and $t$-dependent  
effective couplings, the possibility of  
constructing an Improved Born Approximation in the LEP2 energy  
range has been studied~\cite{dbd}, with the aim of providing simple and transparent  
formulae to be eventually implemented in Monte Carlo's for off-shell  
$W$-pair production. 

Considering the realistic case of $4f$ production, 
in the following a brief account is given on  
the state-of-the-art about the three main classes of radiative  
corrections of interest for LEP2 physics:  
weak, QCD and QED corrections.  

\begin{figure}[hbt]
\begin{center}
\epsfig{file=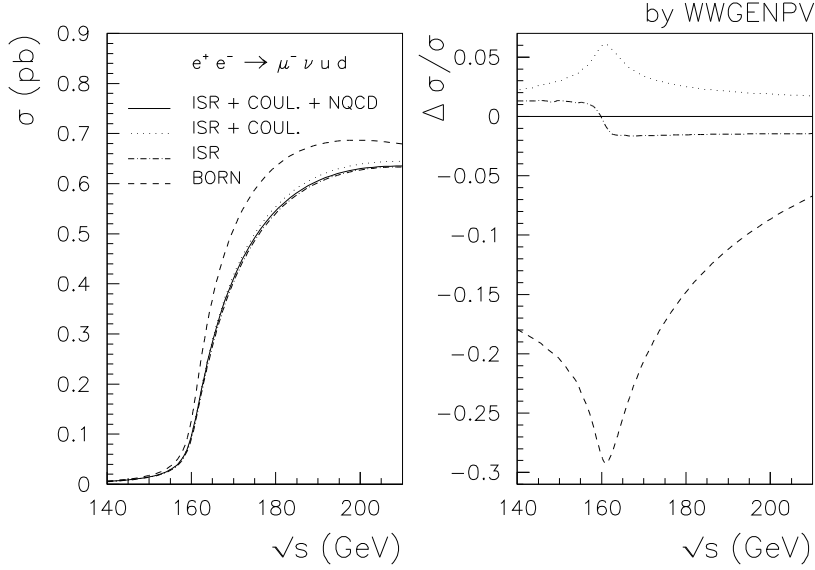, width=12truecm}
\end{center}
\caption{The contribution of various radiative corrections. The left  window
shows the  cross section for the channel $e^+ e^- \to \mu^-
\bar{\nu_{\mu}} u \bar d $ at the tree  level (dashed  line) and including ISR
(dash-dotted line), the coulombic correction (dotted line) and naive  QCD
corrections (continuous  line). The right window shows the relative effect of
each correction.  $\Delta \sigma  / \sigma  $ is defined as $(\sigma_{ISR}  -
\sigma_{Born}) / \sigma_{Born} $ for the dashed line (the relative effect of
ISR), $(\sigma_{ISR+COUL.}  - \sigma_{ISR}) / \sigma_{ISR} $ for the dotted
line (the relative effect of the Coulomb correction) and 
$(\sigma_{ISR+COUL.+NQCD}  - \sigma_{ISR+COUL.}) / \sigma_{ISR+COUL.} $ for the
dash-dotted line  (the relative effect of naive QCD corrections). Numerical 
results produced by {\tt  WWGENPV}~\cite{wwgenpv}.}
\label{fig:wwcorrections}
\end{figure}

As already pointed out, at LEP2 the $W$-boson mass $M_W$
should be a fitting parameter. By simply adding it to the three data points 
used as input parameters at LEP1 (see Sect.~\ref{sect:rsch}) leads to  a 
 four data points scheme, based on $\alpha$,  
$G_{\mu}$, $M_Z$ and $M_W$, that is overcomplete, so that  
one of the typical LEP1 input  parameters has to be replaced with $M_W$.  
Considering  
eq.~(\ref{eq:gmwr}) with $\Delta r = 0$, at least two tree-level 
schemes are possible: 
\begin{eqnarray} 
s_W^2 &=& 1 - {{M_W^2}\over {M_Z^2}} , \nonumber \\ 
g^2 &=& 4 \sqrt{2} G_{\mu} M_W^2 ,  \nonumber 
\end{eqnarray} 
and  
\begin{eqnarray} 
s_W^2 &=& {{\pi \alpha }\over {\sqrt{2} G_{\mu} M_W^2}} ,\nonumber \\ 
g^2 &=& {{4 \pi \alpha }\over {s_W^2}} .  \nonumber  
\end{eqnarray} 
However, for practical calculations, the leading  
universal weak effects can be absorbed in the tree-level  
calculations by defining the input parameters in terms of 
$G_{\mu}$ and $\alpha (s)$. This choice, applied to the second scheme, is the
one adopted in~\cite{yrwweg96}. 
 
Going beyond this phenomenological attitude, it is necessary to talk about a
complete one-loop calculation.  
By adopting the fermion loop scheme, all  
one-particle irreducible fermionic one-loop corrections are included,  
and a theoretically appealing strategy in this direction would be to calculate  
the remaining bosonic corrections, at least within a suitable approximation. 
The pole  
scheme~\cite{v63,stuart91,avow} offers such a framework, provided that the c.m.
energy is sufficiently far away from the threshold region  (a few $W$-boson
widths above).  Actually, as stated  
in  Sect.~\ref{sect:ginv}, this scheme is gauge invariant,  
contains the corrections for the production and decay  
of on-shell bosons as building blocks, and the number of Feynman  
diagrams to be evaluated is considerably reduced with respect to  
the complete set,  
because the potentially relevant corrections in the LEP2 energy  
region are given by the ones affecting the double-resonant diagrams.  
However, a simpler approximate procedure to  
take into account the leading bosonic corrections in the on-shell $W$'s limit
has been recently suggested~\cite{kks97}, which  
amounts to perform the substitution  
\begin{eqnarray} 
G_{\mu} \to G_{\mu} / (1 + \Delta y_{bos}^{SC}) , \nonumber  
\end{eqnarray} 
with $\Delta y_{bos}^{SC} = 11.1 \cdot 10^{-3}$. This replacement   
is equivalent to use the $SU(2)$ gauge coupling $g(M_W^2)$ at  
the high energy scale of $M_W$, defined by the theoretical value  
of the radiatively corrected leptonic $W$-boson width,  
$g^2(M_W^2) = 48 \pi \Gamma_l^W / M_W$.\footnote{The full 
one-loop electroweak and QCD corrections to the decay width  
of the $W$ boson have been calculated  
in ref.~\cite{wgcorr}. An Improved Born  Approximation absorbing the  
bulk of the radiative corrections can be obtained by writing the  
lowest order width in terms of $G_{\mu}$ and $M_W$~\cite{denner93}. }  
The essence of the approach is to use $\Gamma_l^W$ instead of  
$G_{\mu}$ as input parameter, and it has been checked to work well  
in the case of on-shell $W$-pair production, providing results with  
an accuracy better than 1\% with respect to the full one-loop 
results. Moreover, in ref.~\cite{kks97} it has been shown that strong  
cancellations occur between fermionic and bosonic corrections,   
at least at the level of on-shell $W$-pair production. Since the  
same behaviour can be reasonably expected also in the case of  
off-shell $W$-pair production~\cite{bhp2}, these results  confirm that 
the fermionic loop contributions to $e^+  e^- \to 4f$ need  
to be completed with some approximation of the bosonic corrections. 

As far as the $W$-boson mass determination by the threshold method is
concerned, it is worth mentioning a specific bosonic correction arising from  
the  
exchange of a relatively light Higgs boson between the slowly moving  
$W$ bosons. Given the presently allowed range of variation for the Higgs-boson  
mass, it induces a theoretical uncertainty on the total cross section  
near the threshold of $W$-pair production. 
In the limit $m_H \ll M_W$ the leading behaviour in  
the on-shell case is given by~\cite{dittcrac96}  
\begin{eqnarray} 
\delta \sigma \simeq {{\alpha }\over {2 s_W^2}} {{M_W}\over {m_H}}  
\sigma_{Born} . \nonumber
\end{eqnarray} 
In ref.~\cite{beegj} this correction has been computed for off-shell  
$W$-boson production neglecting the $s$-channel vertex  
contributions, due to the dominance of the $\nu$ exchange  
diagram at threshold. For an Higgs-boson mass of $60$~GeV the  
correction amounts to about $0.8-0.9$\% and decreases for higher  
Higgs-boson mass values (being less than $0.1$\% for $m_H = 300$~GeV).  
This effect translates into an uncertainty on the determination  
of $M_W$ of $15$~MeV from the LEP2 threshold run. 
 
Concerning QCD corrections,  
when at least two quarks are present in the final state, \idest\  in  
semileptonic and fully hadronic channels, also  
the strong radiative corrections to the $4f$ processes have to be  
considered.  
In the case of an inclusive set-up and in the CC03  
approximation, the QCD corrections amount to use the QCD corrected  
values of the total and partial $W$-boson widths in the cross section.  
These have been  
calculated in the literature~\cite{QCDwidth} and can be expressed  
in factorized form by the following replacements:  
\begin{eqnarray} 
&&\Gamma_{Wu_id_j}  \to  \Gamma_{Wu_id_j}\left( 1 +  
{{\alpha_s}\over{\pi}}   \right) , \nonumber \\ 
&&\Gamma_W  \to  \Gamma_W \left( 1 + {{2 \alpha_s}\over  
{3 \pi}}  \right) . \nonumber  
\end{eqnarray} 
The impact of QCD corrections on the angular distribution of the  
decay products of a $W$ boson and their application to the on-shell $W$-pair 
production has been discussed in ref.~\cite{ablampe}.  
The QCD corrections for the inclusive set up and in the double-resonant 
approximations can be used as estimates of the corrections  
for more realistic cases. For this reason the above replacements  
are referred to in the literature as {\it naive} QCD corrections. 
The quality of the approximation   
when cuts on the jet directions are imposed has been discussed in  
ref.~\cite{mp96}. 
The effect of naive QCD corrections on the  cross section of a typical $4f$ process 
is shown in Fig.~\ref{fig:wwcorrections}. 
 
Recently the exact ${\cal O}(\alpha_s)$ corrections to  
semileptonic and hadronic   processes have been  
calculated~\cite{mpp96} and compared with the naive approximation both in a  
fully inclusive situation and in the presence of  
ADLO-TH cuts~\cite{yrwweg96}. The differences are  below the  
foreseen experimental error at LEP2 for the semileptonic channels, while  for fully hadronic final
 states the naive approach can give unreliable results. 
The analysis has been performed  
also at a c.m. energy of $500$~GeV, typical of future  
$e^+ e^-$ linear colliders, where the  
naive approximation deteriorates at the percent level  for NLC/TH  
cuts~\cite{physrepnlc}. 

Concerning at last QED corrections,  
in principle they  should consist of the virtual  
photonic corrections and the real-photon bremsstrahlung in  
order to have an infrared safe result. A separation between  
initial- and final-state radiation respecting the $U(1)$  
electromagnetic gauge invariance, as in the case of LEP1 $s$-channel 
processes,  is not possible because the  
$\nu$-exchange diagram involves a non-conserved charge flow  
in the initial state. The problem is circumvented if  
the treatment of electromagnetic radiation is restricted  
at the LL level. The first theoretical calculations taking into account  such
effects  in $4f$ productions can be found in 
refs.~\cite{bpk94,bpk94b,mnppw95,sjmpw96,jpsw96,blr96,br96}.  
In this case the enhanced logarithmic terms coming from final-state  
radiation (FSR)  cancel by virtue of the KLN theorem (when sufficiently  
inclusive cuts are considered) and the initial-state  
corrections reveal the presence of infra-red and collinear singularities,  
which are process-independent.  
Being phenomenologically relevant, these can be resummed  
to all orders by means of one of the methods described in Appendix A.  
Given the steep slope of the born cross section in the threshold  
region, the numerical effect of ISR  is to reduce  
the cross section by several percent (see Fig.~\ref{fig:wwcorrections}),  
depending on the c.m. energy~\cite{cdmn91}.  
By means of various procedures (for instance the variation of the scale in the
LL theoretical predictions) it is possible to state that the overall accuracy
of the LL QED corrections is of the order of 1\%. 
Another possible way of estimating non-leading QED effects is 
the one followed in ref.~\cite{currsplit}, where  
a gauge invariant definition of ISR is given  
according to the current splitting technique. The electrically  
neutral neutrino of the $t$-channel diagram is split into two  
oppositely flowing charged leptons. One of them is attributed to the  
initial state to build a continuous charge flow, while the  
other one is attributed to the final state. In this way the  
${\cal O} (\alpha)$ ISR form factor for $s$-channel processes used at 
LEP1 (see eq.~(\ref{eq:alfak})) is recovered, 
plus non-factorizing terms numerically small in the LEP2 energy range. 
 
However, when aiming to study the impact of exclusive photon radiation   
on particular observables, such as the photon spectrum or the differential 
angular distributions  
of final-state particles in the presence of realistic experimental  
cuts, some improvements on the LL   
approximation have to be introduced. For example, in ref.~\cite{wwgenpv}  
the transverse degrees of freedom of emitted photons have been 
introduced by means of the $p_T$ dependent SF approach, 
in order to take into account the kinematical effects of the photonic 
radiation from  all external charged particles. Alternatively, 
in ref.~\cite{jadach97} the exclusive photonic contributions, including 
also multiple photon radiation from the $W$ bosons, are implemented 
by means of a ${\cal O}(\alpha )$ YFS exponentiation. In this  
formulation the ${\cal O}(\alpha )$ corrections are implemented using the  
results for $W$-pair production in the on-shell limit. 

\begin{figure}[hbt]
\begin{center}
\epsfig{file=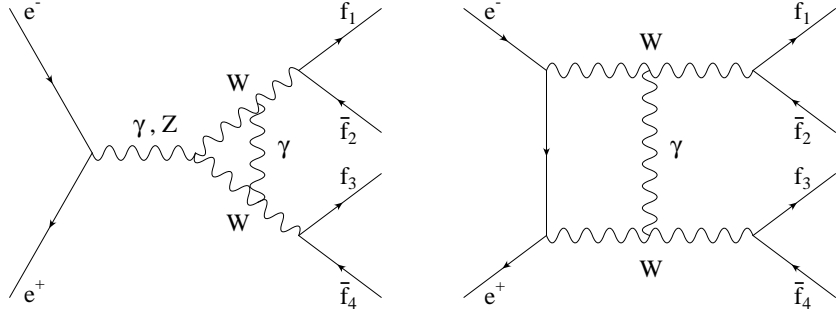, width=11truecm}
\end{center}
\caption{The Feynman diagrams  for the Coulomb correction.}
\label{fig:coulomb}
\end{figure}

In addition to the LL effects, a particularly sizeable 
photonic correction in the $W$-pair production region is the so-called 
{\it Coulomb singularity} (see  Fig.~\ref{fig:coulomb}). 
It originates from the electromagnetic interaction  
between the slowly moving $W$'s. This effect is known since a long  
time~\cite{sommerfeld} for stable particles to give a correction  
factor $\delta \sigma_{Coul}$ of the form  
\begin{eqnarray} 
\delta \sigma_{Coul} = {{\pi \alpha}\over {2 \beta}} \sigma_{Born} , 
\nonumber 
\end{eqnarray} 
where $\beta = \sqrt{1 - M_W^2/E^2}$ represents the velocity of  
the $W$ bosons.  
The off-shellness and the width of the $W$ bosons modify radically  
the Coulomb interaction because they act as an effective  
cut off on the range of the electromagnetic interaction. Only for  
high enough energies, such that  
the typical interaction time between the $W$'s  
is smaller than the $W$-boson lifetime, the effect derived for stable particles  
is substantially unchanged. In terms of Feynman diagrams the  
Coulomb singularity arises from  three- and four-point  
scalar functions contained in the  graphs  
with a photon line connecting the two virtual $W$ bosons  (see  Fig.~\ref{fig:coulomb}). Being the  
coefficients of these scalar functions gauge invariant, they can be  
worked out to give the following correction factor to the lowest  
order cross section resulting from the double-resonant  
diagrams~\cite{coulomb}: 
\begin{eqnarray} 
\delta \sigma_{Coul} = \sigma_{Born}^{CC03} {{\alpha \pi}\over  
{2 {\bar \beta}}} \left[ 1 - {2\over {\pi}}\arctan \left(  
{{\vert \beta_M - {\bar \beta} \vert}\over  
{2 {\bar \beta {\rm Im} \beta_M}}} \right) \right], \nonumber  
\end{eqnarray} 
with  
\begin{eqnarray} 
&& {\bar \beta } = {1\over s} \sqrt{s^2 - 2s (k_+^2 + k_-^2)  
+ (k_+^2 - k_-^2)} ,   \nonumber \\ 
&& \beta_M = \sqrt{1 - {{4 M^2}\over s}} , \quad  
M^2 = M_W^2 - i M_W \Gamma_W - i \varepsilon  .   \nonumber  
\end{eqnarray} 
where $k_+, k_-$ are the four-momenta of the $W^+$ and $W^-$, respectively.
By performing the on-shell limit on this formula the correction  
factor for stable particles is recovered.  
Higher-order effects due to the Coulomb correction 
are unimportant~\cite{coulomb}.  
Numerically the Coulomb singularity amounts to a correction of the  
order of 6\% at threshold and decreases smoothly to about 2\% at  
$\sqrt{s} = 190$~GeV  (see Fig.~\ref{fig:wwcorrections}). 
 
\begin{figure}[hbt]
\begin{center}
\epsfig{file=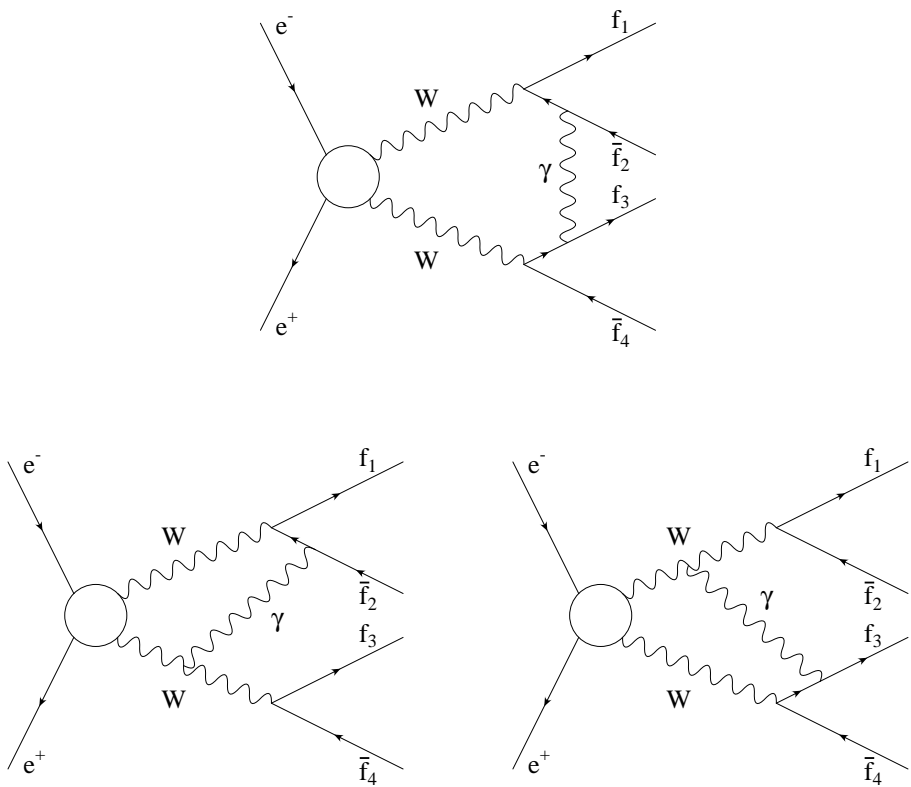, width=12truecm}
\end{center}
\caption{Examples of Feynman diagrams  for non-factorizable virtual  QED
 corrections.}
\label{fig:nonfactor}
\end{figure}

The QED corrections discussed above are phe\-no\-me\-no\-lo\-gi\-cal\-ly relevant,
gauge-in\-va\-riant and include, besides the exponentiation of LL 
contributions, part of the full ${\cal O} (\alpha)$ electroweak 
corrections. The natural development is to consider the 
remaining non-leading ${\cal O} (\alpha)$ corrections.  
As stated previously, a step towards the calculation of the  
bosonic contributions to the ${\cal O(\alpha )}$ radiative  
corrections to $4f$ processes  
can be performed within the pole scheme~\cite{v63,stuart91,avow},  
even if this gives reliable results only a few $\Gamma_W$'s  
above the $W$-pair production threshold.  
In the threshold scan the required theoretical accuracy is, on the  
other hand, more relaxed with respect to the high energy points.  
Given the dominance of the  
double-resonant diagrams in the LEP2 energy region, a good  
approximation is to consider the radiative corrections in the  
double pole approximation. These can be divided in factorizable  
and non-factorizable corrections (see
Fig.~\ref{fig:nonfactor} for an  example of non-factorizable virtual  QED
corrections). 

The former contain manifestly  
two resonant $W$-boson propagators and can be distinguished between  
corrections to $W$-pair production and decay, while the latter do not. 
At present, the factorizable corrections to $e^+ e^- \to 4f$ can be 
taken into account by using the known ${\cal O} (\alpha)$ corrections for on-shell $W$-pair
production and decay. This is a first reasonable approximation 
to the full one-loop calculation, since the neglected terms are of the 
order of $\alpha \Gamma_W / \pi M_W$. This 
approach has been already followed in ref.~\cite{jadach97}.
 Recently, the effects of non-factorizable QED diagrams  in the  
soft-photon limit have been  
investigated~\cite{fkm94,melyak94,melyak96,bbc97,ddr97}. As a result of this 
calculations, the non-factorizable corrections vanish in the case of initial-final  
state interference, and in all cases when the integrations over both  
invariant masses of the virtual $W$ bosons are performed. Hence the  
$W$-boson production angle is insensitive to these corrections, so that  
the studies of anomalous triple gauge-boson couplings at LEP2 are  
not affected by them.  
The case of the $W$-boson invariant mass distribution is different, 
since it receives a relative contribution of the order of 1\%. The same  
corrections vanish for $ZZ$-mediated and $ZH$-mediated $4f$ processes. 
  
\subsubsection{Computational tools}
\label{sect:ffct}
As done in Sects.~\ref{sect:sabct} and  \ref{sect:z0ct}, for small-angle Bhabha 
scattering and $Z^0$ physics, respectively, the basic features of the computer 
codes available for the calculation
of observables of $4f$ production processes
are briefly summarized. The aim of the discussion is to show how 
the theoretical results described in Sects.~\ref{sect:tlcalc} and \ref{sect:ffrc} 
are in practice implemented in computational 
tools used for the experimental analysis of $W$-pair production at LEP2.
Since, as it will be discussed in Sect.~\ref{sect:higgs}, 
Higgs-boson physics at LEP requires the study of 
$4f$ final states, some of the programs developed for $W$-boson 
physics at LEP2 can provide also predictions for Higgs-boson production and decay. 
The goal of this discussion is to make an  inventory of the theoretical  approaches 
to the problem and of their realizations in the form of {\tt FORTRAN} codes, rather then  
to give an exhaustive description of the programs. For a more detailed account of the 
programs,  the reader is referred to refs.~\cite{yrwweg96,yrdpeg96,ohlcrad96}, and to 
the authors of the packages for the most recent developments.  

The most of the $4f$ codes developed for LEP2 are 
Monte Carlo programs. Indeed, since the phase-space integration over the 
$4f$ final state involves seven dimensions, it can be more efficiently 
performed, especially in the presence of arbitrary kinematical cuts, 
by means of standard or adaptive Monte Carlo techniques. However, other 
numerical 
algorithms are available in the literature. For instance, semi-analytical 
programs, where a number of phase-space variables is 
integrated analytically, ending with two integrations over invariant 
masses (see eq.~(\ref{eq:mnw86})) to be performed numerically, and deterministic 
tools have been developed. 
In particular, semi-analytical codes are useful as
benchmark for the Monte Carlo's and for fitting purposes.
Independent of the particular numerical algorithm 
employed, variance reduction techniques (such as importance sampling) 
are demanded and used in practice in order to take care of the 
peaking behaviour of the integrand.

Although different in many theoretical and technical aspects, 
most of the programs include 
the following common ingredients, necessary to match the LEP2 
experimental precision (see the discussion in Sect.~\ref{sect:ffrc}):
\begin{itemize}

\item the full or the numerically relevant set of Feynman diagrams for the 
processes $e^+ e^- \to 4$~fermions;

\item ISR in the collinear approximation, usually in 
the form of QED SF's;

\item Coulomb correction;

\item QCD corrections, in the naive realization, to the total and 
partial $W$-boson widths;

\item possibility of anomalous couplings (see Sect.~\ref{sect:ac});

\item interface to hadronization packages.

\end{itemize}    
Some codes implement additionally  QED  FSR from fermionic 
legs. 
A few programs generate $p_T$-carrying photons, at the level of ISR and/or 
FSR, while others are able to calculate the matrix element for the radiative 
process $e^+ e^- \to 4f + \gamma$. Finally, some programs include one or 
more of the gauge restoration mechanisms discussed in Sect.~\ref{sect:ginv}. 

\noindent
\underline{\tt ALPHA}\cite{alpha} --- This program is based upon an original 
theoretical algorithm recently proposed in the literature~\cite{alpha}, and 
particularly powerful for the treatment of  processes with a high number of 
final-state particles. Differently from all the other $4f$ codes, {\tt 
ALPHA} 
computes automatically the tree-level $2 \to 4$ scattering amplitudes without 
making use of Feynman diagrams, but computing iteratively the saddle point of the
path integral for given external momenta. All $4f$ final states can be 
treated. It can be used to predict 
the rate for the radiative process $e^+ e^- \to 4f + \gamma$~\cite{alphag}. 

\noindent
\underline{\tt CompHEP 3.0}\cite{comphep} --- It is a package of symbolical and 
numerical modules giving in output cross sections and distributions for 
processes with up to five particles in the final state, with a high level of 
automation. It uses the {\tt BASES\&SPRING} package for adaptive Monte Carlo 
integration and unweighted event generation. All $4f$ 
and 4 fermions + 1 photon final states can be treated. ISR is implemented using 
SF's in the collinear approximation. 

\noindent
\underline{\tt ERATO}\cite{erato} --- It gives results for any $4f$ 
final 
state, using the ``E-vector'' formalism for the calculation of the $2 \to 4$ 
helicity amplitudes~\cite{evform}. ISR in the collinear approximation and QCD  
corrections are allowed. A qualifying feature  
of {\tt ERATO} is the incorporation of all $CP$ conserving anomalous couplings. 
The fermion-loop scheme as gauge restoration mechanism is active. 
The program is a Monte Carlo that can run as an event generator and 
as an integrator. It has been interfaced to {\tt JETSET}/{\tt HERWIG}.

\noindent
\underline{\tt EXCALIBUR}\cite{excalibur} --- It is a Monte Carlo 
integrator that can give predictions for 
all possible $4f$ final states (excluding Higgs-boson exchange), with the 
possibility of selecting the contributions of subsets of diagrams and of 
particular spin configurations. The helicity amplitudes are computed using the 
Weyl-van der Waerden formalism~\cite{wvdwform}. ISR in the collinear 
approximation,
Coulomb and QCD corrections, as well as anomalous couplings, are available. 

\noindent
\underline{\tt GENTLE/4fan 3.0}\cite{gentle} --- It is a semi-analytical package 
designed to compute selected $4f$ production cross sections and 
invariant
mass distributions for $CC$ and $NC$ mediated processes. SM Higgs-boson 
production in the
$NC$ case is included. All final states that do not contain identical particles, 
electrons or electron neutrinos can be studied. ISR in the collinear 
approximation is available both in the radiator  and SF approach. 
Moreover, non-universal QED corrections to some classes of diagrams can be 
included. Coulomb and QCD corrections, as well as anomalous couplings, are 
implemented.   

\noindent
\underline{\tt grc4f}\cite{grc4f} --- It is a Monte Carlo generator for all 
$4f$ final states automatically generated by the 
package {\tt GRACE}~\cite{grace}. 
All relevant radiative corrections and anomalous couplings are supplied. In
particular, QED radiation is implemented by means of 
SF's for ISR, but 
a QED PS ({\tt QEDPS})~\cite{qedps} is also available both for ISR 
and FSR. 
Fermion masses can be kept nonzero everywhere. The numerical Monte Carlo 
integration
and event generation is performed with the help of the package {\tt 
BASES\&SPRING}.
Interface to hadronization is provided. 

\noindent
\underline{\tt KORALW 1.03}\cite{koralw} --- It is a Monte Carlo that can treat 
any $4f$ final state {\it via} an interface to the {\tt GRACE} library. 
All the phenomenologically relevant radiative corrections, anomalous couplings 
and interface
to {\tt JETSET} are provided. In particular, ISR is formulated according to the 
YFS exclusive exponentiation, thereby including the 
effect of transverse photon momenta, while FSR is treated within
the SF approach. Recently, ${\cal O} (\alpha)$ non-leading QED corrections 
have been implemented~\cite{jadach97} using the known results for 
on-shell $W$-pair production.

\noindent
\underline{\tt LEPWW}\cite{lepww} --- This generator contains the CC03 and 
NC02 tree-level diagrams, supplemented with ${\cal O}(\alpha)$ ISR and 
FSR treated according to the package {\tt PHOTOS}~\cite{photos}. QCD corrections, 
anomalous couplings and interface to {\tt JETSET} are provided. 
The program aims at a 1-2\% precision.   

\noindent
\underline{\tt LPWW02}\cite{lpww02} --- It is a Monte Carlo code containing the 
Feynman diagrams with two resonant $W$'s and $Z$'s, initial- and final-state 
radiation with SF's in the collinear approximation. Coulomb
singularity and QCD corrections are incorporated. It is interfaced to {\tt JETSET}. 

\noindent
\underline{\tt PYTHIA/JETSET}\cite{pythia} --- It is a general-purpose event 
generator
for a multitude of processes in $e^+ e^-$, $ep$ and $pp$ collisions. It is 
especially
designed for detailed modelling of hadronic final states, rather than precision 
electroweak studies. It includes the CC03 diagrams for the Born matrix element 
and a hybrid SF+PS treatment of ISR. FSR is implemented according to a PS 
description. The Coulomb correction is available. Hadronization is built-in.

\noindent
\underline{\tt WOPPER 1.5}\cite{wopper} --- It is a Monte Carlo generator for  
unweighted $4f$ events. It includes the CC11 set of diagrams, ISR 
according to a PS algorithm, with finite $p_T$ for photons generated 
according to the $1/ (p \cdot k)$ pole. Coulomb and QCD corrections, anomalous 
couplings as well as interface to hadronization are provided. 

\noindent
\underline{\tt WPHACT 1.0}\cite{wphact} ---  This Monte Carlo 
program can compute all processes 
with four fermions in the final state. In particular, by virtue of the 
helicity formalism adopted that allows a semi-automatic calculation of the 
matrix elements~\cite{bm95,phact}, finite $b$-quark mass effects are 
fully taken into account
for NC processes. Higgs-boson production is included. ISR is included using SF 
in the collinear approximation. The Coulomb term and
anomalous couplings are incorporated. Interface to hadronization is present.  

\noindent
\underline{\tt WTO}\cite{wto} --- It is a deterministic code implementing the 
largest part of the $4f$ Feynman diagrams (including Higgs-boson 
signal) obtained with the helicity method 
described in \cite{covform}. ISR using SF's in the collinear
approximation, anomalous couplings, QCD and Coulomb corrections are available. The fermion-loop scheme 
as gauge restoration mechanism is active. Some options for the estimate of the 
theoretical error are provided. Interface to hadronization is present. 

\noindent
\underline{\tt WWF 2.2}\cite{wwf} --- This Monte Carlo generator is a kind of 
merger of an explicit $e^+ e^- \to 4f + \gamma$ matrix element with  
SF's for the description of ISR beyond the LL approximation. 
However, non-leading corrections are implemented in an approximate way. The 
CC20
set of tree-level diagrams is included. FSR is implemented by using an explicit 
one-photon matrix element. All the other relevant radiative corrections, as well 
anomalous couplings and {\tt JETSET} interface, are provided.  

\noindent
\underline{\tt WWGENPV  2.1/HIGGSPV}\cite{wwgenpv} --- They are Monte Carlo
generators computing the largest part of $4f$ tree-level matrix elements 
(including Higgs-boson production) derived according to the helicity formalism of 
ref.~\cite{covform}. ISR and FSR are treated within the SF approach, 
including $p_T/p_L$ effects. Coulomb and QCD corrections are implemented. 
Anomalous couplings and interface to {\tt JETSET} are also available. 

Before  the start of the LEP2 operations, an extensive comparison of all the 
Monte Carlo and semi-analytical programs available for the analysis of 
$4f$ processes was carried out 
in the context of the working group ``Event generators for $WW$ physics'' 
of the CERN workshop ``Physics at LEP2''~\cite{yrwweg96}. The predictions for 
both total cross sections and more exclusive observables of primary importance 
for
the measurement of the properties of the $W$ boson (such as $W$-boson 
production 
angle, invariant masses, radiative energy loss etc.) were compared in detail. 
``Tuned'' comparisons generally showed very satisfactory agreement between the 
codes 
(at the level of 0.1\%), pointing out a correct implementation of the advertised 
features and a high technical precision. Other highlights of the study can be 
summarized as follows
\begin{itemize}

\item the contribution of background processes is in general not negligible and 
therefore dedicated $4f$ codes are more suitable for precision 
measurements 
of the $W$-boson properties than general-purpose programs;

\item for several observables, the effect of photon $p_T$ is important on both 
ISR and FSR;

\item the available estimates of the present theoretical error show that it is 
not 
much smaller than the expected experimental uncertainty; as a consequence, 
a more reliable estimate of the complete one-loop corrections 
to $e^+ e^- \to 4f$ could give some relevant piece of information.

\end{itemize}

\subsection{The $W$-boson Mass and Anomalous Couplings}
\label{sect:wmassac}

As discussed in Sect.~\ref{sect:ffproc}, the basic aims of the study of $W$-pair 
production 
in $4f$ final states at LEP2 are the precise direct measurement of the 
mass of the $W$ boson, $M_W$, 
 and the detection of possible anomalies in the $\gamma W W$ and $Z W W$ 
vertices 
({\it anomalous couplings}). All the decay channels of $W$-pair production are 
used in the 
experimental analysis: the hadronic channel $WW \to q {\bar q} q {\bar q}$ 
(45.6\% decay fraction), 
the semi-leptonic channel $WW \to l \nu q {\bar q}$ (43.8\%) and the leptonic 
channel 
$WW \to l \nu l \nu$ (10.6\%). 

\subsubsection{The $W$-boson mass}
\label{sect:wmass}

Two different methods are employed to measure the mass of the $W$ 
boson at LEP2~\cite{yrwmass96}:
\begin{itemize}

\item the threshold method;

\item the direct reconstruction method.

\end{itemize}
The former determination exploits the strong sensitivity of the $W^+ W^-$ 
production cross section to the $W$-boson mass near the nominal threshold; 
in the latter 
method, the Breit-Wigner resonant shape of the invariant mass  distribution of 
 the $W^{\pm}$-boson decay products
is directly reconstructed, thus yielding $M_W$. The statistically most 
precise determination comes from the direct reconstruction method, since most of 
the 
LEP2 data is and will be collected at energies well above the threshold. 
Furthermore, 
very different systematic errors affect the two strategies, that therefore can 
be 
considered as complementary methods able to provide an internal cross-check on 
the 
$W$-boson mass measurement at LEP2. 

\begin{figure}[hbt]
\begin{center}
\epsfig{file=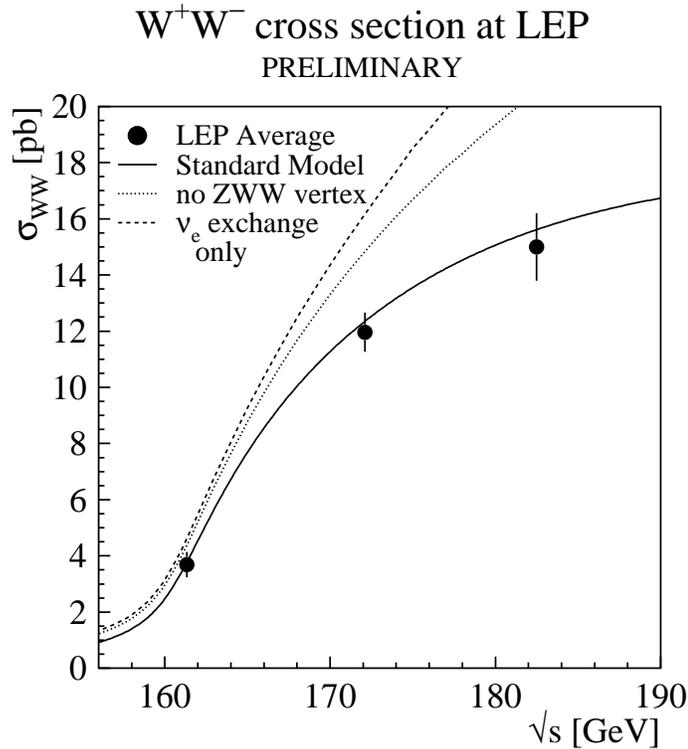, width=9truecm}
\end{center}
\caption{The $W$-pair production cross section  (from~\cite{dward97}).  }
\label{fig:sww_ene}
\end{figure}

As can be seen from Fig.~\ref{fig:wwonoffshell}, the cross section for $W$-pair production 
increases very rapidly near the nominal kinematic threshold at 
$\sqrt{s} = 2 M_W$, although the finite $W$-boson 
width and ISR significantly smear out 
the sharp rise of the (unphysical) cross section for producing two on-shell
$W$'s (see Fig.~\ref{fig:wwcorrections}). This 
means that for a given $\sqrt{s}$ near threshold the $W$-pair production cross 
section 
is very sensitive to the $W$-boson mass, so that a 
measurement of the cross section in this 
energy region directly yields a measurement of $M_W$. In the 
threshold determination, the error on the $W$-boson 
mass due to signal statistics is 
given by
\begin{eqnarray}
\Delta M_W \, = \, \sqrt{\sigma_{WW} } \left\vert { {d M_W} \over {d \sigma_{WW} } } 
\right\vert 
{{1} \over {\sqrt{\varepsilon_{WW} L } } },
\end{eqnarray}
where $\sigma_{WW}$ is the total $W$-pair production cross section, 
$\varepsilon_{WW}$ 
is the overall signal efficiency and $L$ is the integrated luminosity. By 
study\-ing the 
sen\-si\-ti\-vi\-ty factor given by $\sqrt{\sigma_{WW} } \, |d M_W / d 
\sigma_{WW} |$ as a 
function of $\sqrt{s} - 2 M_W$, one finds that there is a minimum located at 
\begin{eqnarray}
\sqrt{s} \, \simeq \, 2 M_W + 0.5~{\hbox {\rm GeV}},
\end{eqnarray}
corresponding to around 161~GeV as the c.m. energy providing the optimal 
sensitivity to $M_W$ (see Fig.~\ref{fig:sww_ene}). 
Since the statistical uncertainty is the most relevant source 
of error for the threshold method, the optimal strategy for data taking consists 
therefore in operating at $\sqrt{s} \simeq 161$~GeV in order to minimize the 
statistical error. The systematic errors on $M_W$ in the threshold method are 
due 
to the uncertainties in the $W$-pair production cross section: the luminosity 
error, 
unknown corrections to the theoretical cross section, as those discussed in 
Sect.~\ref{sect:ffrc}, the uncertainty in the 
beam energy, the background subtraction and 
others. However, as already stressed, the overall error in the threshold
measurement is largely dominated by the limited statistics. The present LEP 
average 
for $M_W$ from the $WW$ production cross section at $\sqrt{s} \simeq 161$~GeV 
is~\cite{dward97}
\begin{eqnarray}
M_W \, = \, 80.40 \pm 0.22 {\hbox {\rm GeV} } \, .
\end{eqnarray} 

The direct reconstruction method consists in reconstructing the 
peak in the invariant mass distributions of the $W$-boson decay products, by 
using either 
the semi-leptonic decay mode ($WW \to l \nu q \bar q$) or the hadronic one 
($WW \to q {\bar q} q {\bar q}$). The measurements are carried out 
at energies above the threshold ($\sqrt{s} > 170$~GeV) in order to reduce the 
statistical error thanks to the larger $WW$ cross section. 
The accuracy of the jet energy measurement is too poor to allow a precision 
measurement of 
$M_W$. Therefore, kinematic fit techniques, 
using the constraints of energy and momentum conservation, together with the 
equality of the two $W$-boson masses in an event, are 
used by the LEP Collaborations to improve the mass resolution~\cite{yrwmass96}. 
This implies that a 
good knowledge of the c.m. energy is essential. As already pointed out,
systematics affecting this method 
are largely independent of those present in 
the threshold method. In particular, the 
$q {\bar q} q {\bar q}$ channel turns out to be more problematic than the 
$l \nu q \bar q$ one: indeed, besides a larger background contamination, the 
phenomena 
of {\it colour reconnection} and {\it Bose-Einstein correlations} (see below) introduce 
additional systematic uncertainties that are absent in the semi-leptonic 
channel.
Systematic errors that are common to both channels are
\begin{itemize}

\item error from the LEP beam energy;

\item errors from the theoretical description (ISR, background, $4f$ 
diagrams and fragmentation);

\item errors from the detector. 

\end{itemize}
In order to estimate the uncertainties due to $4f$ diagrams and ISR, 
the theoretical ingredients discussed in Sects.~\ref{sect:tlcalc} 
and \ref{sect:ffrc}, and 
the corresponding computational tools based on them, are very useful tools. Indeed, 
the effect of $4f$ subprocesses other than the ones belonging to the CC03 class, 
induces an uncertainty of 20~MeV and 
25~MeV
on $M_W$ for the $q {\bar q} q {\bar q}$ and $l \nu q \bar q$ channel, 
respectively, and 
is evaluated by using MC samples generated with one or more of the $4f$ 
generators described in Sect.~\ref{sect:ffct}. 
Moreover, precision 
calculations of the radiative energy loss (see eq.~(\ref{eq:enloss})) 
in $4f$ production at LEP2 are mandatory. Actually, the relatively large 
average energy ($<E_\gamma> \simeq 1.1, 2.1, 3.2$~GeV at $\sqrt{s} = 175, 190, 
205$~GeV,
respectively) carried away by the radiated photons leads to a significant 
mass-shift 
if it is not taken into account in rescaling or constraining the energies 
of the final-state $W$-boson decay products to the 
beam energy. Indeed, the average mass-shift caused by ISR is of the order of 
 $<E_\gamma> M_W / \sqrt{s}$, \idest\  about 500~MeV at $\sqrt{s} = 175$~GeV. 
However, 
since a fit to the mass distribution gives more weight to the peak, the actual 
mass shift at $\sqrt{s} = 175$~GeV is about 200~MeV, anyway important in view 
of the envisaged experimental precision on $M_W$. Since, as discussed in 
Sect.~\ref{sect:ffrc}, the exact treatment of ${\cal O}(\alpha)$ corrections to 
off-shell $W$-pair production is not yet available, the theoretical predictions 
for $<E_\gamma>$, including the presently under control LL and 
Coulomb corrections, are affected by an intrinsic uncertainty, that 
translates into an error of 10-15~MeV on the $W$-boson mass.
This theoretical uncertainty is due to non-leading  
photonic 
terms, for instance associated with 
the radiation of photons off the intermediate $W$ bosons, and should be reduced 
soon 
in the light of the recent progress in the calculation of non-leading QED 
corrections discussed in Sect.~\ref{sect:ffrc}. 

{\it Colour reconnection} and {\it Bose-Einstein effects}~\cite{yrwmass96}, that are systematics 
specific
to the $q {\bar q} q {\bar q}$ channel, are interconnection 
phenomena that may obscure the separate identities of the two $W$ bosons, so 
that the 
$4f$ final state may no longer be considered as consisting of two 
separate $W$-boson decays. Indeed, in the LEP2 energy regime the average distance
between the $W^+$ and $W^-$ decay vertices is smaller than 0.1~fm, \idest\  less 
than a typical hadronic size. Therefore the 
fragmentation of the $W^+$ and the $W^-$ 
may not be independent and may seriously affect the mass reconstruction.  
The colour-reconnection problem arises from the fact that the $W$-boson pair, 
decaying into quark-antiquark pairs $q \bar q$ and $Q \bar Q$, respectively, 
can either fragment into two strings stretched between $q \bar q$ and $Q \bar 
Q$, or 
into two strings stretched between $q \bar Q$ and $Q \bar q$. Owing to the 
colour 
charges, the probability and the properties of the two possible configurations 
are 
very different. Of the two phases characterizing the fragmentation of the 
initial 
quark-antiquark pairs into hadrons, \idest\  perturbative parton cascade and 
non-perturbative hadronization,  the perturbative QCD interconnection effects 
have been 
shown not to be very important~\cite{sk94,abm95}, whereas the influence of 
non-perturbative 
fragmentation can be only estimated by comparing different models. The 
interesting aspect
is that quite different approaches give uncertainties of the same order, \idest\  
about 100~MeV, conservatively. Also Bose-Einstein correlations are due to the 
overlapping of the $W^+$ and $W^-$ hadronization regions: consequently, 
low-momentum 
bosons coming from different $W$'s can experience coherence effects. Their 
influence on the $W$-boson mass-shift can be 
estimated by using theoretical models. 
However, it is worth noticing that experimental studies at LEP2 show no evidence 
yet 
for Bose-Einstein correlations and that recent theoretical investigations 
suggest that 
they may not be a serious systematic (see for instance~\cite{beth}). The present experimental attitude consists 
in trying to measure interconnection effects in data through detailed comparison 
of $q {\bar q} q {\bar q}$ and $l \nu q {\bar q}$ channels. 
  
The present LEP average for $M_W$ from the direct reconstruction method 
at $\sqrt{s} = 172$~GeV is~\cite{dward97}
\begin{eqnarray}
M_W \, = \, 80.53 \pm 0.18~{\hbox {\rm GeV} } \, .
\end{eqnarray} 
It is worth noticing that good consistency between experiments and between 
$q {\bar q} q {\bar q}$ and $l \nu q \bar q$ channels is observed.

By combining the $\sqrt{s} = 161$~GeV (threshold) and 
the $\sqrt{s} = 172$~GeV (direct reconstruction) results, the present 
$M_W$ value from LEP2 is~\cite{dward97}
\begin{eqnarray}
M_W \, = \, 80.48 \pm 0.14~{\hbox {\rm GeV} } \, .
\end{eqnarray}  
The LEP2 result for $M_W$ can be combined with the measurement by CDF/D0 at 
the TEVATRON, 
in order to compile a world average. The $W$-boson 
mass measurements at the TEVATRON are based 
on fits to the transverse mass distribution of $l\nu$ produced in the process 
$q \bar q \to W \to l\nu (l=e,\mu)$. By combining the CDF/D0 results with the 
UA2 measurement, the average $M_W$ from hadron machines is~\cite{dward97}
\begin{eqnarray}
M_W \, = \, 80.41 \pm 0.09~{\hbox {\rm GeV} } \, .
\end{eqnarray} 
The present world average $M_W$ from LEP2 and $p \bar p$ colliders is~\cite{dward97}
\begin{eqnarray}
M_W \, = \, 80.43 \pm 0.08~{\hbox {\rm GeV} } \, .
\end{eqnarray} 
By the end of LEP2 and TEVATRON run 2, both experiments expect to achieve an 
error on 
$M_W$ of  around 0.03-0.04~GeV, comparable with the present error from indirect 
determination 
by means of precision data.  

\subsubsection{Anomalous couplings}
\label{sect:ac}
In addition to the $W$-boson mass, the second important piece of 
information that
$W$-pair production at LEP2 can provide is the structure of the triple 
gauge-boson 
couplings (TGC). Contrary to LEP1 physics, where TGC only enter through loop 
corrections 
to two-fermion production, at LEP2 these couplings are responsible of the 
behaviour of the 
tree-level $W$-pair cross section. Since the specific form for the TGC in the SM 
is a consequence of the Yang-Mills character of the theory, the study of 
$W$-pair production
at LEP2 allows to test directly the non-abelian nature of the $SU(2) \times 
U(1)$ gauge 
theory. Indeed, in spite of the precision tests of the electroweak interaction at 
LEP1 and SLC, 
the non-abelian sector of the SM remains poorly measured to date, even if the very existence of
non-abelian coupling can be considered experimentally proved by the recent LEP2 data (see
Fig.~\ref{fig:sww_ene}). 

$WWZ$ and $WW\gamma$ couplings different from the form predicted by the SM are 
called 
{\it anomalous couplings}. The latter will in general lead to a bad high-energy 
behaviour 
of cross sections violating unitarity. However, since at LEP2 the c.m. energy is 
too low 
to point out such effects, the presence of anomalous couplings can be 
eventually established
through the study of the angular distributions of the $W$ bosons and their decay 
products. 
As for the $W$-boson mass measurement, the control of radiative corrections and 
background 
contributions is necessarily required, since they induce deviations from 
the tree-level angular distributions.   

Possible anomalies in the $WWV$ ($V = Z$ or $\gamma$) vertex can be parameterized 
in terms of a purely phenomenological effective Lagrangian~\cite{gg79,hhpz87} 
\begin{eqnarray}
i {\cal L}_{eff}^{WWV} \, = \, & & g_{WWV} \, \Big[ g_1^V V^{\mu} \left( 
W^-_{\mu\nu} 
W^{+\nu} - W^+_{\mu\nu} W^{-\nu} \right) + k_V W^+_\mu W^-_\nu V^{\mu\nu} + 
 \nonumber \\
& & {\lambda_V \over M^2_W} V^{\mu\nu} W^{+\rho}_\nu W^-_{\rho\mu} 
+ i g_5^V \varepsilon_{\mu\nu\rho\sigma} \left( (\partial^{\rho} W^{-,\mu}) 
W^{+,\nu} - 
W^{-,\mu} ( \partial^{\rho} W^{+,\nu} ) \right) V^{\sigma} +  \nonumber \\
& & i g_4^V W^-_\mu W^+_\nu (\partial^{\mu} V^\nu + \partial^\nu V^\mu) 
- {{\tilde k}_V \over 2} W^-_\mu W^+_\nu
\varepsilon^{\mu\nu\rho\sigma} V_{\rho\sigma} - \nonumber \\
&  & {{\tilde \lambda}_V \over {2 M_W^2} } W^-_{\rho\mu} W^{+\mu}_\nu 
\varepsilon^{\nu\rho\alpha\beta} V_{\alpha\beta} \Big] \, , 
\label{eq:effl}
\end{eqnarray}
that provides the most general Lorentz invariant $WWV$ vertex, observable in 
processes 
where the vector bosons couple to effectively massless fermions. 
The definitions in eq.~(\ref{eq:effl}) read as follows
\begin{eqnarray}
& & g_{WW\gamma} = e  \quad \qquad \qquad \qquad g_{WWZ} = e \cot \vartheta_W 
\nonumber \\ 
& & W_{\mu\nu} = \partial_{\mu} W_\nu - \partial_{\nu} W_{\mu} \qquad 
V_{\mu\nu} = \partial_{\mu} V_\nu - \partial_{\nu} V_{\mu} .
\end{eqnarray}
Therefore, $2 \times 7$ independent parameters are needed to describe 
the most general Lorentz invariant
$WWV$ vertex. Within the SM and at the tree level the couplings in 
eq.~(\ref{eq:effl}) are 
given by $g_1^Z = g_1^\gamma = k_Z = k_\gamma = 1$, with all other couplings 
vanishing. 
It is worth noticing that $g_1^V$, $k_V$ and $\lambda_V$ conserve $C$ and $P$ 
separately, while $g_5^V$ violates $C$ and $P$ but conserves $CP$. Furthermore 
$g_V^4, {\tilde k}_V$ and ${\tilde \lambda}_V$ parameterize a possible $CP$ 
violation in the 
bosonic sector. In particular, the $C$ and $P$ conserving parameters in ${\cal 
L}_{eff}^{WW\gamma}$
can be linked to the static e.m.  moments of the $W^+$ boson 
as follows~\cite{aronson69}
\begin{eqnarray}
&& {\hbox {\rm charge} } \quad  Q_W = e g_1^\gamma \, , \nonumber \\
&& {\hbox {\rm magnetic dipole moment} } \quad   \mu_W = {e \over {2 M_W}} 
(g_1^\gamma+k_\gamma+\lambda_\gamma) \, , \nonumber \\
&& {\hbox {\rm electric quadrupole moment} } \quad  q_W = -{e \over {M_W^2}} 
(k_\gamma-\lambda_\gamma) \, .
\end{eqnarray}

In practice it is impossible to set limits on all the above couplings, so that a 
number of 
assumptions have to be made in order to reduce the number of parameters. In the 
literature, 
two different theoretical strategies are followed. On the one side, the number 
of parameters can be 
reduced by advocating symmetry arguments~\cite{hhpz87,bkrs93}. On the other 
hand, the dimensional 
analysis of the operators involved in the parameterization allows to establish a 
hierarchy 
of the operators themselves according to a scaling law of the kind $(E / 
\Lambda)^{d-4}$, where
 $E$ is the energy of interest, $\Lambda$ is the typical energy scale of 
non-standard physics
and $d$ is the dimension of the operators. The latter procedure allows to single 
out a 
reduced set of parameters able to produce phenomenologically relevant effects at 
the 
energy scale of interest~\cite{dmop}. 

For instance, if $C$, $P$ 
and electromagnetic gauge invariance are imposed, the number of parameters 
reduces to 
five, namely $\lambda_\gamma, \lambda_Z, k_\gamma, k_Z$ and $g_1^Z$. 
Furthermore, 
since the effective Lagrangian of eq.~(\ref{eq:effl}) contains as a particular 
case 
the triple gauge-bosons Lagrangian of the electroweak theory, it is possible and 
convenient 
to introduce deviations from the SM predictions for the five parameters as
\begin{eqnarray}
\Delta g_1^Z = g_1^Z - 1 , \quad \Delta k_{\gamma,Z} = k_{\gamma,Z} - 1 , 
\quad \lambda_\gamma,  \quad \lambda_Z  \, .
\end{eqnarray}
A set of parameters widely used by the LEP Collaborations is given by the 
following three 
combinations, which do not affect the tree-level gauge boson propagators and 
are not yet constrained by low energy and $Z^0$ peak data~\cite{yragc96}
\begin{eqnarray}
&& \Delta k_\gamma - \Delta g_1^Z \cos^2 \vartheta_W  =  \alpha_{B\phi} , 
\nonumber \\
&& \Delta g_1^Z \cos^2 \vartheta_W  =  \alpha_{W\phi} , \nonumber \\
&& \lambda_Z = \lambda_\gamma  =  \alpha_W ,
\label{eq:setlep}
\end{eqnarray}
with the constraint $\Delta k_Z = \Delta g_1^Z - \Delta k_\gamma \tan^2 
\vartheta_W$. The above 
combinations are all zero according to the SM. 

The experimental strategy followed at LEP2 for the determination of anomalous 
couplings relies 
upon the measurement, in $4f$ production processes, of the differential 
cross sections 
$d^n \sigma = d^n \sigma (\Theta, \vartheta_{\pm}^*, \varphi_{\pm}^*; \alpha)$ 
with 
$n = 1,3,5$, where $\Theta$ is the polar production angle of the $W^-$ boson, 
$\vartheta^*_{\pm}$ 
and $\varphi^*_{\pm}$ are the $W^\pm$ decay angles in $W$-boson rest 
frames. Among the 
$W$-pair decay channels, the strongest information comes from the semi-leptonic 
channel because 
it allows unambiguous charge assignment. Also single $W$-boson and single 
$\gamma$ production processes,
that are sensitive to the non-abelian $WW\gamma$ coupling  are studied by the 
LEP Collaborations, 
even if they are less powerful than the $WW \to 4f$ reactions. Present 
combined results 
from the four LEP experiments are~\cite{dward97}
\begin{eqnarray}    
&& \alpha_{W\phi} = 0.02^{+0.16}_{-0.15} \qquad \qquad  
-0.28 < \alpha_{W\phi} < 0.33 \quad (95\% \, 
{\hbox{\rm CL limit}}) \nonumber \\
&& \alpha_W \, \, \, = 0.15^{+0.27}_{-0.27} \qquad \qquad  
-0.37 < \alpha_{W} \, \, < 0.68 \quad (95\% 
\, {\hbox {\rm CL limit}} ) \nonumber \\
&& \alpha_{B\phi} \, = 0.45^{+0.56}_{-0.67} \qquad \qquad 
\, \, -0.81 < \alpha_{B\phi} < 1.50 \quad (95\% \, 
{\hbox{\rm CL limit}} )
\end{eqnarray}
Therefore no discrepancy with the SM is seen in the data. Moreover, an important 
result of the 
anomalous couplings analysis is that the existence of the $ZWW$ vertex is firmly 
established, namely $g_{WWZ} = 0$ is excluded at $ > 95$\% CL (see Fig.~\ref{fig:sww_ene}).

\subsection{Higgs-boson Searches}
\label{sect:higgs}

\begin{figure}[hbt]
\begin{center}
\epsfig{file=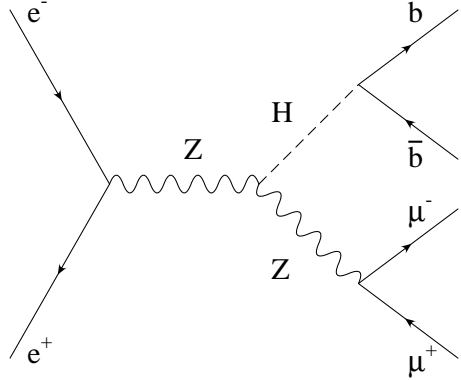, width=6truecm}
\end{center}
\caption{Tree-level Feynman diagram for Higgs-boson production in the  
channel $e^+ e^- \to \mu^+ \mu^-  b
\bar{b} $  (Higgs-strahlung).}
\label{fig:hmmbb}
\end{figure}

As shown in Sect.~\ref{sect:hm}, the electroweak precision data can be 
conveniently used to obtain indirect determinations of the Higgs-boson 
mass through radiative corrections to the physical observables. 
Although the weak dependence of the theoretical predictions 
on the Higgs-boson mass as well as the their intrinsic uncertainties 
prevent to derive stringent predictions for $m_H$, 
there is a slight preference in the precision measurements for a 
relatively  light SM Higgs boson, with an upper limit  
$ m_H < 420 \, {\hbox {\rm GeV}}$ @ $95\%$~CL~\cite{quast97}. Future 
more precise direct measurements of the {\it top}-quark mass at the TEVATRON and 
of the $W$-boson mass at LEP2 and the TEVATRON, associated with 
recent progress in the calculation of two-loop electroweak graphs, 
should improve the present constraints on the mass of this yet elusive 
particle. Indeed, the search for the Higgs boson certainly constitutes 
one of the main tasks of present-day experiments in high-energy physics. 
The discovery of this neutral scalar particle 
would establish the validity of the electroweak theory in 
its standard formulation, and  shed light on the mechanism of electroweak symmetry breaking and mass
generation.  

By searching directly for the SM Higgs boson, the LEP experiments have set a
lower bound on the Higgs-boson mass, namely 
$m_H > 77 \, {\hbox {\rm GeV}}$ @ $95\%$~CL (four LEP experiments 
combined)~\cite{janot97,murray97}. This lower limit comes from negative 
searches of Higgs-boson production in electron-positron collisions around and 
above the $Z^0$ peak. The dominant production mechanism for the SM Higgs boson
in the LEP energy range is the $s$-channel 
Higgs-strahlung process $e^+ e^- \to Z H$, where the intial states annihilate 
into a virtual $Z^0$ boson that converts into a $Z^0$ and a Higgs boson. 
Other 
production mechanisms, \idest\  the fusion processes, where the Higgs boson 
is formed in $WW, ZZ$ $t$-channel collisions, with the $W,Z$'s radiated off 
the incoming electron and positrons, have smaller cross section at LEP
energies. However, it is worth keeping in mind that for $\nu_e {\bar\nu}_e$ 
and $e^+ e^-$ in the final state, the two amplitudes for the 
Higgs-strahlung and fusion process both contribute and interfere. 

\begin{figure}[hbt]
\begin{center}
\epsfig{file=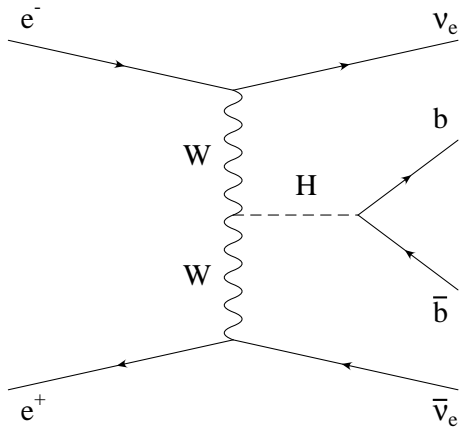, width=6truecm}
\end{center}
\caption{Tree-level Higgs-boson  production {\it via} $WW$ fusion.}
\label{fig:higgsfusion}
\end{figure}

\begin{figure}[hbt]
\begin{center}
\epsfig{file=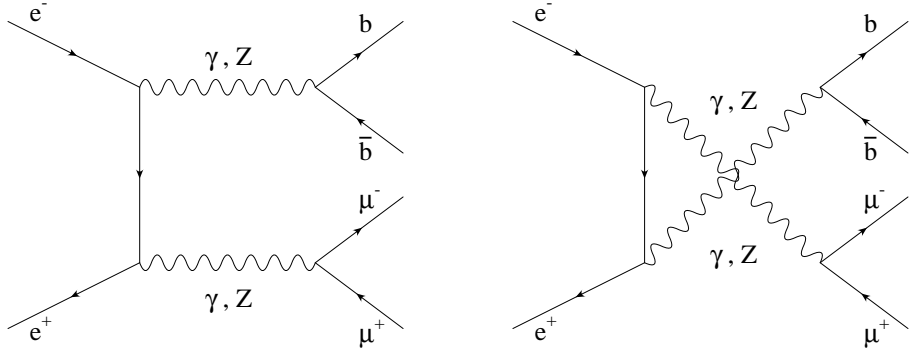, width=12truecm}
\end{center}
\caption{Tree-level Feynman diagrams for the main background processes in the channel 
$e^+ e^- \to \mu^+ \mu^-  b \bar{b} $.}
\label{fig:higgsbckg}
\end{figure}

In view of the LEP2 physics potential, the reachable Higgs-boson mass limit 
is roughly given by $m_H \simeq 
\sqrt{s} - 100$~GeV. Even assuming the highest c.m. energy in the LEP2 
operation ($\sqrt{s} = 205$~GeV), the discovery potential of the 
LEP2 experiments will not exceed a Higgs-boson 
mass of around 100~GeV. For a SM Higgs boson with 
a mass in the range between 77 (present lower limit) and 100 GeV, the 
dominant decay mode is by far $H \to b \bar b$. Branching ratios for 
the decays $H \to \tau^+ \tau^-, c \bar c, gg$ are suppressed by an order 
of magnitude or more. The corresponding total decay width is predicted in the 
electroweak theory to be very narrow, \idest\   less than a few~MeV for 
$m_H$ less than 100~GeV.  
As a consequence of the fact that the Higgs boson predominantly decays 
into $b \bar b$ pairs when $m_H < 100$~GeV, the observed final states in 
Higgs-boson searches at LEP consists of four fermions. In the case of the 
Higgs-boson signal these $4f$ final states are mainly achieved 
through the Higgs-strahlung process $e^+ e^- \to ZH$, with the subsequent 
decays $Z \to f \bar f$ and $H \to b \bar b$.  
In general, whenever excluding Higgs-boson decay channels 
other than $H \to b \bar b$ (such as $H \to \tau^+ \tau^-$ that is 
actually considered in Higgs-boson searches at LEP), the interesting final states
for the physics of the Higgs particle at LEP are $b \bar b \mu^+ \mu^-, 
b \bar b e^+ e^-$ (leptonic channels), $b \bar b \nu \bar\nu$ (missing 
energy channel) and $b \bar b q \bar q$ (four-jet channel). Therefore, 
Higgs-boson physics at LEP2 requires, as for $W$-pair production, the 
calculation of $2 \to 4$~fermions scattering amplitudes, including 
signal and backgrounds in the SM. 
The  Feynman diagram for the Higgs-boson signal in  
the channel $e^+ e^- \to \mu^+ \mu^- b \bar b$ is depicted 
in Fig.~\ref{fig:hmmbb}. The additional Higgs-boson production
mechanism  in  the case  of $\nu_e {\bar\nu}_e b \bar b$ final  state
can be found in  Fig.~\ref{fig:higgsfusion}; an 
analogous diagram with $ZZ$ fusion is present for  the $e^+ e^- b \bar b$ final  state. 
For the channel 
$\mu^+ \mu^- b \bar b$, that provides the cleanest event signature, 
background events are generated, among others, by double vector-boson 
production $e^+ e^- \to ZZ, Z\gamma$ and $\gamma\gamma$ with the 
virtual $Z^0$ and $\gamma$ decaying to $\mu^+ \mu^-$ and $b \bar b$ pairs (see
Fig.~\ref{fig:higgsbckg}). 
 Actually, just to give an idea of the degree of 
complexity of the 
full calculation of $2 \to 4$ amplitudes for Higgs-boson physics, 
25 diagrams are present in the $\mu^+ \mu^- b \bar b$ channel, 
50 diagrams in $e^+ e^- b \bar b$ and 68 in the $b \bar b b \bar b$ channel 
(unitary gauge). Full $4f$ calculations 
for Higgs-boson searches are certainly less demanded with respect to the case of 
precision studies of $W^+ W^-$ production. Indeed, given the small 
Higgs-boson production cross sections (of the order of tens of $fb$, two-three orders of magnitude
smaller than the $e^+ e^- \to W^+ W^-$ cross section) and the low 
statistics, precise predictions are not strictly necessary for discovery 
physics. However, the theoretical and technical expertise developed 
for $4f$ production in $W$-boson physics and described in the previous 
sections allows to apply the corresponding 
``machinery'' to a comprehensive analysis of Higgs-boson production as well. 
The main motivations for complete $4f$ calculations for Higgs-boson 
physics at LEP is that no approximations are introduced in the tree-level 
matrix element, differential distributions for the final-state products 
are directly available and the background can be precisely evaluated. 
Therefore, a full calculation allows to test the degree of validity 
of the usual approximation of computing production cross section $\times$ 
branching ratios and of summing squared amplitudes incoherently, thus 
quantifying off-shellness and interference effects. Furthermore, once the 
Higgs  boson is discovered, complete $4f$ calculations and relative 
generators could be of great help in the extraction of the Higgs-boson 
properties (mass, spin, parity, etc.) that require a good control of 
exclusive distributions.   

Complete $4f$ calculations for Higgs-boson physics at LEP
 have been carried out in the last few years by several 
 groups~\cite{yrwweg96,yrhiggs96,yrdpeg96,hzgen,gp97,blr95,mnph95}.
The typical calculation strategy consists in computing all tree-level 
diagrams for a given channel and including the best available set of radiative
corrections, \idest\  running electroweak couplings, 
initial-state QED radiation, naive QCD final-state
corrections and running heavy-quark masses. This framework is implemented,
although with some minor differences, in $4f$ programs such as 
{\tt compHEP}, {\tt GENTLE/4fan}, {\tt HIGGSPV}, {\tt HZGEN}~\cite{hzgen},  
{\tt WPHACT} and {\tt WTO}, 
that have been described in Sect.~\ref{sect:ffct}. The programs 
{\tt compHEP}, {\tt GENTLE/4fan}, {\tt HZGEN}
and {\tt WPHACT} can evaluate matrix elements taking into account the finite 
$b$ quark mass, while {\tt HIGGSPV} and {\tt WTO} consider massless 
final-state fermions. {\tt WPHACT} includes also the contribution of 
SUSY neutral Higgs-boson production. Some of the codes provide unweighted events 
with the four-momenta of all final-state particles, that can be 
used to process the event through the detector and to apply cuts for 
data analysis. 
The above $4f$ generators are described in more detail in \cite{yrdpeg96}, where 
also reference to other programs for Higgs-boson physics at LEP, such as 
{\tt PHYTIA} and {\tt HZHA}, based on production times decay approximation 
and with an
incomplete treatment of the SM amplitudes, can be found. In ref.~\cite{yrdpeg96} 
some comparisons between the predictions of different programs are shown for a
sample of $4f$ final states, with the conclusions that the agreement 
between the dedicated $4f$ generators is systematically at the level of 1\% or
better, pointing out the high technical precision achieved, and that there are
situations, such as the $b \bar b \nu_e \bar\nu_e$ final state, where the 
difference between $4f$ codes and standard computational tools for Higgs-boson
searches at LEP, such as {\tt HZHA}, not including the full set of 
SM diagrams, can reach 20-40\%, depending on the c.m. energy and 
the Higgs-boson mass value. Thanks to the availability of complete 
$4f$ calculations, some topics previously not investigated 
but of interest for Higgs-boson searches, 
such as the amount of signal-background interference and the effect of finite 
$b$-quark mass corrections, received the due attention.  In the case of massless 
final-state fermions, the interference between 
signal and background diagrams is exactly vanishing, because massless fermions
are coupled to spin-vectors in $\gamma,Z$ decays and to spin-scalar in Higgs-boson 
decay, so that the two amplitudes do not interfere as a consequence of the 
different helicity pattern. If the mass of the $b$ 
quark is kept different from zero in the full matrix element, then a finite 
interference does develop. However, it was found that this effect never 
exceeds the percent level. Indeed, since for $m_H < 100$~GeV the Higgs-boson width is of
the order of a few MeV, it can be expected {\it a priori} that the 
non-vanishing signal-background interference is highly suppressed at LEP, 
as demonstrated by explicit calculations. Therefore, signal-background
interference may be neglected in Higgs-boson search experiments at LEP. Other 
conclusions that comes from these analyses is that ISR is large, varying between
10-20\% on the total cross section, and must be included in Higgs-boson event 
generators, especially in the light of the distortions induced on 
exclusive distributions that are relevant for the determination of the 
properties of the Higgs particle. 

\begin{figure}[hbt]
\begin{center}
\epsfig{file=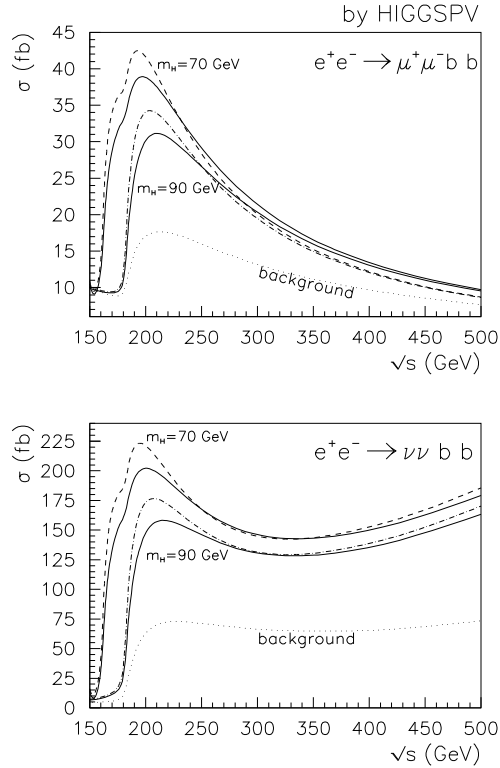, width=9truecm}
\end{center}
\caption{The cross section for Higgs-boson production in the channels 
$e^+ e^- \to  b \bar{b} \mu^+ \mu^-$ and 
$e^+ e^- \to   b \bar{b} \nu \bar\nu$. The dotted and dash-dotted lines show 
tree-level predictions; the solid lines include ISR. The background cross
section is also
 shown. Numerical results by {\tt HIGGSPV}~\cite{wwgenpv}. }
\label{fig:hxsect}
\end{figure}

\begin{figure}[hbt]
\begin{center}
\epsfig{file=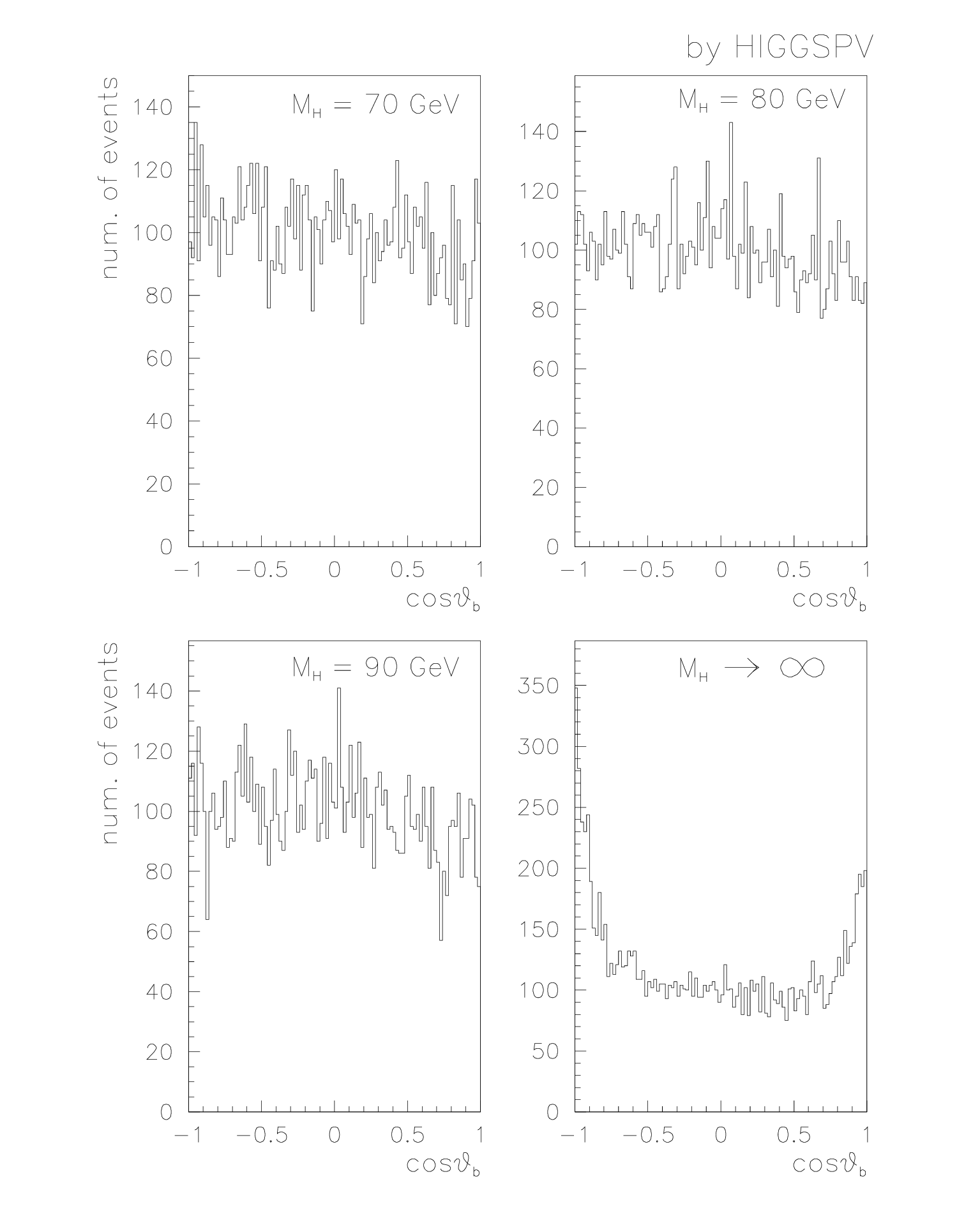, width=9truecm}
\end{center}
\caption{The $b$-quark scattering angle distribution in the 
laboratory frame at $\protect\sqrt{s} = 192$~GeV for the 
process $e^+ e^- \to b \bar b \nu \bar\nu$. Numerical 
results by {\tt HIGGSPV}~\cite{wwgenpv}.}
\label{fig:anb}
\end{figure}

The predictions obtained with the help of 
the $4f$ generator {\tt HIGGSPV} for the Higgs-boson production cross section 
in the channels  $e^+ e^- \to  b
\bar{b} \mu^+ \mu^-$ and $e^+ e^- \to b \bar b \nu \bar\nu$ can be seen in 
Fig.~\ref{fig:hxsect}, for two representative values of the Higgs-boson mass. The 
effect 
of backgrounds and ISR are shown as a function of the c.m. energy up to 
0.5~TeV. It is worth observing that in the LEP2 energy range the size of the 
total cross sections in both channels can give by itself clear evidence 
of Higgs-boson production. In the case of the 
$b \bar b \nu \bar\nu$ final state the rise with the c.m. energy of the cross 
section is due to the logarithmic enhancement introduced by the 
$t$-channel $WW$ fusion processes that are absent in the 
$b \bar b \mu^+ \mu^-$ final state. As said above, in order to extract  
information on the quantum numbers assignment of the Higgs particle, the 
angular distributions are known to be the most sensitive observables. An 
example of such quantities is given in Fig.~\ref{fig:anb}, showing the 
distribution, obtained by means of {\tt HIGGSPV}, of the $b$-quark scattering angle 
in the laboratory frame, 
at $\sqrt{s} = 192$~GeV and for three Higgs-boson mass values and background alone
($e^+ e^- \to b \bar b \nu \bar\nu$ channel). 
As can be seen, in the presence of the Higgs-boson signal as dominant contribution
to the $b \bar b \nu \bar\nu$ cross section, a clearly isotropic spin zero 
behaviour is present, whereas it disappears whenever considering the background 
only. Since this differential distribution can be meaningfully and more 
extensively analyzed by means 
of complete $4f$ calculations, the above example should clarify 
the usefulness of dedicated precision tools for the measurement of the Higgs-boson 
properties.

To conclude the discussion on Higgs-boson searches in electron-positron 
collisions, it is 
worth noticing that, 
if LEP does not discover the Higgs boson and its mass lies in the intermediate 
range ($140 \, {\hbox {\rm GeV}} < m_H < 2 M_Z$), the search for the Higgs 
particle at future $e^+ e^-$ colliders 
will require the analysis of six-fermion production processes, due to the decay 
chain $H \to W^+ W^- \to 4f$. Also in this case, in view of the expected
experimental precision and particularly for a precise determination of the
Higgs-boson  
properties, complete six-fermion calculations of Higgs-boson 
production and relative 
computational tools should be
desirable. Actually, the first results along this direction  very recently 
appeared in the literature~\cite{higgs6fpv,higgs6fto}.  

%% file: concl.txi
\section{Conclusions}
\label{sect:concl}
Since 1989 the four experiments ALEPH, DELPHI, L3 and OPAL at LEP and the SLD 
Collaboration at SLC collected a large amount  of precision data by studying 
the processes of two-fermion production in electron-positron 
collisions at centre of mass energies close to the mass of the 
$Z^0$ boson. About 16 millions of $Z^0$ events have been recorded.  
The LEP phase around the $Z^0$ resonance (LEP1) was terminated in 1995, while SLC 
is continuing in data taking on the $Z^0$ peak. During 1996, the 
LEP energy was increased in order to allow the production of $W$-boson 
pairs for the first time in high-energy $e^+ e^-$ collisions. The LEP2 phase  
is in progress and will continue until 1999. Meanwhile, 
almost final results of the analysis of the LEP1/SLC precision data 
have become available, although the data analyses, including the final 
LEP energy calibrations, are not yet finished.

Looking at the high accuracy reached by the most recent measurements 
of the electroweak observables, the program of 
{\it precision physics at LEP/SLC} can be considered as a complete 
success, thanks to a combined experimental and theoretical effort.
Actually, besides the very successful performance of the LEP machine 
and important experimental/technological achievements, the effort undertaken 
on the theoretical side in precision calculations for $Z^0$ physics 
significantly contributed to the progress of precision 
tests of the Standard Model (SM). 

A key ingredient of the striking success of precision physics at LEP/SLC is
undoubtedly the high precision determination of the machine luminosity. This
has been possible thanks to the experimental achievements on luminometers, and
to  the high-precision calculations  of the small-angle 
Bhabha process, today accurate at the level of 0.1\%. 

The extremely accurate theoretical predictions for observables of 
$e^+ e^-$ collisions into two-fermion final 
states at large scattering angles provided the necessary theoretical 
background for the precise determination of the electroweak parameters
performed by the high-precision experiments.  

Concerning the comparison between theory and experiment, the electroweak 
precision data confirm the validity of the SM with an impressive accuracy. 
The standard theoretical framework is able to accomodate all experimental 
facts, providing predictions in agreement with precision measurements. 
Just to mention 
some of the major achievements, the number of light neutrino species has been 
unambiguously determined to be equal to 3; the $Z$-boson mass is at present
known with a relative precision of $2 \times 10^{-5}$, comparable with the
precision of the muon decay constant $G_\mu$; 
lepton universality has been tested 
with an unprecedented precision; 
the   stringent constraints on the neutral 
current couplings allow the determination of the weak effective mixing angle
with an absolute error at the level of $3 \times 10^{-4}$; the $W$-boson mass is
indirectly determined with a relative error below  0.1\%. 

By virtue of their high accuracy, the precision measurements 
show a clear evidence of pure weak radiative 
corrections and they can be used to infer valuable information about 
the fundamental parameters of the SM. Actually, the measurements are sensitive, 
{\it via} the virtual effects of radiative corrections, to the {\it top}-quark 
mass $m_t$, the Higgs-boson mass $m_H$ and the strong coupling constant $\alpha_s$.  
The indirect constraints obtained from precision data on the 
mass of the particles not energetically accessible 
clearly illustrate the r$\hat{ \hbox{\rm o}}$le of quantum loops 
in raising {\it precision physics} to the level of {\it discovery physics}.

From all the available data (excluding the direct determination of $m_t$ and $M_W$ 
from the TEVATRON and LEP2) the best current estimate of the 
{\it top}-quark mass is~\cite{quast97,dward97}
\begin{eqnarray*}
m_t = 157^{+10}_{-9}~\hbox{\rm GeV}.
\end{eqnarray*}
This indirect determination is in beautiful agreement with the measurement of 
the {\it top}-quark mass from direct production at the 
TEVATRON~\cite{topexp97}, $m_t = 
175.6 \pm 5.5$~GeV, illustrating the constraining power of the precision 
measurements.  

The value of $\alpha_s(M_Z)$ derived from a fit to all electroweak data 
is~\cite{quast97,dward97} 
\begin{eqnarray*}
\alpha_s(M_Z) = 0.120 \pm 0.003,
\end{eqnarray*}   
in very good agreement with the world average and of similar precision.
This indirect determination, when associated with the $\alpha_s$ measurements 
from other processes and at different energy scales, 
significantly proves the running of $\alpha_s$ as predicted by the non-abelian 
structure of QCD. 

The Higgs-boson is the still missing block of the SM. Although the constraints 
inferred on the mass of the Higgs boson from precision data are not 
conclusive, as a consequence of the weak logarithmic dependence of radiative
corrections on $m_H$, 
electroweak measurements imply an indicative mass window, 
thus providing valuable information in 
view of the planned searches at the LHC and future $e^+ e^-$ colliders. A fit 
to all data gives~\cite{quast97,dward97}
\begin{eqnarray*}
&& m_H = 115^{+116}_{-66}~{\rm GeV} ,  \nonumber \\
&& m_H < 420~{\rm GeV} \quad (95\%~{\rm CL}) ,
\end{eqnarray*}
where also the theoretical uncertainty due to missing higher-order corrections 
is taken into 
account in the one-sided 95\%~CL upper limit. A general {\it caveat} is in
order here. The upper limit quoted above depends heavily  on the presence of
the left-right asymmetry from SLC in the data set. Therefore, the only reliable
conclusion is that a very heavy Higgs boson, that 
would be rather problematic for future direct searches, 
is excluded by precision data. Future improved 
measurements of the $W$-boson mass at LEP2 and the TEVATRON 
and of the {\it top}-quark mass at the TEVATRON, together with 
recent progress in the calculation of higher-order radiative corrections, 
are expected to sharpen the existing constraints on $m_H$.   

The first experimental results of LEP2 are available. As in the LEP1 case, the
theoretical tools, noticeably the accurate predictions 
concerning reactions of the kind $e^+ e^-
\to $~4-fermion final states, provide the necessary theoretical scheme for the
data-theory comparison. The recent LEP2 measurements of 
the $W$ pair production cross section already show clear evidence for 
the existence 
of the gauge bosons self-interactions, thus testing a crucial prediction 
of the non-abelian structure of the theory. Future measurements of the $W$-boson
mass are expected to reach an accuracy comparable with the one obtained in the
direct determination performed at the TEVATRON and the indirect determination at
 LEP1/SLC as calculated {\it via } radiative  corrections. 

Looking beyond the SM, the possible scenarios of new physics are strongly 
constrained by the delicate agreement between the SM and the precision data. 
Since no significant anomalies are present, the natural candidates 
are represented by models that do not modify the structure of the SM
significantly. For this reason, theories with fundamental Higgs particles, 
such as supersymmetry, are generally considered more favourite than models 
implying composite Higgs bosons and new strong interactions. Given 
the large value of the {\it top}-quark mass, a fundamental 
Higgs-boson with a relatively low mass, as inferred by precision electroweak   
data, can be nicely accommodated in a supersymmetric theory (such as the 
Minimal Supersymmetric Standard Model) broken at some high energy scale.  
Supersymmetry provides also a viable scenario to realize the unification of the 
coupling constants in a grand unified theory (GUT). Actually, the determination 
of $\sin^2\vartheta_{eff}$ and $\alpha_s(M_Z)$ as obtained from precision data 
is compatible with gauge coupling unification at a large energy scale, provided 
the mass spectrum of the GUT includes supersymmetric particles. 

Although new elementary particles have not been directly discovered by 
experiments at LEP and SLC, it can be said that {\it precision physics} 
provided important pieces of information to our present knowledge in 
particle physics and contributed significantly to obtain hints on how new 
physics beyond the SM is realized in Nature.

%% file: upc.txi
\section{Universal Photonic Corrections}
\label{sect:upc}
A common set of radiative corrections to $s$-channel annihilation and $t$-channel scattering
processes is represented by QED radiative corrections. In the so called leading 
logarithmic (LL) approximation, they are dominated by long-distance contributions, and hence 
are  process independent. Going beyond the LL approximation, it is necessary to take into
account process-dependent corrections, which are no more universal and can be computed by
means of standard diagrammatic techniques. In the following, an overview of the most popular
algorithms developed for the computation of QED corrections in the LL approximation, namely the
Structure Function (SF) method, the Parton Shower (PS) method and 
Yennie-Frautschi-Suura (YFS) exclusive
exponentiation, will be given.\footnote{It is worth noticing that at least two additional 
frameworks for the computation of QED corrections in the LL approximation have been also developed, 
namely the Coherent States  approach~\cite{cohstates} and the Unitary Method~\cite{d91}. 
The  interested reader is referred to the original literature.}
Just for definiteness, only $s$-channel processes will be considered, the generalization
to $t$-channel processes being almost straightforward.  It is worth noticing that for two of these
methods, namely the SF method and YFS exclusive exponentiation, proper procedures for matching the
all-orders LL results with finite-order exact diagrammatic results have been developed.  

\subsection{The Structure Function Method}
\label{sect:sfm}

\begin{figure}[hbt]
\begin{center}
\epsfig{file=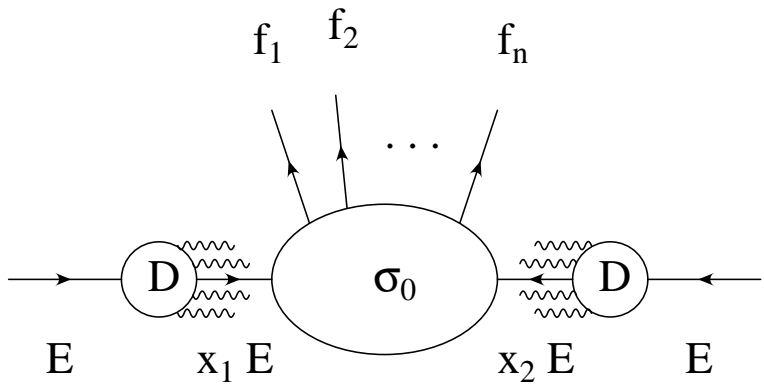, width=10truecm}
\end{center}
\caption{Graphical representation of eq.~(\ref{eq:masterd}). }
\label{fig:strucfun}
\end{figure}

Let us consider the annihilation process $e^- e^+ \to  X$, where $X$ is some given final
state, and let $\sigma_0 (s)$ be its lowest order cross section, possibly including all the
short-distance process dependent corrections, as those discussed in Sect.~\ref{sect:z0phys}. 
Initial-state (IS) QED radiative corrections can be described according to the following 
picture.  Before arriving at the annihilation point, the incoming electron (positron) of
four-momentum $p_{-(+)}$ radiates real and virtual photons (bremsstrahlung). These 
photons, due to the dynamical features of QED, are mainly  radiated along the direction of
motion of the radiating particles, and their effect is mainly to reduce 
the  original four-momentum of the incoming electron (positron) to $ x_{1(2)} p_{-(+)}$. 
After this ``pre-emission'',  the ``hard'' subprocess $e^- (x_1 p_-) e^+ (x_2 p_+) \to  X$
takes place, at a reduced centre of mass energy squared ${\hat s}  = x_1 x_2 s$. The 
resulting cross section, corrected for IS QED radiation, can be represented as 
follows~\cite{strucfun}: 
\begin{equation}
\sigma (s) = \int_0^1 d x_1 d x_2  D(x_1, s) D(x_2, s) \sigma_0 (x_1 x_2 s) 
\Theta(\tx{cuts}), 
\label{eq:masterd}
\end{equation}
where $D(x,s)$ are the electron structure functions, representing the probability that an
incoming electron (positron) radiates a collinear photon, retaining a fraction $x$ of its
original momentum, and $\Theta(\tx{cuts})$ represent the rejection algorithm implemented in
order to take care of experimental cuts (see Fig.~\ref{fig:strucfun} for  a graphical representation). 

Equation~(\ref{eq:masterd}) takes into account only IS photonic radiation, for the sake of
simplicity.  It has however to be noticed that  
in the literature more accurate LL results  can be found, taking into account also kinematical
effects due to IS  radiation and/or the effect of final-state radiation  for the small-angle Bhabha
process~\cite{mnpi93,sabsyr95}. 

Actually, when considering  photonic radiation
only the non-singlet part of the SF is of interest. If the running of the QED
coupling constant $\alpha (s)$ is neglected (which corresponds to neglecting the corrections
due  to additional pair production, that can be taken into account properly by means of
dedicated formulae~\cite{pairs}), the non-singlet part of the SF is the solution of the
following Renormalization Group (RG) equation:
\begin{equation}
s  {{\partial} \over {\partial s}} D(x,s) = {{\alpha} \over {2 \pi}} \int_x^1 {{dz} \over {z}}
P_+(z) D \left( {{x} \over {z}} , s \right) ,  
\label{eq:apeqdiff}
\end{equation}
where $P_+(z) $ is the so called Altarelli-Parisi (AP) splitting function~\cite{ap77}, given by 
\begin{eqnarray}
&& P_+(z) = P(z) - \delta (1-z) \int_0^1 dx P(x) , \nonumber \\
&& P(z) = {{1 + z^2} \over {1-z}} . 
\label{eq:apvert}
\end{eqnarray}
Equation (\ref{eq:apeqdiff}) can be also transformed into an integral equation, subject to the
boundary  condition $D(x, m_e^2) = \delta(1-x)$: 
\begin{equation}
D(x,s)  = \delta (1-x) + {{\alpha} \over {2 \pi}} \int_{m_e^2}^{s} {{d Q^2} \over {Q^2}}
\int_x^1  {{dz} \over {z}} P_+(z) D \left( {{x} \over {z}} , Q^2  \right).  
\label{eq:apeq}
\end{equation}

Equation (\ref{eq:apeq}) can be solved exactly by making use of numerical methods. One of these
is the so called inverse Mellin transform method, which is briefly described in the following. 
The Mellin transformation is defined as 
\begin{eqnarray}
&& F(n) = \int_0^1 dx x^{n-1} f(x) , \\
\label{eq:mellin}
&& f(x) = \int_{\gamma - i \infty}^{\gamma + i \infty} {{dn} \over {2 \pi i}} x^{-n} F(n) . 
\label{eq:amellin}
\end{eqnarray}
By taking the Mellin transform of eq.~(\ref{eq:masterd}), switching to the differential form of the evolution
equation  and solving for the Mellin moments, one finds
\begin{equation}
D(n,s) = \exp \left[ {{\eta} \over {4}}  C(n) \right] , 
\label{eq:dmom}
\end{equation}
where $C(n)$ are the Mellin moments of the AP splitting function 
\begin{equation}
C(n) = \int_0^1 dz z^{n-1} P_+(z) 
\label{eq:pmom}
\end{equation}
and 
\begin{equation}
\eta = {{2 \alpha} \over {\pi}} \ln \left( {{s} \over {m_e^2}} \right) . 
\label{eq:eta}
\end{equation}
By computing the moments of eq.~(\ref{eq:pmom}), 
inserting them into (\ref{eq:dmom}), and performing numerically the anti-transformation
given by  (\ref{eq:amellin}), the exact numerical solution of (\ref{eq:apeq}) is
obtained. 
There is only a technical remark at this point: the integrand of the anti-transformation is
strongly oscillating in the asymptotic region, so that a proper regularization procedure must be adopted
in order to obtain numerical convergence. 

For practical purposes, it is also of great interest to obtain approximate analytical representations of
the solution of the evolution equation. A first analytical solution  can be obtained in the so called
{\it soft photon approximation}, \idest\ in the limit $x \simeq 1$. In such a limit, the dominant
contribution to eq.~(\ref{eq:amellin}) comes from the large $n$ region, so that one can approximate the
moments of the AP splitting function by means of their asymptotic expansion, namely
\begin{equation}
C(n) \simeq {{3} \over {2}}  - 2 \gamma_E -2 \ln n + O \left( {{1} \over {n}} \right) , 
\label{eq:pmoma}
\end{equation}
$\gamma_E $ being the Euler constant. By inserting eq.~(\ref{eq:pmoma}) into 
eq.~(\ref{eq:dmom}) and computing
the anti-transformation one obtains
\begin{equation}
D_{GL}(x,s) = {{\exp \left[ {{{1} \over {2}} \eta \left( {{3} \over {4}} - \gamma_E \right)} \right] } 
\over { \Gamma \left( 1 + {{1} \over {2}} \eta \right) }} {{1} \over {2}} \eta (1 - x)^{{{1} \over {2}} 
\eta - 1} ,  
\label{eq:glsol}
\end{equation}
$\Gamma$ being the Euler gamma-function. The solution shown in eq.~(\ref{eq:glsol}) 
is known as the {\it
Gribov-Lipatov} (GL) approximation~\cite{gl72}. Its main feature is that it 
exponentiates at all perturbative orders the
large logarithmic contributions of the kind $\eta \ln (1-x)$. Its main drawback is that it is valid only
in the soft limit, \idest\ it does not take into account properly hard-photon effects.  
 
The evolution equation (\ref{eq:apeq}) can also be solved iteratively. At the $n$-th step of the iteration, 
one obtains the ${\cal O} (\alpha^n)$  contribution to the structure function. The iterative solution of the
evolution equation is, in some sense, complementary to the exponentiated GL solution. Actually, on the one hand 
the GL solution is exponentiated at all perturbative orders, whereas the iterative solution must be truncated 
at a given finite perturbative order.  On the other hand, the GL solution is valid in the limit $x \simeq 1$,
whereas any given iterative  contribution can be computed exactly.  

In order to go beyond the soft-photon approximation, the following general strategy has been adopted. Given an
iterative solution up to a given perturbative order, for each perturbative contribution it is possible to
isolate the part which is contained in the GL solution.  Then, by 
combining the GL  solution with the iterative one, in which that part has been eliminated in order to avoid 
double counting, one can build a {\it hybrid } solution of the evolution equation, which exploits all the
positive features of the two kinds of solutions and does not present anymore the limitations intrinsic to each 
of them. As a matter of fact, in the literature it is possible to find two classes of hybrid solutions, namely
the additive and factorized ones. The additive solutions have the following general structure: they are built by
simply adding to the GL solution the finite order terms computed by means of the iterative solutions. The
factorized solutions, on the contrary, are obtained by multiplying the GL solution by finite order terms, in
such a way that, order by order, the iterative contributions are exactly recovered. In the following,  the third
order exponentiated additive and factorized solutions are reported. The additive one 
reads~\cite{cdmn92}:  
\begin{eqnarray}
&& D_{A} (x,s)  = \sum_{i=0}^3 d_A^{(i)} (x,s) , \nonumber \\
&& d_A^{(0)} (x,s) = D_{GL} (x,s) , \nonumber \\
&& d_A^{(1)} (x,s) = - {1 \over 4 }  \eta (1+x) , \nonumber \\
&& d_A^{(2)} (x,s) = {1 \over 32 } \eta^2 \left[  (1+x) \left( -4 \ln (1-x) + 3 \ln x 
\right) - 4 {{\ln x } \over  {1-x}} - 5 - x  \right] ,  \nonumber \\ 
&&  d_A^{(3)} (x,s) = {1 \over 384 } \eta^3 \left\{ (1+x) \left[ 18 \zeta (2) - 6 \hbox{\rm Li}_2 (x) 
-12  \ln^2 (1-x) \right]  \right.  \nonumber \\
&& \qquad \qquad + {1 \over {1-x}} \left[ - {3 \over 2} (1 + 8 x + 3 x^2) \ln x 
- 6 (x+5) (1-x) \ln (1-x) \right. 
\nonumber \\
&& \qquad \qquad -12 (1+ x^2) \ln x \ln (1-x) + {1 \over 2} (1 + 7 x^2 ) \ln^2 x \nonumber \\
&& \qquad \qquad \left. \left. - {1 \over 4} (39 - 24 x - 15 x^2 ) \right] \right\} , 
\label{eq:strucfuna}
\end{eqnarray}
where $\zeta $ is the Riemann $\zeta$-function. The factorized one reads~\cite{sj91,s92}: 
\begin{eqnarray}
&& D_F (x,s) = D_{GL} (x,s) \sum_{i=0}^2 d_F^{(i)} , \nonumber \\
&& d_F^{(0)} = {{1} \over {2}} (1 + x^2) , \nonumber \\
&& d_F^{(1)} = {{1} \over {4}} {{\eta} \over {2}} \left[ - {{1} \over {2}} (1 + 3 x^2) \ln x
- (1-x)^2 \right] , \nonumber \\
&& d_F^{(2)} = {1 \over 8} \left( {\eta \over 2} \right)^2 \left[ (1-x)^2 + {1 \over 2} (3 x^2 - 4 x + 1 ) 
\ln x \right. \nonumber \\
&& \left. + {1 \over 12} (1 + 7 x^2) \ln^2 x + (1 - x^2) \tx{Li}_2 (1-x) \right] . 
\label{eq:strucfunf}
\end{eqnarray}
It is worth noting that also higher order
solutions are known, numerically and/or analytically, for both the classes of solution. However, LEP
phenomenology is not sensitive to hard photonic terms beyond the third order. 

If only cuts on the invariant mass of the event after ISR are considered, eq.~(\ref{eq:masterd}) can be
simplified by performing explicitly one of the integrations. Actually, for $s' \ge x_0 s$, it can be shown 
that the QED corrected cross section can be written as
\begin{equation}
\sigma(s) = \int_0^{1 - x_0} dx H(x,s) \sigma_0 \left( (1-x) s \right) , 
\label{eq:masterh}
\end{equation} 
where $H(x,s)$ is the so called {\it radiator}, or {\it flux function}, defined as
\begin{equation}
H (x,s) = \int_{1-x}^1 {{dz} \over {z}} D(z,s) D \left( {{1-x} \over {z}} , s \right) . 
\label{eq:radh}
\end{equation}
By defining
\begin{equation}
K(x,s) = H(1-x,s)
\label{eq:radk}
\end{equation} 
and taking the Mellin moments of eq.~(\ref{eq:radk}), the following identity can be shown:  
\begin{equation}
K(n,s) = \int_0^1 dx x^{n-1} K(x,s) = D^2 (n,s) = \exp \left[ {{\eta} \over {2}}  C(n) \right] ,
\label{eq:kmom} 
\end{equation}
where $C(n)$ are the Mellin moments of the AP splitting function, given by eq.~(\ref{eq:pmoma}). 
In virtue of eq.~(\ref{eq:kmom}), the following identity holds: 
\begin{equation}
H_\eta (x,s) = D_{2 \eta}(1-x , s) . 
\label{eq:reldh}
\end{equation} 

It is worth noticing that in QED  there are some non-leading corrections that behave much
like the leading logarithmic ones. Actually, it can be verified by inspection that after the substitution 
\begin{equation}
\eta \to \beta = {{2 \alpha} \over {\pi}} \left[ \ln \left( {{s} \over {m_e^2}} \right)
-1  \right] 
\label{eq:etabeta}
\end{equation}
the ${\cal O} (\alpha)$  distribution of the invariant mass of the event after ISR as extracted from
eq.~(\ref{eq:masterh}) reproduces exactly the standard diagrammatic calculation.  

Moreover, it has to be noticed that it is possible to  match the all-orders LL results  of the SF
method with exact finite-order diagrammatic results. For the $Z^0$ line shape, the  match has been
performed  in the soft-photon approximation up to ${\cal O} (\alpha^2)$, according to the procedure
described in Sect.~\ref{sect:z0qed}. Also a more general procedure for implementing the matching
beyond the soft-photon approximation, and hence taking into  account hard-photon effects, is known
and can be found in ref.~\cite{a2l96}. 
\subsection{The Parton Shower Method}
\label{sect:psm}
The Parton Shower  method  is substantially a method for providing a Monte Carlo iterative
solution of the evolution equation, at the same time generating the four-momenta of the
electron and photon at a given step of the iteration. It  has been  studied  within the context of 
QCD (see  for instance  refs.~\cite{o80} and~\cite{mw84}) and  subsequently developed also for
QED (see refs.~\cite{km89}, \cite{fsm93}, \cite{m95} and references  therein). 
In order to implement the algorithm, it
is first necessary to assume the existence of an upper limit for $x$, in such a way that the
AP splitting is regularized as follows: 
\begin{equation}
P_+(z) = \theta(x_+ - z) P(z) - \delta (1-z) \int_0^{x_+} dx P(x) . 
\label{eq:apvertmod}
\end{equation}
Of course, in the limit $x_+ \to 1$ eq.~(\ref{eq:apvertmod}) recovers the usual definition of
the AP splitting function given in eq.~(\ref{eq:apvert}). By inserting the modified AP vertex into
eq.~(\ref{eq:apeqdiff}), one obtains
\begin{equation}
s {{\partial} \over {\partial s}}  D(x,s) = {{\alpha} \over {2 \pi}} \int_x^{x_+} 
{{dz} \over {z}} P(z) D \left( {{x} \over {z}} , s \right) 
 - {{\alpha} \over {2 \pi}} D(x,s) \int_x^{x_+} dz P(z) . 
\label{eq:apeqmod}
\end{equation}
Separating now the variables and introducing the Sudakov form factor
\begin{equation}
\Pi (s_1, s_2) = \exp \left[ 
- {{\alpha} \over {2 \pi}} \int_{s_2}^{s_1} {{d s'} \over {s'}} \int_0^{x_+} dz P(z) 
\right] , 
\label{eq:sudakov}
\end{equation}
which is the probability that the electron evolves from virtuality $-s_2$ to $-s_1$ without
emitting photons of energy fraction larger than $1-x_+$, 
eq.~(\ref{eq:apeqmod}) can be recast into integral form as follows: 
\begin{equation}
D(x,s) = \Pi (s,m_e^2) D(x, m_e^2) + {{\alpha} \over {2 \pi}} \int_{m_e^2}^{s} 
{{ds'} \over {s'}} \Pi (s,s') \int_{x}^{x_+} {{dz} \over {z}} P(z) 
D \left( {{x} \over {z}} , s' \right) . 
\label{eq:apeqmodi}
\end{equation}
The formal iterative solution of eq.~(\ref{eq:apeqmodi}) can be represented by the following
infinite series: 
\begin{eqnarray}
D(x,s) = && \sum_{n=0}^{\infty} \prod_{i=1}^{n} \left\{ 
\int_{m_e^2}^{s_{i-1}} {{d s_i} \over {s_i}} \Pi (s_{i-1}, s_i) {{\alpha} \over {2 \pi}}
\int_{x/(z_1 \cdots z_{i-1})}^{x_+} {{d z_i} \over {z_i}} P(z_i) \right\} \nonumber \\
&&\Pi (s_n, m_e^2) D \left( {{x} \over {z_1 \cdots z_n}} , m_e^2 \right) . 
\label{eq:dmc}
\end{eqnarray}
The particular form of (\ref{eq:dmc}) allows to use a Monte Carlo method for building the
solution iteratively. The algorithm is standard, and can be described as follows: 
\begin{itemize}
\item[1 --]{set $Q^2 = m_e^2$, and fix $x=1$ according to the boundary condition $D(x,m_e^2) =
\delta(1-x)$; }
\item[2 --]{generate a random number $\xi$ in the interval $[0,1]$; }
\item[3 --]{if $\xi < \Pi (s,Q^2)$} stop the evolution; otherwise
\item[4 --]{compute $Q'^2$ as a solution of the equation $\xi = \Pi (Q'^2, Q^2)$}; 
\item[5 --]{generate  a random number $z$ according to the probability density $P(z)$  in the
interval $[0,x_+]$; }
\item[6 --]{substitute $x \to xz$ and $Q^2 \to Q'^2$}; go to 2. 
\end{itemize}
An important feature of this algorithm is that it can be used  to generate exclusive events
containing the  complete information on the four-momenta of the particles. This  fact is well
known for the QCD analog of the algorithm, and relies upon the kinematical rules of the so 
called {\it jet calculus}.  For the case under consideration, in which at the $i$-th step of
the iteration a virtual parent electron branches into a virtual electron plus a real photon,
the space-like  kinematics apply. In particular, one has
\begin{equation}
e  (p) \to e (p' ) + \gamma(q) , 
\label{eq:branching}
\end{equation}
where $p^2 = -K^2$, $p'^2 = - K'^2$ and $q^2 = 0$. Assuming that the parent electron moves
along the $z$ axis, the four-momenta of the particles can be parameterized as
follows: 
\begin{eqnarray}
&&p = ( E, \vec{0}, p_z ) , \nonumber \\
&&p'  = (E', \vec{p}_\perp, z p_z) , \nonumber \\
&&q = (E_\gamma , - \vec{p}_\perp, (1-z) p_z) . 
\label{eq:kine}
\end{eqnarray}
From the kinematics of eq.~(\ref{eq:kine}), in the limit $p_z \to \infty$, the following relations
can be derived: 
\begin{eqnarray}
&& E \simeq p_z - {{K^2} \over {2 p_z}} , \nonumber \\
&& E' \simeq z p_z + {{p_\perp^2} \over {2 z p_z}} - {{K'^2 } \over {2 z p_z}} , \nonumber \\
&& E_\gamma \simeq (1-z) p_z + {{p_\perp^2} \over {2 (1-z) p_z}} , 
\label{eq:energies}
\end{eqnarray}
from which, by imposing energy conservation, one obtains
\begin{equation}
- K^2 = - {{K'^2} \over {z}} + {{p_\perp^2} \over {z (1-z)}} . 
\label{eq:ptdet}
\end{equation}
Given $K^2$, $K'^2$ and $z$, eq.~(\ref{eq:ptdet}) allows one to compute the absolute  value of
the transverse momentum at the branching. 

\subsection{YFS Exponentiation}
\label{sect:yfs}

The Yennie-Frautschi-Suura  exponentiation procedure is essentially a technique for
summing up all the infra-red (IR) singularities present in any given process accompanied by
photonic radiation~\cite{yfs}. It is inherently exclusive, \idest\ all the 
summations of the IR-singular
contributions are done before any phase-space integration over the virtual- or real-photon
four-momenta are performed~\cite{jw90}. 
In the following, the general idea underlying the procedure will
be shortly described  (see for  instance~\cite{wascern}  for a more detailed analysis). 

\begin{figure}[hbt]
\begin{center}
\epsfig{file=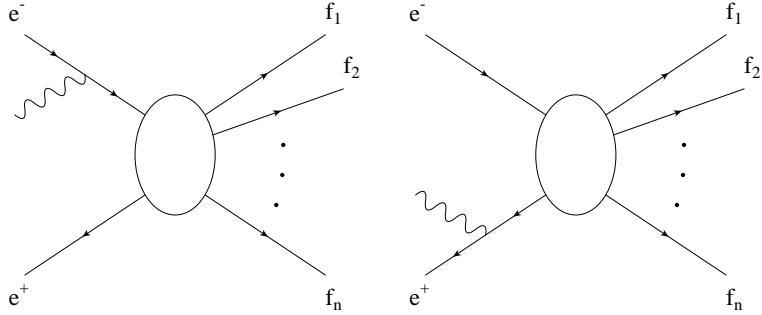, width=10truecm}
\end{center}
\caption{The Feynman diagrams for the initial-state ${\cal O} (\alpha)$  bremsstrahlung correction. }
\label{fig:realg}
\end{figure}

Let us consider the process $e^+ (p_1) e^- (p_2) \to f_1 (q_1) \cdots f_n (q_n)$, 
where $f_1 (q_1) \cdots f_n (q_n)$ represents a given arbitrary final state, and let  
${\cal M}_0$ be its matrix element. 
By using standard Feynman-diagram techniques, it is possible
to show that the  same process, when accompanied by $l$ 
additional real photons radiated by the
initial-state particles, and under the assumption that the $l$ additional photons are soft,
\idest\  their energy is much smaller that any energy scale involved in the process, can be
described by the factorized matrix element built up by the lowest order one,  ${\cal M}_0$,
times the product of $l$ eikonal currents, namely 
\begin{equation} 
{\cal M} \simeq {\cal M}_0 \prod_{i=1}^{l} \left[ e \left( 
{{\varepsilon_i (k_i) \cdot p_2} \over {k_i \cdot p_2}} - 
{{\varepsilon_i (k_i) \cdot p_1} \over {k_i \cdot p_1}}
\right) \right] , 
\label{eq:lrphot}
\end{equation}
where $e$ is the electron charge, $k_i$ are the momenta of the photons and 
$\varepsilon_i (k_i)$ their polarization vectors (see Fig.~\ref{fig:realg} for a 
representation of the
Feynman diagrams relative to the initial-state ${\cal O} (\alpha)$  bremsstrahlung correction). 
By taking the square of the matrix element 
in eq.~(\ref{eq:lrphot}), multiplying for the proper flux factor and the Lorentz-invariant
phase space volume element,  neglecting the four-momenta of the radiated photons in the
overall energy-momentum conservation, summing over the final state photons polarizations 
and combining  properly all the factors, the cross section for the process  
$e^+ (p_1) e^- (p_2) \to f_1 (q_1) \cdots f_n (q_n) + l \tx{ real photons}$ can be written as
\begin{equation}
d \sigma^{(l)}_{r} = d \sigma_0 {{1} \over {l!}} \prod_{i=1}^{l} \left[
k_i d k_i d \cos \vartheta_i d \varphi_i {{1} \over {2 (2 \pi)^3}}
\sum_{\varepsilon_i} e^2 \left( 
{{\varepsilon_i (k_i) \cdot p_2} \over {k_i \cdot p_2}} - 
{{\varepsilon_i (k_i) \cdot p_1} \over {k_i \cdot p_1}}
\right)^2 \right] . 
\label{eq:lreal}
\end{equation}
By summing now on the number of final-state photons, one obtains the cross section for the
original process accompanied by an arbitrary number of real photons, namely
\begin{eqnarray}
d \sigma^{(\infty)}_{r} =&& \sum_{l=0}^{\infty} d \sigma^{(l)}_{r} \nonumber \\
=&& d \sigma_0  \exp \left[
k d k d \cos \vartheta d \varphi {{1} \over {2 (2 \pi)^3}}
\sum_{\varepsilon} e^2 \left( 
{{\varepsilon (k) \cdot p_2} \over {k \cdot p_2}} - 
{{\varepsilon (k) \cdot p_1} \over {k \cdot p_1}}
\right)^2 \right] . 
\label{eq:expr}
\end{eqnarray}
Equation (\ref{eq:expr}), being limited to real radiation only, is IR divergent once the phase
space integrations are performed down to zero photonic energy. This problem, as well known,
finds its solution in the matching between real and virtual photonic radiation. At any rate,
eq.~(\ref{eq:expr}) already shows the key feature of exclusive exponentiation, \idest\ summing
up all the perturbative contributions before performing any phase space integration. 

\begin{figure}[hbt]
\begin{center}
\epsfig{file=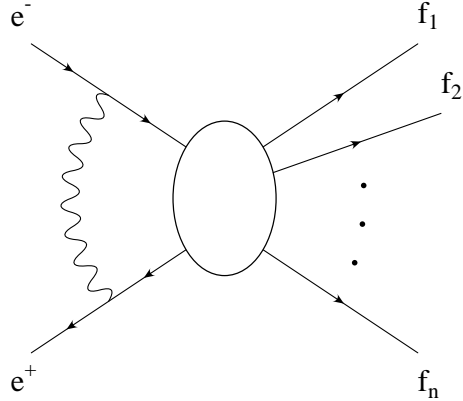, width=6truecm}
\end{center}
\caption{The Feynman diagram for the photonic vertex-like correction. }
\label{fig:virtualg}
\end{figure}

In order to get meaningful radiative corrections, besides initial-state 
real photon corrections it is  necessary to  consider also initial-state virtual photon 
corrections, \idest\ the corrections due to additional  internal photon lines connecting the
initial-state electron and positron (see  Fig.~\ref{fig:virtualg}).  For a vertex-type 
amplitude, the result can be written as
\begin{eqnarray}
{\cal M}_{V_1} = - i {{e^2} \over {(2  \pi)^4}} \int d^4 k &&{{1} \over {k^2  + i \varepsilon}}
{\bar v} (p_1) \gamma^\mu 
{{- ( {\rlap / {p_1}} + {\rlap / {k}} ) + m } \over {2 p_1 \cdot k + k^2 + i 
\varepsilon}}  \nonumber \\
&\times &  \Gamma 
{{ ( {\rlap / {p_2}} + {\rlap / {k}} ) + m } \over {2 p_2 \cdot k + k^2 + i \varepsilon}}
\gamma_\mu u(p_2) , 
\label{eq:virt}
\end{eqnarray}
where $\Gamma$ stands for the Dirac structure competing to the lowest order process, in such a
way that ${\cal M}_0 = {\bar v} (p_1) \Gamma u(p_2)$. The soft-photon part  of the amplitude
can be extracted by taking $k^\mu \simeq 0 $ in all the numerators. In this approximation, the
amplitude of eq.~(\ref{eq:virt}) becomes
\begin{eqnarray} 
&& {\cal M}_{V_1} = {\cal M}_0 \times V \nonumber \\
&& V = {{2 i \alpha} \over {(2 \pi)^3}} \int d^4 k {{1} \over {k^2 + i \varepsilon}}
{{4 p_1 \cdot p_2} \over {(2 p_1 \cdot k + k^2 + i \varepsilon) 
(2 p_2 \cdot k + k^2 + i \varepsilon)}} . 
\label{eq:onevirt}
\end{eqnarray}
Some comments are in order here. First, as in the real case, the IR virtual correction
factorizes off the lowest order matrix element, so that it is universal, \idest\ independent
of the details of the lowest order   process under consideration. Moreover, it is also  free
of ultraviolet (UV) infinities; this, of course, does not mean that the UV behaviour of the
complete amplitude is irrelevant, but that the IR part can be treated independently of
renormalization problems. Last, as  in the real case, it is divergent in the IR portion of the
phase space.  

The correction given by $n$  soft virtual photons can be seen to factorize with an additional 
$1 / n!$ factor~\cite{yfs}, namely 
\begin{equation}
{\cal M}_{V_n}  = {\cal M}_0  \times {{1} \over {n!}} V^n , 
\label{eq:nvirt}
\end{equation}
so that by summing over all  the additional soft virtual photons one obtains
\begin{equation}
{\cal  M}_V = {\cal M}_0 \times \exp [V] . 
\label{eq:virtexp} 
\end{equation}

As  already noticed, both the real  and virtual factors are IR divergent. In order to obtain
meaningful expressions, one has to adopt  some regularization procedure. One possible
regularization procedure is to give the photon a (small) mass $\lambda$ and modifying
eqs.~(\ref{eq:lreal}) and (\ref{eq:onevirt}) accordingly. Once all the expressions are
properly regularized, one can write down the YFS master formula, which takes  into account
real and virtual photonic corrections to the lowest order process. In virtue of the
factorization properties discussed above, the master formula can be obtained from
eq.~(\ref{eq:expr}) with the substitution $d \sigma_0 \to d \sigma_0 \vert \exp (V) \vert^2$,
\idest\
\begin{equation}
d \sigma  = d \sigma_0  \vert \exp (V) \vert^2 \exp \left[
k d k d \cos \vartheta d \varphi {{1} \over {2 (2 \pi)^3}}
\sum_{\varepsilon} e^2 \left( 
{{\varepsilon (k) \cdot p_2} \over {k \cdot p_2}} - 
{{\varepsilon (k) \cdot p_1} \over {k \cdot p_1}}
\right)^2 \right] . 
\label{eq:yfsmaster}
\end{equation} 
As a last step, it is possible to perform analytically the IR cancellation between virtual 
and very soft  real photons. Actually, since very soft real photons do not affect the
kinematics of the process, the real photon exponent can be split into a contribution coming
from photons with energy less than a cutoff $k_{min}$ plus a contribution  coming from photons
with energy above the same cutoff. The first contribution can be integrated over all its phase
space and then combined with the virtual exponent. The physical meaning of this procedure is
to sum over all the degenerate final states: in fact, a very soft real photon produces a
signature that is indistinguishable from the signature typical of the elastic event. After
this step it is possible to remove the regularizing photon mass by taking the limit 
$\lambda \to 0$, so that eq.~(\ref{eq:yfsmaster}) becomes
\begin{equation}
d \sigma  = d \sigma_0   \exp (Y)  \exp \left[
k d k d \Theta (k - k_{min})\cos \vartheta d \varphi {{1} \over {2 (2 \pi)^3}}
\sum_{\varepsilon} e^2 \left( 
{{\varepsilon (k) \cdot p_2} \over {k \cdot p_2}} - 
{{\varepsilon (k) \cdot p_1} \over {k \cdot p_1}}
\right)^2 \right] , 
\label{eq:yfsfinal}
\end{equation} 
where  $Y$ is given  by 
\begin{equation}
Y = 2 V + \int  k d k d \Theta ( k_{min} - k )
\cos \vartheta d \varphi {{1} \over {2 (2 \pi)^3}}
\sum_{\varepsilon} e^2 \left( 
{{\varepsilon (k) \cdot p_2} \over {k \cdot p_2}} - 
{{\varepsilon (k) \cdot p_1} \over {k \cdot p_1}}
\right)^2  . 
\end{equation}
The explicit form of $Y$  can be derived by performing all the details of the calculation, and
reads
\begin{eqnarray}
&&Y = \beta \ln {{k_{min}} \over {E}} + \delta_{YFS} , \nonumber \\
&&\delta_{YFS} = {{1} \over {4}} \beta + {{\alpha} \over {\pi}} \left( 
{{\pi^2}  \over {3}} - {{1} \over {2}} \right) , \nonumber \\
&&\beta = {{2 \alpha} \over {\pi}} \left[ \ln \left( {{s} \over {m_e^2}} \right) 
- 1 \right]  . 
\end{eqnarray}

As in the SF method, the method  of  YFS exclusive exponentiation has been  refined  in order 
to take  into account all-orders corrections on top of finite-order exact results  (see for
instance ref.~\cite{bhlumi95} and  references therein). 

%% file: vp.txi
\section{Vacuum Polarization Corrections}
\label{sect:vacpol}
\begin{figure}[hbt]
\begin{center}
\epsfig{file=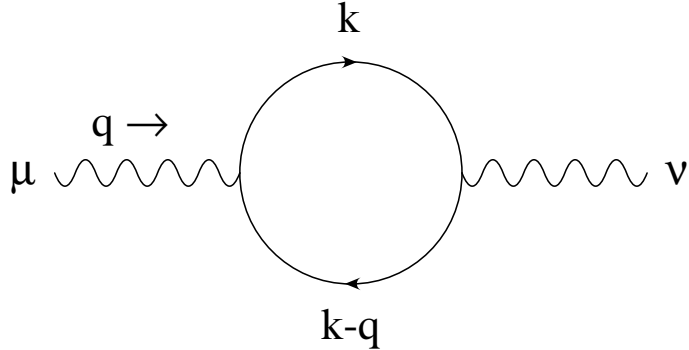, width=90mm}
\end{center}
\caption{The fermionic vacuum polarization Feynman graph.  }
\label{fig:vacpolgraph}
\end{figure}
The input parameter adopted for the analysis of the precision electroweak data collected at
LEP are the QED coupling constant at zero momentum transfer $\alpha$, the Fermi constant 
$G_\mu$ and the $Z$-boson mass $M_Z$. 
Given these input parameters, all the other unknown (or poorly known)
parameters, such as the {\it top}-quark mass, the Higgs-boson mass and the QCD coupling
constant $\alpha_s$, are determined by means of best fits. 

Actually, the QED coupling constant
$\alpha$ is known with great precision at zero momentum transfer from measurements 
such as the electron/muon $g-2$, the Lamb shift, the muonium hyperfine splitting, the neutron
compton wavelength, the quantum Hall  and the Josephson 
effects~\cite{kinoshita90,kinoshita95}; 
but for precision physics at LEP it must be evolved up to
the $Z$-boson mass scale, so that the relevant parameter is rather the running QED coupling
constant $\alpha (s)$. As is well known, the running of $\alpha$ is largely dominated by the
contribution of fermionic vacuum polarization diagrams, which represent by themselves a
gauge-invariant and universal subset of radiative corrections for two-fermion production processes
(see Fig.~\ref{fig:vacpolgraph}). The bosonic contribution to
vacuum polarization is, on the contrary, gauge-dependent. Its contribution is 
small, at least around the $Z^0$ resonance, 
and can be taken into account together with other gauge- and process-dependent 
radiative corrections at a given perturbative order.  
From now on, only the fermionic contribution to vacuum polarization will be discussed. 
\begin{figure}[hbt]
\begin{center}
\epsfig{file=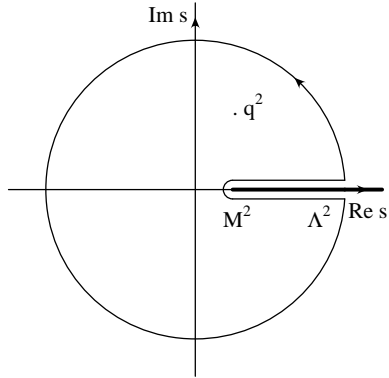, width=50mm}
\end{center}
\caption{The integration contour ${\cal C}$ to be used for the integral of 
eq.~(\ref{eq:cauchy}) in order to obtain the dispersion relation of eq.~(\ref{eq:subds}). }
\label{fig:contour}
\end{figure}

The QED running coupling constant is given by 
\begin{eqnarray}
&&\alpha (s) = {{\alpha(0)} \over {1 - \Delta \alpha (s)}} , \nonumber \\
&&\Delta \alpha (s) = - 4 \pi \alpha(0) 
{\tx{Re}} \left[ \Pi_\gamma (s) - \Pi_\gamma (0) \right] ,  
\label{eq:adef}
\end{eqnarray}
where $\Pi (q^2) $ is the two-point electromagnetic correlator, given by 
\begin{equation}
( q^\mu q^\nu - q^2 g^{\mu \nu}) \Pi_\gamma (q^2) = i \int d^4 x e^{i q \cdot x}
\langle 0 \vert T ( j^\mu (x) j^\nu (0) ) \vert 0 \rangle . 
\label{eq:pigrec}
\end{equation}

The vacuum polarization contribution can be naturally split into the contributions coming from
the leptons, the {\it  light} hadrons and the {\it top} quark  according to the following
relation: 
\begin{equation}
\Delta \alpha (s) = \Delta \alpha_{l} (s) +  \Delta \alpha_{h}^{(5)} (s) 
+ \Delta \alpha_t (s) .  
\label{eq:dacontr} 
\end{equation}
\begin{figure}[hbt]
\begin{center}
\epsfig{file=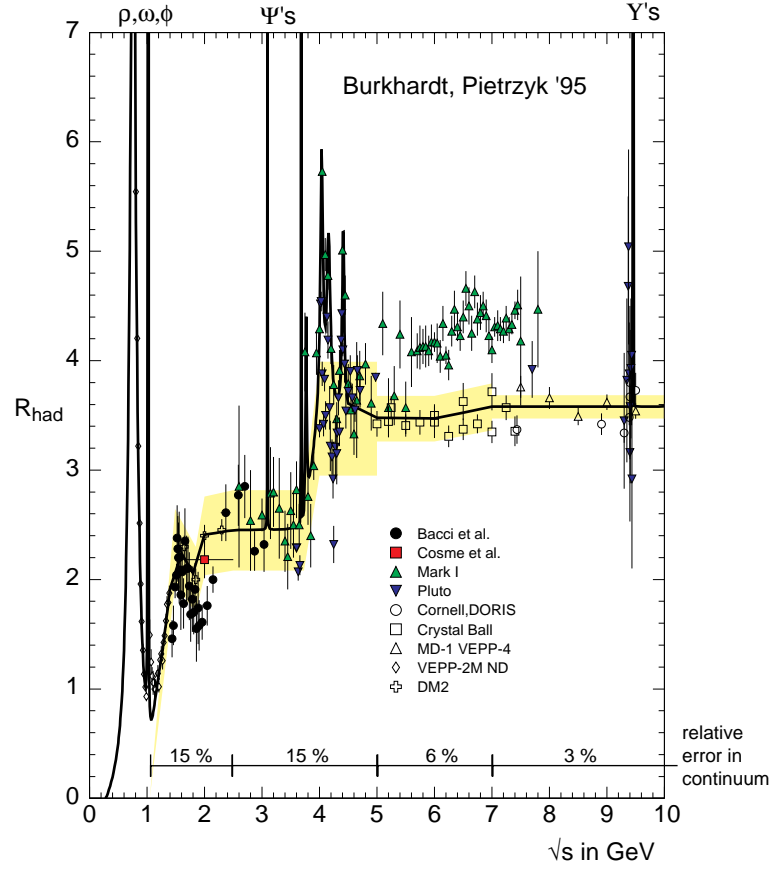,width=100mm}
\end{center}
\caption{The total cross section for  the process  $e^+ e^- \to hadrons$,   normalized  to the total 
cross section for  the  process  $e^+ e^- \to  \mu^+ \mu^-$ at low centre of mass energy 
(from~\cite{bolek95}). }
\label{fig:rhad_lowen}
\end{figure}

The contributions from leptons and {\it top}-quark loops can be safely computed by means of
ordinary perturbation theory. In particular, the leptonic contributions can be very well
approximated by the expression of the loop-integrals for $m_l^2 << s$, and read
\begin{equation}
\Delta \alpha_{l} (s) = {{\alpha (0)} \over {3 \pi }} \sum_l 
\left( \ln {{s} \over {m_l^2}} - {5 \over 3}  \right) . 
\label{eq:dalep}
\end{equation}
On the other hand, the {\it top}-quark contribution, due to the decoupling properties of QED,
can be represented by 
\begin{equation}
\Delta \alpha_t (s) = - {{\alpha(0)} \over {3 \pi }} {4 \over 15} {{s} \over {m_t^2}} . 
\label{eq:datop}
\end{equation}
In particular, eqs.~(\ref{eq:dalep}) and (\ref{eq:datop}), when evaluated at $s = M^2_Z$ give
the values
\begin{eqnarray}
&& \Delta \alpha_{l} (M^2_Z) = 0.03142 ,  \nonumber \\
&& \Delta \alpha_{t} (M^2_Z) = -0.6 \times 10^{-4} . 
\end{eqnarray}
\begin{figure}[hbt]
\begin{center}
\epsfig{file=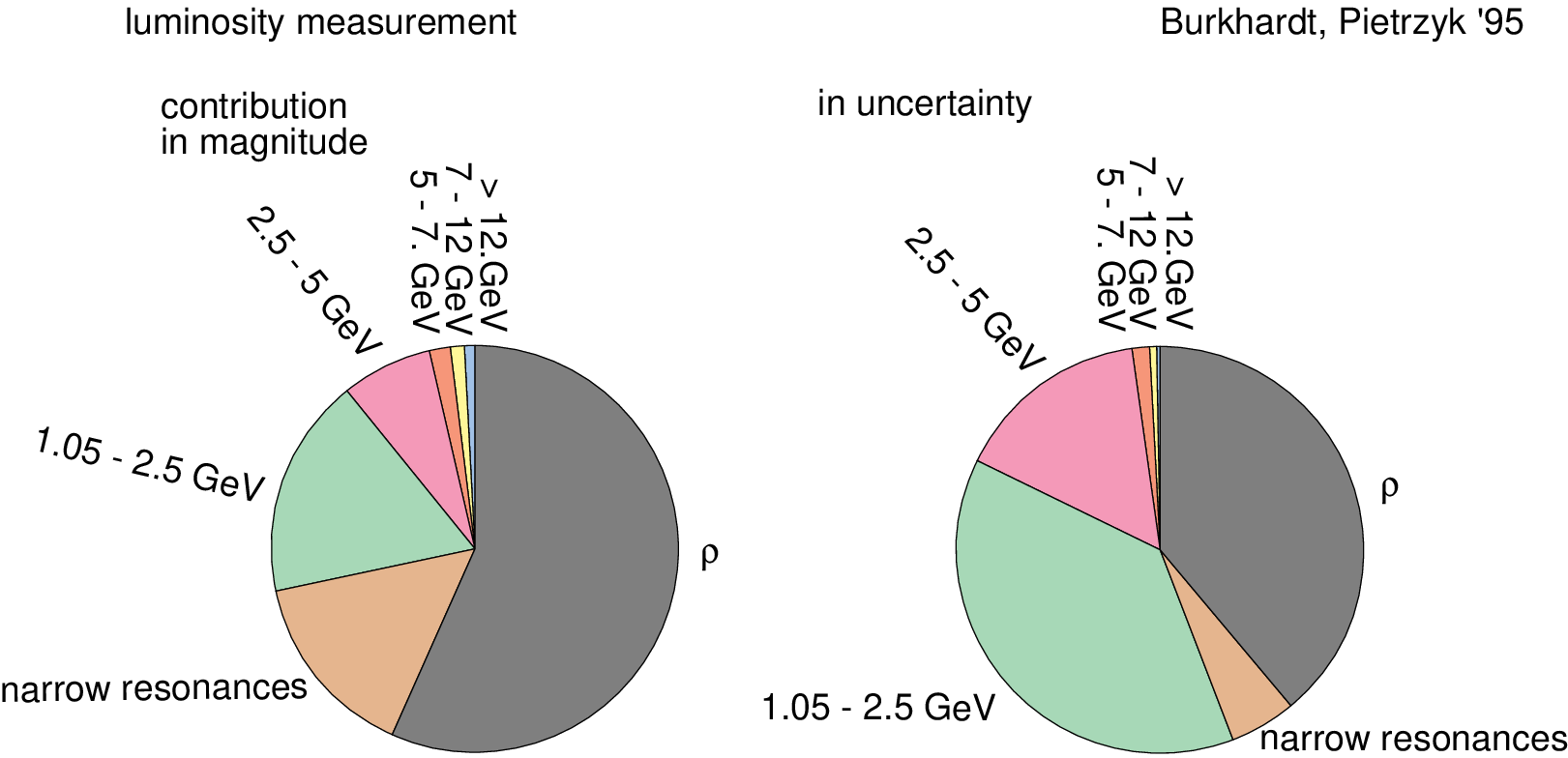,width=115mm}
\end{center}
\caption{The relative contributions to $\Delta \alpha (t= -1.424$~GeV$^2)$ in magnitude and
uncertainty (from~\cite{bolek95}). }
\label{fig:vacpolpiebha}
\end{figure}

For the {\it light}-quark contributions, on the contrary, the perturbative expression of
eq.~(\ref{eq:dalep})
cannot be used. In fact, due to the ambiguities inherent in the definition of the 
{\it light}-quark masses, the answer provided by eq.~(\ref{eq:dalep}) is affected by very large
uncertainties, proportional to $\delta m_q / m_q$. Moreover, one can {\it a priori} expect
very large  non-perturbative QCD corrections in the region of the hadronic resonances. 
The standard procedure adopted in order  to circumvent  this problem is to exploit the
analyticity properties of the vacuum polarization amplitudes by making use of dispersion
relations (DR) techniques (see for instance~\cite{k96}). 

\begin{figure}[hbt]
\begin{center}
\epsfig{file=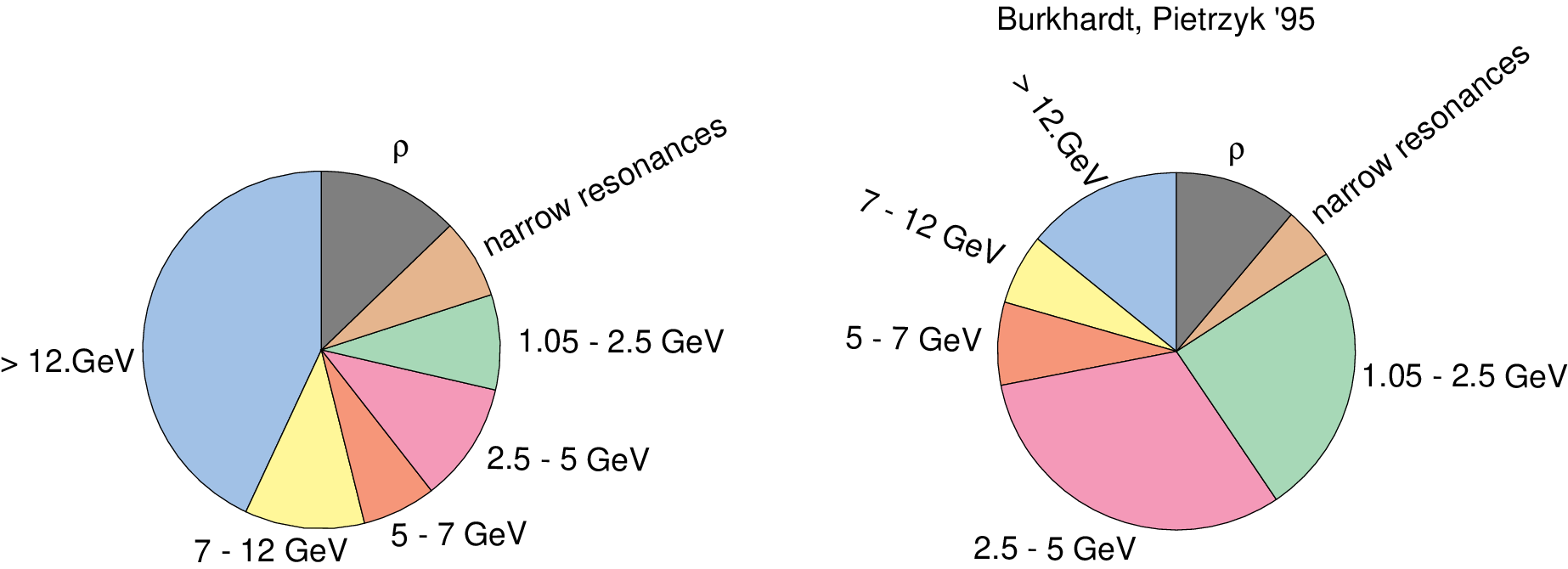,width=130mm}
\end{center}
\caption{The relative contributions to $\Delta \alpha (s= M_Z^2)$ in magnitude and
uncertainty (from~\cite{bolek95}). }
\label{fig:vacpolpie}
\end{figure}

Given a complex-valued function $F$ of complex argument $s$, 
if $F$ is olomorphic in a region ${\cal R}$ and on its boundary ${\cal C}$, the value 
of $F$ at any point $q^2$ inside ${\cal R}$ can be computed, according to the Cauchy's 
integral representation, as a contour integral:  
\begin{equation}
F (q^2) = {{1} \over {2  \pi i}} { \oint}_{\cal C} ds {{F (s)} \over {s - q^2}} . 
\label{eq:cauchy} 
\end{equation}
Under the assumption that 
\begin{itemize}
\item $F(s)$ is real for real $s$, up to a threshold $M^2$,  
\item $F(s)$ has a branch cut for real $s > M^2$, 
\item $F(s)$ is olomorphic except along the branch cut,  
\end{itemize}
and taking as integration contour the  one shown in Fig.~\ref{fig:contour}, 
one  can derive the following once-subtracted DR
\begin{equation}
F(q^2) = F(q_0^2) + {{q^2 - q_0^2} \over {\pi }} \int_{M^2}^{\infty}
{{ds} \over {s  - q_0^2} } {{{\tx{Im}} F(s)} \over {s - q^2 - i \varepsilon}} . 
\label{eq:subds}
\end{equation}
\begin{figure}[hbt]
\begin{center}
\epsfig{file=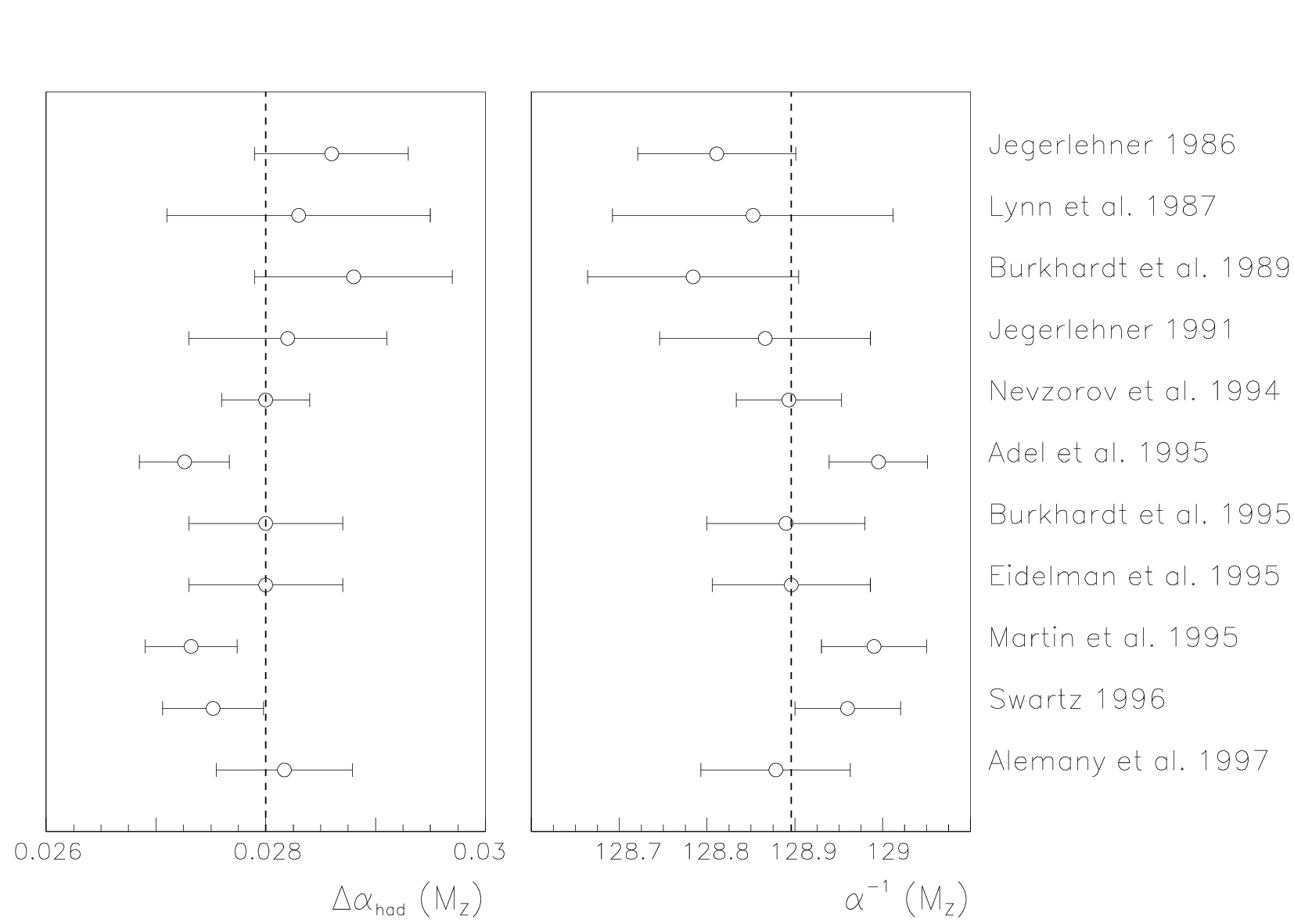,width=130mm}
\end{center}
\caption{The determinations of $\Delta \alpha_h^{(5)} (M_Z)$ and $\alpha^{-1} (M_Z)$. The
dashed vertical line selects the value  used by the LEP Collaborations. }
\label{fig:alphadet}
\end{figure}

Equation (\ref{eq:subds}) allows one to compute the value of $F$ at any point $q^2$ with  the
only knowledge of the imaginary part of $F$ along the branch cut. 
The optical theorem provides the link between the absorptive part of the hadronic vacuum 
polarization amplitude and the total cross section for the reaction $e^+ e^- \to 
{\tx{hadrons}}$, namely 
\begin{equation}
\sigma (s) = {{16 \pi^2 \alpha^2 (s)} \over {s}} {\tx{Im}} \Pi_\gamma (s) . 
\label{eq:opticalth}
\end{equation}
Hence, from the experimental data for $e^+ e^- \to  {\tx{hadrons}}$ (see
Fig.~\ref{fig:rhad_lowen})  one can compute the
hadronic contribution of the hadronic vacuum polarization as follows: 
\begin{equation}
\Delta \alpha_h^{(5)} (s) = - {{s} \over {4 \pi^2 \alpha}} {\tx{Re}}
\int_{4 m_\pi^2}^{\infty} d s' {{\sigma_h (s')} \over {s'  - s - i \varepsilon}} . 
\label{eq:dah}
\end{equation}

Several determinations of the hadronic contribution to the running of the QED coupling
constant $\alpha $ have
been performed~\cite{jeger86,lpv87,betal89,jeger91,nn94,adyn95,
bolek95,ej95,mz95,swartz96,adh97}. 
The differences between the various determinations 
can be traced back to the
procedures adopted for fitting the data, treating the experimental errors and  
performing
the numerical integrations. The differences are also determined by the different 
thresholds
chosen to start the application of perturbative QCD at large $s$, and to the value 
used for $\alpha_s (M_Z)$. Figures~\ref{fig:vacpolpiebha} and \ref{fig:vacpolpie} show the
relative contributions to $\Delta \alpha$ in magnitude  and uncertainty at a scale typical
of   the luminosity measurement (Fig.~\ref{fig:vacpolpiebha}) and  at  the $Z^0$ peak  
(Fig.~\ref{fig:vacpolpie}). In Fig.~\ref{fig:alphadet} several  determinations  of $\Delta
\alpha (M_Z)$ and   $\alpha^{-1} (M_Z)$ are  shown, together with the value used by the LEP
Collaborations.\footnote{Two new determinations of $\alpha (M_Z)$ have recently
appeared~\cite{anew1,anew2} giving $\alpha^{-1} (M_Z) = 128.923 \pm 0.036$ and 
$\alpha^{-1} (M_Z) = 128.98 \pm 0.06$, respectively. After the completion of this work, a further analysis of
$\alpha (M_Z)$ has been performed in~\cite{ks98}, yelding the result 
$\alpha^{-1} (M_Z) = 128.928 \pm 0.023$. } 

For practical  purposes, the running of  the QED coupling constant for large time-like 
momenta  can be computed by means of the following effective formula, in terms of
fermion masses and effective {\it light}-quark masses: 
\begin{equation}
\Delta \alpha (s) = {{\alpha (0)} \over {3 \pi }} \sum_f Q_f^2 N_{cf}
\left( \ln {{s} \over {m_f^2}} - {5 \over 3}  \right) . 
\label{eq:daeff}
\end{equation}
In eq.~(\ref{eq:daeff}), the sum is extended to all {\it light} fermions (charged
leptons  and {\it light} quarks), $Q_f$ is the fermion charge in units of  the electron
charge and $N_{cf}$ is 1 for leptons and 3 for quarks. The effective {\it light}-quark 
masses $m_d = m_u  =  47$~MeV, $m_s   = 150$~MeV, $m_c = 1.55$~GeV  and  $m_b =
4.7$~GeV insure that the hadronic vacuum polarization contribution of 
refs.~\cite{bolek95,ej95} is reproduced.

%% file: scalint.txi
  
\section{One-loop Integrals and Dimensional Regularization}  
\label{sect:scalint}
  
The evaluation of one-loop diagrams generally leads to   
the problem of ultraviolet divergences, so that a   
regularization procedure has to be adopted in order to deal   
with well defined objects before implementing a renormalization program. 
As said in Sect.~\ref{sect:olfd} the   
method commonly used in the case of gauge theories is the   
dimensional regularization~\cite{bg72,ashm72,hv72},   
where the singularities arise as   
poles of the form $1/(4-n)$, $n$ being the number of   
space time dimensions where the integrals are convergent.   
In the literature a general   
method has been developed~\cite{pv79}, which allows to   
write every loop diagram in terms of certain combinations of   
scalar form factors~\cite{hv79}.   
In this appendix some technical details concerning this   
procedure are provided for the simple case of the two-point   
form factors.   
In the evaluation of a self-energy diagram the following   
integrals are encountered:  
\begin{eqnarray}  
B_0; B_{\mu}; B_{\mu \nu}(p^2, m_1, m_2) = {1\over {i \pi^2}}   
\int d^n q {{1; q_{\mu}; q_{\mu \nu }}\over   
{(q^2 + m_1^2) ((q+p)^2 + m_2^2)}} .  
\end{eqnarray}  
The vector and tensor integrals can be decomposed into Lorentz   
covariants and scalar coefficients:  
\begin{equation}  
B_{\mu}(p^2, m_1, m_2) = p_{\mu} B_1(p^2. m_1, m_2) ,  
\label{eq:bmu}  
\end{equation} 
 
\begin{equation}  
B_{\mu \nu}(p^2, m_1, m_2) = p_{\mu} p_{\nu}   
B_{21}(p^2, m_1, m_2) + \delta_{\mu \nu} B_{22}(p^2, m_1, m_2).   
\label{eq:bmunu}  
\end{equation}  
The functions $B_1$, $B_{21}$ and $B_{22}$ are algebraically   
related to the two-point scalar integral $B_0$ and to the   
one-point scalar integral $A(m)$, defined by   
\begin{eqnarray}  
A(m) = {1\over {i \pi^2}} \int d^n q {1\over {q^2 + m^2}} .  
\label{eq:amdef}  
\end{eqnarray}  
The expression for $B_1$ can be obtained multiplying   
eq.~(\ref{eq:bmu}) by $p_{\mu}$ and taking into account of the   
identity   
\begin{eqnarray}  
q \cdot p = {1\over 2} \left[ (q+p)^2 + m_2^2 - (q^2 + m_1^2)   
- p^2 - m_2^2 + m_1^2 \right] .  
\end{eqnarray}  
It follows that   
\begin{eqnarray}  
B_1 = {1\over {2 p^2}} \left[ A(m_1) - A(m_2) -   
(p^2 + m_2^2 - m_1^2)B_0 \right] ,  
\end{eqnarray}  
where the arguments have been omitted for simplicity of notation.  
The expressions for $B_{21}$ and $B_{22}$ are analogously obtained   
multiplying eq.~(\ref{eq:bmunu}) by $p_{\mu} p_{\nu}$ and   
$\delta_{\mu \nu}$:    
\begin{eqnarray}  
B_{22} = {1\over 3}\left\{ -m_1^2 B_0 + {1\over 2} [A(m_2)   
- (m_1^2 + m_2^2 + {1\over 3}p^2)   
- (m_1^2 - m_2^2 - p^2)B_1 ] \right\} ,  
\end{eqnarray}  
\begin{eqnarray}  
B_{21} = {1\over {3 p^2}} \left[ A(m_2)   
+ 2 (m_1^2 - m_2^2 - p^2)B_1 +   
{1\over 2} (m_1^2 + m_2^2 + {1\over 3}p^2) + m_1^2 B_0 \right] .  
\end{eqnarray}  
In so doing only the basic scalar integrals $A(m)$ and   
$B_0(p^2, m_1, m_2)$ need to be calculated.   
  
Let us consider the one-point integral $A(m)$ in four   
dimensions:  
\begin{eqnarray}  
A(m) = {1\over {i \pi^2}} \int_0^{\infty} dq_0 \int d^3 q   
{1\over {q^2 + m^2 - i\varepsilon}} .  
\end{eqnarray}  
In the complex $q_0$ plane the integrand has poles for   
\begin{eqnarray}  
q_0 = \pm \left( \sqrt{\vert q\vert^2 + m^2} - i\varepsilon   
\right) .  
\end{eqnarray}  
The introduction of the Wick rotation allows to transform the   
Minkowski space in a Euclidean space and perform the   
integration by means of polar coordinates.   
  
Working in Euclidean space in the case of $n$ dimensions,   
a set of generalized polar coordinates can be   
introduced~\cite{hv72,v94}  
\begin{eqnarray}  
\int d^n q = \int_0^{\infty} \omega^{n-1} d\omega   
\int_0^{2 \pi} d \vartheta_1 \int_0^{\pi} \sin  
\vartheta_2 d\vartheta_2   
\ldots \int_0^{\pi} \sin^{n-2} \vartheta_{n-1} d  
\vartheta_{n-1} .  
\end{eqnarray}  
  
Since the integrand of $A(m)$ depends only on $q^2$,   
the angular integrations are easily performed by means of   
the relation  
\begin{eqnarray}  
\int_0^{\pi} \sin^m \vartheta d \vartheta = \sqrt{\pi} {{\Gamma   
\left( {{m+1}\over 2}\right) }\over {\Gamma \left(   
{{m+2}\over 2}\right) }} ,  
\end{eqnarray}  
yielding   
\begin{eqnarray}  
A(m) &=& {{2 \pi^{{n\over 2}-2}}\over {\Gamma \left(   
{n\over 2} \right) }} \int_0^{\infty} \omega^{n-1} d\omega   
{1\over {\omega^2 + m^2}} \\  
&=& {{2 \pi^{{n\over 2}-2}}\over {\Gamma \left(   
{n\over 2}\right) }} \cdot {1\over 2} \cdot   
{{\Gamma \left( {n\over 2} \right) \Gamma \left(   
1 - {n\over 2} \right) }\over {\Gamma (1) m^{2 \left(   
1 - {n\over 2} \right) }}} \\   
&=& {{\pi^{{n\over 2}-2}}\over {m^{2 - n}}}   
\Gamma \left( 1 - {n\over 2} \right) ,  
\label{eq:am}  
\end{eqnarray}  
where $\Gamma (x)$ is the Euler $\Gamma$ function with the   
property $x \Gamma (x) = \Gamma (x+1)$.   
Since the physical results are recovered for $n$ approaching   
$4$, it is convenient to introduce the notation   
$n = 4 -\varepsilon $, and expand the expression of $A(m)$ around   
$\varepsilon = 0$:  
\begin{eqnarray}  
A(m) &=& m^2 \pi^{-{{\varepsilon}\over 2}} m^{-\varepsilon}   
{{\Gamma \left( {{\varepsilon}\over 2} \right) }\over {-1 +   
{{\varepsilon}\over 2}}} \\  
&\simeq & m^2 \left( -\Delta + \ln  m^2 - 1 \right) ,  
\end{eqnarray}  
where   
\begin{eqnarray}  
\Delta = {2\over {4-n}} - \gamma - \ln \pi   
\end{eqnarray}   
contains the ultraviolet divergence,   
and the expansion of the $\Gamma $ function around the zero of   
its argument has been used   
\begin{eqnarray}  
\Gamma (x) = {1\over x} - \gamma + {\cal O}(x) ,  
\end{eqnarray}  
with $\gamma $ indicating the Euler constant.  
  
The calculation of the two-point scalar integral $B_0$ is more   
involved than $A(m)$ because of the presence of two factors   
in the denominator. As a first step it is necessary to use   
the Feynman parameterization to combine the denominators:  
\begin{eqnarray}  
{1\over {a b}} = \int_0^1 dx {1\over {[ax + b(1-x)]^2}},   
\end{eqnarray}  
where   
\begin{eqnarray}  
b &=& q^2 + m_1^2 , \nonumber \\  
a &=& (q+p)^2 + m_2^2 .\nonumber  
\end{eqnarray}  
With a shift in the integration variable $q \to q + xp$ the term   
linear in the integration variable disappears, yielding    
\begin{eqnarray}  
B_0(p^2, m_1, m_2) = {1\over {i \pi^2}} \int_0^1 dx \int d^n q   
{1\over {(q^2 + \chi - i \varepsilon)^2}} ,  
\end{eqnarray}  
with $\chi = -p^2 x^2 + (p^2 + m_2^2 - m_1^2) x + m_1^2$.   
The angular integration is now straightforward and the last   
integral is again worked out in terms of the $\Gamma $ function:  
\begin{eqnarray}  
B_0(p^2, m_1, m_2) = {{\pi^{{n\over 2}-2}}\over   
{\chi^{2 - {n\over 2}}}} {{\Gamma \left( 2 - {n\over 2}   
\right) }\over {\Gamma (2)}} .  
\end{eqnarray}  
By means of an expansion of $n$ around $4$ as for the case of   
the one-point scalar integral, the expression for $B_0$ reads:  
\begin{eqnarray}  
B_0(p^2, m_1, m_2) = \Delta - \int_0^1 dx \ln \chi .  
\label{eq:b0}  
\end{eqnarray}  
  
For arbitrary values of momentum and masses the integral in 
eq.~(\ref{eq:b0}) can be worked out by means of the method outlined in   
ref.~\cite{gv80}. The integration becomes straightforward   
for particular values of the arguments. For example, to obtain the 
asymptotic expressions of the fermionic self-energies written in 
Sect.~\ref{sect:olfd}, the following asymptotic expressions 
for the $B_0$ scalar function are useful:   
\begin{eqnarray}  
& & B_0(p^2, m, m)_{ \vert p^2 \vert \gg m^2} \simeq \Delta - \ln (-p^2) + 2  + i \pi , \nonumber \\
& & B_0(p^2, m, m)_{ \vert p^2 \vert \ll m^2} \simeq \Delta - \ln m^2 - {{p^2}\over {6 m^2}} + 
{{p^4} \over {60 m^4}}, \nonumber \\  
& & B_0(p^2, 0, m)_{ \vert p^2 \vert \gg m^2} \simeq \Delta - \ln (-p^2) + 2 + i \pi , \nonumber \\
& & B_0(p^2, 0,m)_{ \vert p^2 \vert \ll m^2} \simeq \Delta - \ln m^2 + 1 
- {{p^2} \over {2 m^2}}. \nonumber 
\end{eqnarray}  

Concerning the vertex correction diagrams, the following   
three-point integrals are encountered:  
\begin{equation}  
C_0; C_{\mu}; C_{\mu \nu}(p_1,p_2,m_1,m_2,m_3) = {1\over {i \pi^2}}   
\int d^n q {{1; q_{\mu}; q_{\mu \nu }}\over   
{(1) (2) (3)}}, \nonumber
\end{equation}
where 
\begin{eqnarray}
(1) &=& q^2 + m_1^2 , \nonumber \\
(2) &=& (q + p_1)^2 + m_2^2 , \nonumber \\
(3) &=& (q + p_1 + p_2)^2 + m_3^2 . \nonumber
\end{eqnarray}  

The vector and tensor integrals can be decomposed into Lorentz   
covariants and scalar coefficients in the following way:  
\begin{eqnarray}  
& & C_{\mu} = p_1^{\mu} C_{11} + p_2^{\mu} C_{12} , \nonumber \\
& & C_{\mu \nu} = p_1^{\mu} p_1^{\nu} C_{21} + p_2^{\mu} p_2^{\nu} C_{22} 
+ (p_1^{\mu} p_2^{\nu} + p_2^{\mu} p_1^{\nu} ) C_{23} + 
\delta_{\mu \nu} C_{24} ,  \nonumber 
\end{eqnarray}
where the arguments in the functions $C_{ij}$ have been neglected for 
simplicity of notations. 
According to the procedure illustrated in ref.~\cite{pv79}, the three-point 
scalar form factors can be expressed as linear combinations of two-point 
functions and of the fundamental three-point scalar integral $C_0$. It is 
worth noticing that only the 
function $C_{24}$ contains the UV divergences, while the other form 
factors are UV finite.